\documentclass[a4paper,12pt]{article}
\usepackage[francais]{babel}
\usepackage{amssymb}
\usepackage{latexsym}
\usepackage{amsmath}

\begin{document}
\title {\bf  L'Univers en expansion et probl\`{e}mes d'\'energie}
\author{{\bf\large Moukaddem Nazih}
\hspace{2mm}\vspace{3mm}\\
{\it D\'{e}partement de Math\'{e}matiques},\\
{\it Universit\'{e} Libanaise}, {\it Tripoli-Liban}}
\date{Janvier $2007$-Mars $2009$}

\maketitle \vskip -15cm \vskip 14cm \vspace{1cm}

\textbf{Abstract} In this paper we first construct a mathematical
model for the universe expansion that started up with the original
Big Bang. Next, we discuss the problematics of the mechanical and
physical laws invariance regarding the spatial frame exchanges. We
then prove the (theoretical) existence of a variable metric $g_t$,
depending on time and satisfying a simplified Einstein equation,
so that all free ordinary trajectories are geodesics. This is done
by considering the classical Galileo$-$Newtonian space and time
notions, by using generalized Newtonian principles and adding the
approved physical new ones (as covariance principle, Mach
principle, the Einstein equivalence principle \ldots) in order to
establish a new cosmological model of the dynamical universe as
being $(U(t))_{t > 0}=(B_e(O,R(t)),g_t)_{t > 0}$, where
$B_e(O,R(t))$ is the Euclidean ball of radius $R(t)$ in $\mathbb
R^3$ and $R(t) \sim t$ when \emph{t} $\gg$ 0 and $c=1$. The
cosmological metric $g_t$ is totally determined, at time $t,$ by
the mass$-$energy distribution $E_t(X)$ on $B_e(O,R(t))$. We also
study the black holes phenomenon and we prove that the total and
global cosmological energy distribution $E_t(X)$ satisfies a wave
equation whose solutions are characterized by pseudo-frequencies
depending on time and related to the spectrum of the Dirichlet
problem on the unit ball $B_e(O,1)$ for the Laplace$-$Beltrami
operator $-\Delta$. Our model is consistent in the sense that all
Newtonian and classical physical laws are valid as particular
cases in classical situations. We end this construction by
introducing, possibly, the most important feature of the
expansion$-$time$-$energy triangle that is the
temperature$-$pressure duality factor and so achieving the
construction of our real physical model of the expanding universe.
Then, we show that all basic results of modern Physics are still
valid without using neither the erroneous interpretation of the
special relativity second postulate nor the uncertainty principle.
Moreover, we give a mathematical model that explains
the matter$-$antimatter duality and classifies the fundamental particles and we conclude that there exist only two privileged fundamental forces.\\
We then show that our model results in a well posed initial value
formulation for the most general Einstein's equation and leads to
a well determined solution to this equation by using a constraint
free Hamiltonian system that reduces, according to our model, to
twelve equations relating twelve independent
unknown functions.\\
We also adapt the Einstein's general relativity theory to our
setting thus freeing it from several obstacles and constraints and
leading to the unification of general relativity with quantum
Physics and Newton - Lagrange - Hamilton's Mechanics.\\
We end this paper by determining (within the framework of our
model) the age, the size and the total energy of our universe and
proving that only the energy \emph{E}, the electromagnetic
constant $ke^2$, the Boltzmann characteristic $K_B T$ (where $T$ is the cosmic temperature) and the
speed of light \emph{c} (to which we add a quantum Statistics'
constant \emph{A}) are time - independent universal constants. The
other fundamental constants (such as \emph{G}, $\overline{h}$,
$K$, $\alpha$...) are indeed time - dependent and naturally related
to the previous ones proving, in that way, the unity of the
fundamental forces and that of all Physics' notions. This
essentially is done by adapting the Einstein - de Sitter model
(for the Hubble homogeneous and
isotropic Cosmology) and the Einstein - Friedmann equations to our setting.\\\\\\\\

{\bf{\large{0$\hskip 0.20cm$ Introduction, Sommaire et Table de mati\`eres}}}\\\\

\normalsize{\`{A} l'aube du 21$^{\grave{e}me}$ si\`{e}cle
r\'{e}appara\^{i}t une crise, qui semble \^{e}tre structurelle, au
sein de la Physique moderne. Celle-ci semblait pourtant avoir
r\'{e}solu, au cours de la premi\`{e}re moiti\'{e} du si\`{e}cle
dernier, tous les probl\`{e}mes qui ont surgi \`{a} la fin du
19$^{\grave{e}me}$ si\`{e}cle avec la d\'{e}couverte de quelques
ph\'{e}nom\`{e}nes et lois qui semblaient \^{e}tre en
contradiction avec la M\'{e}canique et la Physique
galil\'{e}o-newtoniennes classiques. De nos jours, la Cosmologie
moderne est bas\'{e}e sur la th\'{e}orie de l'expansion de
l'Univers (le Big Bang) avec une multitude de preuves \`{a}
l'appui. Plus r\'{e}cemment notre compr\'{e}hension de la
structure de la mati\`{e}re est fond\'{e}e sur le mod\`{e}le
atome-noyau-nucl\'{e}ons-quarks d'un c\^{o}t\'{e} et sur la
classification hadrons-leptons et la dualit\'{e}
mati\`{e}re-antimati\`{e}re de l'autre c\^{o}t\'{e}. Le tout
ob\'{e}\"{\i}t \`{a} une quantisation rigoureuse des charges et
des niveaux d'\'{e}nergie et \`{a} des r\`{e}gles pr\'{e}cises
dont les plus importantes sont les lois de conservation de
l'\'{e}nergie et de la quantit\'{e} de mouvement et la loi
d'exclusion de Pauli. Avec la fameuse formule d'Einstein \emph{E =
mc$^2$}, on comprend mieux l'\'{e}quivalence de toutes les formes
de l'\'{e}nergie. La quantisation et l'unification des formes
d'\'{e}nergie se comprennent encore mieux avec la d\'{e}couverte
du photon et de l'effet photo\'{e}lectrique \'{e}galement due
\`{a} Einstein. D'un autre c\^{o}t\'{e}, la M\'{e}canique
quantique et les \'{e}quations de Schr\"{o}dinger ont permis de
r\'{e}aliser un grand progr\`{e}s dans la compr\'{e}hension de la
nature ondulatoire de la mati\`{e}re et d'un grand nombre de
ph\'{e}nom\`{e}nes naturels sans pour autant fournir des
explications (th\'{e}oriques et exp\'{e}rimentales)
pr\'{e}cises.\\

\`{A} la lumi\`{e}re de ces grandes d\'{e}couvertes dont une
partie peut \^{e}tre mise au profit de la th\'{e}orie quantique et
l'autre partie au profit de la th\'{e}orie de la relativit\'{e},
deux grandes questions, parmi tant
d'autres, s'imposent :\\

1. Y a-t-il une compatibilit\'{e} ou une compl\'{e}mentarit\'{e}
entre ces deux th\'{e}ories qui semblent avoir toutes les deux
contribuer \`{a} faire sortir la Physique classique de ses impasses?\\

2. Apr\`{e}s la compr\'{e}hension d\'{e}finitive du
ph\'{e}nom\`{e}ne \'{e}lectromagn\'{e}tique et les progr\`{e}s
dans la th\'{e}orie de l'unification de la force
\'{e}lectromagn\'{e}tique avec les deux forces d'interaction
(forte et faible), est-ce que toutes les forces (en ajoutant celle
de la gravitation aux trois premi\`{e}res) pourraient \^{e}tre
unifi\'{e}es au sein d'une th\'{e}orie globale?\\

La r\'{e}ponse \`{a} la premi\`{e}re question semble \^{e}tre
n\'{e}gative. Plusieurs th\'{e}ories d'unification se heurtent
\`{a} une r\'{e}alit\'{e} qui para\^{i}t inexplicable. Les raisons
invoqu\'{e}es de l'\'{e}chec de telles tentatives sont vari\'{e}es
comme, \`{a} titre d'exemple, notre incapacit\'{e} technique
d'effectuer des mesures infiniment petites ou infiniment grandes
ou l'existence de plusieurs dimensions insaisissables, mais aussi
l'inexistence de r\'{e}alit\'{e}s objectives gouvern\'{e}es par
des lois et des r\`{e}gles pr\'{e}cises ou, en cas d'une telle
existence, notre incapacit\'{e} de comprendre leur vraie nature et
leur vrai fonctionnement.\\

\`{A} l'aide de notre mod\`{e}le global, nous nous proposons de
r\'{e}soudre un grand nombre de probl\`{e}mes ouverts et de lever
les contradictions apparentes quant \`{a} l'interpr\'{e}tation des
r\'{e}sultats et faits nouveaux, sans pour autant mettre en cause
les principes fondamentaux de la M\'{e}canique et de la Physique,
aussi bien classiques que modernes tant qu'ils soient
scientifiquement (th\'{e}oriquement et exp\'{e}rimentalement)
valides. Parmi ces derniers, on peut citer, \`{a} titre d'exemple,
le principe de  conservation de l'\'{e}nergie et de la
quantit\'{e} de mouvement, les lois de Maxwell, la loi de
Mach-Einstein sur l'\'{e}quivalence entre la mati\`{e}re,
l'\'{e}nergie, et la courbure de l'espace, le principe de la
constance de la vitesse de propagation des ondes
\'{e}lectromagn\'{e}tiques dans le " vide absolu ", la nature
ondulatoire de la mati\`{e}re et la nature quantis\'{e}e des ondes
et enfin l'indissociabilit\'{e} des notions mati\`{e}re-energie,
ondes, expansion, temp\'{e}rature, pression, forces d'interaction
et \'{e}quilibres \'{e}nerg\'{e}tiques. Un certain nombre de ces
principes se trouve reconfirm\'{e} dans le cadre de notre
mod\`{e}le.\\

Cependant notre mod\`{e}le remet en cause certains principes qui
ont \'{e}t\'{e} introduits et utilis\'{e}s avec comme seule
justification (mi-intellectuelle, mi-exp\'{e}rimentale) la
volont\'{e} de faire sortir la Physique classique de certaines
impasses apparentes. En effet \`{a} la base de notre mod\`{e}le se
trouve la remise en cause (parfaitement justifi\'{e}e)
l'interpr\'{e}tation erronn\'{e}e du deuxi\`{e}me postulat de la
relativit\'{e} restreinte qui consiste \`{a} supposer que la
vitesse de la lumi\`{e}re ne d\'{e}pend pas de la vitesse du
r\'{e}f\'{e}rentiel inertiel utilis\'{e} pour mesurer cette
vitesse, sans pour autant remettre en cause l'ind\'{e}pendance de
cette m\^{e}me vitesse vis-\`{a}-vis du mouvement de la source. Un
autre aspect fondamental de notre mod\`{e}le est de bien situer
les \'{e}quations de Schr\"{o}dinger et les Statistiques
quantiques dans leur propre contexte et \`{a} l'int\'{e}rieur de
leurs justes limites. Elles consistent en effet en une sorte
d'approche approximative et pr\'{e}dictive, aussi importante
qu'elle soit, des ph\'{e}nom\`{e}nes \'{e}tudi\'{e}s et des
r\'{e}sultats exp\'{e}rimentaux obtenus ; \`{a} ceci on
associe la remise en cause du principe d'incertitude.\\
\'{E}videmment, la remise en cause de ces postulats est bas\'{e}e
sur un raisonnement logique et math\'{e}matique rigoureux tout en
offrant une alternative coh\'{e}rente pour expliquer des
ph\'{e}nom\`{e}nes dont la contradiction apparente avec les
principes bien \'{e}tablis de la Physique \'{e}tait \`{a} la base
de leur
adoption.\\

En ce qui concerne la relativit\'{e} restreinte, on montre
qu'aucune des exp\'{e}riences (r\'{e}elles ou imaginaires) qui ont
conduit \`{a} recourir \`{a} la notion relativiste de
l'espace-temps, ne justifie l'alt\'{e}ration de la relation
naturelle (galil\'{e}o-newtonienne) de l'espace avec le temps.
Toutes ces exp\'{e}riences admettent des explications
cons\'{e}quentes et simples. C'est le cas, par exemple, de
l'exp\'{e}rience du train, des deux observateurs, de
l'\'{e}metteur et du miroir, de celle de Michelson-Morley ou de
celle de l'\'{e}metteur au milieu d'un camion avec les deux
miroirs sur les deux c\^{o}t\'{e}s...\\
De plus, on montre que la loi de covariance est tout \`{a} fait
respect\'{e}e par les \'{e}quations de Maxwell en montrant que
l'\'{e}quation des ondes se transforme d'une fa\c{c}on canonique
vis-\`{a}-vis de tout changement de rep\`{e}re inertiel. Ceci
n\'{e}cessite simplement l'utilisation d'une notion, tout \`{a}
fait naturelle, de d\'{e}rivation qui int\`{e}gre le mouvement
relatif des rep\`{e}res. Une notion de d\'{e}rivation plus
g\'{e}n\'{e}rale a servi en fait \`{a} d\'{e}montrer les
propri\'{e}t\'{e}s de covariance (de tensorialit\'{e}) pour
diff\'{e}rents types de rep\`{e}res mobiles et qui co\"{\i}ncide,
dans les cas habituels, avec la notion de d\'{e}rivation usuelle.
On montre aussi que le clivage arbitraire entre des particules
relativistes et non relativistes et les formules relativistes
elles-m\^{e}mes conduisent \`{a} des contradictions
\'{e}videntes.\\

De m\^{e}me, en se basant sur une logique math\'{e}matique
difficilement contestable, on montre que des exp\'{e}riences  du
genre de celles des particules \'{e}nerg\'{e}t-\\iques arrivant
sur un \'{e}cran, apr\`{e}s avoir travers\'{e} deux minces fentes
l\'{e}g\`{e}rement espac\'{e}es, ne nous autorisent pas de
conclure que le fait de reconna\^{i}tre \`{a} partir desquelles
des deux fentes sont pass\'{e}es les particules suffit en
soi-m\^{e}me d'alt\'{e}rer le r\'{e}sultat physique r\'{e}el et
objectif. Ces r\'{e}sultats pourraient \^{e}tre alt\'{e}r\'{e}s
uniquement par les moyens techniques et conjoncturels utilis\'{e}s
pour aboutir \`{a} cette reconnaissance. Nous montrons
\'{e}galement que les incertitudes constat\'{e}es r\'{e}ellement,
partout dans la nature, sont dues en r\'{e}alit\'{e} \`{a} la
nature dynamique et \'{e}volutive complexe des ph\'{e}nom\`{e}nes
naturels (trajectoires des mouvements, interactions,
\'{e}quilibres \'{e}nerg\'{e}tiques…) ainsi qu'\`{a}
l'imperfection et les limites impos\'{e}es par nos moyens
techniques d'une fa\c{c}on conjoncturelle. Ces derniers sont
heureusement de plus en plus performants et pr\'{e}cis ; ce qui
explique les progr\`{e}s permanents r\'{e}alis\'{e}s au niveau de
notre compr\'{e}hension de l'Univers et de la structure de la
mati\`{e}re. Les \'{e}quations de Schr\"{o}dinger offrent
\'{e}videmment un moyen puissant pour d\'{e}terminer, par exemple,
la probabilit\'{e} de trouver les particules dans une r\'{e}gion
donn\'{e}e de l'espace et pour expliquer des ph\'{e}nom\`{e}nes
qui semblent \^{e}tre inattendus, mais cela ne nous autorise pas
\`{a} \'{e}riger les Statistiques quantiques \`{a} un statut de
principe th\'{e}orique ou exact.\\
Ainsi, on montre en consid\'{e}rant  l'exemple du pendule simple
en \'{e}quilibre vertical stable et l'exemple d'une balle au repos
dans une caisse (suppos\'{e}e \'{e}galement au repos), que la
nature ondulatoire de la mati\`{e}re ne nous autorise pas de
parler de fr\'{e}quence ni de longueur d'onde correspondant \`{a}
des oscillations dans l'espace et par suite on ne peut pas parler
de leur \'{e}nergie minimale qui serait non nulle, en
contradiction flagrante avec les principes de Newton. Par
cons\'{e}quent le fait m\^{e}me d'invoquer le principe
d'incertitude n'a aucun lieu d'\^{e}tre. De m\^{e}me l'\'{e}nergie
minimale de l'\'{e}lectron dans un atome d'hydrog\`{e}ne (ground
state energy) et le rayon (approximatif) de Bohr qui lui est
associ\'{e} sont les r\'{e}sultats d'un \'{e}quilibre
\'{e}nerg\'{e}tique (entre plusieurs formes d'\'energie et
plusieurs forces d'interaction interne et externe) bas\'{e} sur le
fait que l'\'{e}nergie potentielle minimale de l'\'{e}lectron au
sein de l'atome doit \^{e}tre finie et ceci n'a rien \`{a} voir
avec le principe d'incertitude. L'\'{e}quilibre
\'{e}nerg\'{e}tique se traduit, d'une fa\c{c}on naturelle, par
l'aspect nuageux de
l'orbite.\\

Par ailleurs, nous montrons que l'attribution d'une longueur
d'onde (d\'{e}finie par la relation de De Broglie
$\lambda\;=\;\frac{h}{p}$) \`{a} toute particule mat\'{e}rielle
ponctuelle, comme les \'{e}lectrons au sein d'un atome, ne peut
\^{e}tre qu'une approximation utile et peut conduire dans certains
cas \`{a} des contradictions. De m\^{e}me, nous consid\'{e}rons
que l'utilisation de la M\'{e}canique quantique et de la
Statistique quantique n'est justifi\'{e}e que dans les cas
infinit\'{e}simaux subatomique l\`{a} o\`{u} nos capacit\'{e}s
d'effectuer des mesures exactes (ou m\^{e}me approximatives) sont
tr\`{e}s limit\'{e}es rendant inefficace l'analyse de ces cas dans
le cadre de la M\'{e}canique classique. La co\"{i}ncidence
(approximative) des r\'{e}sultats obtenus par les m\'{e}thodes
bas\'{e}es sur la M\'{e}canique quantique avec la M\'{e}canique
newtonienne, lagrangienne et hamiltonienne par l'interm\'{e}diaire
de l'\'{e}quation de Hamilton-Jacobi, ne doit servir qu'\`{a} une
justification (ou \`{a} une l\'{e}gitimisation) de l'utilisation
des m\'{e}thodes quantiques (essentiellement probabilistes) dans
les cas infinit\'{e}simaux et ne permet pas de conclure que les
lois de la Nature ob\'{e}issent uniquement aux r\`{e}gles
(approximatives) de la M\'{e}canique quantique. Bien au contraire,
les incertitudes inh\'{e}rentes \`{a} ces m\'{e}thodes (et qui en
sont en fait des r\'{e}sultats l\'{e}gitimes) refl\`{e}tent
l'aspect approximatif de cette approche et il n'est pas exclu,
qu'en utilisant d'autres moyens d'analyse, exp\'{e}rimentaux ou
th\'{e}oriques, on arrivera \`{a} mieux optimiser ces
approximations et ces incertitudes et \`{a} une meilleure
compr\'{e}hension de ces
ph\'{e}nom\`{e}nes physiques.\\

Nous allons donner dans la suite quelques id\'{e}es  forces de
notre mod\`{e}le. Notons, pour commencer, que celui-ci est
bas\'{e} sur toutes les lois et principes de la Physique
math\'{e}matique dont la validit\'{e} s'est av\'{e}r\'{e}e
(th\'{e}oriquement et exp\'{e}rimentalement) indiscutable tout en
soumettant ceux qui ont \'{e}t\'{e} partiellement admis (plus ou
moins conjoncturellement) dans le seul but de r\'{e}soudre
quelques probl\`{e}mes de parcours, \`{a} un r\'{e}examen
attentif. Ceux qui n'ont pas r\'{e}sist\'{e} \`{a} la logique
math\'{e}matique ont \'{e}t\'{e} abandonn\'{e}s, ainsi que toutes
leurs cons\'{e}quences, apr\`{e}s avoir fourni les justifications
n\'{e}cessaires et \'etabli des alternatives nettement plus
naturelles. Ensuite le mod\`{e}le a \'{e}t\'{e} construit sur la
base de quelques id\'{e}es simples mais qui sont loin d'\^{e}tre
simplistes. On peut les r\'{e}sumer par la th\'{e}orie de
l'expansion (qui a fait son chemin depuis Hubble) et l'utilisation
d'une m\'{e}trique riemannienne variable avec le temps et la
position refl\'{e}tant le principe de Mach repris par Einstein :
Mati\`{e}re = Courbure ; auxquelles on ajoute le principe \`{a} la
fois scientifique et philosophique qui consiste en l'unit\'{e} et
la coh\'{e}rence des lois de la nature qui englobe, entre autres :
les lois de conservation, la loi de covariance, les lois
d'\'{e}quivalence et la dualit\'{e} attraction-r\'{e}pulsion
originelle
des forces.\\

Ainsi, l'Univers \`{a} l'instant $t>0$ consiste, selon notre
mod\`{e}le, en une boule $B_e(O,R$($t$)$)$ de $\mathbb{R}^3$ (avec
$R(t) \sim t$ lorsque $t \gg$ 0) munie d'une m\'{e}trique
riemannienne $g_t$($X$). Celle-ci refl\`{e}te, \`{a} tout instant,
par l'interm\'{e}diaire de sa courbure variable, la distribution
de la mati\`{e}re-\'{e}nergie ainsi que tous ses effets. Cette
m\'{e}trique contracte les distances et volumes autour des
agglom\'{e}rations mat\'{e}rielles \`{a} forte densit\'{e} et
surtout autour des trous noirs caract\'{e}ris\'{e}s par une
densit\'{e} \'{e}nerg\'{e}tique extr\^{e}me. Par contre, cette
m\'{e}trique mesure les distances selon notre \'{e}chelle
(euclidienne) conventionnelle dans les endroits quasiment loin de
toute influence (essentiellement gravitationnelle) de la
mati\`{e}re. Toutes les trajectoires d\'{e}crivant des mouvements
libres (i.e. sous la seule action des forces naturelles) dans
l'Univers seraient (relativement \`{a} cette m\'{e}trique) des
g\'{e}od\'{e}siques comme le sont les trajectoires associ\'{e}es
aux mouvements newtoniens libres (i.e. non soumis \`{a} des forces
ext\'{e}rieures) dans l'espace plat. Ces derni\`eres ne sont
autres que les lignes droites parcourues \`{a} vitesse constante
ou, en d'autres termes, les g\'{e}od\'{e}siques relatives \`{a}
notre
m\'{e}trique euclidienne plate.\\
La chute libre, \`{a} une distance raisonnable de la terre,
d\'{e}crit une g\'{e}od\'{e}sique $X$(\emph{t}) pour une
m\'{e}trique $g_t$ (i.e.$\hskip 0.1cm$ $\nabla_
{X^{'}(t)}^{g_t}X^{'}(t)$ = 0) qu'on peut d\'{e}terminer
facilement dans les deux cas o\`{u} l'on suppose soit que la
gravitation est uniforme soit qu'elle est centrale. Cette notion
nous permet de r\'{e}soudre num\'{e}riquement le probl\`{e}me des
$n$ corps, par exemple.\\

Toutefois, au lieu d'essayer de d\'{e}terminer la m\'{e}trique en
question \`{a} l'aide de l'\'{e}quation tensorielle d'Einstein, on
a opt\'{e} de suivre toute une autre d\'{e}marche. En effet la
d\'{e}pendance de cette \'{e}quation d'un grand nombre de
facteurs, en plus du temps, rend la r\'{e}solution inextricable
malgr\'{e} toutes les simplifications et les r\'{e}ductions
possibles. Notre d\'{e}marche sera progressive, commen\c{c}ant par
une mod\'{e}lisation math\'{e}matique purement th\'{e}orique de
l'expansion virtuelle de l'espace, suivie de l'introduction
progressive des r\'{e}alit\'{e}s physiques passant de
l'id\'{e}alisation \`{a} la r\'{e}gularisation et \`{a} la
quasi-lin\'{e}arisation pour finir par int\'{e}grer tous les
facteurs qui fa\c{c}onnent notre Univers r\'{e}el d'une fa\c{c}on
essentiellement simultan\'{e}e et
indissociable.\\

Dans un premier temps, on montre, en utilisant les principes
fondamentaux (g\'{e}n\'{e}ralis\'{e}s) de la M\'{e}canique
newtonienne, que la cr\'{e}ation et l'expansion de l'espace dans
lequel vit l'Univers devrait (\`{a} partir d'un certain temps) se
produire \`{a} une vitesse quasi-constante qui tend vers \emph{c}
(suppos\'{e}e \'{e}gale \`{a} 1). On introduit ensuite la
distribution de la masse mat\'{e}rielle $m_t(X)$ et celle de
l'\'{e}nergie potentielle g\'{e}n\'{e}ralis\'{e}e $E_t(X)$ sur la
boule \emph{B}$_e$(0,\emph{R}(\emph{t})) pour tout \emph{t}.
Ensuite on leur associe successivement les mesures $\rho_t$ =
$m_t(X)dX$ et $\nu_t$ = $E_t(X)dX$ et on consid\`{e}re la mesure
$\mu_t$ associ\'{e}e \`{a} la m\'{e}trique physique $g_t$ qui est
d\'{e}termin\'{e}e par la distribution $E_t$(\emph{X}) de la
fa\c{c}on suivante:
$$\mu_t = dv_{g_t} = v_t(X)dX = dX-\nu_t(X) = dX - E_t(X)dX $$
La mesure $\nu_t$ mesure le d\'{e}faut caus\'{e} par la
mati\`{e}re-\'{e}nergie pour que le volume soit euclidien et
$\mu_t$ mesure le volume r\'{e}el physique dans l'Univers (\`{a}
l'instant \emph{t}) en tenant compte de toutes les manifestations
de la mati\`{e}re-\'{e}nergie. On consid\`{e}re ensuite le
demi-c\^{o}ne de l'espace-temps
$$C^{'} = \{(x,y,z,t) \in \mathbb{R}^4; x^2+y^2+z^2 \leq R^2(t); t \geq
0\}= \bigcup_{t \geq 0}B(0,R(t)) \times \{t\},$$qu'on peut
consid\'{e}rer, tout au long de notre construction (par souci de
simplicit\'{e} et de clart\'{e}), comme \'{e}tant:
$$C = \{(x,y,z,t) \in \mathbb{R}^4; x^2+y^2+z^2 \leq t^2; t \geq
0\}= \bigcup_{t \geq 0}B(0,t) \times \{t\}$$Ceci revient en fait
\`{a} supposer que la vitesse des ondes \'{e}lectromagn\'{e}tiques
\'{e}tait de tout temps \'{e}gale \`{a} la vitesse de la
lumi\`{e}re dans le vide absolu (i.e. \emph{c} = 1) et que la
vitesse de l'expansion \'{e}tait de tout temps celle de
l'expansion de l'espace g\'{e}om\'{e}trique vide qui est
d\'{e}termin\'{e}e, selon notre mod\`{e}le, par la propagation
\'{e}lectromagn\'{e}tique. Le cas g\'{e}n\'{e}ral sera discut\'{e}
\`{a} la fin de notre article.\\
L'Univers \`{a} l'instant $t_0$ serait alors l'intersection de ce
demi-c\^{o}ne avec l'hyperplan d'\'{e}quation \emph{t} = $t_0$ de
$\mathbb{R}^4$ munie de la m\'{e}trique riemannienne $g_{t_0}$. On
applique alors le th\'{e}or\`{e}me de Stokes sur ce demi-c\^{o}ne
muni de la m\'{e}trique plate de Minkowski (consid\'{e}rant ainsi
l'espace virtuel vide au sein duquel \'{e}volue l'espace
g\'{e}om\'{e}trique avec le d\'{e}roulement du temps) d'une part
et sur ce m\^{e}me demi-c\^{o}ne muni de la m\'{e}trique
$$h_t = dt^2 - g_t$$
d'une autre part pour montrer que l'\'{e}nergie
g\'{e}n\'{e}ralis\'{e}e (englobant la mati\`{e}re)
$E(t,X):=E_t(X)$ v\'{e}rifie une \'{e}quation d'onde:
$$\square E(t,X) = \frac{\partial^2}{\partial t^2}E(t,X) -
\Delta E(t,X) = 0 \mbox { pour } X \in B(0,t)$$avec
$$E(t,X) |_{S(0,t)} = 0 \mbox { pour tout}\hskip 0.2cm t,$$dont les
solutions sont des fonctions pseudo-p\'{e}riodiques admettant des
pseudo-fr\'{e}quences
d\'{e}croissant avec le temps.\\
S'appuyant sur le principe de Planck-Einstein, on \'{e}crit tout
au long de la propagation:
$$E_\mu(t,X) = g_\mu(t)\psi(\frac{X}{t}) = h_\mu(t)f_\mu(t)$$
o\`{u} $\psi$ et $\mu$ sont respectivement les fonctions propres
et les valeurs propres associ\'{e}es au probl\`{e}me de Dirichlet
sur la boule unit\'{e} \emph{B$_e$}(0,1), $f_\mu(t)$ est la
fr\'{e}quence de
la solution et $h_\mu(t)$ est une sorte de constante de Planck.\\
En introduisant le facteur temp\'{e}rature, qui est (avec la
pression) indissociable de l'expansion de l'Univers, on montre
que, pour tout mouvement libre (g\'{e}od\'{e}sique pour $g_t$)
X(t), l'\'{e}nergie $E_\mu(t,X(t))$ est une fonction
d\'{e}croissante du temps (via la d\'{e}croissance de la
temp\'{e}rature cosmique) d\'{e}pendant de $\mu$ d'une mani\`{e}re purement conventionnelle.\\

Finalement, on retrouve, dans le cadre de notre mod\`{e}le, la
fameuse relation $E = mc^2 (=m)$ et on montre que, pour toute
particule mat\'{e}rielle de masse initiale au repos $m_0 = m(0)$
circulant \`{a} une vitesse $v(t) <$ 1, l'\'{e}nergie totale
$E(t)$ est \'{e}gale \`{a} $\gamma(t) m_0 c^2 + \frac{1}{2}
\gamma_1(t) m_0 v^2$ o\`{u} $\gamma(t)$ est le facteur de Lorentz
et $\gamma_1(t)$ est un facteur qui provient de la perte de
l'\'{e}nergie de masse par l'interm\'{e}diaire des radiations
li\'{e}es aux fluctuations de la vitesse et de la temp\'{e}rature.
Ce facteur pourrait \^{e}tre calcul\'{e} th\'{e}oriquement ou
exp\'{e}rimentalement de plusieurs fa\c{c}ons. On montre qu'il
d\'{e}cro\^{i}t de 1 \`{a} 0 lorsque la vitesse varie de 0 \`{a}
1. On montre ensuite que notre formulation concernant
l'\'{e}nergie et la quantit\'{e} du mouvement des particules
mat\'{e}rielles sont approximativement les m\^{e}mes que les formulations relativistes.\\

On pourrait continuer dans ce sens et r\'{e}examiner toutes les
formules (dont la validit\'{e} est approuv\'{e}e
exp\'{e}rimentalement) de la Physique moderne, o\`{u} l'on utilise
soit la notion relativiste de l'espace-temps soit les Statistiques
quantiques ou le principe d'incertitude, pour leur redonner une
interpr\'{e}tation plus solide et pourquoi pas plus pr\'{e}cise,
du moment qu'il ne s'agissait que de r\'{e}sultats approximatifs
\'{e}tablis \`{a} partir des exp\'{e}riences. Ceci rel\`{e}ve en
fait d'un travail collectif laborieux et assidu. Toutefois ce
r\'{e}examen n\'{e}cessite le r\'{e}ajustement de quelques notions
et de r\'{e}tablir la d\'{e}pendance du temps pour quelques
notions et constantes. Citons, \`{a} titre d'exemple, que le
ph\'{e}nom\`{e}ne du d\'{e}calage vers le rouge (redshift)
s'explique en fait par l'accroissement, avec le temps et la
distance, des longueurs
d'onde et non pas par la vitesse de la source des ondes.\\

Notons, finalement, que notre mod\`{e}le est tout \`{a} fait
cons\'{e}quent dans le sens o\`{u} il est compatible avec la
Physique classique dans les situations newtoniennes et quasi-
newtoniennes, o\`{u} la m\'{e}trique $g_t$ serait tr\`{e}s proche
de $g_e$ et la mesure $\mu_t$ serait tr\`{e}s proche de la mesure
de Lebesgue d\`{e}s que la distribution $E(t,X)$ serait \`{a} peu
pr\`{e}s nulle dans une r\'{e}gion donn\'{e}e de l'espace. La
m\'{e}trique $g_t(X)$ int\`{e}gre et explique toutes les
situations approximatives, r\'eelles et m\^eme singuli\`{e}res
(trous noirs) et exprime que la r\'{e}alit\'{e} physique est
continue sans \^{e}tre diff\'{e}rentiable (\`{a} l'exception de la singularit\'{e} originelle).\\

Par ailleurs, il est clair que, dans le cadre de notre mod\`{e}le,
on retrouve, tout en les pr\'{e}cisant, un grand nombre de
r\'{e}sultats dans le domaine de la Cosmologie moderne. Ceux-ci
recouvrent les r\'{e}sultats qui sont \'{e}tablis en se basant sur
les travaux de Hubble, Friedmann et sur le mod\`{e}le
d'Einstein-de Sitter. En particulier, notre mod\`{e}le est
conforme \`{a} la deuxi\`{e}me affirmation du principe fondamental
de la Cosmologie qui stipule que la vitesse relative des galaxies
est proportionnelle \`{a} leur distance sans pour autant
adh\'{e}rer \`{a} la premi\`{e}re confirmation de ce principe
stipulant que l'Univers para\^{i}trait (\`{a} une grande
\'{e}chelle) exactement le m\^{e}me pour n'importe quel
observateur
indiff\'{e}remment situ\'{e} au sein de l'Univers.\\

On continue notre \'{e}tude en \'{e}tablissant un mod\`{e}le
math\'{e}matique qui conduit \`a une classification de toutes les
particules fondamentales (mat\'erielles et antimat\'erielles).
Ceci est r\'ealis\'e en utilisant une \'{e}quation d'onde
v\'{e}rifi\'{e}e par l'op\'{e}rateur de Dirac \emph{D}, d\'{e}fini
par la structure spinorielle de l'espace riemannien
(\emph{B}(0,1),$g_e$) et on conclut \`{a} l'existence de deux
forces fondamentales privil\'{e}gi\'{e}es seulement, qui sont
essentiellement li\'{e}es \`{a} l'unit\'{e} originelle de la
mati\`{e}re-\'{e}nergie, au mouvement d'expansion originel et
\`{a} l'unit\'{e} naturelle de l'Univers, d'une part, et sur les
deux aspects antagonistes fondamentaux des forces naturelles
li\'{e}es intrins\`{e}quement \`{a} la mati\`{e}re, qui sont
l'attraction et la r\'{e}pulsion, d'une autre part. Ce sont la
force
gravitationnelle et la force \'{e}lectromagn\'{e}tique.\\

On montre ensuite que, dans le cadre de notre mod\'{e}lisation, la
solution de l'\'{e}quation d'Einstein la plus g\'{e}n\'{e}rale
peut \^{e}tre obtenue au moyen d'un probl\`{e}me d'\'{e}volution
bien pos\'{e}, \`{a} valeurs initiales libres de toute condition,
conduisant \`{a} un d\'{e}veloppement de Cauchy maximal bien
d\'{e}fini. De m\^{e}me, on montre que la r\'{e}solution de
l'\'{e}quation d'Einstein est \'{e}quivalente \`{a} celle d'un
syst\`{e}me hamiltonien libre de toute contrainte qui se
r\'{e}duit \`{a} douze \'{e}quations \`{a} douze fonctions inconnues
ind\'{e}pendantes $g_{ij}$ et $\pi^{ij}$. Cette solution
correspond \`{a} une m\'{e}trique "initiale" $g_{t_0}$ d\'{e}finie
sur une surface de Cauchy quelconque $\Sigma_{t_0}$ (ou, d'une
fa\c{c}on \'{e}quivalente, sur l'Univers \emph{B}(\emph{O},$t_0$))
\`{a} un instant initial $t_0$ telles que les d\'{e}riv\'{e}es de
ses composantes, par rapport au temps, ${\dot{g}}_{ij}$($t_0$)
s'identifient aux doubles des composantes de la courbure
extrins\`{e}que $K_{ij}$($t_0$) de ($\Sigma_{t_0}, g_{t_0}$) \`{a}
l'int\'{e}rieur de l'espace - temps \emph{M} = \emph{C}(\emph{t})
muni de sa m\'{e}trique lorentzienne $h_t = dt^2 - g_t.$\\

On finit cet article en r\'{e}adaptant la th\'{e}orie de la
relativit\'{e} g\'{e}n\'{e}rale d'Einstein \`{a} notre mod\`{e}le.
Pour cela, on utilise, \`{a} la fois, le mod\`{e}le macroscopique
de la cosmologie homog\`{e}ne isotropique (consid\'{e}rant
l'Univers en tant que poussi\`{e}re de galaxies), les valeurs
exp\'{e}rimentales fiables actuelles de quelques constantes
fondamentales et les deux \'{e}quations de Friedmann - Einstein
pour \'{e}tablir correctement l'\^{a}ge, l'\'{e}tendue et
l'\'{e}nergie totale de notre Univers. En se basant ensuite sur
des r\'{e}sultats issus de la Statistique quantique, on montre que
seules l'\'{e}nergie \emph{E}, la constante
\'{e}lectromagn\'{e}tique $ke^2$, la caract\'{e}ristique de
Boltzmann $K_B T$ et la vitesse de la lumi\`{e}re \emph{c}
(auxquelles on ajoute une constante quantique \emph{A}) sont des
constantes universelles ind\'{e}pendantes du temps; les autres
constantes fondamentales (La constante gravitationnelle \emph{G},
la constante de Planck $\overline{h}$, le facteur de la force
\'{e}lectromagn\'{e}tique $\alpha$ et le param\`{e}tre de courbure
\emph{K}(\emph{t})) sont en fait d\'{e}pendantes du temps. Notons,
en passant, que ce fait ajoute une perspective nouvelle au
processus de quantification indiquant ses limites et son aspect
relatif. Finalement, on \'{e}tablit des relations impliquant ces
diff\'{e}rentes constantes montrant clairement l'unit\'{e} de
toutes les th\'{e}ories de la Physique:
L'\'{e}lectromagn\'{e}tisme, la relativit\'{e} g\'{e}n\'{e}rale,
la Physique quantique, la thermodynamique et la M\'{e}canique de
Newton - Lagrange - Hamilton et conduisant \`{a} l'unification des
forces fondamentales de la Nature. Tous nos r\'{e}sultats sont
conformes aux r\'{e}sultats bien confirm\'{e}s de la Physique
classique et moderne. Cependant, on enregistre certaines
d\'{e}viations par rapport \`{a} des r\'{e}sultats approximatifs
pr\'{e}vus, jusqu'ici, d'une fa\c{c}on g\'{e}n\'{e}rale (et
parfois hypoth\'{e}tique) sans, pour autant, \^{e}tre
rigoureusement \'{e}tablis et qui sont loins de faire
l'unanimit\'{e} de la communaut\'{e} scientifique.\\
Notre mod\`{e}le confirme clairement que les lois de l'Univers et
de son expansion sont gouvern\'{e}es par la th\'{e}orie
(l\'{e}g\`{e}rement revis\'{e}e) de la relativit\'{e}
g\'{e}n\'{e}rale d'Einstein.\\

En fin de compte, nous pensons que, pour aller plus loin dans
notre compr\'{e}hension de l'Univers nous devrions marier la
th\'{e}orie \`{a} la pratique, les Math\'{e}matiques \`{a} la
Physique en y ajoutant un peu d'imagination, de philosophie et de
confiance en nos aptitudes dans l'avenir.\\\\\\\\\\\\
\begin{center}
{\bf\Large{Table de mati\`eres}}
\end{center}
\vspace{0.5cm}
$$
\begin{array}{lll}
{1.}&{\mbox{Isom\'etries et rep\`eres mobiles}.............................................................}&{12}\\
{2.}&{\mbox{Mod\'{e}lisation math\'{e}matique de l'expansion de l'Univers}...................}&{19}\\
{3.}&{\mbox{Tensorialit\'{e} g\'{e}n\'{e}rale du changement de rep\`{e}re}.................................}&{35}\\
{4.}&{\mbox{Mod\'{e}lisation physico-math\'{e}matique de l'Univers en expansion}........}&{42}\\
{5.}&{\mbox{Mati\`{e}re, \'{E}nergie, Masse et Trous noirs}.............................................}&{68}\\
{6.}&{\mbox{\'{E}nergie, Pseudo-ondes et Fr\'{e}quences}................................................}&{84}\\
{7.}&{\mbox{R\'{e}percurssions sur la Physique moderne}..........................................}&{94}\\
{8.}&{\mbox{The limits of Quantum theory}..........................................................}&{111}\\
{9.}&{\mbox{Mati\`{e}re, antimati\`{e}re et forces fondamentales}...................................}&{140}\\
{10.}&{\mbox{La relativit\'{e} g\'{e}n\'{e}rale revue et
    simplifi\'{e}e}..........................................}&{160}\\
{11.}&{\mbox{Introduction \`{a} une Cosmologie revis\'{e}e}.............................................}&{177}\\
{12.}&{\mbox{Constantes fondamentales de la Physique moderne}.........................}&{192}\\
{13.}&{\mbox{Commentaires et probl\`{e}mes ouverts}.................................................}&{201}\\
\end{array}
$$
\vspace{1cm}
\section{Isom\'etries et rep\`eres mobiles}
Signalons pour commencer que les trois premiers paragraphes de cet
article sont consacr\'es \`{a} \'etablir certaines propri\'et\'es
de tensorialit\'e concernant les changements de rep\`{e}res
mobiles et \`{a} construire un mod\`{e}le math\'ematique (purement
th\'eorique) de la cr\'eation de l'espace. Cet espace est en
r\'ealit\'e rempli simultan\'ement par l'Univers physique (r\'eel)
dont la mod\'elisation va \^{e}tre accomplie progressivement au
cours des paragraphes suivants. Le mod\`{e}le d\'efinitif
caract\'eris\'e par la m\'etrique physique r\'eelle $g_t(X)$ ne
sera achev\'e qu'au cours du septi\`{e}me paragraphe lorsqu'on
introduit le facteur qui consiste en la dualit\'e
temp\'erature-pression. La m\'etrique introduite auparavant
constitue une approximation raisonnable de la vraie
m\'etrique sur des intervalles de temps finis $[$\emph{t$_0$},\emph{t}$]$ pour \emph{t$_0$} $\gg$ 1.\\
On suppose qu'il existe sur $\mathbb R^3$ une famille de
m\'etriques riemanniennes $g_t$, contin\^ument diff\'erentiable
par rapport \`a $t\in]0,\;\infty[$ (ce qui sera le cas pour tous
les objets math\'ematiques index\'es par $t$ dans la suite) et
que, pour un $t_0$ fix\'e, il existe une famille d'isom\'etries
$\phi _{(t_0,\;t)}=:\phi _t$ de $(\mathbb R^3,\;g_{t_0})$ sur
$(\mathbb R^3,\;g_t)$. Supposons, qu'\`a l'instant $t=t_0$,
$\mathbb R^3$ est rep\'er\'e \`a l'aide d'un rep\`ere ${\cal
R}_{0}(t_0)$et consid\'erons un rep\`ere mobile ${\cal {R}}(t)$
co\"incidant \`a l'instant $t=t_0$ avec un rep\`ere ${\cal
{R}}(t_0)$ ayant la m\^eme origine que ${\cal R}_{0}(t_0)$ et dont
l'origine d\'ecrit une courbe $ a_{0}(t)$ (dans le rep\`ere ${\cal
R}_{0}(t_0)$) et effectuant en m\^eme temps une famille continue
de transformations $A_t$ (par rapport \`a ${\cal R}_{0}(t_0)$). Un
mod\`ele de cette situation sera \'etudi\'e avec plusieurs
cons\'equences dans le deuxi\`eme paragraphe de cet article.
Consid\'erons enfin une particule ponctuelle mobile
co\"{\i}ncidant \`a l'instant $t=t_0$ avec l'origine du rep\`ere
${\cal R}_{0}(t_0)$ et supposons que sa trajectoire est d\'ecrite
(dans le rep\`ere ${\cal R}_{0}(t_0)$) par $x_0(t)$, pour $t\geq t_0$.\\
Pour $t_1 > t_0$ et $t_0\leq t\leq t_1$, posons
$$y_0(t)=\phi_{t_1}(x_0(t)),$$
$$b_0(t)= \phi _{t_1}(a_0(t)),$$
$$\hskip 4.5cm u_0(t)=\phi_t(x_0(t)) \;\;\;\;\;\;\;\;\;\;\;\;\;\;\;\;\;\;\;\;\;\;\;\;\;\;\;\;\;\;\;\;\;\;({\cal{R}}_0)$$ et
$$\alpha _0(t)=\phi _t(a_0(t))$$ et notons $x_1(t)$,
$y_1(t)$, $\alpha _1(t)$ et $u_1(t)$ les nouvelles coordonn\'ees
des courbes $x_0(t)$, $y_0(t)$, $\alpha _0(t)$ et
$u_0(t)$ dans le rep\`ere ${\cal {R}}(t_1)$ d'origine $b_0(t_1)= \alpha _0(t_1)=\phi _{t_1}(a_0(t_1))$ (voir figure 1). \\
Ainsi, pour $t_0 \leq t\leq t_1$, on a :
$$x_0(t)-b_0(t_1)=A_{t_1}.x_1(t),$$
$$\hskip 3.6cm y_0(t)-b_0(t_1)=A_{t_1}.y_1(t), \;\;\;\;\;\;\;\;\;\;\;\;\;\;\;\;\;\;\;\;\;\;\;\;\;\;\;\;({\cal{R}}_1)$$
d'o\`u $$y_1(t)-x_1(t)=A_{t_1}^{-1}(y_0(t)-x_0(t))$$ et
$$y_1(t_1)-x_1(t_1)=A_{t_1}^{-1}(y_0(t_1)-x_0(t_1))=
A_{t_1}^{-1}.u_0(t_1)-A_{t_1}^{-1}.x_0(t_1)$$et par suite \\
$$
u_1(t_1)=x_1(t_1)+ A_{t_1}^{-1}.u_0(t_1)-A_{t_1}^{-1}.x_0(t_1)$$
$$=x_1(t_1)+ A_{t_1}^{-1}.u_0(t_1)-A_{t_1}^{-1}.b_0(t_1)-x_1(t_1)$$
$$=A_{t_1}^{-1}(u_0(t_1)-b_0(t_1)),$$puisque  l'on a (en utilisant (${\cal{R}}_0$) et (${\cal{R}}_1$)): $u_0$($t_1$) = $y_0$($t_1$),$\;\;\;\;\;$ $u_{1}$($t_{1}$) =
$y_{1}$($t_{1}$) $\;\;\;\;\;\;\;\;\;$ et $\;\;\;\;\;\;\;$
$x_{0}$($t_{1}$) = $b_{0}$($t_{1}$) + $A_{t_{1}}.
x_{1}$($t_{1}$).\\
Cette relation s'\'ecrit aussi $$u_1(t_1)-\alpha _1(t_1)=A_{t_1}^{-1}(u_0(t_1)-\alpha _0(t_1))$$ puisque $\alpha _1(t_1)= 0$.\\
Ainsi, pour tout $t\geq t_0$, on a la formule de changement de
coordonn\'ees d\'ecrivant le passage du rep\`ere ${\cal
R}_{0}(t_0)$ au rep\`ere mobile ${\cal R}(t)$:
\begin{equation}\label {r1}
u(t)-\alpha (t)=A_{t}^{-1}(u_0(t)-\alpha _0(t)).
\end {equation}
Ici, $u_0(t)$ et $\alpha _0(t)$ d\'esignent respectivement les
trajectoires dans ${\mathcal R}_{0}(t_0)$ de la particule et de
l'origine du rep\`ere mobile ${\cal R}(t)$ dans l'espace $\mathbb
R^3$ muni, \`a tout instant $t\geq t_0$, de la m\'etrique variable
$g_t$ i.e. dans $(\mathbb R_t^3, g_t)_{t\geq t_0}$, tandisque
$u(t)$ et $\alpha (t)=0$ d\'esignent les vecteurs coordonn\'ees
des trajectoires $u_0(t)$ et $\alpha _0(t)$ dans le rep\`ere
mobile le long de $\alpha_0(t)$ ${\cal R}(t)$. Ainsi $u_0(t)$ et $\alpha_0(t)$
mod\'{e}lisent des trajectoires, rep\'{e}r\'{e}es \`{a} l'aide du
rep\`{e}re fixe $\mathcal{R}_0(t_0)$, dans un Univers en \'{e}volution
permanente et qui est muni \`{a} chaque instant $t$ d'une
m\'{e}trique \`{a} courbure variable $g_t$, tandis que $u(t)$
mod\'{e}lise la trajectoire de la particule, dans le rep\`{e}re
variable $\mathcal {R}(t)$ circulant le long de $\alpha_0(t)$, dans ce m\^{e}me Univers.\\
On en d\'eduit (en utilisant (1))
\begin{equation} \label {r2}
u'(t)-\alpha '(t)= A_{t}^{-1}(u'_0(t)-\alpha
'_0(t))+(A_{t}^{-1})'(u_0(t)-\alpha _0(t))
\end{equation}
ou aussi (en utilisant (1) de nouveau):

$$
u'(t)=A_{t}^{-1}(u'_0(t)-\alpha '_0(t))+(A_{t}^{-1})'\circ
A_t(u(t))\hskip 3cm (2')$$  . En particulier, si le mouvement du
rep\`ere ${\cal R}(t)$ est uniforme par rapport \`a ${\cal
R}_{0}(t_0)$ (i.e. $\alpha '_0(t)=\overrightarrow {V_0}$ et $A_t
=A_{t_0}=:A$ pour tout $t\geq t_0$), alors on a:
$$u'(t)=A^{-1}(u'_0(t)-\overrightarrow {V_0})$$ et si de plus $\phi _t=Id _{\mathbb R^3}$, alors on obtient $$x'(t)=
A^{-1}(x'_0(t)-\overrightarrow {V_0})$$
et$$x''(t)=A^{-1}.x''_0(t)$$i.e.$$\Gamma (t)=A^{-1}.\Gamma
_0(t),$$o\`u l'on a d\'esign\'e par $x(t)$ les coordonn\'ees de
$x_0(t)$ dans le rep\`ere mobile ${\cal R}(t)$. En prenant $A=Id
_{\mathbb R^3}$, on obtient$$x'(t)=x'_0(t)-\overrightarrow
{V_0}$$et
$$x''(t)=x''_0(t)$$i.e.$$\Gamma (t)=\Gamma _0(t).$$
D'autre part, si $a_0(t)=0=\alpha _{0}(t)$ pour tout $t\geq t_0$,
alors on obtient d'apr\`{e}s (2$^{'}$) et (1):
$$u'(t)=A_{t}^{-1}.u'_0(t)+(A_{t}^{-1})'.u_0(t), \;\;\;\;\;\;\;\;\;\;\;\;\; \hskip 2cm (2^{''})$$ et si de plus $A_t \equiv A$, alors on a ${\cal R}(t)=A{\cal R}_{0}(t_0)$ pour tout $t\geq t_0$ et on obtient
$$u'(t)=A^{-1}.u'_0(t),$$
$$u''(t)=A^{-1}.u''_0(t),$$
et finalement, si on suppose de plus que $\phi _t=Id _{\mathbb R^3}$, on a $$x'(t)=A^{-1}.x'_0(t)$$
et bien \'{e}videmment $$\Gamma(t) = A^{-1}.\Gamma_0(t).$$

\subsection*{Un nouvel op\'{e}rateur de d\'{e}rivation par
rapport au temps}

Dans le cas g\'en\'eral, posons
$$v_0(t)=u_0(t)-\alpha _0(t)$$ et
$$v(t)=u(t)-\alpha (t)=u(t),$$ ainsi on a
$$\frac {d}{dt} v_0(t)=\frac {d}{dt}u_0(t)-\frac {d}{dt}\alpha _0(t).$$
Posons
$$\hskip 2cm \frac{d_1}{dt}v(t)=
\frac {d}{dt} v(t)-(A_{t}^{-1})'(u_0(t)-\alpha _0(t))
\;\;\;\;\;\;\;\;\;\;\;\;\;\;\hskip 2cm(d_1)$$
$$(=\frac
{d}{dt} v(t)-(A_{t}^{-1})' \circ A_t.v(t)\mbox {, d'apr\`es la
relation (1)) }.$$ La relation (2) montre alors que l'on a:
\begin{equation}\label {r3}
\frac{d_1}{dt}v(t)= A^{-1}_{t}(\frac
{d}{dt}v_0(t))=A^{-1}_{t}(\frac {d_1}{dt}v_0(t)),
\end{equation}
puisque $\frac{d_1}{dt}$ co\"{\i}ncide avec $\frac{d}{dt}$ pour
les coordonn\'{e}es dans $R_{0}$($t_{0}$).\\
Cette formule montre la tensorialit\'e du changement de
coordonn\'ees relatif au changement de rep\`ere pour le "vecteur
vitesse" d\'efini par cette d\'erivation qui tient compte de la
vitesse de
l'origine du rep\`ere mobile et de sa rotation.\\\\
De m\^eme, concernant le "vecteur acc\'el\'eration" d\'efini \`a
partir de ce vecteur vitesse et de la m\'etrique variable $g_t$,
on a:
$$\nabla^{g_t}_{\frac{d_1}{dt}v(t)}\frac{d_1}{dt}v(t)= \nabla ^{g_t}_{A_{t}^{-1}.\frac {d_1}{dt}v_0(t)}A_{t}^{-1}.\frac {d_1}{dt}v_0(t)
=\nabla ^{g_t}_{A_{t}^{-1}.\frac {d}{dt}v_0(t)}A_{t}^{-1}.\frac
{d}{dt}v_0(t) .$$ Si $A_t=A$ pour $t\geq t_0$, on obtient
(d'apr\`{e}s la d\'{e}finition ($d_1$))
$$\hskip 4.5cm \frac {d_1}{dt}v(t)=\frac
{d}{dt}v(t) \hskip 5.1cm (3')$$et
\begin{equation}\label{r4}
\nabla^{g_t}_{\frac{d}{dt}v(t)}\frac{d}{dt}v(t)=\nabla
^{g_t}_{A^{-1}.\frac {d}{dt}v_0(t)}A^{-1}.\frac {d}{dt}v_0(t)
\end{equation}
et si de plus $\alpha'_{0}(t)=0$, alors on a (en utilisant (3) et
(3'))
$$\frac{d_1}{dt} u(t) = A^{-1} . \frac{d_1}{dt} u_0(t) =
A^{-1}.\frac{d}{dt} u_0(t)$$ et
$$\nabla^{g_t}_{\frac{d}{dt}u(t)}\frac{d}{dt}u(t)=\nabla ^{g_t}_{A^{-1}.\frac {d}{dt}u_0(t)}A^{-1}.\frac {d}{dt}u_0(t)$$
qui s'\'ecrit
\begin{equation}\label {r5}
\nabla^{g_t}_{u'(t)}u'(t)=\nabla
^{g_t}_{A^{-1}.u'_0(t)}A^{-1}.u'_0(t).
\end{equation}
Si l'on suppose maintenant que les m\'etriques $g_t$ sont plates,
alors on obtient, en utilisant (4) et (5) successivement:
$$v''(t)=\nabla^{g_t}_{v'(t)}v'(t)=\nabla ^{g_t}_{(A^{-1}.v_0(t))'}(A^{-1}.v_0(t))'=A^{-1}.v''_0(t)$$et

\begin{equation}\label{r6}
u''(t)=A^{-1}.u''_0(t)
\end{equation} et finalement,
pour $\phi _t=Id _{\mathbb R^3}$ et $a^{'}_0(t) \equiv
\overrightarrow{V}$ (un vecteur constant non n\'ec\'essairement nul), on obtient
\begin{equation}\label{r7}
x''(t)=\nabla^{g_{t_0}}_{x'(t)}x'(t) = A^{-1}.x''_0(t),
\end{equation}
relation qui s'\'ecrit $$ \Gamma (t)= A^{-1}.\Gamma _0(t).$$
Remarquons que les relations (6) et (7) restent vraies si l'on
suppose seulement que $A_t\equiv A$, $g_t$ est \`a courbure nulle
et que $\alpha _0(t)$ (resp. $a_0(t)$) est une g\'eod\'esique. Par
cons\'equent si l'on suppose de plus que $u_0(t)$ (resp. $x_0(t)$)
est une
g\'eod\'esique, alors il en est de m\^eme pour $u(t)$ (resp. $x(t)$).\\\\
Plus g\'en\'eralement, consid\'erons deux rep\`eres mobiles ${\cal
R}_1(t)$ et ${\cal R}_2(t)$ provenant tous les deux du rep\`ere
${\cal R}_{0}(t_0)$ de la m\^eme fa\c con que le rep\`ere ${\cal
R}(t)$ ci-dessus. En utilisant les notations \'evidentes suivantes
(semblables \`a celles utilis\'ees plus haut),
$$v_0(t)=u_0(t)-\alpha _0(t),$$  $$w_0(t)=u_0(t)-\beta _0(t),$$
$$u_1(t) = v_1(t) \mbox{ (coord. de } v_0(t) \mbox{ dans} R_1(t)
\mbox{)},$$
$$u_2(t) = w_2(t) \mbox{ (coord. de } w_0(t) \mbox{ dans} R_2(t)
\mbox{)},$$

$$\frac {d_1}{dt}u_1(t)=\frac {d}{dt}u_1(t)-(A_{t}^{-1})'\circ A_t.u_1(t),$$

$$\frac {d_2}{dt}u_2(t)=\frac {d}{dt}u_2(t)-(B_{t}^{-1})'\circ B_t.u_2(t),$$
on obtient
$$\frac {d_1}{dt}u_1(t)=A_t^{-1}.\frac {d}{dt}v_0(t)$$ et

$$\frac {d_2}{dt}u_2(t)=B_t^{-1}.\frac {d}{dt}w_0(t)= B_t^{-1}.\frac {d}{dt}(u_0(t)-\beta _0(t))
 = B_t^{-1}.\frac {d}{dt}(v_0(t)+\alpha _0(t)-\beta _0(t)) $$
 $$\hskip 3 cm=B_t^{-1}.\frac {d}{dt}(v_0(t)) +B_t^{-1}.\frac {d}{dt}(w_0(t)-v_0(t)).$$
 D'o\`u (en utilisant (3), qui implique $\frac{d}{dt}v_0(t) =
 A_t.\frac{d_1}{dt}u_1(t)$)

$$\frac {d_2}{dt}u_2(t)=B_t^{-1}\circ A_t.\frac
{d_1}{dt}u_1(t)+B_t^{-1}. \frac {d}{dt}(w_0(t)-v_0(t))$$
\begin{equation}\label{r8}
=B_t^{-1}\circ A_t.\frac {d_1}{dt}u_1(t)+B_t^{-1}. \frac
{d_1}{dt}(\alpha_{0}(t) - \beta_{0}(t))
\end{equation}
Pour $\alpha'_0(t)=\beta'_0(t),$ on obtient
$$\frac {d_2}{dt}u_2(t)=B_t^{-1}\circ A_t.\frac {d_1}{dt}u_1(t)$$et
$$\nabla ^{g_t}_{\frac{d_2}{dt}u_2(t)}\frac{d_2}{dt}u_2(t)=
 \nabla ^{g_t}_{B_t^{-1}\circ A_t.\frac {d_1}{dt}u_1(t)}B_t^{-1}\circ A_t.\frac {d_1}{dt}u_1(t).$$
 Si on suppose toujours que $\alpha' _0(t)= \beta '_0(t)$ et que $A_t=A$ et $B_t=B$ pour tout $t$, on obtient
 $$\frac {d}{dt} u_2(t)=B^{-1}\circ A.\frac {d}{dt}u_1(t)$$i.e.$$u'_2(t)=B^{-1}\circ A.u'_1(t)$$et
 $$\nabla ^{g_t}_{u'_2(t)}u'_2(t)=\nabla ^{g_t}_{(B^{-1}\circ A.u_1(t))'}(B^{-1}\circ A.u_1(t))'.$$
Si $g_t$ est de plus \`a courbure nulle, on
obtient$$u''_2(t)=B^{-1}\circ A.u''_1(t). $$ Finalement, si on
suppose que $A _t = A$, $B_t=B$, $a' _0(t)= b'_0(t)$ et $\phi
_t=Id _{\mathbb R^3}$ pour tout $t \geq t_0$, alors on obtient
\'evidemment$$x'_2(t)=B^{-1}\circ A.x'_1(t)$$ et
$$x''_2(t)=B^{-1}\circ A.x''_1(t)$$ i.e.$$\Gamma _2(t)=B^{-1}\circ
A.\Gamma _1(t),$$ o\`u l'on a d\'esign\'e par $x_1(t)$ et $x_2(t)$
respectivement les coordonn\'ees de $x_0(t)$ dans les deux
rep\`eres mobiles ${\cal R}_1(t)$ et ${\cal R}_2(t)$. Remarquons
que, pour obtenir les relations $u''_2(t)=B^{-1}\circ A.u''_1(t)$
(resp. $ x''_2(t)=B^{-1}\circ A.x''_1(t)$), il suffit que $\alpha
_0(t)$ et $\beta _0(t)$ (resp. $a_0(t)$ et $b_0(t)$) soient des
g\'eod\'esiques pour la m\'{e}trique plate. Si $u_1(t)$ (resp.
$x_1(t)$) est de plus une g\'eod\'esique, il en est de m\^eme pour
$u_2(t)$ (resp. $x_2(t)$).\\\\
Ces derni\`eres relations montrent la tensorialit\'e du vecteur
acc\'el\'eration vis \`a vis de changements de rep\`eres \`a
vitesses \'egales ou \`a vitesses constantes
et se d\'eduisant l'un de l'autre \`a l'aide d'une transformation lin\'eaire constante.\\\\
Si l'on suppose maintenant que les transformations $A_t$ et $B_t$
sont des isom\'etries de $(\mathbb R^3,\;g_{t})$ et si $\alpha
'_0(t)=\beta '_0(t)$, on obtient (d'apr\`{e}s la relation (8))
$${\parallel \frac {d_2}{dt} u_2(t)\parallel}_{g_t}=
{\parallel \frac{d_1}{dt} u_1(t)\parallel}_{g_t}.$$ Si de plus, on
a $A_t=A$ et $B_t=B$ pour $t\geq t_0$, alors on a n\'ecessairement
$$g_t=g_{t_0},$$
$${\parallel u'_2(t)\parallel}_{g_{t_0}}={\parallel u'_1(t)\parallel}_{g_{t_0}}$$et
$$||\widetilde {\Gamma }_2(t)||_{g_{t_0}}=||\widetilde {\Gamma
}_1(t)||_{g_{t_0}}$$o\`u, pour une m\'etrique plate, 
$$\widetilde {\Gamma }_2(t):=\nabla^{g_{t_0}}_{u'_2(t)}u'_2(t)=u''_2(t) \hskip 1cm \mbox{ et }
\hskip 1cm  \widetilde {\Gamma }_1(t):=\nabla^{g_{t_0}}_{u'_1(t)}u'_1(t)=u''_1(t).$$ 
Et finalement, dans le cas
o\`u $ \phi _t=Id _{\mathbb R^3}$, on obtient bien \'evidemment
$${\parallel x'_2(t)\parallel}_{g_{t_0}}={\parallel x'_1(t)\parallel}_{g_{t_0}}={\parallel x'_0(t)\parallel}_{g_{t_0}}$$ et
$${\parallel x''_2(t)\parallel}_{g_{t_0}}={\parallel x''_1(t)\parallel}_{g_{t_0}}=
{\parallel x''_0(t)\parallel}_{g_{t_0}}$$i.e., pour une m\'etrique plate,
$${\parallel{\Gamma}_{2}(t)\parallel}_{g_{t_0}}={\parallel{\Gamma}_{1}(t)\parallel}_{g_{t_0}}={\parallel{\Gamma}_{0}(t)\parallel}_{g_{t_0}}.$$

\section {Mod\'elisation math\'ematique de l'expansion de l'Univers}
Consid\'erons une fonction $\lambda \in C^0([0,\;\infty[)$ qui
v\'erifie les propri\'et\'es suivantes:
\begin{enumerate}
\item $\lambda \in C^2(]0,\;\infty[)$, \item $\lambda(t)\not= 0$,
pour $t$ $\in [0,\;\infty[$.
\end{enumerate}
consid\'erons ensuite, pour $t\geq 0$, la m\'etrique d\'efinie sur l'espace virtuel (th\'eorique) $\mathbb R^3$ par:\\
$$g_t:=\frac {1}{\lambda ^2(t)}g_e,$$ et l'application de $\mathbb R^3$ dans $\mathbb R^3$  d\'efinie
par: $$\phi _t=\lambda (t)Id_{\mathbb R^3}.$$ On a alors
$$^{t}\phi _t\circ g_t\circ \phi _t= g_e$$et par
suite $\phi _t:\;(\mathbb R^3,\;g_e)\longrightarrow (\mathbb R^3,\;g_t)$ est une isom\'etrie.\\
Consid\'erons enfin, dans $(\mathbb R^3,\;g_e)$ muni d'un rep\`ere
orthonorm\'e euclidien $(O,\vec{i},\vec {j},\vec{k})$, la
g\'eod\'esique ${exp}_O(t \overrightarrow {V_0})$, pour
$\overrightarrow {V_0} \in \mathbb R^3$, qu'on peut consid\'erer
comme \'etant l'axe des abscisses $Ox$ param\'etr\'e par
$t\longrightarrow x(t)=v_0 t$, o\`u $v_0=
{\parallel\overrightarrow {V_0}\parallel}_{g_e}$. Posons, pour
$t\geq 0$, $U(t)= \phi _t (t\overrightarrow {V_0})$, relation
qu'on va noter abusivement $$ u(t)=\phi _t (tv_0).$$ En supposant
de plus que $v_0=1$, on va noter
$$u(t)=\phi _t (t)=t \lambda(t)$$ la trajectoire dans (${\mathbb R^3}_t,g_t$)$_{t\geq 0}$ d'une particule se trouvant \`a l'instant
$t=0$ \`a l'origine $O$.\\ Dans ces conditions on a, pour $t > 0$:
$$u'(t)=t
\lambda'(t)+\lambda(t)$$ et
$$\widetilde {\Gamma}(t):=\nabla ^{g_t}_{u'(t)}u'(t)=\frac {d}{dt}u'(t)=u''(t)= t\lambda ''+\lambda'+\lambda'=
t\lambda''+2\lambda'.$$ D\'esignons par $m$ la masse d'une
particule mat\'{e}rielle fondamentale, qu'on peut supposer (pour
le moment) ind\'ependante du temps (bien que le volume d\'epend
\'evidemment de $g_t$).\\\\\textbf{Remarque:} Signalons quand
m\^{e}me qu'une particule de masse \emph{m} non nulle aussi petite
qu'elle soit ne peut circuler en r\'{e}alit\'{e} qu'\`{a} une
vitesse euclidienne \emph{v} $<$ 1 bien que sa vitesse
(suppos\'{e}e constante) peut atteindre toute valeur aussi proche
de 1 que l'on veut ($1$ est ici la vitesse de la lumi\`ere dans le vide absolu).\\\\Posons ensuite $F(t)=m\widetilde
{\Gamma}(t)$ et d\'esignons par $E(t)$ une primitive de la
fonction $ g_t(m\widetilde {\Gamma}(t),u'(t))$ i.e.
$$\frac {dE}{dt}=E'(t)=g_t(m\widetilde {\Gamma}(t),u'(t)).$$ Consid\'erons
finalement l'\'equation diff\'erentielle
$$g_t(m\widetilde
{\Gamma}(t),u'(t))= g_e(m\Gamma(t),x'(t))\equiv 0,$$ o\`u
$\Gamma(t)=x''(t)=0.$ Cette \'equation s'\'ecrit $$
\frac{1}{\lambda ^2(t)}m u''(t)u'(t)\equiv 0$$ ou
$$u''(t)u'(t)\equiv 0$$ ou encore
$$
(t\lambda ''+2\lambda ')(t\lambda'+\lambda)=0 \hskip 3cm (E)
$$
Toute solution $\lambda$ de $(E)$ v\'erifiant les conditions
pr\'erequises fournira, en quelque sorte, une g\'en\'eralisation
\`a l'espace $\mathbb R^3$ muni de la m\'etrique variable $g_t$,
des lois fondamentales de la M\'ecanique classique.\\
Montrons que l'ensemble de ces solutions n'est autre que l'ensemble des constantes non nulles.\\
En effet, la relation $u''(t)u'(t)= 0$ implique n\'ecessairement
$\frac{1}{2}\frac{d}{dt}u'^2(t)=0$ et par suite $u'^2(t)=C^2$ et
$|u'(t)|=C$. Or $C$ ne peut pas \^etre nulle, puisque
$$u'=0\Rightarrow
t\lambda'+\lambda=0\Rightarrow\frac{\lambda'}{\lambda}=\frac{-1}{t},\mbox{
pour $t >0$ }$$
$$\Rightarrow ln \frac
{\lambda}{C_1}=-lnt=ln\frac{1}{t}\Rightarrow\lambda=\frac
{C_1}{t},$$ ce qui contredit notre hypoth\`ese sur la
r\'egularit\'e \`a l'origine de $\lambda$. L'\'equation (\emph{E})
est donc \'equivalente, sur $[0,\;\infty[$, \`a
$$t\lambda''+2\lambda'=0 \hskip 3 cm (E').$$
Supposons donc que l'on a une solution locale qui n'est pas
identiquement \'egale \`a une constante non nulle et supposons que
l'on a, par exemple, $\lambda(1)=a>0$ et $\lambda'(1)=b>0$.
consid\'erons alors la solution analytique $\lambda$ de
l'\'equation ($E^{'}$) v\'erifiant ces m\^emes conditions
initiales et d\'efinie sur un voisinage de $t=1$. Cette solution
est en fait d\'efinie sur $]0,\;\infty[$, puisque pour $t_0>0$,
l'\'{e}quation ($E^{'}$) montre que $\lambda
'(t_0)=0\Leftrightarrow \lambda ''(t_0)=0$; ce qui montre que s'il
existe un tel $t_0$, alors la seule solution de ($E^{'}$) sur
$]0,\;\infty[$ est $\lambda\equiv\lambda(t_0)=\lambda(1)$ qui
implique $\lambda'(1)=0$, ce qui est absurde. Ceci prouve aussi
que $\lambda'$ et $\lambda''$ ne peuvent pas s'annuler sur
$]0,\;\infty[$. Ainsi l'\'equation ($E^{'}$) est \'equivalente sur
$]0,\;\infty[$ \`a
$$t\lambda ''=-2\lambda'\Leftrightarrow\frac{\lambda''}{\lambda'}=
\frac {-2}{t}\Leftrightarrow \int _{1}^{t}
\frac{\lambda''}{\lambda'}ds=-\int _{1}^{t}\frac {2}{s}ds$$ $$\Rightarrow ln\lambda'(t)-lnb=-2lnt=ln\frac{1}{t^2}\\
$$$$\Rightarrow \lambda'(t) =\frac{b}{t^2}\Rightarrow
\lambda(t)=\frac{-b}{t}+c$$ avec $c=a+b$.\\ $b$ \'etant $\not =0$,
cette solution $\lambda$(\emph{t}) n'est pas prolongeable, non
seulement en une fonction continue sur $[0,\;\infty[$ , mais aussi
en une distribution sur $[0,\;\infty[$. On obtient la m\^eme
contradiction
en supposant que $\lambda'(1)=b<0$.\\
Par cons\'equent, les seules familles $g_t$ et $\phi _t$
v\'erifiant ces conditions sont d\'efinies par $
g_t=\frac{1}{\lambda ^2}g_e$ et $\phi _t=\lambda Id_{\mathbb R^3}$ pour $t\geq 0$ et une constante $\lambda \not = 0.$\\
En posant $\lambda=1$, d\'eterminant ainsi une \'echelle
privil\'egi\'ee et une m\'etrique riemannienne sp\'ecifique, on obtient
$g_t=g_e$ et $\phi _t= Id_{\mathbb R^3}$ pour tout $t\geq 0.$ A
l'aide de ce choix naturel, on obtient $u'(t)=1$ pour tout $t\geq
0.$ Cette derni\`ere propri\'et\'e s'\'ecrit $u'=H$ (fonction de
Heaveside).\\
Ainsi, on montr\'{e} que toute m\'{e}trique riemannienne $g$ sur
$\mathbb{R}^3$ appartenant \`{a} la classe conforme de la
m\'{e}trique euclidienne $g_e$ et v\'{e}rifiant les lois
fondamentales de la M\'{e}canique ($F = m \widetilde{\Gamma}$ et
$\frac{dE}{dt} = g(F,u^{'})$) est, \`{a} une constante positive
pr\`{e}s. la m\'{e}trique euclidienne elle m\^{e}me. Choisir cette
constante \'{e}gale \`{a} 1 revient \`{a} fixer une \'{e}chelle
donn\'{e}e pour toutes les grandeurs physiques. Cette m\'{e}trique
caract\'{e}rise toute r\'{e}gion de l'Univers physique r\'{e}el
suppos\'{e} vide et \`{a} l'abri de toute influence de la
mati\`{e}re. Celle-ci modifie les distances et les volumes
euclidiens et cr\'{e}e la m\'{e}trique physique (r\'{e}elle) qui
contracte ces derniers et qui sera construite le long des
chapitres suivants.\\
Remarquons que le choix du facteur de conformit\'{e} $\lambda$
comme \'{e}tant une fonction de la seule variable $t$ refl\`{e}te la propri\'{e}t\'{e} d'homog\'{e}n\'{e}it\'{e} et
d'isotropie globales de l'Univers et le choix de $\lambda$ = 1
correspond \`{a} une densit\'{e} d'\'{e}nergie $\rho$ quasiment
nulle dans un Univers de volume euclidien quasiment infini au sein
duquel on mesure conventionnellement les distances \`{a} l'aide de
la m\'{e}trique euclidienne.\\

\subsection*{Cons\'equences}
D'apr\`es ce qui pr\'ec\`ede, on peut assimiler l'Univers spatial
\`a l'instant $t_0 > 0$ en tant qu'espace g\'eom\'etrique virtuel
vide, \`a la boule euclidienne de rayon $ R(t_0)=\int
_{0}^{t_0}{\parallel u'(r)\parallel}_{g_e}dr=t_0$, muni de la
norme euclidienne $g_e$:$$ U(t_0):=(B(O,t_0),g_e),$$ et on peut
prendre la boule euclidienne $B(0,1)$ munie de la m\'etrique
variable $ g_t:=t^*g_e:=t^2g_e$ comme mod\`ele de l'Univers en
expansion en tant qu'espace g\'{e}om\'{e}trique virtuel (\`a
l'instant $t > 0$):
$$U_1(t):=(B(O,1),t^2 g_e).$$
Ici l'application $X\longrightarrow t_0X$ est une isom\'etrie de $U_1(t_0)$ sur $U(t_0)$ pour tout $t_0 > 0.$\\
Les propri\'et\'es $u=tH$, $u'=H$ et $\Gamma = u^{''} = \delta$
(mesure de Dirac sur ${\mathbb R}_{+}$) se g\'en\'eralisent \`a
$\mathbb R^3$, pour $X(t)=t{\overrightarrow {V}}$ ($t\geq 0$ et
${\overrightarrow {V}}\in \mathbb R^3$ avec ${\parallel
{\overrightarrow {V}}\parallel}_{g_e}=1$), sous la forme
$$t=|X(t)|:={\parallel X(t)\parallel}_{g_e}=d(O,X(t)),$$
$$u'=1 \mbox { sur la demi-droite }t{\overrightarrow {V}},$$ $$u=Id_{t{\overrightarrow {V}}}\;\mbox{ et }\;
u(t)=|X(t)| = t,$$ $$\Gamma _{t{\overrightarrow {V}}} =\delta
_{t{\overrightarrow {V}}} \mbox { (mesure de Dirac sur la
demi-droite } t{\overrightarrow {V}}).$$
Elles peuvent \^etre formul\'ees ainsi:\\

Le temps est la distance euclidienne.\\

La vitesse est l'unit\'e de temps et de distance.\\

L'acc\'el\'eration $\Gamma _{t{\overrightarrow {V}}}$ est le
potentiel du mouvement dans la direction
$t{\overrightarrow {V}}$ concentr\'e \`a l'origine du temps et de l'espace.\\

$\emph{E}_{0} = \emph{m}_{0}$ est l'\'energie (ou la masse) potentielle originelle, perp\'etuelle et \'eternelle de la particule mat\'erielle ponctuelle.\\

$\emph{F} = m_0 \Gamma _{t{\overrightarrow {V}}} = m_0
\delta_{t\overrightarrow{V}}$ est l'\'energie (ou la masse)
originelle $m_0$ anim\'ee du potentiel du mouvement dans la
direction de $\overrightarrow {V}.$
\\\\
Notons que ces \'enonc\'es sont valables dans le cadre d'une
approche purement th\'eorique de l'expansion r\'eelle de l'Univers
physique. Ils d\'ecrivent seulement la cr\'eation de l'espace
g\'eom\'etrique vide et ils peuvent \^etre, en fait, utilis\'es en
tant qu'une macro-approximation de l'Univers physique newtonien.
Dans la suite de ce paragraphe, nous allons d\'evelopper des
r\'esultats
allant dans ce sens plut\^{o}t virtuel que r\'{e}el.\\
Notons aussi que, pour \'etudier (dans le contexte pr\'ecis\'e
ci-dessus) un \'ev\`enement ou un mouvement se d\'eroulant entre
un temps $t_1$ et un temps $t_2$ , on peut utiliser l'un ou
l'autre des mod\`eles suivants:
$$(B(O,1),t^2g_e)\hskip 2 cm t_1\leq t \leq t_2,$$
$$(B(O,t_1),\frac {t^2}{t_1^2}g_e)\hskip 2 cm t_1\leq t \leq t_2,$$
$$(B(O,t_2),\frac {t^2_2}{t^2}g_e)\hskip 2 cm t_1\leq t \leq t_2.$$
Ainsi, si le mouvement r\'eel d'une particule de masse $m$ se
traduit, pour $0 < t\leq T$, par la trajectoire $X(t)$, dans
$(B(O,T),g_e)$, alors il correspond au mouvement dont la
trajectoire est $u(t)=\frac{1}{T}X(t)$ dans $ (B(O,1),g_T)$:=
$(B(O,1),T^2g_e)$ qui est, \`a l'instant $T$ et d'un point de vue
statique (i.e. consid\'erant l'Univers
comme \'etant fig\'e \`a l'instant $T$), isom\'etrique \`a $(B(O,T),g_e)$.\\
Par isom\'etrie, la vitesse dans $(B(O,1),T^2g_e)$ est
$u'(t)=\frac {X'(t)}{T}$, de norme $\sqrt{g_T(u'(t),u'(t))}=
{\parallel X'(t)\parallel}_{g_e}$ et d'acc\'el\'eration $u''(t)=\frac {X''(t)}{T}.$\\
La variation de son \'energie cin\'etique, entre $t=t_{1}$
et$\hskip 0.25cm$
 $t=t_{2}$, est
$$ \triangle E_{c} = \int _{t_1}^{t_2} g_{T}(mu''(t),u'(t))dt=\int
_{t_1}^{t_2}mX''(t) \hskip 0.05cm . \hskip 0.05cm X'(t)dt$$
$$=\frac {1}{2} \int _{t_1}^{t_2} m\frac {d}{dt}{(X'(t))}^2dt =
\frac {1}{2}m({X'(t_2)}^2-{X'(t_1)}^2).$$ En particulier, si
$X(t)=t{\overrightarrow {V_1}}$, alors on a
$u(t)=\frac{t}{T}{\overrightarrow {V_1}}$ et
$u'(t)=\frac{{\overrightarrow {V_1}}}{T}.$ Les coordonn\'ees (dans
le rep\`ere euclidien de la boule unit\'e) de la particule sont
les composantes du vecteur $\frac{t}{T}{\overrightarrow
{V_1}}=:\frac {t}{T}(x_1,y_1,z_1)$, mais sa distance, dans
$(B(O,1),g_T)$, \`a l'origine est donn\'ee, \`a l'instant $t$,
par:$$\int _{0}^{t} \sqrt {g_T(u'(s),u'(s))} ds= \int
_{0}^{t}{\parallel {\overrightarrow
{V_1}}\parallel}_{g_e}ds=:v_1t;$$ son acc\'el\'eration est
$u''(t)=0$ et son \'energie cin\'etique est constante. Il est
clair que l'on doit avoir ${\parallel {\overrightarrow
{V_1}}\parallel}_{g_e}=v_1 < 1$. Quand au mouvement de la
particule correspondant \`a l'expansion traduit par
$X(t)=t{\overrightarrow {V}}$ avec ${\parallel {\overrightarrow
{V}}\parallel}_{g_e}=1,$ alors on constate que celle-ci se trouve,
\`a l'instant $t=T$, sur $S(O,1)$ \`a une distance $T$ de
l'origine (lorsqu'elle est mesur\'{e}e par $g_T$) et sa vitesse
est de norme
$$\sqrt {g_T(\frac {{\overrightarrow {V}}}{T},\frac
{{\overrightarrow {V}}}{T})}= {\parallel {\overrightarrow
{V}}\parallel}_{g_e}=1.$$ Remarquons aussi que l'on a (pour
l'Univers statique \`{a} l'instant $\emph{t}$):
$$vol(U_1(t))=\frac{4\pi t^3}{3}=vol (U(t)).$$
Par ailleurs, lorsque le mouvement est rectiligne et \`a vitesse
constante dans $U(T)$, alors il est donn\'e par $$ X(t)=a+(t-t_0)
{\overrightarrow {V_1}}$$et on a $X'(t)= {\overrightarrow {V_1}}$
et le mouvement dans $U_1(T)$ s'\'ecrit sous la forme $$u(t)=\frac
{a}{T}+\frac{t-t_0}{T} {\overrightarrow {V_1}}$$ et la vitesse est
$u'(t)
=\frac { {\overrightarrow {V_1}}}{T}.$\\\\
D'un point de vue dynamique (i.e. tenant compte de l'expansion
bas\'ee sur l\'evolution du temps), la situation n'est pas la
m\^eme. L'Univers en mouvement est alors assimil\'e \`a la
vari\'et\'e (riemannienne) $B(O,1)$ muni, pour $t > 0$, de la
famille de m\'etriques $g_t=t^2g_e$:
$${(U_t(t))}_{t > 0}={(B(O,1),g_t)}_{t > 0}$$
qui exprime le fait qu'\`a chaque instant $t_0 > 0$, l'Univers
$U_t(t)$ se r\'eduit \`a $$
U_{t_0}(t_0):=U_1(t_0)=(B(O,1),t_0^2g_e)\simeq B(O,t_0),g_e).$$ Le
mouvement correspondant \`a $X(t)=t\overrightarrow {V_0}$, pour
$t\geq 0$, se traduit alors dans ${(U_t(t))}_{t > 0}$ par
$u(t)=\overrightarrow {V_0}$ pour tout $t$, d\'efinissant ainsi un
point fixe dans $B(O,1)$
qui se trouverait sur $S(O,1)$ lorsque ${\parallel {\overrightarrow {V_0}}\parallel}_{g_e}=1.$\\
Pour d\'eterminer le rayon $R_1(t)$ et le volume $V_1(t)$
correspondant \`a l'Univers en expansion ${(U_s(s))}_{0 < s \leq
t}={(B(O,1),g_s)}_{0 < s\leq t}$, consid\'erons les courbes
param\'etr\'ees par
$$X:[0,t]\longrightarrow B(O,t)$$  $$\hskip 2 cm s\longrightarrow X(s)=s\overrightarrow {V}$$ et
$$u:[0,t]\longrightarrow B(O,1)$$ $$\hskip 2 cm s\longrightarrow u(s)=
\frac {s}{t}\overrightarrow {V}$$ \`a valeurs dans $B(O,t)$ et
$B(O,1)$ respectivement, avec ${\parallel {\overrightarrow
{V}}\parallel}_{g_e}=1.$ On a:
$$u'(s)=\frac {\overrightarrow {V}}{t}$$et
$${\parallel u'(s)\parallel}_{g_t}={\parallel \frac{\overrightarrow {V}}{t}\parallel}_{g_t}=
{\parallel {\overrightarrow {V}}\parallel}_{g_e}=1,$$en utilisant
l'isom\'etrie $\overrightarrow {U}\longrightarrow t\overrightarrow
{U}$ de $(B(O,1),g_t)$ sur $(B(O,t),g_e).$ Or en utilisant
l'isom\'etrie $\overrightarrow {U}\longrightarrow s\overrightarrow
{U}$ de $(B(O,1),g_s)$ sur $(B(O,s),g_e)$ pour tout $s$; $0 <
s\leq t$, on a ${\parallel u'(s)\parallel}_{g_s}= {\parallel
\frac{\overrightarrow {V}}{t}\parallel}_{g_s}=\frac {s}{t}.$
Ainsi, le rayon $R_1(t)$ de la boule unit\'e $B(O,1)$ par rapport
\`a la m\'etrique variable ${(g_s)}_{0 < s\leq t}$ (i.e. la
distance de $O$ au point de la sph\`ere euclidienne $S(O,1)$
d\'etemin\'e par $\overrightarrow {V}$) est donn\'ee par
$$R_1(t)=\int _{0}^{t}\frac {s}{t}ds=\frac {1}{t}{[\frac
{s^2}{2}]}_{0}^{t}=\frac {t}{2}.$$ Quand au volume de la boule
$B(O,1)$ par rapport \`a la m\'etrique variable ${(g_s)}_{0 <
s\leq t}$, alors on a:
$$V_1(t)=\int _{0}^{t}vol (S(O,\frac {s}{t}),g_s) ds =
\int _{0}^{t}s^2vol (S(O,\frac {s}{t}),g_e) ds$$
$$\hskip 3 cm=\int _{0}^{t}4 {\pi}\frac {s^2}{t^2}s^2ds=
 \frac {4 {\pi}}{t^2}{[\frac {s^5}{5}]}^{t}_{0}=\frac {4 {\pi}t^3}{5}.$$
 (Remarquons que $$vol(S(O,s),g_e)=vol(S(O,1),g_s)=4 {\pi}s^2$$et que$$
 vol(S(O,\frac {s}{t}),g_t)=t^24 {\pi} \frac{s^2}{t^2}=4 {\pi}s^2=vol(S(O,s),g_e)).$$
 Notons que la d\'ependance de cette famille de m\'etriques en fonction du temps $t$ se traduit par une
 d\'ependance radiale d'une m\'etrique unique s'exprimant \`a l'aide de la
 distance \`a l'origine \'egale au temps $t$.\\ Pour un mouvement quelconque se traduisant, dans un rep\`{e}re d'origine $O_1 \in B(O,T)$, par la
 trajectoire $X(t)$ pour $t_1\leq t\leq t_2\leq T$, dans $(B(O,T),g_e)$, alors on a, dans
 ${(U_t(t))}_{0 < t\leq T}$, $$u(t)=\frac {X(t)}{t}$$et par suite $$u'(t)=-\frac {1}{t^2}X(t)+\frac {
 X'(t)}{t}$$ et $$ \nabla ^{g_t}_{u'(t)}u'(t)=u''(t)=\frac {2}{t^3}X(t)-\frac {1}{t^2}X'(t)-
 \frac {1}{t^2}X'(t)+\frac{X''(t)}{t}.$$D'o\`u
 $${{\parallel u'(t)\parallel}_{g_t}}^2=g_t(u'(t),u'(t))=t^2(\frac {1}{t^4}X(t)^2-
 \frac {2}{t^3}X(t).X'(t)+\frac {X'(t)^2}{t^2})$$$$=\frac {1}{t^2}X(t)^2- \frac {2}{t}X(t).X'(t)+
 X'(t)^2.$$La longueur de cette trajectoire est donc
 $$\int _{t_1}^{t_2}\sqrt {g_t(u'(t),u'(t))} dt \approx \int _{t_1}^{t_2}{\parallel X'(t)\parallel}_{g_e}dt,$$
 lorsqu'on suppose que $t_1$ est tr\`es grand et
 ${\parallel X(t)\parallel}_{g_{e}}<< t_1$.\\
 La variation de l'\'energie cin\'etique (pour une particule test de masse 1) est donn\'ee par
 $$\int _{t_1}^{t_2}g_t(u''(t),u'(t)) dt=\int _{t_1}^{t_2} t^2 g_e(-\frac {1}{t^2}X(t)+\frac {X'(t)}{t}
 , \frac {2}{t^3}X(t)-\frac {2}{t^2}X'(t)+\frac{X''(t)}{t})dt$$
 $$\hskip 3cm \approx \int _{t_1}^{t_2} X'(t).X''(t)dt=\frac {1}{2}\int _{t_1}^{t_2}
 \frac {d}{dt}X'(t)^2dt = \frac{1}{2}(X^{'}(t_2)^2 - X^{'}(t_1)^2),$$pour $t_1$ tr\`es grand.\\
 Si le mouvement est d\'ecrit par une g\'eod\'esique
 dans $(B(O,t_2),g_e)$:$$X(t)=a+(t-t_1)\overrightarrow {V},$$alors on a
 $$u(t)=\frac {a}{t}+\frac {t-t_1}{t}\overrightarrow {V},$$
 $$u'(t)=-\frac{a}{t^2}+\frac {t_1}{t^2}\overrightarrow {V},$$
 $$u''(t)=\frac {2a}{t^3}-\frac {2t_1}{t^3}\overrightarrow {V}.$$
 Ainsi, bien que la variation de l'\'energie cin\'etique
 $\int _{t_1}^{t_2} X'(t).X''(t)dt$ est nulle dans $(B(O,t_2),g_e)$, elle ne l'est pas dans
 ${(U_t(t))}_{0 < t \leq t_2}$; elle est \'egale \`a
 $$\int _{t_1}^{t_2}t^2(-\frac {2a^2}{t^5}+\frac{2t_1 a. \overrightarrow {V}}{t^5}+
 \frac{2t_1 a. \overrightarrow {V}}{t^5}-\frac {2 t_1^2 V^2}{t^5})dt$$
 $$\hskip 3 cm \approx -2 \int _{t_1}^{t_2}\frac {t_1^2}{t^3}V^2dt = -2 t_1^2 {[-\frac{1}{2 t^2}]}_{t_1}^{t_2}V^2$$
 $$\hskip 3 cm = t_1^2(\frac {1}{t_1^2}-\frac {1}{t_2^2})V^2= (1- \frac {t_1^2}{t_2^2})V^2\approx 0,$$
 pour $t_1$ et $t_2$ tr\`es grands et assez proches l'un de l'autre.\\\\
 Signalons que pour les petites valeurs
 de $t$ et pour $t$ infiniment petit (correspondant au d\'ebut de la formation de l'Univers), de telles approximations
 sont aberrantes et une \'etude sp\'eciale s'impose.\\
 Cependant, pour un hypoth\'etique mouvement originel d'une particule
 mat\'erielle ponctuelle de masse $m$ d\'ecrit par $X(t)=t\overrightarrow {V}$ pour ${\parallel {\overrightarrow
{V}}\parallel}_{g_e}=V<1$, on a
 $X'(t)=\overrightarrow {V}$ et $X''(t)=0$ et par suite
 $$E(t)-E(0)=\int _{0}^{t}\frac {d}{ds}E(s)ds=\int _{0}^{t}g_e(mX''(s),X'(s))ds=0.$$Par
 cons\'equent on a
 $$E(t)=E(0) \mbox { pour tout } t $$ et l'\'{e}nergie d'une particule effectuant un tel mouvement est
 constante.\\\\
 Consid\'erons ensuite le cas o\`u un \'emetteur de lumi\`ere
 effectue une trajectoire s'exprimant par $Y(t)$ dans $\mathbb R^3$ muni du rep\`ere ${\cal R}_0(t_0)$, par exemple, tandisqu'un
 r\'ecepteur d\'ecrit une trajectoire s'exprimant par $X(t)$. (fig. 2)\\
 Le rayon \'emis par l'\'emetteur qui arrive au r\'ecepteur \`a un instant $t_1$ est en fait celui qui a \'et\'e
 \'emis par le premier,
 \`a un instant $t_0<t_1$, \`a partir d'un point $Y(t_0)$ tel que
 $${\parallel X(t_1)-Y(t_0)\parallel}_{g_e}=d=t_1-t_0.$$

 Notons enfin, qu'\'etant donn\'e que les grandeurs
 m\'ecaniques et physiques ne sont essentiellement perceptibles et mesurables qu'\`a partir de rep\`eres mouvant
 par rapport au rep\`ere originel et les uns relativement aux autres, on peut \'evacuer le premier
 genre de mobilit\'e de la fa\c con suivante:\\
 Si le mouvement d'une particule par rapport au rep\`ere originel $(O,\vec {i},\vec{j},\vec{k})$ est d\'ecrite par le
 vecteur $\overrightarrow {OM}=X(t)$, celui des deux rep\`eres par $a(t)$ et $b(t)$, alors le mouvement
 de cette particule est d\'ecrit dans le premier rep\`ere ${\cal R}_1(t)$ par le vecteur
 $\overrightarrow {O_1M}=Y(t)=X(t)-a(t)$ et dans le second ${\cal R}_2(t)$ par le vecteur
 $\overrightarrow {O_2M}=Z(t)=X(t)-b(t).$ La relation entre $Y(t)$ et $Z(t)$ est donn\'ee par
 $$Y(t)=Z(t)+b(t)-a(t)=Z(t)+\overrightarrow {O_1O_2}$$qui donne les relations entre les coordonn\'ees
 de $M$ dans ${\cal R}_2(t)$ et celles de $M$ et de $O_2$ dans ${\cal R}_1(t)$. De m\^eme ceci nous permet,
 lorsqu'il s'agit de v\'erifier qu'une loi m\'ecanique ou physique s'exprime invariablement dans l'un
 ou l'autre des deux rep\`eres, de supposer que l'un d'eux est fixe ou au repos. Ainsi on peut rep\'erer l'espace ambiant
 \`a l'aide de n'importe quel rep\`ere euclidien $(O_1,\vec {i_1},\vec{j_1},\vec{k_1})$ o\`u $O_1$
 est un point fictif co\"{\i}ncidant \`a un instant donn\'e $t_0> 0$ (l'instant pr\'esent par exemple) avec un
 point donn\'e (sur la terre par exemple). On peut \'etudier ainsi les ph\'enom\`enes se
 produisant jusqu'\`a l'instant $t> t_0$ dans la partie de l'Univers $(B(O_1,t-t_0),g_e)$ dans le
 rep\`ere $(O_1,\vec {i_1},\vec{j_1},\vec{k_1})$ et le mod\'eliser par $(B(O_1,1),{(t-t_0)}^2g_e).$

\subsection*{Demi-c\^one de l'espace-temps}
 Soit $C$ le demi-c\^one d\'efini dans $\mathbb R^4$ par
 $$C=\{ (x,y,z,t)\in \mathbb R^4;\;x^2+y^2+z^2\leq t^2,\;t\geq 0\}= B(O,t) \times \bigcup _{t\geq 0} \{t\}.$$
Ainsi, si $B(O,t)$ est l'espace g\'eom\'etrique dans lequel
''vit'' l'Univers physique (r\'eel) \`a l'instant $t$, alors $C$
constitue l'espace g\'eom\'etrique \`a l'int\'erieur duquel se
d\'eroule le processus de l'expansion et la cr\'eation de l'espace
physique qui est intrins\`equement li\'ee au
d\'eroulement du temps.\\
Pour un \'ev\`enement $P_0=(x_0,y_0,z_0,t_0) \in C$, $U(t_0)$ est
consid\'er\'e comme \'etant l'ensemble de tous les
 \'ev\`enements $P$ qui se r\'ealisent simultan\'ement avec $P_0$ i.e. $P=(x,y,z,t_0)\in U(t_0)\times \{t_0\}$. Le futur
 de $P_0$, d\'esign\'e par ${\cal F}(P_0)$, est l'ensemble de tous les \'ev\`enements $Q\in U(t)$ avec
 $t\geq t_0$ qui pourraient \^etre situ\'es, \`a un instant donn\'e $t\geq t_0$, sur une trajectoire
 d'origine $P_0$. L'ensemble ${\cal F}(P_0)$ est le demi-c\^one droit ''sup\'erieur'' de sommet $P_0$
 et d'angle d'ouverture $\frac {\pi}{2}.$ Les points $P'$ de la surface conique de ce demi-c\^one
 ne peuvent \^etre atteints que par des trajectoires form\'ees de lignes droites parcourues \`a la vitesse 1 joignant le point
 $P_0 = (x_0,y_0,z_0)_{t_0}$ de $B(O,t_0)$ au point $P^{'}(x_1,y_1,z_1)_{t_1}$ de $B(O,t_1)$ (voir fig.3).
 Un point $Q\in U(t_1)\times \{t_1\} \simeq B(O,t_1)$ int\'erieur au demi-c\^one ne peut pas
\^etre joint \`a $P_0 \in B(O,t_0)$ que par une trajectoire
$\gamma$ de longueur euclidienne (dans $\mathbb{R}^3$),
$l(\gamma)$, inf\'erieure \`a $t_1-t_0$ (fig.3).\\
 On d\'esigne par ${\cal P}(P_0)$ (le pass\'e de $P_0$)l'ensemble des \'ev\`enements
 $Q\in U(t)\times \{t\}$ avec $t\leq t_0$
 qui peuvent \^etre joints \`a $P_0$ par une trajectoire d'origine $Q$.
 Cet ensemble n'est autre que l'intersection du demi-c\^one ''inf\'erieur'' de sommet $P_0$ avec le demi-c\^one de
 l'espace-temps comme indiqu\'e sur la figure 3. Une
 trajectoire $\gamma$ joignant un \'ev\`enement $Q\in U(t_2)\times \{t_2\}$ \`a $P$, pour $t_2\leq t_0$, doit v\'erifier
 $l(\gamma)\leq t_0-t_2$. Les seules trajectoires \`a vitesse 1 passant par $P_0$ proviennent de la surface conique de cette intersection.\\
 
 \subsection*{L'exp\'erience du train, du miroir et des deux observateurs}
 Quant au fameux exemple du train, du miroir et des deux observateurs, alors si le rep\`ere virtuel
 $(O_1,\vec {i_1},\vec{j_1},\vec{k_1})$ est fixe et le rep\`ere fix\'{e} au train $(O'_1,\vec {i'_1},\vec{j'_1},\vec{k'_1})$ est
 ''inertiel'' par rapport au premier (i.e. se d\'eplace d'une fa\c{c}on rectiligne et uniforme dans la direction $O_1x$ par exemple) et
 co\"{\i}ncidant avec $(O_1,\vec {i_1},\vec{j_1},\vec{k_1})$ \`a l'instant $t=0$, l\`a o\`u se trouve le passager,
 l'\'emetteur et le miroir plus haut sur l'axe $O_1z$, alors le rayon, ou plus exactement le photon \'emis \`a l'instant $t=0$ qui se dirige verticalement
 (i.e. dans la direction $O_1z$) continue son chemin et se trouve \`a l'instant $2t$ \`a la distance $2h$
 sur $O_1z$ ($h$ est la distance de l'\'emetteur au miroir) puisque le miroir est cens\'{e} \^{e}tre aussi petit
 et aussi haut que l'on veut de sorte que le rayon vertical pourrait bien ne pas le croiser. D'un autre c\^ot\'e le photon \'emis
 au m\^eme instant $(t=0)$
 et qui se dirige dans une autre direction pour rencontrer le miroir qui, entre temps, s'est
 d\'eplac\'e d'une distance horizontale $d$
 (i.e. dans la direction $O_1x$) et se r\'efl\'echit pour retrouver le passager, qui s'est
 d\'eplac\'e entre temps d'une distance horizontale de $2d$, aurait parcouru la distance de
 $2\sqrt{d^2+h^2}$.
 Le passager retrouve donc le deuxi\`eme photon \`a l'instant $2t_1=2\sqrt{d^2+h^2}$, tandis que le premier
 photon
 se retrouverait \`a la m\^eme distance $2t_1$ au m\^eme instant $2t_1$. Il aurait fallu mettre un miroir fictif fixe par
 rapport \`a $(O_1,\vec {i_1},\vec{j_1},\vec{k_1})$ un peu plus haut pour que le premier
 photon
 se retrouve au point $O_1$ au m\^eme instant du retour du premier jusqu'au passager. Par ailleurs, si l'on suppose que l'\'emission
 de lumi\`ere est strictement instantan\'ee (ne s'effectuant pas en une petite fraction de seconde) et uni-directionnelle (i.e.
 produisant un seul rayon vertical) et si le miroir est suffisamment haut et suffisamment petit, alors notre passager n'aurait
 jamais re\c cu de rayon r\'efl\'echi. \\
 La m\^eme situation se produit lorsqu'on consid\`ere le rep\`ere originel fixe et deux rep\`eres
 terrestres, l'un se d\'epl\c cant avec une vitesse relative constante par rapport \`a l'autre (
 le calcul des distances devient un peu plus compliqu\'e) (voir figure
 4).\\
 D'autre part, si on suppose que l'\'{e}metteur et le passager
 sont tous les deux continuellement situ\'{e}s au point $O^{'}_1$, origine du rep\`{e}re pour lequel ils sont au repos, tout en
 supposant que le miroir au dessus d'eux est macroscopique, alors
 le photon \'{e}mis verticalment \`{a} l'instant $t=0$ croise le
 miroir en un point qui n'est pas le m\^{e}me que celui qui
 \'{e}tait originellement situ\'{e} \`{a} la verticale de
 $O^{'}_1$. De plus, le photon ne reviendra pas, apr\`{e}s
 r\'{e}flexion, exactement au point $O^{'}_1$ qui s'est, entre
 temps, d\'{e}plac\'{e} horizontalement. Ainsi, ce photon n'est
 rest\'{e} \`{a} aucun instant \`{a} la verticale de $O^{'}_1$.\\
 Quant au passager et son rep\`{e}re d'origine $O^{'}_1$, alors
 il ne pourra pas d\'{e}cider que le photon (ou le signal) qu'il a
 re\c{c}u au temps $2t$ (apr\`{e}s r\'{e}flexion) a effectu\'{e} la
 distance $2h$ que si le photon est rest\'{e} tout le long du
 trajet verticalement au dessus de lui. Or, \`{a} une vitesse
 donn\'{e}e $v$, la hauteur $h$ est d\'{e}termin\'{e}e par le
 temps pr\'{e}vu du croisement entre le photon et le miroir, qui
 est \'{e}gal \`{a}
 $t$, et d\'{e}termine la direction du photon \'{e}mis
 \`{a} l'instant $t=0$ afin qu'il puisse v\'{e}rifier la
 propri\'{e}t\'{e} ci-dessus. De m\^{e}me un autre photon \'{e}mis
 juste apr\`{e}s celui-ci et dans la m\^{e}me direction que lui ne
 v\'{e}rifiera pas cette propri\'{e}t\'{e} et ne croisera le miroir \`{a} l'instant
 $t$ qu'\`{a} condition de
 diminuer ad\'{e}quatement la hauteur $h$ et dans ce cas le croisement ne se produira \'{e}videmment pas au dessus du passager. Un autre photon \'{e}mis
 \`{a} l'instant $t=0$ dans une autre direction n\'{e}cessitera
 aussi une modification de cette hauteur et le point de rencontre ne sera pas au dessus du passager. Ainsi, \`{a} $v$, $h$ et
 $t$ donn\'{e}s, le photon qui croiserait le miroir au dessus du
 passager serait probablement \'{e}mis \`{a} un temps
 ult\'{e}rieur au temps $t=0$ et ne serait probablement pas le
 m\^{e}me que celui qui serait capt\'{e} par le passager au temps
 $2t$. Peut-on alors dire que, dans le rep\`{e}re du passager, le
 rayon lumineux a effectu\'{e} la distance $2h$ pendant le temps
 $2t$ ?\\
 La notion d'un rep\`{e}re propre, qui est en mouvement par
 rapport \`{a} un autre rep\`{e}re inertiel, est \'{e}videmment
 intrins\`{e}quement li\'{e}e \`{a} la notion de la progression du
 temps et elle conduit, dans le cas de notre exp\'{e}rience, \`{a}
 la situation suivante:\\
La vitesse $v$ du passager par rapport au rep\`{e}re fixe
 \'{e}tant arbitrairement choisie alors le choix d'un temps
 de rencontre $t$ implique qu'un photon unique, ayant
 une direction bien d\'{e}termin\'{e}e, pourrait rester \`{a} la
 verticale du passager le long de son trajet pour une hauteur bien pr\'{e}cise $h$ du miroir. Les autres photons
 du rayon lumineux ayant cette m\^{e}me direction ainsi que ceux
 qui appartiennent aux autres rayons du faisceau ne v\'{e}rifient
 pas cette propri\'{e}t\'{e}. Par suite ils parcourent, par
 rapport au rep\`{e}re propre du passager, des distances
 diff\'{e}rentes les unes des autres. Donc, le rep\`{e}re propre
 du passager ne peut pas \^{e}tre utilis\'{e} d'une fa\c{c}on
 canonique pour mesurer les distances parcourues par les photons
 et encore moins par les rayons qui sont form\'{e}s, pour chacun
 d'entre eux, d'une succession "infinie" de photons et par suite
 ne pourra pas \^{e}tre utilis\'{e} afin de mesurer la distance
 parcourue par la "lumi\`{e}re" et encore moins de la vitesse
 intrins\`{e}que de la "lumi\`{e}re". La canonicit\'{e} de la
 constance de la vitesse de la lumi\`{e}re i.e. de tous les
 photons d'un rayon lumineux quelconque, va \^{e}tre \'{e}tablie
 plus loin dans le contexte qui lui est propre i.e. par rapport
 \`{a} un rep\`{e}re virtuellement fixe (en utilisant
 l'op\'{e}rateur de d\'{e}rivation $\frac{d_*}{dt}$) et par
 rapport \`{a} tout autre rep\`{e}re inertiel (en utilisant la
 transformation galil\'{e}enne et l'op\'{e}rateur
 $\frac{d_1}{dt}$).\\

\subsection*{Remarque sur la notion relativiste de l'espace-temps}
\bigskip

Il est bien connu que les notions pr\'{e}relativistes
(adopt\'{e}es par Euclide, Descartes, Galil\'{e}, Newton et bien
d'autres) conf\`{e}rent \`{a} l'espace \`{a} trois dimensions et
au temps, qui progresse contin\^{u}ment, un caract\`{e}re absolu.
La distance euclidienne $\Delta x$ s\'{e}parant deux objets
ponctuels \`{a} un instant donn\'{e} $t_0$ et l'intervalle de
temps $\Delta t$ entre deux \'{e}v\`{e}nements non simultan\'{e}s
ont une r\'{e}alit\'{e} intrins\`{e}que ind\'{e}pendante des
observateurs inertiels. On va montrer dans la suite que
l'introduction de la notion relativiste d'espace-temps n'a aucune
raison d'avoir
lieu.\\
Pour cela, consid\'{e}rons deux observateurs inertiels $O_1$ et
$O_2$ situ\'{e}s respectivement \`{a} l'origine de deux
rep\`{e}res euclidiens $R_1=(O_1, e_1, e_2, e_3)$ et $R_2=(O_2,
e_1, e_2, e_3)$ qui co\"{\i}cident \`{a} un instant $t$ = 0 et
tels que $O_2$ se d\'{e}place sur l'axe $O_1 x$ de $R_1$ \`{a} une
vitesse relative constante $v$. Les deux observateurs se mettent
facilement d'accord, \`{a} chaque instant $t \geq 0$, sur la
mesure de la distance euclidienne s\'{e}parant deux points
quelconques $A_1$ et $A_2$ de l'espace $\mathbb{R}^3$. En effet,
on peut supposer, sans r\'{e}duire la g\'{e}n\'{e}ralit\'{e} de ce
probl\`{e}me que le rep\`{e}re $R_1$ est fixe, que les points
$A_1$ et $A_2$ sont fix\'{e}s sur l'axe $O_1 x$ et que le
rep\`{e}re $R_2$ glisse sur l'axe $O_1 x$ avec une vitesse
constante $v$ par rapport \`{a} $R_1$. Ainsi, si $x_1$ et $x_2$
d\'{e}signent les abscisses des points $A_1$ et $A_2$ dans $R_1$,
alors la longueur euclidienne du segment $A_1A_2$ est donn\'{e}e
par $x_2-x_1$ lorsqu'elle est mesur\'{e}e par les deux
observateurs \`{a} l'aide des deux rep\`{e}res $R_1$ et $R_2$
\`{a} chaque instant $t \geq 0$. En effet, \`{a} chaque instant $t
\geq 0$, la distance $A_1A_2$ mesur\'{e}e par le rep\`{e}re $R_2$
est toujours $(x_2-tv)-(x_1-tv)=x_2-x_1.$ Ainsi une barre
m\'{e}tallique, par exemple, d'extr\'{e}mit\'{e}s $A_1$ et $A_2$ a
une longueur $l = A_1A_2=x_2-x_1$ lorsqu'elle est mesur\'{e}e
\`{a} l'aide du rep\`{e}re $R_1$ et de tout rep\`{e}re inertiel
$R$ circulant \`{a} n'importe quelle vitesse constante $v < 1$ par
rapport \`{a} $R_1$. De m\^{e}me, si cette m\^{e}me barre se
d\'{e}place en glissant sur l'axe $O_1 x$ \`{a} une vitesse
constante $v_0$ par rapport \`{a} $R_1$, alors, \`{a} tout instant
\emph{t}, sa longueur mesur\'{e}e par $R_1$ est $x_2+v_0
t-(x_1+v_0 t)=x_2-x_1$, et elle est \'{e}gale \`{a} $x_2+v_0
t-vt-(x_2+v_0 t-vt) = x_2-x_1=l$ lorsqu'elle est mesur\'{e}e \`{a}
l'aide de n'importe quel rep\`{e}re $R_2$ circulant le long de
l'axe $O_1 x$ \`{a} la vitesse constante (quelconque) $v$ par
rapport \`{a} $R_1$. De plus, lorsqu'on consid\`{e}re
l'espace-temps galil\'{e}o-newtonien $\mathbb{R}^4$ muni du
rep\`{e}re $R^{'}_1=(O_1, e_1, e_2, e_3, e_4)$ o\`{u} $e_4$
correspond \`{a} l'axe du temps $O t$ qui est orthogonal \`{a}
l'hyperplan rep\'{e}r\'{e} par $R_1=(O_1, e_1, e_2, e_3)$, alors
la barre m\'{e}tallique a toujours la m\^{e}me longueur $l
=x_2-x_1$ lorsque celle-ci est mesur\'{e}e \`{a} l'aide des deux
rep\`{e}res euclidiens $R_1$ et $R^{'}_1$ \`{a} un instant
donn\'{e} quelconque $t$. Or, lorsqu'on introduit la
quatri\`{e}me dimension, i.e. le temps repr\'{e}sent\'{e} par
l'axe $O_1 t$, les points $A_1$ et $A_2$ devront \^{e}tre
rep\'{e}r\'{e}s, \`{a} tout instant $t \geq 0$, dans le rep\`{e}re
$R^{'}_1$ par les coordonn\'{e}es $(x_1, 0,0,t)$ et $(x_2,0,0,t)$
qu'on va noter $(x_1,t)$ et $(x_2,t)$ en n\'{e}gligeant les deux
autres dimensions. La longueur de la barre est alors donn\'{e}e
dans ce rep\`{e}re par $\sqrt{(x_2-x_1)^2+(t-t)^2}=x_2-x_1=\Delta
x$. Lorsque l'observateur $O_2$ mesure cette longueur dans
l'espace-temps, \`{a} l'instant $t$, il faut tenir compte du fait
que $O_2$ serait alors rep\'{e}r\'{e} par $R^{'}_1$ \`{a} l'aide
des coordonn\'{e}es $(vt,t)$ et qu'il est lui m\^{e}me situ\'{e}
dans l'hyperplan de hauteur $t$ dans $\mathbb{R}^4$, qui est le
m\^{e}me hyperplan o\`{u} se trouvent les points $A_1$ et $A_2$
\`{a} l'instant $t$, c'est \`{a} dire aux points $(x_1,t)$ et
$(x_2,t)$ dans le rep\`{e}re $R^{'}_1$. Ainsi l'observateur $O_2$
obtient la m\^{e}me longueur $\Delta x = l$
en utilisant aussi bien son rep\`{e}re $R_2$ que son rep\`{e}re \`{a} quatre dimensions $R^{'}_2 = (O_2, e_1, e_2, e_3, e_4)$.\\
Le faux probl\`{e}me qui a justifi\'{e} l'introduction de la
notion relativiste de l'espace-temps est le fait que l'intervalle
spatial $\Delta x$ s\'{e}parant deux \'{e}v\`{e}nements non
simultan\'{e}s $E_1$ ayant les coordonn\'{e}es $(x_1,t_1)$ dans
$R^{'}_1$ et $E_2$ ayant les coordonn\'{e}es $(x_2,t_2)$ \`{a} des
instants $t_1 < t_2$ d\'{e}pend de l'observateur inertiel. En
effet, l\'{e}v\`{e}nement $E_1$ est rep\'{e}r\'{e} par $O_2$ dans
$R^{'}_2$ \`{a} l'instant $t_1$ par $(x_1-vt_1,t_1)$ et $E_2$ est
rep\'{e}r\'{e} par $O_2$ dans $R^{'}_2$ \`{a} l'instant $t_2$ par
$(x_2-vt_2,t_2)$ si l'on suppose que $O_2$ reste sur l'axe $O_1
x$. D\'{e}signons respectivement par $A^{'}_1$, $A^{'}_2$,
$A^{''}_1$ et $A^{''}_2$ les points de coordonn\'{e}es
$(x_1,t_1)$,$(x_2,t_1)$, $(x_1,t_2)$ et $(x_2,t_2)$ dans le
rep\`{e}re $R^{'}_1$. L'intervalle spatial de $A^{'}_1$ \`{a}
$A^{'}_2$ mesur\'{e} par $O_2$, \`{a} l'aide de $R^{'}_2$ \`{a}
l'instant $t_1$, est $x_2-vt_1-(x_1-vt_1)=x_2-x_1$ et l'intervalle
spatial de $A^{''}_1$ \`{a} $A^{''}_2$, mesur\'{e} par $O_2$ \`{a}
l'aide de $R^{'}_2$ \`{a} l'instant $t_2$, est
$x_2-vt_2-(x_1-vt_2)=x_2-x_1$. Ce sont les m\^{e}mes que les
longueurs mesur\'{e}es par cet observateur \`{a} l'aide de son
rep\`{e}re $R_2$ lorsqu'il est situ\'{e} dans l'espace-temps aux
points $O^{'}_2(t_1) = (O_2,t_1)$ et $O^{''}_2(t_2) = (O_2,t_2)$
respectivement aux instants $t_1$ et $t_2$ (voir fig.4$^{'}$).\\
Par contre, l'intervalle spatial $\Delta_1 x$ entre $A^{'}_1$ et
$A^{''}_2$ mesur\'{e} par $R^{'}_1$ est $x_2-x_1$ et l'intervalle
spatial $\Delta_2 x$ entre $A^{'}_1$ et $A^{''}_2$ mesur\'{e} par
$R^{'}_2$ est $x_2-vt_2-(x_1-vt_1) \neq x_2-x_1.$ De m\^{e}me
l'intervalle euclidien entre $A^{'}_1$ et $A^{''}_2$ mesur\'{e}
par $R^{'}_1$ est $I_1 = \sqrt{(x_2-x_1)^2+(t_2-t_1)^2}$ et ce
m\^{e}me intervalle mesur\'{e} par $R^{'}_2$ est $I_2 =
\sqrt{((x_2-vt_2)-(x_1-vt_1))^2+(t_2-t_1)^2}$ et par suite $I_2
\neq I_1$.\\
Or, ni les intervalles $\Delta_i x$ ni les intervalles $I_i$
(\emph{i} = 1,2) ne correpondent pas \`{a} une r\'{e}alit\'{e}
physique. En effet, si on consid\`{e}re la barre $A_1A_2$ qui se
trouve \`{a} l'instant $t_1$ en $A^{'}_1A^{'}_2$ dans
l'espace-temps $\mathbb{R}^4$ et en $A^{''}_1A^{''}_2$ dans
l'espace-temps $\mathbb{R}^4$ \`{a} l'instant $t_2$, alors sa
longueur spatiale r\'{e}elle mesur\'{e}e par $O_2$ \`{a} l'aide du
rep\`{e}re $R^{'}_2$ est \'{e}agle \`{a}
$x_2-vt_1-(x_1-vt_1)=x_2-x_1$ \`{a} l'instant $t_1$ et \`{a}
$x_2-vt_2-(x_1-vt_2)=x_2-x_1$ \`{a} l'instannt $t_2$. De m\^{e}me,
les deux intervalles euclidiens dans l'espace-temps $\mathbb{R}^4$
mesur\'{e}s par $O_2$ \`{a} l'aide de $R^{'}_2$ aux instants $t_1$
et $t_2$ sont \'{e}galement \'{e}gaux:
$$
\sqrt{((x_2-vt_1)-(x_1-vt_1))^2+(t_1-t_1)^2}=$$
$$\sqrt{((x_2-vt_2)-(x_1-vt_2))^2+(t_2-t_2)^2}=x_2-x_1$$
Les deux extr\'{e}mit\'{e}s qui sont rep\'{e}r\'{e}es, dans
l'Univers physique, par ($x_1,y_1,z_1$) (resp.($x_2,y_1,z_1$))
\`{a} l'aide du premier rep\`{e}re devront \^{e}tre
rep\'{e}r\'{e}es physiquement, \`{a} l'aide du second rep\`{e}re
"mobile", par ($x_1-vt, y_1, z_1$) (resp. ($x_2-vt, y_1, z_1$))
puisque ce dernier a effectu\'{e} physiquement un d\'{e}placement
de $vt$ dans la direction des $x$ positives pendant le temps
$t$.\\
Les mesures des intervalles $A^{'}_1A^{''}_2$ et $A^{'}_2A^{''}_1$
dans les rep\`{e}res $R^{'}_1$ et $R^{'}_2$ n'ont aucune
signification physique. Elles n'ont rien \`{a} avoir avec la barre
ni \`{a} l'instant $t_1$, ni \`{a} l'intsant $t_2$.\\
On peut dire la m\^{e}me chose pour l'aire et le volume de
n'importe quel objet \`{a} deux dimensions ou \`{a} trois
dimensions dans l'espace r\'{e}el $\mathbb{R}^3$. Un tel corps est
situ\'{e} enti\`{e}rement, \`{a} chaque instant $t$, dans
l'hyperplan de hauteur $t$ dans l'espace-temps $\mathbb{R}^4$. Les
longueurs, les aires et les volumes sont les m\^{e}mes que ce soit
dans $\mathbb{R}^3$ lorsqu'ils sont mesur\'{e}s par $R_1$ et $R_2$
ou dans $\mathbb{R}^4$ lorsqu'ils sont mesur\'{e}s par $R^{'}_1$
et $R^{'}_2$ \`{a} n'importe quel instant $t$.\\
Par ailleurs, si l'Univers en expansion permanente est
repr\'{e}sent\'{e} par $U(t_1)$ \`{a} l'instant $t_1$ et par
$U(t_2)$ \`{a} l'instant $t_2 > t_1$ et si on suppose qu'il est
muni \`{a} chaque instant $t$ d'une m\'{e}trique $g_t$ qui
\'{e}volue avec le temps, alors on ne peut pas mesurer la distance
d'un point $A_1$ de $U(t_1)$ au point $A_2$ de $U(t_2)$ puisqu'on
ne peut pas utiliser ad\'{e}quatement ni $g_{t_1}$ ni $g_{t_2}$
pour effectuer cette mesure. \`{A} aucun instant $t$, l'Univers
n'est fig\'{e} et la m\'{e}trique $g_t$ n'est pas la m\^{e}me
\`{a} n'importe quels deux instants diff\'{e}rents $t_1$ et $t_2$.
Donc le probl\`{e}me de mesurer la distance spatiale $\Delta x$
entre deux \'{e}v\`{e}nements $E_1$ et $E_2$ \`{a} des instants
diff\'{e}rents $t_1$ pour $E_1$ et $t_2$ pour $E_2$ ou
l'intervalle euclidien $I$ dans l'espace-temps $\mathbb{R}^4$
entre $E_1$ et $E_2$ ne devrait pas se poser puisqu'il est
physiquement vide de sens.\\\\
Supposons maintenant que le mouvement d'une particule est
rep\'er\'e \`a l'aide de deux rep\`eres tels que la vitesse
relative de l'un par rapport \`a l'autre est constante et que l'un
se d\'eduit de l'autre \`a l'aide d'une transformation orthogonale
$A_t$ \`{a} tout instant \emph{t}.\\ Ainsi, si $\overrightarrow
{OM}=X(t)$ dans le premier rep\`ere, $\overrightarrow
{O'M}=\overrightarrow {O'O}+ \overrightarrow {OM}=\overrightarrow
{OM}-\overrightarrow {OO'}$ et $\overrightarrow {OO'}=a(t)$ dans
le premier rep\`ere, on a
$$\overrightarrow {O'M}=Y(t)=A_{t}^{-1}(X(t)-a(t))$$ d'o\`u
$$Y'(t)=(A_{t}^{-1})'(X(t)-a(t))+A_{t}^{-1}(X'(t)-a'(t))$$ et si $A_{t}^{-1}\equiv A^{-1}$, on obtient
$$Y'(t)=A^{-1}(X'(t)-a'(t))$$
$$Y''(t)=A^{-1}(X''(t)-a''(t))$$
et si on a de plus $a'(t)=0$, alors
$$Y'(t)=A^{-1}.X'(t)$$et$${\parallel Y'(t)\parallel}_{g_e}={\parallel X'(t)\parallel}_{g_e}.$$
De m\^eme si $a''(t)=0,$ alors$$Y''(t)=A^{-1}.X''(t).$$Ainsi, on a
(\`a titre d'exemple):
$${\parallel Y''(t)\parallel}_{g_e}={\parallel X''(t)\parallel}_{g_e}$$et
$$\triangle E_1(t):=\int _{t_0}^{t}X''(r).X'(r)dr=\frac {1}{2}\int _{t_0}^{t}\frac{d}{dr}
{\parallel X'(r)\parallel}^{2}_{g_e}dr$$
$$= \frac{1}{2}({\parallel X'(t)\parallel}^{2}_{g_e}-{\parallel X'(t_0)\parallel}^{2}_{g_e})$$
$$= \triangle E_2(t):=\frac {1}{2}\int _{t_0}^{t}Y''(r).Y'(r)dr$$
$$= \frac{1}{2}({\parallel Y'(t)\parallel}^{2}_{g_e}-{\parallel Y'(t_0)\parallel}^{2}_{g_e}).$$\\

\section{Tensorialit\'e g\'en\'erale du changement de rep\`ere}
Notons que l'hypoth\`ese de l'ind\'ependance de la famille $g_t$
de la position dans $\mathbb R^3$ est parfaitement justifi\'ee
dans la proc\'edure de mod\'elisation th\'eorique de l'expansion
de l'Univers. Par contre, l'\'etude d'un mouvement ou d'un
ph\'enom\`ene physique se d\'eroulant entre deux temps $t_1$ et
$t_2$ dans l'espace r\'eduit \`a $B(O,t_2)$, par exemple, devrait
tenir compte de l'existence dans $B(O,t)$, pour \emph{t} $\in$
$[$\emph{t$_1$},\emph{t$_2$}$]$ de champs de gravitations
cr\'e\'es par des masses, des trous noirs et des ph\'enom\`enes
\'energ\'etiques, \'electromagn\'etiques et quantiques dispers\'es
dans l'Univers. Ceci expliquerait, par exemple, les d\'eviations
subies par la propagation de la lumi\`ere par rapport aux
g\'eod\'esiques
classiques (lignes droites).\\
Peut-on int\'egrer raisonnablement toutes ces donn\'ees variables,
globalement ou m\^eme localement, en construisant une m\'etrique
$g_t(X)$ d\'ependant \`a la fois du temps et de la position
absolue, ou m\^eme
relative, dans l'espace?\\
Aurions-nous suffisemment de donn\'ees num\'eriques ou empiriques
pour pouvoir tenir compte de la dynamique de l'expansion afin de
mod\'eliser localement une partie de l'Univers ambiant \`a l'aide
de boules
$${(B(O,t-t_0),g_t(X))}_{t\geq t_0}\approx {(B(O,1),{(t-t_0)}^2g_t(X))}_{t\geq t_0}?$$
Th\'eoriquement une telle m\'etrique \`a courbure variable existe
(l'Univers est malgr\'e tout fini \`a chaque instant $t$) mais
dans la pratique, il y a beaucoup de travail \`a faire,
n\'ecessitant la d\'etermination de nouveaux objets
g\'eom\'etriques (tels les connexions, les symboles de
Christoffel, les g\'eod\'esiques....) et impliquant des
adaptations convenables des lois de la M\'ecanique et de la Physique.\\\\\\
Dans ce qui suit, on va consid\'erer le cas g\'en\'eral et
quelques cas particuliers (r\'eels, approximatifs ou virtuels)
concernant les familles $g_t$, $a_0(t)$ et $A_t$ qui vont \^etre
pr\'ecis\'es ci-dessous. En effet $g_t$ peut d\'ependre ou non de
la position ou du temps (localement ou globalement), $a_0(t)$ peut
\^etre ou non une g\'eod\'esique et finalement $A_t$ peut
d\'ependre ou non du temps, \^etre ou non une isom\'etrie et
\^etre ou non parall\`ele le long de $a_0(t)$.\\\\
Consid\'erons donc, pour $T>>0$, l'Univers assimil\'e \`a la boule
euclidienne $B(O,T)$ muni de la m\'etrique variable
$g_t(X)=g(t,X)$, o\`u $0 < t < T$ et $X\in B(O,T)$ et supposons
qu'un \'ev\`enement se produit entre deux instants $t_0$ et $t_1$
tels que $t_0<t_1<T$. Soit ${\cal {R}}_0(t_0)=(O_0,\vec{i_0},\vec
{j_0},\vec{k_0})$ un rep\`ere virtuellement fixe et un rep\`ere
mobile ${\cal {R}}(t)$ co\"{\i}ncidant \`a l'instant $t=t_0$ avec
${\cal {R}}(t_0)$. On suppose de plus que l'origine de ce rep\`ere
d\'ecrit une trajectoire rep\'er\'ee dans ${\cal {R}}_0(t_0)$ par
$a_0(t)$ avec $a_0(t_0)=0$. D\'esignons enfin par $X_0(t)$ et
$X(t)$ la trajectoire d'une particule donn\'ee relativement aux
rep\`eres ${\cal {R}}_0(t_0)$ et ${\cal {R}}(t)$ successivement.
Soit $A_t$ la matrice de passage du rep\`ere ${\cal {R}}_0(t_0)$
au rep\`ere ${\cal {R}}(t)$. On a alors:
$$X(t)=A_t(X_0(t)-a_0(t))$$ et $$X'(t)=A_t(X'_0(t)-a'_0(t))+A'_t(X_0(t)-a_0(t)).$$

\subsection*{Un autre op\'{e}rateur de d\'{e}rivation}

En posant
$$\hskip 2cm \frac{d_*}{dt}X(t)=X'(t)+A_t.a'_0(t)-A'_t(X_0(t)-a_0(t)), \hskip 3cm (d_*)$$on
obtient
\begin{equation}\label{r9}
\frac{d_*}{dt}X(t)=A_t.X'_0(t)
\end{equation}
Supposons dans un premier temps que $A_t$ soit une isom\'etrie par
rapport \`a $g_t$ \`a tout instant $t$; alors la d\'erivation
$\frac{d_*}{dt}$ qui, par l'interm\'ediaire de la matrice $A_t$
(qui d\'epend uniquement de la g\'eom\'etrie variable de l'Univers
et du rep\`ere ${\cal {R}}_0(t_0)$), tient compte tout
naturellement de la g\'eom\'etrie de l'espace et du mouvement du
rep\`ere mobile ${\cal {R}}(t)$. Le choix du rep\`ere ${\cal
{R}}_0(t_0)$ n'a pas d'influence sur la nature des r\'esultats qui
vont \^etre obtenus puisqu'il s'agit en fait de l'\'etude du
passage d'un rep\`ere \`a l'autre, tous les deux mobiles par rapport \`a ${\cal {R}}_0(t_0)$.\\\\
Notons que la relation (9) montre que, dans le cas g\'en\'eral, le
vecteur vitesse se transforme d'une fa\c con tensorielle dans le
sens o\`u, si le vecteur $X'_0(t)$ est multipli\'e par la fonction
$f(X_0(t))$ o\`u $f$ est une fonction diff\'erentiable autour de
la trajectoire, alors le vecteur $\frac{d_*}{dt}X(t)$
est multipli\'e par cette m\^eme fonction (de $t$).\\
Etudions ensuite la transformation de l'\'ecriture (induite par ce
changement de rep\`eres) du vecteur acc\'el\'eration d\'efini par
$$\Gamma _0(t)=\nabla _{\frac{d}{dt}X_0(t)}^{g_t}\frac{d}{dt}X_0(t)=\nabla _{X'_0(t)}^{g_t}X'_0(t).$$
Pour cela, posons
$$\Gamma _*(t):=\nabla _{\frac{d_*}{dt}X(t)}^{g_t}\frac{d_*}{dt}X(t)=\nabla _{A_t.X'_0(t)}^{g_t}A_t.X'_0(t).$$
Si $A_t$ est ind\'ependante du temps i.e. $A_t\equiv A$ et $A$
quelconque, on obtient
$$\frac{d_*}{dt}X(t)=A.X'_0(t),$$
$$X'(t)=A.(X'_0(t)-a'_0(t)),$$
et
$$\Gamma _*(t)=\nabla _{\frac{d_*}{dt}X(t)}^{g_t}\frac{d_*}{dt}X(t)
=\nabla _{A.X'_0(t)}^{g_t}A.X'_0(t).$$Si de plus $g_t$ est plate,
on aura
$$\Gamma _*(t)=A.X''_0(t)=A.\nabla _{X'_0(t)}^{g_t}X'_0(t)=A.\Gamma _0(t)$$
et$$\Gamma (t):=\nabla _{X'(t)}^{g_t}X'(t)=\nabla
_{A.(X'_0(t)-a'_0(t))}^{g_t}A.(X'_0(t)-a'_0(t))$$
$$\hskip 1cm = X''(t)=A.(X''_0(t)-a''_0(t)).$$
Si on suppose maintenant que $a_0(t)$ est une g\'eod\'esique,
alors on obtient comme dans le cas euclidien
$$\Gamma _*(t)=\Gamma (t)=A.X''_0(t)=X''(t)=A.\Gamma _0(t).$$Evidemment si $X_0(t)$
est de plus une g\'eod\'esique on a:$$\Gamma _*(t)=\Gamma
(t)=\Gamma _0(t)=0.$$ Notons que si on suppose que $A_t\equiv A$
et $a_0(t)=0$, alors on a, dans tous les cas:
$$X'(t)=A.X'_0(t)$$et$$\Gamma (t)=\Gamma _*(t)=\nabla _{A.X'_0(t)}^{g_t}A.X'_0(t)$$et si $g_t$ est plate, on a
$$\Gamma (t)=X''(t)=A.X''_0(t)=A.\Gamma _0(t)$$et par suite si $X_0(t)$ est une g\'eod\'esique, alors
$X(t)$ l'est aussi. Si on suppose de plus que $g_t\equiv g_{t_0}$
et $A$ est une isom\'etrie (locale), alors on a$${\parallel\Gamma
(t)\parallel}_{g_{t_0}}= {\parallel\Gamma
_0(t)\parallel}_{g_{t_0}}.$$Revenons au cas g\'en\'eral o\`u $g_t$
et $A_t$ sont quelconques et multiplions, dans les expressions de
$\Gamma _0(t)$ et $\Gamma _*(t)$, le vecteur $X'_0(t)$ par une
fonction $h(X_0(t))=:h(t)$ o\`u $h$ est une fonction
diff\'erentiable autour de la trajectoire; on obtient alors
$$\nabla _{hX'_0(t)}^{g_t}hX'_0(t)=h(h\nabla _{X'_0(t)}^{g_t}X'_0(t)+dh(X_0(t)).X'_0(t)X'_0(t))$$
$$\hskip 1 cm =h^2\Gamma _0(t)+\frac{1}{2}dh^2(X_0(t)).X'_0(t)X'_0(t)$$
$$\hskip 1 cm =h^2\Gamma _0(t)+\frac{1}{2}\frac{d}{dt}h^2(X_0(t))\frac{d}{dt}X_0(t)$$et
$$\nabla _{A_t.hX'_0(t)}^{g_t}A_t.hX'_0(t)= \nabla^{g_t}_{h
\frac{d_*}{dt} X(t)} h \frac{d_*}{dt} X(T) = \nabla
_{h(X_0(t))A_t.X'_0(t)}^{g_t}h(X_0(t))A_t.X'_0(t)$$
$$\hskip 0.05cm =\nabla _{h(X_0(t))\frac{d_*}{dt}X(t)}^{g_t}h(X_0(t))\frac{d_*}{dt}X(t)$$
$$\hskip 1.2cm =h(h\nabla _{\frac{d_*}{dt}X(t)}^{g_t}\frac{d_*}{dt}X(t)+
dh(X_0(t)).(\frac{d_*}{dt}X(t))\frac{d_*}{dt}X(t))$$
$$\hskip 0.8 cm =h^2\Gamma _*(t)+\frac{1}{2}dh^2(X_0(t)).(\frac{d_*}{dt}X(t))\frac{d_*}{dt}X(t).$$
Notons que, dans le cas o\`u les $A_t$ sont des isom\'etries, le
second terme de cette relation d\'epend de
la fonction $h$ et de la g\'eom\'etrie de l'espace.\\
Dans le cas g\'en\'eral, la comparaison de l'expression des deux
vecteurs $\nabla _{hX'_0(t)}^{g_t}hX'_0(t)$ et $\nabla^{g_t}_{h
\frac{d_*}{dt} X(t)} h \frac{d_*}{dt} X(t)$ = $\nabla
_{hA_t.X'_0(t)}^{g_t}hA_t.X'_0(t)$ montre que les expressions du
vecteur acc\'el\'eration dans les deux rep\`eres se transforment
de la m\^eme mani\`ere lorsqu'on multiplie le vecteur vitesse par
une fonction donn\'ee. Si on suppose que $A_t\equiv A_{t_0}=A$ et
$a^{'}_0(t)=0$, on obtient:
$$\nabla _{hX'_0(t)}^{g_t}hX'_0(t)=h^2(t) \Gamma_0(t) + \frac{1}{2} dh^2(X_0(t)).\frac{d}{dt} X_0(t) \frac{d}{dt} X_0(t) = h^2(t)\Gamma_0(t)+\frac{1}{2}\frac{d}{dt}h^2(t)X'_0(t)$$et
$$ \nabla^{g_t}_{h X'(t)} h X'(t) = \nabla^{g_t}_{h \frac{d_*}{dt}
X(t)} h \frac{d_*}{dt} X(t) = \nabla
_{hA.X'_0(t)}^{g_t}hA.X'_0(t)$$
$$=h^2(X_0(t))\Gamma (t)+
\frac{1}{2}dh^2(X_0(t)).(A.X'_0(t))A.X'_0(t)
$$
$$=h^2(t)\Gamma(t)+\frac{1}{2}dh^2(X_0(t)).\frac{d_*}{dt}X(t)\frac{d_*}{dt}X(t) $$
$$= h^2(t) \Gamma(t) + \frac{1}{2} dh^2 (X_0(t)).X'(t)X'(t)$$ Si de
plus, la fonction $h$ est suppos\'ee constante sur un voisinage de $X_0(t)$ ou si $h$ est contante sur $X_0(t)$ et $A_t\equiv Id_{\mathbb{R}^3}$, alors les expressions du vecteur
acc\'el\'eration seront toutes les deux multipli\'ees par $h_0^2$
($\Gamma _0(t)$ se transforme en $h_0^2\Gamma _0(t)$ et $\Gamma
(t)$
se transforme en $h_0^2\Gamma (t)$).\\
Dans ces conditions, cette propri\'et\'e implique une
sorte de tensorialit\'e du vecteur acc\'el\'eration dans le sens suivant:\\
Supposons que ${\cal{X}}(X)$ est un champ de vecteurs dans
$B(O,t_1)$ admettant les courbes int\'egrales $X_0(t)$ (i.e.
$X'_0(t)={\cal{X}}(X_0(t))$ pour $t_0\leq t\leq t_1$). Lorsqu'on
multiplie ${\cal{X}}$ par une fonction $h(X)$ suppos\'ee constante
sur un voisinage des courbes int\'egrales $X_0(t)$ ou si $h$ est contante sur $X_0(t)$ avec $A_t\equiv Id_{\mathbb{R}^3}$, alors les deux
expressions du vecteur vitesse sont multipli\'ees par la m\^eme
constante $h_0$ et les deux expressions du vecteur
acc\'el\'eration sont multipli\'ees par
$h_0^2 (h_0=h(X_0)=h(x_0,y_0,z_0)$ lorsque la courbe int\'egrale passe par le point $(x_0,y_0,z_0)$ pour $t=t_0$).\\
On peut aussi exprimer cette propri\'et\'e en disant que lorsque
le vecteur acc\'el\'eration $\Gamma _0(t)$ de la courbe
int\'egrale $X_0(t)$ du champ de vecteurs ${\cal{X}}$ dans le
rep\`ere ${\cal {R}}_0(t_0)$ est multipli\'ee par une constante,
alors l'expression de ce m\^eme vecteur dans le
rep\`ere ${\cal {R}}(t)$ est multipli\'ee par cette m\^eme constante.\\\\\\
Plus g\'en\'eralement, consid\'erons deux rep\`eres mobiles ${\cal
{R}}_1(t)$ et ${\cal {R}}_2(t)$ comme ceux consid\'er\'es au
paragraphe 1. Utilisons les notations \'evidentes suivantes:
$$X_1(t):=A_t(X_0(t)-a_0(t)),$$ $$X_2(t):=B_t(X_0(t)-b_0(t)),$$
$$\frac{d_*}{dt}X_1(t):=X'_1(t)+A_t.a'_0(t)-A'_t(X_0(t)-a_0(t))=A_t.X'_0(t)$$et
$$\frac{d_*}{dt}X_2(t):=X'_2(t)+B_t.b'_0(t)-B'_t(X_0(t)-b_0(t))=B_t.X'_0(t).$$On a alors
\begin{equation}\label{r10}
\frac{d_*}{dt}X_2(t)=B_t\circ A_t^{-1}.\frac{d_*}{dt}X_1(t),
\end{equation}et
$$\Gamma _{1*}(t):=\nabla _{\frac{d_*}{dt}X_1(t)}^{g_t}\frac{d_*}{dt}X_1(t)=
\nabla _{A_t.X'_0(t)}^{g_t}A_t.X'_0(t),
$$et
$$\Gamma _{2*}(t):=\nabla _{\frac{d_*}{dt}X_2(t)}^{g_t}\frac{d_*}{dt}X_2(t)$$
$$=
\nabla _{B_t.X'_0(t)}^{g_t}B_t.X'_0(t) =\nabla _{B_t\circ
A_t^{-1}\frac{d_*}{dt}X_1(t)}^{g_t}B_t\circ
A_t^{-1}\frac{d_*}{dt}X_1(t).$$ Donc si $B_t$ et $A_t$ sont des
isom\'etries, alors la relation (10) ne d\'epend que de la
g\'eom\'etrie (d\'{e}finie par $g_t$) de l'espace et on a:
$${\parallel\frac{d_*}{dt}X_2(t)\parallel}_{g_{t}}
={\parallel\frac{d_*}{dt}X_1(t)\parallel}_{g_{t}}$$et
$${\parallel\Gamma _{2*}(t)\parallel}_{g_{t}}=
{\parallel\Gamma _{1*}(t)\parallel}_{g_{t}}$$et si $B_t=A_t$, on a
(dans le cas g\'en\'eral):
$$\Gamma _{2*}(t)=\Gamma _{1*}(t).$$D'autre part si $A_t\equiv A$ et $B_t\equiv B$, alors on a:
$$\frac{d_*}{dt}X_1(t)= \frac{d}{dt} X_1(t) + A.a'_0(t) = A.X'_0(t)=\frac{d}{dt}(A.X_0(t))$$et
$$\frac{d_*}{dt}X_2(t)= \frac{d}{dt} X_2(t) + B. b'_0(t) = B. X'_0(t) = \frac{d}{dt}(B.X_0(t)),$$par suite on obtient:
$$\Gamma _{1*}(t)=\nabla ^{g_t}_{(AX_0(t))'}(AX_0(t))'$$et
$$\Gamma _{2*}(t)=\nabla ^{g_t}_{(BX_0(t))'}(BX_0(t))'.$$Par cons\'equent, si $g_t$ est plate, on a
$$\Gamma _{1*}(t)=A.X''_0(t)=A.\Gamma _{0}(t)$$et
$$\Gamma _{2*}(t)=B.X''_0(t)=B.\Gamma _{0}(t)$$d'o\`u
$$\Gamma _{2*}(t)=B\circ A^{-1}.\Gamma _{1*}(t).$$
De m\^eme on a (toujours dans le cas o\`u $A_t\equiv A$ et
$B_t\equiv B$ et $g_t$ est plate) les relations suivantes entre
$\Gamma _1(t)$, $\Gamma _2(t)$, $\Gamma _{1*}(t)$, $\Gamma
_{2*}(t)$ et $\Gamma _0(t)$:
$$\Gamma _1(t)=\nabla _{X'_1(t)}^{g_{t}}X'_1(t)=X''_1(t)=A.(X_0(t)-a_0(t))''$$et
$$\Gamma _2(t)=\nabla _{X'_2(t)}^{g_{t}}X'_2(t)=X''_2(t)=B.(X_0(t)-b_0(t))''$$
et si $a_0(t)$ et $b_0(t)$ sont des g\'eod\'esiques, on obtient
$$\Gamma _{1}(t)=A.X''_0(t)=A.\Gamma _{0}(t)=\Gamma _{1*}(t),$$
$$\Gamma _{2}(t)=B.X''_0(t)=B.\Gamma _{0}(t)=\Gamma _{2*}(t),$$
$$\Gamma _{2}(t)=B\circ A^{-1}.\Gamma _{1}(t)$$
et
$$\Gamma_2(t)=\Gamma_1(t)\quad\mbox{pour}\;\; A=B\quad\mbox{ou, en particulier, pour}\quad A=B=\mbox{Id}_{\mathbb{R}^3}.$$
De m\^eme si on a $A_t\equiv A$, $B_t\equiv B$, $a_0(t)=b_0(t)=0$
et $g_t$ quelconque, alors on obtient
$$\Gamma _1(t)=\Gamma _{1*}(t)=\nabla _{A.X'_0(t)}^{g_{t}}A.X'_0(t),$$
$$\Gamma _2(t)=\Gamma _{2*}(t)=\nabla _{B.X'_0(t)}^{g_{t}}B.X'_0(t)$$et si de plus $g_t$ est plate
alors on obtient (\'evidemment) de nouveau$$\Gamma _{2}(t)=B\circ
A^{-1}.\Gamma _{1}(t)$$et si $A$ et $B$ sont de plus des
isom\'etries, alors on a
$${\parallel\Gamma _{2}(t)\parallel}_{g_{t_0}}=
{\parallel\Gamma _{1}(t)\parallel}_{g_{t_0}}.$$ En revenant au cas
g\'en\'eral o\`u $g_t$, $A_t$ et $B_t$ sont quelconques, on montre
qu'en multipliant le vecteur vitesse $\frac{d_*}{dt}X_1(t)$ par
une fonction $h(X_1(t))=:h(t)$, o\`u $h$ est une fonction
diff\'erentiable autour de la trajectoire $X_1(t)$, alors le
vecteur $\frac{d_*}{dt}X_2(t)$ est multipli\'e par cette m\^eme
fonction et on a:
$$\nabla _{h\frac{d_*}{dt}X_1(t)}^{g_t}h\frac{d_*}{dt}X_1(t)=
h^2(t)\Gamma_{1*}(t)+\frac{1}{2}dh^2(X_0(t)).(\frac{d_*}{dt}X_1(t))\frac{d_*}{dt}X_1(t)$$et
$$\nabla _{h\frac{d_*}{dt}X_2(t)}^{g_t}h\frac{d_*}{dt}X_2(t)=
h^2(t)\Gamma_{2*}(t)+\frac{1}{2}dh^2(X_0(t)).(\frac{d_*}{dt}X_2(t))\frac{d_*}{dt}X_2(t).$$A
partir de ces relations, on obtient comme ci-dessus des
propri\'et\'es concernant des cas particuliers qui montrent toutes
une certaine tensorialit\'e du vecteur acc\'el\'eration vis \`a
vis du changement de deux rep\`eres tous les deux mobiles par rapport \`{a} un troisi\`{e}me.\\
En particulier, lorsque $g_t$ est plate et $a_0(t)$ et $b_0(t)$
sont des g\'{e}od\'{e}siques (ou $a_0(t) = b_0(t) = 0$) alors
cette tensorialit\'{e} est valide pour le vecteur
acc\'{e}l\'{e}ration ordinaire $\Gamma(t) = \frac{d^2}{dt^2}X(t) =
X^{''}(t)$ puisque l'on a alors
$$\frac{d_*}{dt} X(t) = \frac{d}{dt} X(t) = A.\frac{d}{dt}
X_0(t) .$$

Notons pour finir que dans le cas g\'en\'eral ($g_t=g_t(X)$), on
peut d\'efinir globalement ou localement tous les objets
g\'eom\'etriques (i.e. d\'ependant uniquement de $g_t$) tels que
les groupes d'isom\'etrie, le gradient et le hessien d'une
fonction, la divergence d'un champ de vecteurs ou d'une forme
diff\'erentielle, le laplacien d'une fonction ou d'une forme
diff\'erentielle (l'op\'erateur de Hodge), l'op\'erateur de
Dirac...\\\\
Quant \`a la tensorialit\'e du changement de rep\`eres dans
l'Univers assimil\'e \`a ${(U_t(t))}_{t > 0}\cong
{(B(O,1),g_t)}_{t
> 0}$, alors on montre que l'on obtient des propri\'et\'es du
m\^eme ordre que celles obtenues auparavant pour des
\'ev\`enements se produisant entre deux instants $t_1$ et $t_2$
suffisemment proches l'un de l'autre et tels que $t_2\geq t_1 \geq
t_0 >>0$. En effet, on pourra alors consid\'erer deux rep\`eres
mobiles ${\cal {R}}_1(t)$ et ${\cal {R}}_2(t)$ ayant le m\^eme
origine $O_1$ \`a l'instant $t=t_0$ et d\'ecrivant des
trajectoires $a(t)$ et $b(t)$ dans le rep\`ere $(O_1,\vec{i},\vec
{j},\vec{k})$ et comparer $u_1(t)=\frac{X_1(t)}{t}$, $u'_1(t)$ et
$u''_1(t)$ \`a $u_2(t)=\frac{X_2(t)}{t}$, $u'_2(t)$ et $u''_2(t)$,
o\`u $X_1(t)$ et $X_2(t)$ sont les expressions d'une m\^eme
trajectoire rep\'er\'ee dans les deux rep\`eres mobiles. On
utilisera alors le fait que l'on a:
$${\parallel a(t)\parallel}_{g_{e}},
{\parallel b(t)\parallel}_{g_{e}},{\parallel
X_1(t)\parallel}_{g_{e}}, {\parallel
X_2(t)\parallel}_{g_{e}}<<t_0.$$

\section{Mod\'elisation physico-math\'ematique de l'Univers en
expansion}

Signalons pour commencer que, dans la suite de cet article, on
utilisera fondamentalement les deux principes d'inertie de Newton,
le principe de la covariance, les deux principes d'\'equivalence
et les principes (dans un sens intuitif) d'homog\'en\'eit\'e et
d'isotropie. Signalons aussi que la majorit\'e des notions et des
r\'esultats utilis\'es dans cet article se trouve dans un grand
nombre d'ouvrages classiques sur la th\'eorie de la relativit\'e
et, en particulier, on peut consulter pour tous les notions et
r\'esultats, le livre de R.Wald:
"General relativity".\\ \\
Il est clair, d'apr\`es ce qui pr\'ec\`ede, que l'exp\'erience du
train, du miroir et des deux observateurs, ne justifie pas
l'introduction de la nouvelle conception relativiste de la
relation espace-temps. Montrons que l'invalidit\'e du principe de
covariance vis \`{a} vis de la loi de Maxwell dans les rep\`eres
inertiels n'est en fait qu'apparente. \\

En effet, consid\'erons deux rep\`eres euclidiens inertiels
$\mathcal{R}_1$ et $\mathcal{R}_2$ tels que le mouvement du second
par rapport au premier soit uniforme. Cette situation peut \^etre
ramen\'e \`a supposer que, si $(x_1(t),y_1(t),z_1(t))$ d\'ecrit
une trajectoire donn\'ee dans le rep\`ere $\mathcal{R}_1,$ alors
cette trajectoire est exprim\'ee dans $\mathcal{R}_2$ par
$(x_2(t),y_1(t),z_1(t))$ avec $x_2(t)=x_1(t)-vt.$ Cependant ces
deux rep\`eres, dont l'\'etude du passage de l'un \`a l'autre a
conduit \`a en d\'eduire l'invalidit\'e du principe de covariance,
sont en fait des rep\`eres mobiles inertiels. Par cons\'equent, si
$v_1$ et $v_2$ sont respectivement les vitesses constantes de ces
deux rep\`eres par rapport \`a un rep\`ere virtuel $\mathcal{R}_0$
suppos\'e fixe (on peut alors supposer que les deux vecteurs
vitesse ont la m\^eme direction que l'axe $Ox$ de $\mathcal{R}_0$)
et si $\varphi$ est une fonction qui peut \^etre suppos\'ee de la
forme $\varphi(x_1,t)$ dans le rep\`ere $\mathcal{R}_1,$ alors
l'\'equation des ondes s'\'ecrit dans $\mathcal{R}_1$ sous la
forme
\begin{equation}\label{11}
\square_1\varphi(x_1,t)\;:=\;\frac{\partial^2\varphi}{\partial
t^2}(x_1,t)\;-\;\frac{\partial^2\varphi}{\partial
x_1^2}(x_1,t)\;=\;f(x_1,t).
\end{equation}
En posant $\varphi(x_2,t)\;=\;\varphi(x_1-vt,t),$ la forme de
cette \'equation, lorsqu'elle est \'ecrite dans le rep\`ere
$\mathcal{R}_2,$ reste la m\^eme. Elle s'\'ecrit, en effet, sous
la forme
\begin{equation}\label{12}
\square_2\varphi(x_2,t)\;:=\;\frac{\partial^2\varphi}{\partial
t^2}(x_2,t)\;-\;\frac{\partial^2\varphi}{\partial
x_2^2}(x_2,t)\;=\;f(x_2,t)
\end{equation}
dans le sens suivant:\\
Lorsqu'on \'ecrit $\varphi(x_1,t)\;=\;\varphi(x_0-v_1t,t),$
l'\'equation des ondes s'\'ecrivant dans $\mathcal{R}_0$ sous sa
forme canonique
\begin{equation}\label{13}
\square_0\varphi(x_0,t)\;:=\;\frac{\partial^2\varphi}{\partial
t^2}(x_0,t)\;-\;\frac{\partial^2\varphi}{\partial
x_0^2}(x_0,t)\;=\;f(x_0,t),
\end{equation}
alors l'\'equation (12) est obtenue \`a partir de l'\'equation
(13) de la m\^eme fa\c{c}on que l'\'equation (11) est obtenue de
la m\^eme \'equation (13). Ceci est r\'{e}alis\'{e} en donnant \`a
la d\'eriv\'ee partielle par rapport \`a la deuxi\`eme variable,
i.e. par rapport \`a $t,$ dans l'\'equation (12) le m\^eme sens
qu'on a donn\'ee \`a $\frac{\partial\varphi}{\partial t}$ dans
(11). En d'autres termes, on d\'efinit la d\'erivation
$\frac{\partial\varphi}{\partial t}(x_1,t)$ en rempla\c{c}ant dans
$$\frac{d}{dt}\varphi(x_0,t)=\frac{\partial\varphi}{\partial
x_0}(x_0,t)x_0^{'}(t)+\frac{\partial\varphi}{\partial t}(x_0,t),$$
$x_0(t)$ par $x_1(t)(=x_0(t)-v_1t$) et $x^{'}_0$(\emph{t}) par
$x^{'}_1$(\emph{t}), consid\'{e}r\'{e} comme \'{e}tant le m\^{e}me
que $x^{'}_0(t)$, et non pas en tant que
$$\frac{\partial\varphi}{\partial
t}(x_1,t)=\frac{d}{dt}\varphi(x_0(t)-v_1t,t)=\partial_1\varphi(x_0-v_1t,t)(x^{'}_0(t)-v_1)+\partial_2\varphi(x_0-v_1t,t)$$
et on proc\`ede de la m\^eme mani\`ere pour la d\'erivation
$\frac{\partial\varphi}{\partial t}(x_2,t)$ en rempla\c{c}ant,
dans $\frac{d}{dt}\varphi(x_0,t),$ $x_0(t)$ par
$x_2(t)(=x_0(t)-v_2t$) et $x^{'}_0$(\emph{t}) par
$x^{'}_2$(\emph{t}), consid\'{e}r\'{e} comme \'{e}tant le m\^{e}me
que $x^{'}_0(t)$, et non pas en tant que
$$\frac{\partial\varphi}{\partial
t}(x_2,t)=\frac{d}{dt}\varphi(x_0(t)-v_2t,t).$$ Ainsi, lorsqu'on
suppose que $\mathcal{R}_1$ est au repos et $\mathcal{R}_2$ est
inertiel par rapport \`a $\mathcal{R}_1,$ on d\'eduit (12) de (11)
en utilisant la d\'eriv\'ee partielle
$\frac{\partial\varphi}{\partial t}(x_2,t)$ qui s'obtient \`a
partir de
$$\frac{d}{dt}\varphi(x_1,t) = \partial_1 \varphi (x_1,t)
x^{'}_1(t) + \partial_2 \varphi (x_1,t)$$ en rempla\c{c}ant
$x_1(t)$ par $x_2(t)(=x_1(t)-vt$)et $x^{'}_1$(\emph{t}) par
$x^{'}_2$(\emph{t}), consid\'{e}r\'{e} comme \'{e}tant le m\^{e}me
que $x^{'}_1(t)$, et non pas en tant que
$$\frac{\partial\varphi}{\partial t}(x_2,t)\;=\;\frac{d}{dt}\varphi(x_1(t)-vt,t).$$
En d'autres termes, lorsqu'on effectue la d\'eriv\'ee partielle
par rapport \`a la variable $t$ de la fonction
$\varphi(x_1-vt,t),$ on suppose que la premi\`ere variable est
$x_1$ et non pas $x_1-vt$ (i.e en consid\'{e}rant que la
d\'{e}pendance du temps de la variable $x_2$ est la m\^{e}me que
celle de la variable $x_1$) et
en rempla\c{c}ant ensuite $x_1 -vt$ par $x_2$.\\
Plus g\'en\'eralement, nous allons utiliser ici les deux notions
de d\'erivation $\frac{d_1}{dt}$ et $\frac{d_*}{dt}$ d\'ej\`a
introduites et quelques propri\'et\'es de tensorialit\'e
concernant le changement de rep\`eres d\'ej\`a \'etablies dans les
paragraphes pr\'ec\'edents.\\Consid\'erons donc un rep\`ere
euclidien virtuel fixe
$\mathcal{R}_0=(O_0,\overrightarrow{i_0},\overrightarrow{j_0},\overrightarrow{k_0})$
de $\mathbb{R}^3$ et deux autres rep\`eres
$\mathcal{R}_1=(O_1,\overrightarrow{i_1},\overrightarrow{j_1},\overrightarrow{k_1})$
et
$\mathcal{R}_2=(O_2,\overrightarrow{i_2},\overrightarrow{j_2},\overrightarrow{k_2}).$
On suppose que $O_1$ et $O_2$ circulent, par rapport \`a
$\mathcal{R}_0,$ avec deux vitesses constantes respectives
$\overrightarrow{v_1}$ et $\overrightarrow{v_2}.$ On suppose de
plus que $\mathcal{R}_1$ et $\mathcal{R}_2$ sont obtenus, pour
$t\geq 0,$ \`{a} partir de $\mathcal{R}_0$ par les deux
transformations lin\'eaires $A_t$ et $B_t.$ Finalement, soient
$X_0(t)=(x_0(t),y_0(t),z_0(t)),$ $X_1(t)=(x_1(t),y_1(t),z_1(t))$
et $X_2(t)=(x_2(t),y_2(t),z_2(t))$ les expressions respectives,
dans ces trois rep\`eres, d'une trajectoire donn\'ee dans
$\mathbb{R}^3.$ On a alors, pour tout $t\geq0:$
\begin{eqnarray*}
  X_1(t)&=&A_t(X_0(t)-t\overrightarrow{v_1})=:A_t.Y_0(t),\\
X_2(t)&=&B_t(X_0(t)-t\overrightarrow{v_2})=:B_t.Z_0(t)
\end{eqnarray*}
et
\begin{eqnarray*}
  X_2(t) &=& B_t.(A_t^{-1}.X_1(t)+t\overrightarrow{v_1}-t\overrightarrow{v_2}) \\
  &=& B_t\circ A_t^{-1}.X_1(t)-B_t.t\overrightarrow{v}
\end{eqnarray*}
o\`u
$\overrightarrow{v}=\overrightarrow{v_2}-\overrightarrow{v_1}$ est
la vitesse relative de $O_2$ par rapport \`a $O_1.$ Ainsi, on a
(d'apr\`{e}s la relation ($d_1$)):
\begin{eqnarray*}
  \frac{d_1}{dt}X_1(t) &=& \frac{d}{dt}X_1(t)-A_t^{'}(X_0(t)-t\overrightarrow{v_1}) = \frac{d}{dt}X_1(t) - A^{'}_tY_0(t)\\
 &=& \frac{d}{dt}X_1(t)-A_t^{'}\circ A_t^{-1}.X_1(t)
\end{eqnarray*}
et, d'apr\`{e}s la relation ($d_*$):
$$\frac{d_*}{dt}X_1(t)=\frac{d}{dt}X_1(t)+A_t.\overrightarrow{v_1}-A_t^{'}(X_0(t)-t\overrightarrow{v_1}).$$
Or
$$\frac{d}{dt}X_1(t)=A_t\left(\frac{d}{dt}X_0(t)-\overrightarrow{v_1}\right)+A_t^{'}(X_0(t)-t\overrightarrow{v_1})$$
qui s'\'ecrit
$$\frac{d}{dt}X_1(t)=A_t.\frac{d}{dt}Y_0(t)+A_t^{'}.Y_0(t).$$
D'o\`u
\begin{equation*}
    \frac{d_1}{dt}X_1(t)=A_t.\frac{d}{dt}Y_0(t)=A_t.\frac{d_1}{dt}Y_0(t)
    \;\;\;\;\;\;\;\;\;\;\;\;\;\;\;\;\;\; (d_1^{'})
\end{equation*}
et
\begin{eqnarray*}
  \frac{d_*}{dt}X_1(t) &=& A_t.\frac{d}{dt}Y_0(t)+A_t.\overrightarrow{v_1}\\
 &=& A_t.\frac{d}{dt}X_0(t)=A_t.\frac{d_*}{dt}X_0(t)
\end{eqnarray*}
puisque, rappelons le, $\frac{d_1}{dt}$ et $\frac{d_*}{dt}$ sont
identiques \`a
$\frac{d}{dt}$ pour les coordonn\'ees dans $\mathcal{R}_0.$\\
On note que l'on a
$$\frac{d_2}{dt}X_2(t):=B_t.\frac{d}{dt}Z_0(t)\qquad\mbox{pour}\qquad
Z_0(t) = X_0(t) - tv_2 =Y_0(t)-t\overrightarrow{v} \;\;\;\;\;
(d_2^{'})$$ et
$$\frac{d_*}{dt}X_2(t):=B_t.\frac{d}{dt}X_0(t),\qquad\qquad\qquad\qquad\qquad\qquad\quad$$
ce qui donne
\begin{equation*}
\frac{d_*}{dt}X_2(t)=B_t\circ A_t^{-1}.\frac{d_*}{dt}X_1(t).
\end{equation*}
Par cons\'equent, lorsque $B_t\equiv A_t,$ on obtient
$$\frac{d_*}{dt}X_2(t)=\frac{d_*}{dt}X_1(t).$$
Par ailleurs, lorsque $\overrightarrow{v_1}=\overrightarrow{v_2}$
(i.e. $\overrightarrow{v}=0$), on obtient (d'apr\`{e}s les
relations ($d_2^{'}$) et ($d_1^{'}$))
\begin{equation*}
\frac{d_2}{dt}X_2(t)=B_t\circ A_t^{-1}.\frac{d_1}{dt}X_1(t)
\end{equation*}
et si de plus $B_t\equiv A_t,$ alors
$$\frac{d_2}{dt}X_2(t)=\frac{d_1}{dt}X_1(t).$$
Rappelons que ces deux d\'erivations tiennent compte d'une
fa\c{c}on naturelle \`a la fois des vitesses relatives et des
rotations des rep\`eres mobiles. La premi\`ere co\"incide, pour
$A_t\equiv A$, avec la d\'erivation classique. la seconde devient,
pour $A_t\equiv A$:
$$\frac{d_*}{dt}X_1(t)=\frac{d}{dt}X_1(t)\;+\;A.\overrightarrow{v_1}.$$
Quelques unes des relations ci-dessus montrent quelques
propri\'et\'es de tensorialit\'e pour le vecteur vitesse vis \`a
vis
des changements de rep\`eres.\\
Dans le cas particulier o\`u $A_t=Id_{\mathbb{R}^3},$ on obtient
$$\frac{d_1}{dt}X_1(t)=X_1^{'}(t)=\frac{d_1}{dt}Y_0(t)=Y_0^{'}(t)=X_0^{'}(t)-\overrightarrow{v_1} ,$$
qui n'est autre que la d\'erivation Galil\'eenne classique, et
$$\frac{d_*}{dt}X_1(t)=X_0^{'}(t).$$

A la lumi\`ere de ce qui pr\'ec\`ede, on peut clarifier plus
pr\'ecisemment le probl\`eme de la covariance de l'\'equation de
Maxwell. En effet, en rempla\c{c}ant $X_0(t),$ $X_1(t)$ et
$X_2(t)$ par $x_0(t),$ $x_1(t)$ et $x_2(t)$ et les vecteurs
vitesses $\overrightarrow{v_1},$ $\overrightarrow{v_2}$ et
$\overrightarrow{v},$ par les vitesses scalaires $v_1,$ $v_2$ et
$v,$ on peut affirmer que la validit\'e du principe de covariance
necessite uniquement une l\'eg\`ere modification de la notion de
la variable spaciale $(x\longrightarrow x-vt)$ et celle de la
diff\'erentiation par rapport au temps
$\left(\frac{d}{dt}\longrightarrow\frac{d_*}{dt}\right).$ Ces
modifications constituent les deux proc\'edures naturelles pour
inclure le mouvement du rep\`ere dans le mouvement g\'en\'eral
dans l'univers, contrairement \`a la conception relativiste de
l'espace-temps qui conduit \`a l'alt\'eration de la relation
naturelle entre l'espace et le temps puisque pour nous, les
distances sont essentiellement proportionnelles au temps.\\

\subsection*{Forme canonique de l'\'{e}quation de Maxwell}

Consid\'erons maintenant la d\'eriv\'ee (par rapport au temps)
classique
\begin{eqnarray*}
   \frac{d}{d t}\varphi(x_1,t)&=&\frac{d}{dt}\varphi(x_0-v_1t,t)\\
   &=&\partial_1\varphi(x_0-v_1t,t)(x_0^{'}(t)-v_1)+\partial_2\varphi(x_0-v_1t,t).
\end{eqnarray*}
En prenant la d\'eriv\'ee $\frac{d_*}{d t}$ \`a la place de
$\frac{d}{dt}$ lors d'un changement de variable $x_1 = x_0 - v_1
t$, on obtient
\begin{eqnarray*}
   \frac{\partial_*\varphi}{\partial t}(x_1,t)&=& \frac{\partial_*\varphi}{\partial t}(x_0-v_1t,t) \\
   &:=& \partial_1\varphi(x_0-v_1t,t)\frac{d_*}{dt}x_1(t)+\partial_2\varphi(x_0-v_1t,t) \\
   &=& \partial_1\varphi(x_0-v_1t,t)x_0^{'}(t)+\partial_2\varphi(x_0-v_1t,t) \\
  &=&\frac{\partial\varphi}{\partial t}(x_0-v_1t,t)
\end{eqnarray*}
dans le sens pr\'ecis\'e auparavant, puisqu'ici  $\frac{d_*}{dt}
x_1(t) = x^{'}_0(t)$, $A_t$ \'{e}tant ici identiquement \'{e}gal
\`{a} $Id_{\mathbb{R}^3}$. Cette d\'erivation est donc effectu\'ee
en consid\'erant que la d\'ependance par rapport \`a la variable
temps de la premi\`ere variable $x_1(t)=x_0(t)-v_1t$ est la m\^eme
que celle de la variable $x_0(t)(=x_1(t)+v_1t),$ i.e. en
n\'egligeant le terme $v_1t$ qui provient du mouvement relatif du
rep\`ere $\mathcal{R}_1$; ce qui revient \`{a} prendre la
d\'{e}riv\'{e}e de la trajectoire rep\'{e}r\'{e}e dans $\mathcal{R}_1$ par
$x_1(t)$ comme \'{e}tant la d\'{e}riv\'{e}e de la m\^{e}me
trajectoire rep\'{e}r\'{e}e par $\mathcal{R}_0$, mais aussi par tout autre
rep\`{e}re fixe isom\'{e}trique \`{a} $\mathcal{R}_0$ et en particulier par
$\mathcal{R}_1$ \`{a} condition de consid\'{e}rer ce dernier comme \'{e}tant
au repos par rapport \`{a} $\mathcal{R}_0$. Dans le m\^eme ordre d'id\'ee
posons
$$\frac{\partial^2_*\varphi}{\partial
t^2}(x_1,t)\;=\;\frac{\partial^2\varphi}{\partial
t^2}(x_0-v_1t,t).$$ En utilisant ces notions et ces notations, on
peut maintenant affirmer que la forme canonique intrins\`eque de
l'\'equation de Maxwell est
\begin{equation*}
\square_*\varphi(x,t)\;:=\;\frac{\partial^2_*\varphi}{\partial
t^2}(x,t)\;-\;\frac{\partial^2\varphi}{\partial x^2}(x,t)\;=\;0
\end{equation*}
qui se r\'eduit, lorsqu'on se ram\`ene au rep\`ere pour lequel
l'origine est au repos, \`a l'\'equation classique
\begin{equation*}
\square\varphi(x,t)\;=\;\frac{\partial^2\varphi}{\partial
t^2}(x,t)\;-\;\frac{\partial^2\varphi}{\partial x^2}(x,t)\;=\;0.
\end{equation*}
Lorsqu'on consid\`ere deux autres rep\`eres $\mathcal{R}_1$ et
$\mathcal{R}_2,$ l'\'equation intrins\`eque s'\'ecrit
successivement
\begin{equation*}
\square_*\varphi(x_1,t)\;:=\;\frac{\partial^2_*\varphi}{\partial
t^2}(x_1,t)\;-\;\frac{\partial^2\varphi}{\partial
x_1^2}(x_1,t)\;=\;0
\end{equation*}
et
\begin{equation*}
\square_*\varphi(x_2,t)\;:=\;\frac{\partial^2_*\varphi}{\partial
t^2}(x_2,t)\;-\;\frac{\partial^2\varphi}{\partial
x_2^2}(x_2,t)\;=\;0.
\end{equation*}
Ces deux \'equations se r\'eduisent \'evidemment \`a l'\'equation
classique lorsqu'on consid\`ere les rep\`eres appropri\'es
respectifs. N\'eamoins, lorsqu'on effectue le changement de
rep\`ere $\mathcal{R}_1\longrightarrow\mathcal{R}_2,$ la variable
$x_2$ dans la deuxi\`eme \'equation doit \^etre consid\'er\'ee
comme \'etant $x_2=x_1-vt,$ o\`u $v$ est la vitesse relative de
$\mathcal{R}_2$ par rapport \`a $\mathcal{R}_1,$ et cette
\'equation ne se r\'eduit pas \`a la forme classique. Le second
membre prend en fait la signification
$$\frac{\partial^2_*\varphi}{\partial
t^2}(x_1-vt,t)\;-\;\frac{\partial^2\varphi}{\partial
x_1^2}(x_1-vt,t).$$ Ceci veut dire que, dans la d\'eriv\'ee de
$\varphi(x_1-vt,t)$ par rapport au temps, le premier terme est
obtenu en consid\'erant que la d\'ependance du temps de la
premi\`ere variable provient uniquement de celle de la variable
$x_1(t).$ La d\'ependance du temps de $x_2(t)=x_1(t)-vt$ provenant
de $vt$ n'est pas prise en compte, ce qui est tout \`a fait
naturel puisque le mouvement relatif du rep\`ere consid\'er\'e par
rapport \`a tout autre rep\`ere ne peut pas constituer une
caract\'eristique canonique de la propagation des ondes
\'el\'ectromagn\'etiques ou, plus g\'en\'eralement, de tout
mouvement physique r\'eel. Si $v=0,$ les deux \'equations
canoniques pr\'ec\'edentes sont identiques et elles se r\'eduisent
\`a la forme classique lorsqu'elle sont
\'ecrites dans n'importe quel rep\`ere au repos.\\

Evidemment, les deux op\'erateurs $\square_1$ et $\square_2$
introduits au d\'ebut de ce paragraphe \`a partir de l'op\'erateur
$\square_0$ se confondent avec l'op\'erateur $\square_0.$ Comme
l'op\'erateur $\square_0,$ ils se confondent, lorsqu'ils sont
\'ecrits chacun dans son propre rep\`ere au repos, avec la forme
classique. L'\'etude de ces deux op\'erateurs avait pour but de
montrer que l'on ne peut pas privil\'egier un rep\`ere inertiel
donn\'e au d\'epens de tous les autres. Tous les rep\`eres
inertiels ont le m\^eme droit d'exprimer une loi physique. Par
cons\'equent, on ne peut pas choisir l'un d'entre eux afin
d'affirmer la non validit\'e du principe de covariance en
invoquant simplement le fait que la m\^eme loi physique s'exprime
diff\'eremment dans un autre rep\`ere inertiel. Ainsi, seul
l'op\'erateur $\square_*$ est canonique et par suite il est le
seul qui peut \^etre utilis\'e canoniquement pour traduire
une loi physique donn\'ee en coordonn\'ees locales.\\

\subsection*{Canonicit\'{e} de la vitesse de la lumi\`{e}re}

On peut maintenant conclure que seul le cadre pr\'ec\'edent
constitue le contexte propre qui permet une interpr\'etation
correcte du principe fondamental de l'invariabilit\'e de la
vitesse de la lumi\`ere dans le vide par rapport \`a tous les
rep\`eres inertiels. En effet, si dans un rep\`ere inertiel
$\mathcal{R}_0$ la trajectoire d'un rayon de lumi\`ere est
d\'ecrite par $x_0(t)=ct,$ alors lorsqu'on consid\`ere un autre
rep\`ere $\mathcal{R},$ circulant uniformement suivant l'axe $Ox$
de $\mathcal{R}_0$ avec une vitesse constante $v$ (voir fig.5), on
a:
$$x(t)\;=\;x_0(t)\;-\;vt\;=\;ct\;-\;vt,$$
o\`u $x(t)$ est l'expression de cette m\^eme trajectoire dans le
rep\`ere $\mathcal{R}$, puisque l'on a ici $x(t)=A_t(x_0(t)-vt)$
et $A_t=Id_{\mathbb{R}^3}.$ D'o\`u
$$\frac{d_*}{dt}x(t)\;=\;\frac{d}{dt}x_0(t)\;=\;c\;=\;1 , $$
qui est la vitesse canonique de la lumi\`ere, et
$$\frac{d_1}{dt}x(t)\;=\;\frac{d}{dt}x(t)\;=\;c-v$$
qui est conforme \`a la notion galil\'eo-newtonienne de la
vitesse.\\

Ceci montre de nouveau l'invalidit\'e du second postulat de la
relativit\'e sp\'eciale lorsqu'elle affirme que la vitesse de la
lumi\`ere ne d\'epend pas de la vitesse du rep\`ere mobile
utilis\'e pour la mesurer. Naturellement le principe fondamental
de l'ind\'ependance de cette vitesse de celle de la source est
parfaitement
valide.\\\\
On peut \'{e}galement r\'{e}sumer ce qui pr\'{e}c\`{e}de en disant
que lorsqu'on \'{e}crit l'\'{e}quation de Maxwell dans un
r\'{e}f\'{e}rentiel galil\'{e}en quelconque $\mathcal{R}$ sous la forme
$$ \square \varphi(x,t) = \frac{\partial^2 \varphi}{\partial t^2}
(x,t) - \frac{\partial^2 \varphi}{\partial x^2} (x,t) = 0$$ on
est, en fait, en train de consid\'{e}rer implicitement que ce
rep\`{e}re est fixe et que la vitesse de la lumi\`{e}re dans ce
rep\`{e}re est 1. L'\'{e}quation de Maxwell devrait s'\'{e}crire
alors
$$ \square_* \varphi(x_1,t) = \frac{\partial^2_* \varphi}{\partial t^2}
(x_1,t) - \frac{\partial^2 \varphi}{\partial x_1^2} (x_1,t) = 0$$
dans tout autre rep\`{e}re galil\'{e}en arbitraire $\mathcal{R}_1$ circulant
\`{a} une vitesse arbitraire donn\'{e}e $v_1$ par rapport \`{a}
$\mathcal{R}$ et la vitesse de la lumi\`{e}re par rapport \`{a} ce
rep\`{e}re est alors $1-v_1$ $(\frac{d_*}{dt}x_1(t)=\frac{d}{dt}x_1(t)+v_1)$.\\
Si l'on \'{e}crit, par exemple, que l'\'{e}quation de Maxwell par
rapport au r\'{e}f\'{e}rentiel propre du soleil sous la
premi\`{e}re forme ci-dessus, on est en train de d\'{e}cider que
la vitesse de la lumi\`{e}re par rapport au soleil est 1.
L'\'{e}quation de Maxwell dans le r\'{e}f\'{e}rentiel propre de la
terre devrait \^{e}tre \'{e}crite sous la deuxi\`{e}me forme
ci-dessus et la vitesse de la lumi\`{e}re, \'{e}mise par le
soleil, par rapport \`{a} ce
r\'{e}f\'{e}rentiel est alors approximativement $1-10^{-4}$.\\\\
Ainsi, on doit noter qu'il y a, pour nous, une diff\'{e}rence
r\'{e}elle (physique) entre le mouvement r\'{e}el (par rapport
\`{a} un r\'{e}f\'{e}rentiel initial) d'un rep\`{e}re inertiel et
le mouvement r\'{e}el d'un corps ou d'une particule. En effet,
lorsqu'on suppose que la particule (par exemple) a une
acc\'{e}l\'{e}ration par rapport au r\'{e}f\'{e}rentiel initial,
alors il est en acc\'{e}l\'{e}ration par rapport au rep\`{e}re
inertiel et il \'{e}met des radiations (physiques) tandis que si
l'om suppose que la particule est au repos (ou anim\'{e}e d'un
mouvement uniforme) par rapport au r\'{e}f\'{e}rentiel initial,
alors quelque soit le mouvement du rep\`{e}re mobile, il est
\'{e}vident que la particule n'\'{e}met pas de radiations.\\\\
\textbf{Remarque:} Lorsque nous affirmons que la vitesse de la
lumi\`{e}re d\'{e}pend du rep\`{e}re inertiel utilis\'{e} pour la
mesurer, ceci est entendu dans le sens suivant:\\
Lorsqu'on suppose qu'une source de lumi\`{e}re est situ\'{e}e
\`{a} une distance $d$, suppos\'{e}e constante, sur l'axe des $x$
d'un rep\`{e}re euclidien $R_1 = (O_1, e_1, e_2, e_3)$ et qu'un
autre rep\`{e}re $R_2 = (O_2, e_1, e_2, e_3)$, qui co\"{\i}ncide
avec $R_1$ \`{a} l'instant $t =0$, se d\'{e}place le long de l'axe
des $x$ avec une vitesse uniforme $v >0 $ relativement au
rep\`{e}re $R_1$ dans la direction de la source, alors nous
affirmons que la lumi\`{e}re \'{e}mise par la source \`{a} un
instant $t> 0$ atteint l'observateur situ\'{e} en $O_1$ un peu
plus tard que celui qui est situ\'{e} en $O_2$. De m\^{e}me, si
$R_1$ est suppos\'{e} fixe et si deux sources sont suppos\'{e}es
situ\'{e}es toutes les deux, \`{a} l'instant $t=0$, \`{a} une
m\^{e}me distance $d$ de $O_1$ et qui s'\'{e}loignent toutes les
deux de $O_1$ l'une \`{a} une vitesse uniforme $v_1$ et l'autre
\`{a} une vitesse uniforme $v_2 > v_1$ par rapport \`{a} $R_1$,
alors la lumi\`{e}re \'{e}mise par la source \`{a} un instant $t
>0$ atteint l'observateur situ\'{e} en $O_1$ avant celle qui est
\'{e}mise au
m\^{e}me instant par la deuxi\`{e}me source.\\

Pour illustrer la port\'{e}e de l'erreur qu'on peut commettre
lorsqu'on \'{e}crit hativement les changements de variables entre
deux rep\`{e}res galil\'{e}ens, voici un exemple extrait de
l'excellent livre sur la relativit\'{e} intitul\'{e}:
Relativit\'{e} - Fondement et applications ([7]), o\`{u} l'on
consid\`{e}re, au chapitre 1, une source lumineuse plac\'{e}e
\`{a} l'origine $O'$ d'un r\'{e}f\'{e}rentiel galil\'{e}en $R' =
O'x'y'z'$, de vitesse $v = \beta c$ (o\`{u} $\beta=\frac{v}{c}$)
par rapport \`{a} un autre r\'{e}f\'{e}rentiel galil\'{e}en $R =
Oxyz$ (voir la figure 12). En appelant $E_1$ l'\'{e}v\`{e}nement
"\'{e}mission de la lumi\`{e}re en $O'$", $E_2$
l'\'{e}v\`{e}nement "reflexion sur un miroir $M'$", situ\'{e} sur
l'axe $O'y'$ \`{a} la distance $l$ de $O'$ et $E_3$
l'\'{e}v\`{e}nement "d\'{e}tection de la lumi\`{e}re en $O'$ \`a l'instant $t=t'=0$" et
en \'{e}crivant les coordonn\'{e}es de $E_1, E_2$ et $E_3$ dans
$R'$ comme \'{e}tant respectivement
$$ E_{1_{R'}} \left\{%
\begin{array}{ll}
    x' = 0 \\
    y' = 0 \\
    z' = 0 \\
    ct' = 0 \\
\end{array}%
\right. \;\;\; E_{2_{R'}} \left\{%
\begin{array}{ll}
    x' = 0 \\
    y' = l \\
    z' = 0 \\
    ct' = l \\
\end{array}%
\right. \;\;\; E_{3_{R'}} \left\{%
\begin{array}{ll}
    x' = 0 \\
    y' = 0 \\
    z' = 0 \\
    ct' = 2l \\
\end{array}%
\right.$$ on a d\'{e}cid\'{e}, d'une fa\c{c}on simpliste, que
$ct'=l$. Or cela est le cas, comme on vient de le signaler, d'un
photon exceptionnel et ce fait ne nous autorise pas de d\'{e}crire
un changement de variable galil\'{e}en arbitraire de cette
mani\`{e}re. Dans le rep\`{e}re $R'$, la distance parcourue par un
photon arbitraire, pendant le temps $t'$, n'est pas \'{e}gal \`{a}
$l$ (i.e. $ct' \neq l$) comme on l'a d\'ej\`a d\'emontr\'e dans l'exp\'erience du train, du miroir et des deux observateurs.\\
De m\^{e}me, lorsqu'on d\'{e}cide qu'en accord avec la
transformation de Galil\'{e}e, les \'{e}v\`{e}nements $E_1, E_2$
et $E_3$ s'\'{e}crivent dans $R$ comme \'{e}tant
$$ E_{1_{R}} \left\{%
\begin{array}{ll}
    x = 0 \\
    y = 0 \\
    z = 0 \\
    ct = 0 \\
\end{array}%
\right. \;\;\; E_{2_{R}} \left\{%
\begin{array}{ll}
    x = \beta l \\
    y = l \\
    z = 0 \\
    ct = ct' = l \\
\end{array}%
\right. \;\;\; E_{3_{R}} \left\{%
\begin{array}{ll}
    x = 2 \beta l \\
    y = 0 \\
    z = 0 \\
    ct = 2l \\
\end{array}%
\right.$$ on commet la m\^{e}me erreur en consid\'{e}rant que $c t
= l$ et que $x = \beta l = v \frac{l}{c}$ puisque, en
g\'{e}n\'{e}ral, $c t' = ct \neq l$ et $\frac{l}{c} \neq t$.
Ainsi, le fait de tirer la conclusion que la distance qui
s\'{e}pare les deux \'{e}v\`{e}nements $E_1$ et $E_2$ n'est pas
invariante par changement de rep\`{e}re galil\'{e}en, en se basant
sur l'\'{e}galit\'{e} $d' = l$ (o\`u $d'$ est suppos\'ee comme \'etant la distance effectu\'ee par la lumi\`ere lorsqu'elle est mesur\'ee \`a l'aide de $\mathcal{R}'$), n'est pas l\'{e}gitime puisque la
distance travers\'{e}e par chaque photon est diff\'{e}rente des
distances travers\'{e}es par les autres photons et le rep\`{e}re
$R'$ ne peut pas \^{e}tre utilis\'{e} ad\'{e}quatement pour
d\'{e}terminer la "distance travers\'{e}e par la lumi\`{e}re".
Cette distance (mesur\'ee dans $R$) $d$ v\'erifie en fait $d = (l^2 + \beta^2 d^2)^{\frac{1}{2}}$, ce qui donne, pour $v\ll c$, $d\simeq l(1+\beta^2){\frac{1}{2}}\simeq\gamma l$.\\
De m\^{e}me, lorsqu'on utilise les transformations relativistes
$$ x = \gamma (x' + \beta c t')$$
$$y = y' \hskip 1.8cm$$
$$ z = z' \hskip 1.8cm$$
$$ ct = \gamma (ct' + \beta x')$$ o\`{u} $\gamma = (1-\beta^2)^{-\frac{1}{2}}$, on \'{e}crit les
coordonn\'{e}es des \'{e}v\`{e}nements $E_1, E_2$ et $E_3$ dans le
rep\`{e}re $R$ sous la forme
$$ E_{1_{R}} \left\{%
\begin{array}{ll}
    0 \\
    0 \\
    0 \\
    0 \\
\end{array}%
\right. \;\;\; E_{2_{R}} \left\{%
\begin{array}{ll}
   \gamma \beta l \\
    l \\
    0 \\
   \gamma l \\
\end{array}%
\right. \;\;\; E_{3_{R}} \left\{%
\begin{array}{ll}
    2 \gamma \beta l \\
    0 \\
    0 \\
    2 \gamma l \\
\end{array}%
\right.$$et on commet la m\^{e}me erreur en d\'{e}cr\'{e}tant que
$ct' = l$.\\\\
Par ailleurs, on peut ajouter \`a ce qui pr\'ec\`ede que le fait
de consid\'erer les temps n\'egatifs ($t<0$) est fondamentalement
incompatible avec la th\'eorie confirm\'ee de l'expansion de
l'Univers qui pr\'esuppose qu'il y ait un temps originel, ou une
origine du temps d\'esign\'ee par $t=0$ \`a partir duquel (ou, plus
pr\'{e}cisement juste apr\`{e}s) le temps s'\'ecoule d'une
fa\c{c}on
croissante homog\`ene ($t>0$) simultan\'ement avec le processus de l'expansion.\\

\subsection*{Interpr\'{e}tation alternative de l'exp\'{e}rience de
Michelson - Morley}

Montrons, pour finir que l'interpr\'etation g\'en\'eralement
acc\'ept\'ee de l'exp\'erience de Michelson-Morley est erronn\'ee.\\
Nous allons commencer par analyser une exp\'erience qui pourrait
servir \`a montrer la d\'eficience de l'interpr\'etation de
l'exp\'erience de l'\'emetteur au milieu d'un camion (avec les
deux miroirs sur les deux cot\'es) d'une part et qui indiquerait
la mani\`ere d'analyser d'autres exp\'eriences tel que
l'\'emetteur
dans un avion se dirigeant vers un observateur donn\'e.\\

Consid\'erons donc un \'emetteur de lumi\`ere qui se trouve au
repos par rapport \`a la terre et qui \'{e}met des rayons de
lumi\`ere parall\`element au mouvement de la terre (suppos\'e
uniforme) vers un miroir situ\'e \`a une distance $L$ de
l'\'emetteur. Soient $0<v<1$ et $c=1$ les vitesses respectives de
la terre et de la lumi\`ere relativement \`a un rep\`ere virtuel
fixe co\"incidant \`a l'instant $t=0$ avec l'\'emetteur et ayant
son axe $Ox$ parallel au mouvement. Soient $t_1$ et $t_2$ les
temps mis respectivement par un rayon (ou plus exactement par un
photon) pour atteindre le miroir et ensuite pour retourner \`a
l'\'emetteur (voir fig.4$^{''}$). On a \'evidemment, en m\'esurant
ces deux distances relativement au rep\`ere fixe:
$$L\;+\;t_1v\;=\;t_1$$
et
$$L\;-\;t_2v\;=\;t_2.$$
D'o\`u
$$t_1\;=\;\frac{L}{1-v}\qquad\mbox{et}\qquad t_2\;=\;\frac{L}{1+v}$$
et parsuite
$$t_1\;+\;t_2\;=\;\frac{2L}{1-v^2}.$$ Si on suppose maintenant,
conform\'ement \`a la deuxi\`eme partie du deuxi\`eme postulat de
la relativit\'e sp\'eciale, que la vitesse de la lumi\`ere par
rapport au rep\`ere pour lequel l'\'emetteur est au repos est
donn\'ee par $1,$ on obtient
$$t^{'}_1\;+\;t^{'}_2\;=\;2L$$
o\`u
$$t^{'}_i\;=\;\gamma(t_i\;-\;xv)\qquad\mbox{pour}\qquad
i=1,2\qquad\mbox{et}\qquad \gamma\;=\;\frac{1}{\sqrt{1-v^2}}.$$
Montrons que cette \'egalit\'e m\`ene \`a une contradiction. En
effet, on a
\begin{eqnarray*}
  2 L = t^{'}_1\;+\;t^{'}_2 &=& \gamma\left[t_1\;-\;(L+t_1v)v\right]+ \gamma [t_2 - (-t_2) v] \\
  &{}&((-t_2)\;\mbox{est la distance alg\'ebrique parcourue par la lumi\`ere}\\
  &{}&\mbox{ par rapport au r\'ef\'erentiel fix\'e})\\
  &=& \gamma(t_1+t_2-Lv-t_1v^2+t_2v) \\
  &=&
  \gamma\left(\frac{2L}{1-v^2}-Lv-\frac{L}{1-v}v^2+\frac{L}{1+v}v\right).
\end{eqnarray*}
D'o\`u
$$\gamma\left(\frac{2}{1-v^2}\;-\;v\;-\;\frac{v^2}{1-v}\;+\;\frac{v}{1+v}\right)\;=\;2$$
et par suite
$$\frac{2-v+v^3-v^2-v^3+v-v^2}{1-v^2}\;=\;2\sqrt{1-v^2}$$
ou
$$\frac{2-2v^2}{1-v^2}\;=\;2\sqrt{1-v^2}$$
ce qui donne $1=\sqrt{1-v^2}$ et parsuite $v=0;$ ce qui est absurde.\\
Par ailleurs, lorsqu'on utilise la transformation
galil\'eo-newtonienne classique on obtient
$$y(t)\;=\;x(t)\;-\;vt\;=\;t\;-\;vt\qquad\qquad\;\;\;\;\quad\quad\mbox{pour}\qquad 0<t<t_1\;\;\;\;\;\;\;$$
et (en prenant $t_1$ comme temps de base)
$$y(t)\;=x(t)-v(t-t_1)\;=\;t_1\;-\;vt_1\;-\;(t-t_1)\;-\;v(t-t_1)\qquad\mbox{pour}\qquad t_1<t<t_1+t_2,$$
o\`u $x(t)$ d\'esigne l'abscisse par rapport au rep\`ere de l'\'emetteur lorsqu'il est situ\'e \`a l'instant $t_1$ et suppos\'e comme $y$ \'etant fix\'e. D'o\`u
$$y(t_1)\;=\;t_1\;-\;vt_1$$
et
$$\qquad \qquad y(t_1+t_2)\;=\;t_1\;-\;vt_1\;-\;t_2\;-\;vt_2.$$
Or  $y(t_1+t_2)=0$  donne
$$t_1\;-\;t_2\;=\;v(t_1+t_2)$$
et
$$v\;=\;\frac{t_1-t_2}{t_1+t_2}$$
comme il se doit (la vitesse du rep\`ere en mouvement par rapport au rep\`ere fixe est le quotient de la distance par le temps).\\
De plus, on a
$$y^{'}(t)\;=\;1-v\qquad\quad\quad\mbox{pour}\qquad 0<t<t_1\qquad$$
et
$$y^{'}(t)\;=\;-(1+v)\qquad\;\;\mbox{pour}\qquad t_1<t<t_1+t_2.$$
D'o\`u
\begin{eqnarray*}
  y^{'}(t)t_1\;+\;(-y^{'}(t)t_2)&=&(1-v)t_1+(1+v)t_2 \\
  &=&L+L\;=\;2L
\end{eqnarray*}
en agr\'ement avec les notions galil\'eo-newtoniennes.\\

Maintenant, \`a la lumi\`ere de l'exp\'erience pr\'ec\'edente et
de l'exp\'erience du train et du miroir, on peut montrer que
l'interpr\'etation des exp\'eriences du genre Michelson-Morley est
erronn\'ee. En effet, supposons (pour simplifier) que l'appareil
est constitu\'{e} (comme dans [7]) de deux miroirs $M_1$ et $M_2$
et d'une lame semi-transparente $L_s$ qui divise le faisceau
lumineux en deux parties d'\'{e}gales intensit\'{e}s et que l'on
r\'{e}alise l'interf\'{e}rence des ondes issues de l'image $S_1$,
donn\'{e}e par $L_s$ et $M_1$, et de l'image $S_2$, donn\'{e}e par
$L_s$ et $M_2$ qui est l\'{e}g\`{e}rement inclin\'{e} (fig. 13).
Pour l'exp\'{e}rimentateur, la variation de l'\'{e}clairement au
point $P$, o\`{u} se trouve le d\'{e}tecteur, d\'{e}pend de la
diff\'{e}rence $\tau$ des dur\'{e}es mises par les ondes issues de
$S$ et d\'{e}tect\'{e}es en $P$. Plus pr\'{e}ci\'{e}ment, l'aspect
interf\'{e}rentiel d\'{e}pend de la diff\'{e}rence de phase $2 \pi
\nu \tau$, $\nu$ \'{e}tant la fr\'{e}quence du rayonnement
monochromatique \'{e}mis par $S$.\\
Ainsi, si $\tau_1$ d\'{e}signe la dur\'{e}e mise par la
lumi\`{e}re pour aller de $I$ \`{a} $I_1$ puis revenir en $I$, et
$\tau_2$ est la dur\'{e}e qu'elle met pour aller de $I$ \`{a}
$I_2$ et revenir en $I$, on a $\tau = \tau_2 - \tau_1$ puisque les
trajets $SI$ et $IP$ sont communs. Exprimons $\tau_1$ et $\tau_2$
en fonction de la longueur $l$ des bras de
l'interf\'{e}rom\`{e}tre ($I I_1 = I I_2 = l$), de la vitesse de
translation $\overrightarrow{v_{e}}$ du laboratoire, i.e. la
vitesse de la terre par rapport \`{a} un r\'{e}f\'{e}rentiel $R$,
qu'on peut consid\'{e}rer comme \'{e}tant fixe, et de la vitesse
de la lumi\`{e}re $v = c$ par rapport \`{a} $R$. La relation
galil\'{e}o-newtonienne entre les vitesses $\overrightarrow{v},
\overrightarrow{v_{e}}$ et $\overrightarrow{v'}$, o\`{u}
$\overrightarrow{v'}$ est la vitesse de la lumi\`{e}re par rapport
au r\'{e}f\'{e}rentiel $R'$ li\'{e} \`{a} la terre, s'\'{e}crit
$$\overrightarrow{v} = \overrightarrow{v'} +
\overrightarrow{v_e}$$ avec $v =c$.\\
Or, on a \'{e}tabli plus haut que l'on a, dans le rep\`{e}re $R$,
$$ l + t_1 v_e = ct_1 \;\;\; \mbox{ et } \;\;\; l - t_2 v_e =
ct_2,$$ce qui implique
$$ t_1 = \frac{l}{c - v_e} \;\;\; \mbox{ et } \;\;\; t_2 =
\frac{l}{c +v_e}$$et par suite
$$ \tau_2 = \frac{l}{c-v_e} + \frac{l}{c+v_e} =
\frac{2lc}{c^2-v_e^2} = \frac{2l}{c} \frac{1}{1-\beta_e^2} =
\frac{2l}{c} \gamma_e^2.$$ De m\^{e}me, en appliquant la relation
galil\'{e}o-newtonienne, on obtient, dans $R'$
$$ \tau^{'}_2 = \frac{l}{c-v_e} + \frac{l}{c+v_e} = \frac{2l}{c}
\gamma_e^2= \tau_2$$ puisque la vitesse de la lumi\`{e}re dans
$R'$ \`{a} l'all\'{e}e est $c-v_e$ et au retour est $c+v_e$.\\
D'autre part, ayant dans $R$
$$\tau_1 = \frac{2d_1}{c} \simeq \frac{2}{c} \sqrt{l^2+\frac{v_e^2l^2}{c^2}}\simeq 2\gamma_e\frac{l}{c}$$
on obtient
$$ \tau = \tau_2 - \tau_1 = \frac{2l}{c} \gamma_e^2 -\frac{2l}{c}\gamma_e.$$ 
Lorsqu'on proc\`{e}de pour
mesurer $\tau_1$ dans $R'$ en utilisant le diagramme de la figure
13, on affirme que la "vitesse de la lumi\`{e}re" dans $R'$ est
$\overrightarrow{c} - \overrightarrow{v_{e}}$; ce qui n'est autre
que la vitesse du photon qui reste continuellement au dessus de
$I$ et, comme on l'a d\'{e}j\`{a} fait remarquer, ceci n'est pas
le cas dans cette exp\'{e}rience et par suite, on ne peut pas
affirmer que
$$ || \overrightarrow{c} - \overrightarrow{v_e}|| = (c^2 + v_e^2 -
2 v_e c\; cos \theta)^{\frac{1}{2}} = (c^2
-v_e^2)^{\frac{1}{2}},$$o\`{u} $\theta$ est l'angle d\'{e}fini par
$cos \theta = \frac{v_e}{c}$. Par cons\'{e}quent, on ne peut pas
continuer le raisonnement et affirmer que
$$ \tau_1 = \frac{2l}{(c^2
-v_e^2)^{\frac{1}{2}}} = \frac{2l}{c} \gamma_e\quad\mbox{et}\quad \tau=\frac{2l}{c}\gamma_e(\gamma_e-1)\simeq \frac{l}{c}\beta^2_e$$
pour conclure que
la diff\'{e}rence de phase est
$$ \varphi = \varphi_2 - \varphi_1 = 2 \pi \nu \tau \simeq
\frac{2 \pi}{\lambda_0} l \beta_e^2$$par contre, on peut affirmer
correctement que, mesur\'{e} dans $R$, on a
$$ \tau = \frac{2l}{c} \gamma_e^2 - \frac{2l}{c}\gamma_e.$$
Ainsi, en faisant tourner l'appareil de
$90^\circ$, les r\^{o}les jou\'{e}s par les miroirs $M_1$ et $M_2$
sont invers\'{e}s de sorte que $|\tau| = |\tau_1 - \tau_2 |$ reste
invariant. Par cons\'{e}quent, dans les deux cas, la
diff\'{e}rence entre les deux distances parcourues par les deux
faisceaux reste la m\^{e}me; ce qui explique l'invariance du
sc\'{e}ma d'interf\'{e}rence.\\\\
\subsection*{Remarque:} Bien que le temps propre $\tau$ joue un
r\^{o}le primordial dans la th\'{e}orie de la relativit\'{e}, il
n'est, dans le cadre de notre mod\`{e}le, qu'un param\`{e}tre qui
est naturellement reli\'{e} au temps universel $t$ par
l'interm\'{e}diare de la relation de changement de variable:
$\frac{d \tau}{dt} = \frac{1}{\gamma}$, o\`{u} $\gamma =
\frac{1}{\sqrt{1-\frac{v^2}{c^2}}}$ est le facteur de Lorentz qui
est apparu dans les travaux de W. Kaufmann comme \'{e}tant un
facteur de proportionalit\'{e} de telle sorte que la quantit\'{e}
$\gamma m_0$ (o\`{u} $m_0$ est la masse au repos d'une particule
mat\'{e}rielle) f\^{u}t appel\'{e} par lui comme \'{e}tant la
masse apparente de la particule en mouvement. Ce facteur est
apparu \'{e}galement dans les calculs de distance ci-dessus et il
va jouer un r\^{o}le important, comme au sein de la
relativit\'{e}, dans le cadre de notre th\'{e}orie et notamment
pour d\'{e}finir l'\'{e}nergie globale et la quantit\'{e} de
mouvement des particules en mouvement plus ou moins rapide.\\

\subsection*{G\'{e}om\'{e}trisation de l'Univers physique}

A la lumi\`ere de tout ce qui pr\'ec\`ede, on a propos\'e dans cet
article de conserver la conception pr\'erelativiste
(Galil\'eo-newtonienne) de l'espace-temps et de consid\'erer le
demi-c\^one $C= \{(x,y,z,t) \in \mathbb{R}^4;\; x^2+y^2+z^2 \leq
t^2\;\mbox{et}\;t\geq0 \}$ du monde des trajectoires dans $\mathbb{R}^4$ avec des
restrictions impos\'ees sur les trajectoires r\'eelles
garantissant la causalit\'{e}. Ainsi, d'apr\`es ce qui
pr\'ec\`ede, on peut d\'eduire que toutes les lois de la
M\'ecanique et de la Physique se reposant sur des relations
impliquant les vecteurs vitesses, les vecteurs acc\'el\'erations
et des champs de vecteurs quelconques (ou plus g\'en\'eralement
des champs de tenseurs) sont invariantes par les changements de
rep\`eres effectu\'es dans l'espace ($\mathbb{R}^3, g_{t_0}$)
impliquant des vitesses relatives constantes et des isom\'etries
fix\'ees ($a'_0(t) = \vec {V_1}, b'_0(t) = \vec{V_2}, A, B \in
O(g_{t_0})$). On peut citer, \`{a} titre d'exemple : les lois
fondamentales de la M\'ecanique, l'\'equation de Maxwell, le
principe de la conservation de
l'\'en\'ergie, le principe de la moindre action \ldots .\\
On a montr\'e aussi, dans le cas g\'en\'eral d'une m\'etrique
variable ( \`a courbure variable et d\'ependante du temps) $g_t$ ,
certaines propri\'et\'es de tensorialit\'e des changements de
rep\`eres et la covariance des lois de la Physique dans certains
d'entre eux, notamment ceux qui sont isom\'etriques  (relativement
\`a $g_t$ ) incluant toutes les isom\'etries et non seulement un
sous-groupe de leur groupe total.
\\ \\
On se propose dans la suite d'aller plus loin dans la direction de
la r\'ealit\'e physique de l'Univers en tenant compte des
ph\'enom\`enes gravitationnels et des diff\'erentes manifestations
de la mati\`ere comme la masse, l'\'electromagn\'etisme et les
diff\'erentes formes de l'\'energie ( \`a l'exception des
ph\'enom\`enes quantiques et des singularit\'es \'energ\'etiques
bien qu'ils sont, dans notre
nouveau contexte, beaucoup plus faciles \`a traiter.).\\ \\
Pour cela supposons, dans un premier temps, qu'une particule de
masse inertielle (quasiment) constante $m_I$, ne portant aucune
charge, est en chute libre dans un champ gravitationnel uniforme (
localement dans ($\mathbb{R}^3, g_e$)) d\'{e}fini par $ \vec{g} =
(0,0,-g)$ dans un
rep\`ere euclidien fixe.\\
Lorsque cette particule est rep\'er\'ee \`a l'aide d'un rep\`ere
euclidien ($0,e_1,e_2,e_3$)qui suit lui-m\^eme la trajectoire d'un
mouvement similaire de chute libre tel que $ e_3=(0,0,1) $, alors la
particule para\^{i}t (comme l'a montr\'e Einstein) comme si elle
\'etait au repos dans ce nouveau rep\`ere. De m\^eme si cette
particule avait une vitesse initiale (horizontale) $\vec{V_0} $
(dans le rep\`{e}re fixe) pour $ t=t_0$, alors son mouvement
para\^{i}t dans le rep\`ere mobile comme s'il \'etait uniforme
tandis-qu'il est d'allure parabolique dans le rep\`ere fixe. Le
rep\`ere mobile \`a acc\'el\'eration uniforme
$\vec{\Gamma}=\vec{g}$ dans le rep\`ere fixe joue par cons\'equent
le r\^ole d'un rep\`ere inertiel pour les deux lois d'inertie de
Newton. Cette m\^eme propri\'et\'e est valable pour tout rep\`ere
effectuant un mouvement de chute libre semblable \`a notre
rep\`ere mobile. Appelons un tel rep\`ere $ \vec{g}$-inertiel. Il
existe (localement dans $\mathbb{R}^3$) un tenseur m\'etrique
dynamique $g_{ab}$ tel que $\bigtriangledown^{g_{ab}}_{x'(t)}
x'(t)=0$ pour tout mouvement de chute libre d\'ecrit par $x(t)$
dans le rep\`ere fixe, c.\`a.d. tel que les trajectoires de la
chute libre (dans ce syst\`eme isol\'e) soient des g\'eod\'esiques
pour cette m\'etrique. En effet, gr\^ace aux sym\'etries
suppos\'ees de notre syst\`eme (et en utilisant les cordonn\'ees
du rep\`ere fixe), on peut supposer que l'on a:
$$ g_{ij}=dx_{1}^2+dx_{2}^2+b(t,x_3,g)dx_{3}^2 $$
avec $b(t,x_3,g)$ d\'ecroissant en fonction de $t$ et de $g$. On a
alors$$||x||^{2}_{g_t(x)}=x_1^2+x_2^2+b(t,x_3,g)x_{3}^2$$et si on
prend $x=x_3e_3$, on a$$||x||^{2}_{g_t(x)}=b(t,x_3,g)x_{3}^2.$$Le
mouvement de chute libre \'etant d\'ecrit
par$$x_3(t)=a_3-\frac{1}{2}gt^2\;\mbox{ avec
}\;x_3(0)=a_3,\;x_3(t_0)=0\;\mbox{ et
}\;a_3=\frac{1}{2}gt_0^2,$$on a, pour $g>0$:
$$t=\sqrt{\frac{2}{g}(a_3-x_3(t))}\;\mbox{ et }\;x'_3(t)=-gt.$$On
a alors les \'equivalences suivantes:
$$\nabla_{x'_3(t)e_3}^{g_t}x'_3(t)e_3 =0\Leftrightarrow g^2t\nabla_{e_3}^{g_t}te_3=0\Leftrightarrow
$$$$g^2t^2\nabla_{e_3}^{g_t}e_3+g^2t\frac{d}{dx_3}\sqrt{\frac{2}{g}(a_3-x_3(t))}e_3=0\Leftrightarrow$$
$$t\frac{\frac{d}{dx_3}b(t,x_3,g)}{b(t,x_3,g)}=-\frac{d}{dx_3}\sqrt{\frac{2}{g}(a_3-x_3(t))}\Leftrightarrow$$
$$\frac{\frac{d}{dx_3}b(t,x_3,g)}{b(t,x_3,g)}=-
\frac{\frac{d}{dx_3}\sqrt{\frac{2}{g}(a_3-x_3(t))}}{\sqrt{\frac{2}{g}(a_3-x_3(t))}}\Leftrightarrow$$
$$b(t,x_3,g)=\frac{k(t)}{\sqrt{\frac{2}{g}(a_3-x_3(t))}}=\frac{k(t)}{t}.$$
Pour \emph{x}(\emph{t}) = $x_3$(\emph{t})$e_3$, on a
$$||x_3(t) e_3||^2_{g_{ij}(t)} = b(t,x_3,g)x_3(t)^2 = x_3(t)^2
||e_3||_{g_{ij}(t)}$$et par suite
$$b(t,x_3,g) = ||e_3||_{g_{ij}(t)} = \frac{k(t)}{t}$$Or,
$$ ||x^{'}(t)||_{g_{ij}(t)} = ||x^{'}_3(t) e_3||_{g_{ij}(t)} = ||g t e_3||_{g_{ij}(t)} = g t
||e_3||_{g_{ij}(t)} = c(t)$$D'o\`{u}
$$||e_3||_{g_{ij}(t)} = \frac{c(t)}{g t}
= b(t,x_3,g).$$ Par cons\'{e}quent, on obtient
$$g_{ij}(t) = dx_1^2 + dx_2^2 + \frac{c(t)}{g t}dx_3^2 = dx_1^2 +
dx_2^2 + \frac{c(t)}{\sqrt{2g(a_3 - x_3(t))}}dx_3^2$$ Cette
m\'{e}trique d\'{e}pend \'{e}videmment du niveau initial choisi
$x_3(0)$ = $a_3$.\\
Dans le cas o\`u le champ de gravitation $\vec{g}$ est central et
de norme (euclidienne) constante $g$ et que son centre $C$ est
fixe, alors on peut int\'egrer (localement) cette gravitation dans
une m\'etrique $g_{ab}(t)$ \`a l'aide de sa connexion canonique $
\bigtriangledown^{g_{ab}}$ en d\'efinissant les g\'eod\'esiques
issues de $C$ par $ \bigtriangledown^{g_{ab}}_{x'(t)} x'(t)=0$
o\`u $x(t)$ d\'esigne les coordonn\'ees de ces courbes dans un
rep\`ere euclidien d'origine $C$ ou tout autre rep\`ere fixe.\\
Gr\^ace \`a la sym\'etrie (principe d'homog\'en\'eit\'e et
d'isotropie locales), on pourrait d\'eterminer la m\'etrique
$g_{ab}(t)$ en utilisant les coordonn\'ees normales riemanniennes
autour de $C$ et en les transformant ensuite en coordonn\'ees
sph\'eriques pour obtenir:
$$g_{ab}(t)=b(t,r,g)dr^2+r^2d\sigma ^2$$avec $b(t,r,g)$ d\'ecroissant
avec $t$ et $g$. Ainsi les distances au centre et les volumes sont
"inversement proportionnels" \`a l'intensit\'e de la gravitation.\\
Lorsqu'on suppose que le centre de gravitation $C(t)$ est mobile,
la chute libre ne se produit pas en ligne droite dirig\'ee vers le
centre, mais suivant une courbe $x(t)$ dont la tangente est, \`a
tout instant $t$, point\'ee vers $C(t)$. Cependant, un champ de
gravitation central n'est jamais uniforme; il d\'{e}pend de la
distance au centre (et par suite du temps de param\'{e}trisation
des trajectoires libres). Si $C(t)$ est mobile, alors ce champ est
radialement constant relativement \`{a} un rep\`{e}re centr\'{e}
en $C(t)$. N\'{e}anmoins, si $C$ est fixe, alors, pour des objets
assez proches les uns des autres et suffisamment loins du centre,
on peut consid\'{e}rer le champ gravitationnel comme \'{e}tant
uniforme. Si l'on suppose maintenant que $C'(t) \ll 1$ et $ g_t
\ll 1$, on retrouve approximativement la m\'etrique euclidiennne.
C'est le cas, par exemple, lorsqu'on se trouve (localement) \`a
une distance raisonnable de la surface de la terre. Mais cela
n'est pas, en g\'en\'eral, la situation qui correspond \`a la
r\'ealit\'e physique. En effet, bien que l'on puisse imaginer un
rep\`ere inertiel tel que la vitesse relative d'un astre donn\'e,
par exemple, est suffisemment petite, le vecteur $\vec{g}$ (ou
plut\^{o}t $\vec{g_t}$ ) d\'epend fortement de la distance
euclidienne du corps en mouvement au centre de gravitation. Mais,
dans un syst\`eme suppos\'e isol\'e, on peut d'une facon empirique
d\'eterminer les g\'eod\'esiques (trajectoires d'objets en chute
libre) et \'etudier leurs vecteurs vitesse et leurs vecteurs
acc\'el\'eration relatives (en d\'eterminant leurs d\'eviations
relatives) et en d\'eduire les symboles de Christoffel associ\'es
\`a la m\'etrique $(g_t)_{ab} $ et son tenseur de courbure $
R^{a}_{bcd} (t) $ \`a l'aide des \'equations de d\'eviation
infinit\'esimale des g\'eod\'esiques (R.Wald:3.3.18), tout en
tenant compte de la relation
$\bigtriangledown^{(g_t)_{ab}}(g_t)_{ab}=0$ ainsi que
des sym\'etries possibles.\\ \\

\subsection*{La m\'{e}trique physique}

Ces m\^emes consid\'erations doivent \^etre prises en compte
lorsqu'il s'agit de d\'eterminer (m\^eme localement) la m\'etrique
globale $g_{ab} (t,x)$ de l'Univers en expansion
$U(t)=(B(O,t),g_{ab}(t,x))$ en tenant compte des autres
ph\'enom\`enes (\'electro-magn\'etiques, thermo-nucl\'eaires,
radio-actives, quantiques, singularit\'es \'energ\'etiques et
autres) qui doivent \^etre int\'egr\'es dans la m\'etrique. Ceci
nous conduit \`a un type d'\'equation d'Einstein dont la
r\'esolution conduit \`a une m\'etrique approximative
$g_{ab}(t,x)$  pouvant d\'ecrire les trajectoires des mouvements
libres i.e. les g\'eod\'esiques de l'Univers dynamique
($\nabla_{x'(t)}^{g_{ab}}x'(t)$=0). La m\'{e}trique $g_{ab}$(\emph{t},\emph{x}) sera dite la m\'{e}trique physique.\\ \\
En effet, signalons pour commencer, qu'on va noter (conform\'ement
au principe d'\'equivalence faible) $m_I=m_g$ la masse inertielle
ou gravitationnelle d'un corps lorsqu'il est situ\'e dans un champ
gravitationnel g\'en\'eral d'intensit\'e $\vec{g_t}$ induit
localement par une m\'etrique $(g_t)_{ab}$. Dans ces conditions on
a $\vec{g_t} =\vec{ \Gamma_t}=\bigtriangledown^{g_e}_{x'(t)}
x'(t)=x''(t)$ pour une trajectoire $x(t)$ dans un rep\`ere
euclidien virtuel fixe quelconque avec
$\bigtriangledown^{(g_t)_{ab}}_{x'(t)} x'(t)=0$ (comme on va le
voir dans la cinqui\`eme partie de cet article). Ainsi la masse
$\vec{g_t}$-inertielle $m_I(t)$ d\'epend du temps $t$ par
l'interm\'ediaire de la m\'etrique $(g_t)_{ab} $ qui refl\`ete
toutes les formes d'\'{e}nergie et toute sorte de fluctuations \'energ\'etiques.\\ \\
Si ${\cal {R}}_0$ d\'esigne un rep\`ere euclidien fixe, ${\cal
{R}}_1$ d\'esigne un autre rep\`ere identique \`a ${\cal {R}}_0$
dont l'origine d\'ecrit la courbe $a(t)$ (par rapport \`a ${\cal
{R}}_0$),
posons$$\widetilde{\Gamma}_0(x(t))=\nabla_{x'(t)}^{g_t}x'(t)$$
et$$\widetilde{\Gamma}_{01}(x(t))=\nabla_{x'(t)}^{g_t}x'(t)-
\nabla_{a'(t)}^{g_t}a'(t)=\widetilde{\Gamma}_0(x(t))-\widetilde{\Gamma}_0(a(t)).$$
$\widetilde{\Gamma}_{01}(t) := \widetilde{\Gamma}_{01}(x(t))$
d\'esigne l'acc\'el\'eration dynamique de la trajectoire $x(t)$
dans le rep\`ere ${\cal {R}}_1$ qui a lui-m\^eme sa propre
acc\'el\'eration dynamique $\widetilde{\Gamma}_0(a(t))$ =:
$\Gamma_0(t)$ et on a
$$\widetilde{\Gamma}_{01}(t)=\widetilde{\Gamma}_{0}(x(t)) \mbox{si et
seulement si le rep\`ere} {\cal {R}}_1 \mbox{est
}g_t-\mbox{inertiel (i.e. }\nabla_{a'(t)}^{g_t}a'(t)=0).$$ Dans
ces conditions on a les
\'equivalences suivantes:\\
$x(t)$ d\'ecrit la trajectoire d'un mouvement libre
$\Leftrightarrow$ $x(t)$ est une courbe g\'eod\'esique pour la
m\'etrique dynamique $g_t \Leftrightarrow
\widetilde{\Gamma}_{0}(x(t))=0$\\et on a alors:
$$ \widetilde{\Gamma}_{01}(t)=0 \mbox{ si et seulement si }a(t) \mbox{ est une
g\'eod\'esique}.$$ Ceci g\'en\'eralise le premier principe
d'inertie de Newton qui stipule qu'un mouvement est uniforme dans
un rep\`ere fixe si, et seulement si, il est uniforme dans tout
autre rep\`ere inertiel. Remarquons que
$\nabla_{x'(t)-a'(t)}^{g_t}x'(t)-a'(t)$ n'est pas n\'ecessairement
nulle lorsque $\nabla_{x'(t)}^{g_t}x'(t)=0$ et
$\nabla_{a'(t)}^{g_t}a'(t)=0$ contrairement \`a la propri\'et\'e
$\nabla_{x'(t)-a'(t)}^{g_e}x'(t)-a'(t)=0$ lorsque
$x''(t)=a''(t)=0.$\\\\
A la lumi\`ere de la r\'ealit\'e physique de l'Univers, du
principe fondamental de Mach repris par Einstein
(Mati\`ere=Courbure) et du principe g\'en\'eral de la
mod\'elisation (Physique=G\'eom\'etrie), il faut tenir compte des
autres aspects de la mati\`ere-\'energie. Cela nous conduit au
(0,2)-tenseur sym\'etrique simplifi\'e (d\'efini  sur
$U(t)=(B(O,t) \subset \mathbb{R}^3, g_{ab}(t,x))$ et modifi\'e
(dans un sens qu'on va pr\'eciser ci-dessous) de la
mati\`ere-\'{e}nergie $T_{ab}^{*}(t)$ dont les
\'el\'ements variables \`a d\'eterminer sont au nombre de $6$. \\
\\
Consid\'erons donc l'Univers physique identifi\'e \`a chaque
instant $t > 0$ \`a $U(t)=(B(O,t) , g_{ab}(t,x))$ o\`u $B(O,t)$
est la boule de rayon euclidien $t$ et  $g_{ab}(t)$ est la
m\'etrique riemannienne r\'{e}gularis\'{e}e associ\'ee \`a
l'Univers \`a l'instant $t$. Cette m\'etrique est variable avec le
temps et la position et elle est \`a courbure g\'en\'eralement
positive, elle aussi variant avec le temps et la position.\\
Cette courbure est d\^ue \`a la distribution des masses-\'energies
\`a chaque instant et par suite elle est d\^ue essentiellement \`a
la globalit\'e des champs gravitationnels et de toutes les forces
d'interaction.\\
Le demi-c\^one de l'espace-temps $C$ est form\'e de toutes les
sections $t=cte$ ($t \geq 0$) qui sont des boules de $\mathbb{R}^3
$ - hypersurfaces de $C \subset \mathbb{R}^4$ (\`{a} l'exception
de \emph{t} = 0), orthogonales \`a l'axe du temps pour la
m\'etrique de Minkowski sur $\mathbb{R}^4$. Chacune de ces boules
$B(I,t_0)$, de centre $(0,0,0,t_0)$ et de rayon euclidien $t_0$,
est munie de la m\'etrique riemanienne $g_{t_0}(x)$ qui d\'epend
de la position $x$ et qui est \`a courbure variable. L'\'equation
d'Einstein s'\'ecrit, en g\'en\'eral, sous la forme:
$$ G_{ab}(t):= R_{ab}(t)- \frac {1}{2}R(t) g_{ab}(t)=  T_{ab}^{*}(t)\hskip 2cm\mbox{    ($\cal{E}$)   }$$
o\`u $R_{ab}(t)$ d\'esigne le (0,2)-tenseur de Ricci associ\'e \`a
la m\'etrique $g_{ab}(t) $ dans $U(t) \subset \mathbb{R}^3$,
$R(t)$ est la courbure scalaire de  $U(t)$. Le tenseur d'Einstein
modifi\'e $T_{ab}^{*}(t)$ d\'epend naturellement de la densit\'e
$\rho (t)$ de la distribution de la masse-\'energie, du champ
gravitationnel ambiant, de la pression $P(t)$ r\'esultant des
ph\'enom\`enes typiquement associ\'es \`a des fluides parfaits, du
champ \'electrique global $ \vec{E(t)}$ et du champ magn\'etique
global $\vec{B(t)}$ et des autres manifestations \'energ\'etiques.
Tous ces objets tensoriels varient avec le temps et la position
suivant que l'espace soit, localement et \`a un instant donn\'e,
\`a dominante mat\'erielle ou radiationnelle. Notons que l'on a:
$$\bigtriangledown^a T_{ab}^{*}(t)=0,$$(ainsi que
$\bigtriangledown^a G_{ab}(t)=0$ d'apr\`es la deuxi\`eme
identit\'e de Bianchi) o\`u $\bigtriangledown=\bigtriangledown
(g_{ab}(t))$ est  la connexion de Levi-Civita associ\'ee \`a la
m\'etrique riemannienne $g_{ab}(t)$, assurant ainsi la validit\'e
de la loi de la conservation de l'\'energie et que, par
construction, les trajectoires des corps soumis uniquement aux
forces naturelles, des astres et des galaxies sont
substanciellement et globalement des g\'eod\'esiques pour la
m\'etrique $g_{ab}(t)$ (sur le plan th\'eorique). La tache
cruciale qui reste est de r\'esoudre (localement) cette \'equation
en se basant sur une banque de donn\'ees dynamiques aussi
pr\'ecises que
possible. Cette r\'{e}solution va \^{e}tre effectu\'{e}e au paragraphe 10. \\\\
Remarquons que le probl\`eme des trois corps (ou plus
g\'en\'eralement des $n$ corps), qu'ils forment ou non un
syst\`eme isol\'e, doit \^etre trait\'e dans ce contexte. Si
$(g_t)_{ab} $ est la m\'etrique globale ambiante et si $x_i(t)
(i=1,2,3)$ d\'ecrivent les trajectoires de ces trois corps dans un
rep\`ere virtuel fixe quelconque, alors on a:$$
\bigtriangledown^{(g_t)_{ab}}_{x_i'(t)} x_i'(t)=0 \quad \mbox{pour
} i=1,2,3.$$ Si l'on suppose que ces trois corps forment un
syst\`eme isol\'e, alors on peut r\'eciproquement utiliser ces
\'equations avec toutes les donn\'ees du probl\`eme pour retrouver
la m\'etrique $(g_t)_{ab}$.\\\\
Plus g\'en\'eralement, on peut d\'eterminer localement la
m\'etrique $(g_t)_{ab}$ d'une fa\c con empirique: Si $X(t)$
d\'esigne la trajectoire d'une particule ou d'un corps mobile sous
la seule action des forces naturelles (i.e. des forces qui ne sont
pas issues de ph\'{e}nom\`{e}nes \'{e}nerg\'{e}tiques
singuli\`{e}res; un tel mouvement sera qualifi\'{e} de mouvement
libre)dans un rep\`ere fixe, alors on a $\nabla
_{X'(t)}^{g_t}X'(t)=0$ \`a tout instant $t$. Ainsi la
d\'etermination empirique d'un certain nombre de g\'eod\'esiques
$X(t)$ permet de d\'eterminer approximativement les symboles de
Christoffel et la m\'etrique $g_t$ en tout point de $X(t)$. Ainsi
la m\'etrique $g_t$ est d\'etermin\'ee soit \`a l'aide de ses
g\'eod\'esiques soit en d\'eterminant le tenseur $T_{ab}^{*}$ et
en r\'esolvant l'\'equation d'Einstein simplifi\'ee. Signalons que
lorsque $T^{*}_{ab}=0$ sur une r\'egion $D\subset B(O,t)$, cela
signifie d'apr\`{e}s la d\'{e}finition m\^{e}me de ce tenseur, que
la r\'{e}gion \emph{D} est non seulement d\'{e}pourvue de
mati\`{e}re mais aussi qu'elle est en dehors de toutes les
influences \'{e}nerg\'{e}tiques de la mati\`{e}re et par
cons\'{e}quent on a $g_{ab}$ = $g_e$ sur \emph{D} ainsi que
$R_{ab}$ = 0 et $R$ = 0. Dans les cas particuliers (Fluide
parfait, champ \'electro-magn\'etique et champ scalaire de
Klein-Gordon), les cas asymptotiques et les cas quasi-newtoniens,
la r\'esolution de l'\'equation ($\cal{E}$) est beaucoup plus
facile dans le cadre de notre mod\`ele que dans celui de la
relativit\'e g\'en\'erale standard. On reviendra sur ces sujets
dans la
suite.\\\\
\textbf{Remarque:} Notre tenseur $T^*_{ab}$ et notre m\'{e}trique
physique $g_t$ qui lui est associ\'{e}e int\`{e}grent dans leur
d\'{e}finition toutes les formes de la mati\`{e}re-\'{e}nergie
ainsi que tous leurs effets contrairement au tenseur d'Einstein et
de la m\'{e}trique de l'espace-temps qui lui est associ\'{e}e qui
refl\`{e}te le champ gravitationnel caus\'{e} par une masse
donn\'{e}e en pr\'{e}sence d'un champ de mati\`{e}re (du genre
fluide parfait avec ou sans pression) et \'{e}ventuellement d'un
champ \'{e}lectromagn\'{e}tique ou en leur absence. L'\'{e}quation
d'Einstein dans le vide
$$ R_{ab} - \frac{1}{2} R \; g_{ab} = 0$$ est caract\'{e}ris\'{e}e
par $\rho$ = 0, $T_{ab}$ = 0, $R_{ab}$ = 0 et $R$ = 0 n'implique
pas que la m\'{e}trique de l'espace-temps se r\'{e}duit \`{a} la
m\'{e}trique plate. Par ailleurs, lorsqu'on introduit la constante
cosmologique $\Lambda$, l'\'{e}quation d'Einstein dans le vide
devient
$$ R_{ab} = \frac{1}{2} R \; g_{ab} - \Lambda \; g_{ab}$$et on a
alors $R = 4 \Lambda$ et si $\Lambda \neq$ 0, alors $R \neq$ 0 et
les g\'{e}od\'{e}siques dans le vide d'Einstein ne sont pas celles
de l'espace-temps plat. Ainsi, \`{a} la lumi\`{e}re de notre mod\`{e}le,
$\Lambda$ refl\`{e}te l'influence de la mati\`{e}re cosmique
l\`{a} o\`{u} il n'y a pas de mati\`{e}re (bien qu'il y ait de la
gravit\'{e} intergalactique, des radiations cosmiques et des
neutrinos) et en toute logique elle d\'{e}pendrait du temps et
probablement des r\'{e}gions dans l'espace. Pour nous, une
r\'{e}gion de l'espace est dans le vide absolu (i.e. $T^*_{ab}$ =
0) si, et seulement si, cette r\'{e}gion est (quasiment) \`{a}
l'abri de toutes les manifestations de la mati\`{e}re ainsi que de
leurs influences.\\\\
Disons un dernier mot sur l'horizon des particules ou la partie de
l'Univers \`a partir de laquelle un observateur
isotropique (se d\'epla\c{c}ant dans le sens de l'expansion \`{a} la vitesse de la lumi\`{e}re approximativement) peut recevoir un emetteur de lumi\`ere qui y est situ\'e \`a un instant donn\'e :\\
Si $P_0$ est un tel observateur, alors l'horizon des particules
pour lui est la r\'eunion de toutes les boules de centre $P_t$
situ\'e sur la ligne isotropique $(OP_0)$, par suite cet horizon
n'est autre que le demi-espace situ\'e au del\`a du plan
perpendiculaire en $P_0$ \`a cette ligne (voir figure 6).
\'{E}videmment cet horizon est purement th\'{e}orique \`{a} cause
des singularit\'{e}s (essentiellement les trous noirs) qui se
trouvent dans l'Univers physique r\'{e}el. Un observateur
ordinaire ne peut voir qu'une petite partie de l'Univers en
expansion.\\\\
\subsection*{ Comparaison avec l'invariance relativiste de la
vitesse de la lumi\`{e}re}

Dans le cadre de la relativit\'{e} restreinte, la vitesse de la
lumi\`{e}re est ind\'{e}pendante de l'observateur galil\'{e}en qui
est g\'{e}n\'{e}ralement repr\'{e}sent\'{e}, dans l'espace-temps
relativiste ($\mathbb{R}^4,\eta$) (o\`{u} $\eta$ est la
m\'{e}trique de Minkovsky $-dt^2+dx_1^2+dx_2^2+dx_3^2$), par une
droite de type temps $D$. En consid\'{e}rant une
param\'{e}trisation normale $c: I \subset \mathbb{R}
\longrightarrow \mathbb{R}^4$ de $D$ (i.e. $\eta (c^{'}(t),
c^{'}(t))$ = -1) et une param\'{e}trisation quelconque
$\widetilde{c}: J \subset \mathbb{R} \longrightarrow \mathbb{R}^4$
d'un autre observateur galil\'{e}en $\widetilde{D}$, on \'{e}crit
$$ {\widetilde{c}}\;^{'}(t) = \overrightarrow{k} + \alpha c^{'}(t)$$
pour $\overrightarrow{k} \in c^{'}(t)^{\perp}$ et $\alpha \in
\mathbb{R}$, o\`{u} $c^{'}(t)^{\perp}$ est l'hyperplan orthogonal,
pour la m\'{e}trique $\eta$, du vecteur $c^{'}(t)$, qui n'est
autre que l'ensemble des points simultan\'{e}s \`{a} $c(t) \in D$
\`{a} l'instant $t$. On d\'{e}finit alors le vecteur vitesse
relative de l'observateur $\widetilde{D}$ par rapport \`{a} $D$
par
$$ {\overrightarrow{v}}_{\widetilde{D}/D} =
\frac{\overrightarrow{k}}{\alpha}$$ et si $\widetilde{c}$ est une
param\'{e}trisation normale, par rapport \`{a} $D$, de
$\widetilde{D}$ alors on a $\alpha$ =1 et
${\overrightarrow{v}}_{\widetilde{D}/D} =
\overrightarrow{k}$.\\
On d\'{e}montre alors que la vitesse relative de la lumi\`{e}re
par rapport \`{a} l'observateur galil\'{e}en arbitraire $D$ est
$c$=1. Ceci est \'{e}tabli en consid\'{e}rant une droite de type
lumi\`{e}re $L$ param\'{e}tr\'{e}e par $\widetilde{c}(t) = M + tl$
o\`{u} $\eta(l,l)$ =0 et $M$ est un point quelconque du c\^{o}ne
de type temps et en prenant $A = c(t) \in D$ et $B =
\widetilde{c}(t)$ le point de $L$ simultan\'{e} \`{a} $c(t)$ pour
$D$ et en \'{e}crivant
$$ {\widetilde{c}}\;^{'}(t) = l = \overrightarrow{k} + \alpha
c^{'}(t)$$o\`{u} $\alpha \in \mathbb{R}$ qu'on peut supposer
positif et $\overrightarrow{k} \in c^{'}(t)^{\perp}$ et finalement
en consid\'{e}rant le vecteur vitesse relative comme \'{e}tant
$$ \overrightarrow{v}_{L/D} =
\frac{\overrightarrow{k}}{\alpha}$$qui est cens\'{e}
repr\'{e}senter le vecteur vitesse de la lumi\`{e}re par rapport
\`{a} l'observateur $D$.\\
Ainsi la relation $\eta(l,l) =
\eta(\overrightarrow{k},\overrightarrow{k}) + \alpha^2
\eta(c^{'}(t),c^{'}(t)) =
\eta(\overrightarrow{k},\overrightarrow{k}) - \alpha^2 = 0$
implique
$$\alpha = \sqrt{\eta(\overrightarrow{k},\overrightarrow{k})} \;\;\;\; \mbox{
et} \;\;\;\; \parallel \overrightarrow{v}_{L/D} \parallel_\eta =
\sqrt{\eta
(\frac{\overrightarrow{k}}{\alpha},\frac{\overrightarrow{k}}{\alpha})}
=
\frac{\sqrt{\eta(\overrightarrow{k},\overrightarrow{k})}}{\alpha}
= 1.$$

Pour nous, ce raisonnement es valable uniquement dans le cas d'un
observateur stationnaire i.e. lorsque $D$ se confond avec l'axe du
temps o\`{u} l'on a alors $c(t) = N + t(1,0,0,0) \hskip 0.1cm$,
$\hskip 0.1cm$ $c^{'}(t) = (1,0,0,0) \hskip 0.1cm$, $\hskip 0.1cm$
$c^{'}(t) ^{\perp}$ est un hyperplan parall\`{e}le \`{a}
$(O,x_1,x_2,x_3)$ pour $x_1,x_2,x_3 \in \mathbb{R}$ et
$\widetilde{c}(t) = M +
t(1,\frac{1}{\sqrt{3}},\frac{1}{\sqrt{3}},\frac{1}{\sqrt{3}})$
v\'{e}rifiant $\eta
({\widetilde{c}}\;^{'}(t),{\widetilde{c}}\;^{'}(t)) =0$.\\
Lorsqu'on prend un observateur galil\'{e}en quelconque, $c^{'}(t)
^{\perp}$ ne sera plus n\'{e}cessairement, dans le cadre
relativiste, l'hyperplan - espace consid\'{e}r\'{e}\\
pr\'{e}c\'{e}demment.\\
La relation \'{e}tablie ci-dessus pour deux observateurs
galil\'{e}ens $D$ et $\widetilde{D}$ qui a conduit \`{a} la
d\'{e}finition de la vitesse relative $\overrightarrow{v}_{
\widetilde{D}/D } = \overrightarrow{k}$ dans le cadre relativiste
n'a pour nous aucune signification physique. Cette d\'{e}finition,
ainsi que la notion relativiste de l'espace - temps, a \'{e}t\'{e}
con\c{c}u uniquement pour justifier l'aspect erronn\'{e} du
deuxi\`{e}me postulat de la relativit\'{e} restreinte qui stipule
que la vitesse de la lumi\`{e}re est la m\^{e}me pour tous les
observateurs galil\'{e}ens. Cet aspect de ce m\^{e}me postulat a
\'{e}t\'{e} adopt\'{e} par Einstein pour se conformer avec le
principe relativiste de Galil\'{e} qui stipule que toutes les lois
de la Physique (et en particulier, les lois de Maxwell) devraient
\^{e}tre les m\^{e}mes pour tous les observateurs inertiels. Or,
nous venons de prouver que ce coup de force est compl\`{e}tement
inutile en donnant un contenu physique \`{a} la notion de
canonicit\'{e} de
l'\'{e}quation de Maxwell.\\

Pour nous (comme on va le montrer au paragraphe 5), l'Univers
physique r\'{e}el n'est autre qu'une partie de $\mathbb{R}^3$ qui
est mod\'{e}lis\'{e}e, \`{a} chaque instant $t$, par la boule
euclidienne $B(O,R(t))$ de $\mathbb{R}^3$ munie de la m\'{e}trique
physique $g_t$ (qui refl\`{e}te la consistance physique de
l'Univers) et, dans ces conditions, la mesure de la vitesse de
tout observateur ou trajectoire dans $B(O,R(t))$,
param\'{e}tr\'{e}e dans le semi-c\^{o}ne de l'espace - temps
$$C = \{(x,y,z,t); x^2+y^2+z^2 \leq R^2(t) \sim t^2, t > 0\} =
\bigcup_{t>0} B(O,R(t)) \times \{t\}$$par $c(t) =
(t,x_1(t),x_2(t),x_3(t))$et dans $B(O,R(t))$ par $X(t) =
(x_1(t),x_2(t),x_3(t))$, n'est autre que $||X^{'}(t)||_{g_t}$.
Cette vitesse est mesur\'{e}e donc dans $B(O,R(t))$ \`{a} l'aide
de la m\'{e}trique physique $g_t$ et non pas dans le semi-c\^{o}ne
$C$ de $\mathbb{R}^4$ et si l'on d\'{e}signe par $h_t := dt^2 -
g_t$ la m\'{e}trique \`{a} l'instant $t$ du semi-c\^{o}ne, qui est
associ\'{e}e \`{a} $g_t$, on a
$$ 0 < ||c^{'}(t)||_{h_t} < 1 \;\;\; \mbox{ et } \;\;\; ||X^{'}(t)||_{g_t} <
1$$pour tout observateur tandis que
$$ ||c^{'}(t)||_{h_t} = 0 \;\;\; \mbox{ et } \;\;\;
||X^{'}(t)||_{g_t} = 1$$pour les trajectoires caract\'{e}risant
les rayons de lumi\`{e}re (ceux-ci sont caract\'{e}ris\'{e}s aussi
par $\nabla_{X^{'}(t)}^{g_t} X^{'}(t) = 0$).\\
Ainsi, lorsqu'une partie de l'Univers peut \^{e}tre assimil\'{e}e
au vide absolu, on a alors (dans cette partie) $g_t = g_e$ et $h_t
= dt^2 - g_e = - \eta$ et on peut consid\'{e}rer des observateurs
galil\'{e}ens $D_1$ et $D_2$ qui auront chacun une vitesse
relative par rapport \`{a} un observateur stationnaire $D_0$
repr\'{e}sent\'{e} par l'axe du temps (ou une droite verticale
quelconque) et si $c_1(t)$ et $c_2(t)$ sont les deux
param\'{e}trisations normales de $D_1$ et $D_2$ alors on a
$$ c^{'}_1(t) = \overrightarrow{k_1} + (a,0,0,0)$$
$$ c^{'}_2(t) = \overrightarrow{k_2} + (b,0,0,0)$$o\`{u}
$\overrightarrow{k_1}$ et $\overrightarrow{k_2}$ sont les
projections sur $B(O,R(t))$, parall\`{e}lement \`{a} l'axe du
temps (qui lui est orthogonal) de $c^{'}_1(t)$ et $c^{'}_2(t)$, et
la vitesse relative de $D_2$ par rapport \`{a} $D_1$ n'est autre
que $\overrightarrow{v}_{D_2 / D_1} =: \overrightarrow{k_r} =
\overrightarrow{k_2} - \overrightarrow{k_1}$. De m\^{e}me la
vitesse de la lumi\`{e}re par rapport \`{a} $D_0$ est, dans ce
cadre pr\'{e}cis, $|| \overrightarrow{k}||_{g_e}$ = 1 lorsque
$\widetilde{c}(t)$ d\'{e}signe la trajectoire de la lumi\`{e}re et
${\widetilde{c}}\;^{'}(t) = l$ s'\'{e}crit sous la forme $l =
\overrightarrow{k} + (1,0,0,0)$ avec $h(l,l) := (dt^2 - g_e)(l,l)$
=0. De plus, la vitesse relative de la lumi\`{e}re par rapport
\`{a} l'observateur $D_1$ n'est autre que
$\overrightarrow{k}-\overrightarrow{k_1}$. Ainsi, si $D_0, D_1$ et
$L$ se trouvent dans le m\^{e}me plan et si $v_{D_1 / D_0} = v$,
alors la vitesse relative de la lumi\`{e}re par rapport \`{a}
$D_1$ est $1 -v$.\\
Dans l'Univers physique r\'{e}el $(B(O,R(t),g_t)$, on peut
d\'{e}finir le vecteur vitesse relative instantan\'{e} de la
lumi\`{e}re par rapport \`{a} un obseravteur $D_1$ uniquement en
un point o\`{u} $D_1$ et la trajectoire $L$ de la lumi\`{e}re se
croisent \`{a} l'instant $t$ par
$\overrightarrow{k}-\overrightarrow{k_1}$ ayant pour norme
physique $||\overrightarrow{k}-\overrightarrow{k_1}||_{g_t}$
o\`{u} $\overrightarrow{k}$ et $\overrightarrow{k_1}$ sont les
vecteurs de type espace associ\'{e}s \`{a} $L$ et $D$. Cette norme
se r\'{e}duit \`{a} la norme euclidienne dans le vide absolu.

\subsection*{Simultan\'eit\'e}
D'autre part, illustrons, \`a l'aide d'un exemple classique (c.f. [2]), que la simultan\'eit\'e a un caract\`ere universel. En effet, supposons que dans rep\`ere (suppos\'e fixe) $R$, une particule est produite avec une vitesse constante $v$ et que, lorsqu'elle est situ\'ee au point $x=0$, elle se d\'esint\`egre en deux photons $p_1$ et $p_2$ qui se propagent le long de l'axe des $x$ dans deux directions oppos\'ees. Les d\'etecteurs $D_1$ et $D_2$ \'etant situ\'es aux points $x=-L$ et $x=L$ et la vitesse des deux photons dans $R$ \'etant $c$, ils arrivent \`a  $D_1$ et $D_2$ en m\^eme temps $t=\frac{L}{c}$. Lorsqu'on analyse cet \'ev\`enement \`a l'aide du rep\`ere $R'$, pour lequel la particule est au repos, en utilisant les formules de transformations relativistes alors $p_2$ arrive \`a $D_2$ avant que $p_1$ arrive \`a $D_1$. Ceci conduit au fait que deux \'ev\`enements qui sont simultan\'es dans le rep\`ere $R$ ne le sont pas dans le rep\`ere $R'$. Notre interpr\'etation est totalement diff\'erente. Pour nous, lorsque les deux photons sont \'emis ils se dirigent chacun de son cot\'e avec la m\^eme vitesse $c$ (dans $R$) ind\'ependamment de la particule qui leur a donn\'e naissance et du rep\`ere $R'$.\\
Si l'on insiste \`a utiliser le rep\`ere $R'$, ayant la vitesse $v$ par rapport \`a $R$, on aboutit, en utilisant la transformation galil\'eenne, au constat suivant:\\
Dans $R'$, la vitesse de $p_2$ est $c-v$ et $p_2$ effectue la distance $L'-vt_2-0=L-vt_2$, o\`u $t_2$ est le temps (dans $R'$) mis par ce photon pour arriver \`a $D_2$; ce qui donne $t_2=\frac{L-vt_2}{c-v}$ et par suite $ct_2=L$. De m\^eme la vitesse de $p_1$ dans $R'$ \'etant $c+v$ et la distance effectu\'ee par $p_1$ \'etant $L'+vt_1-0=L+vt_1$, o\`u $t_1$ est le temps (dans $R'$) mis par ce photon pour arriver \`a $D_1$, alors on a $t_1=\frac{L+vt_1}{c+v}$ et par suite $ct_1=L$. Ainsi, on obtient $ct_2=ct_1$ et $t_1=t_2=\frac{L}{c}=t$.

\subsection*{Remarques finales}
Notons aussi que le ph\'enom\`ene relativiste de la contraction des longeurs n'est pas concevable ni logiquement ni physiquement. En effet, si l'on consid\`ere deux barres m\'etalliques identiques $b$ et $b'$ de m\^eme longueur $l$ et si la barre $b$ est fix\'ee le long de l'axe $Ox$ d'un rep\`ere suppos\'e fixe $R$ et si la barre $b'$ est en mouvement uniforme le long de l'axe $Ox$, alors l'utilisation des formules de transformation relativiste, reliant les coordonn\'ees \'ecrites dans $R$ \`a celles \'ecrites dans le rep\`ere $R'$ pour lequel la barre $b'$ est au repos, conduit \`a une aberration physique flagrante:\\
Lorsqu'on utilise le rep\`ere $R$ pour analyser la situation, on verrait que lorsque l'origine de la barre $b'$ co\"incide avec celui de $b$ l'extr\'emit\'e de $b'$ ne co\"incide pas avec celle de $b$ et lorsque l'extr\'emit\'e de $b'$ co\"incide avec celle de $b$ alors l'origine de $b'$ ne co\"incide pas avec celui de $b$ (c.f. [7]). Cette situation qui pourrait \^etre admise th\'eoriquement n'est pas concevable physiquement.\\
Ceci prouve, encore une fois, que l'hypoth\`ese de l'invariabilit\'e de la vitesse de la lumi\`ere vis \`a vis de tous les r\'ef\'erentiels galil\'eens et l'introduction de l'espace-temps relativiste conduisent \`a des situations physiques irr\'ealisables tout en paraissant avoir r\'esolu quelques probl\`emes th\'eoriques mal pos\'es sans oublier les complications th\'eoriques insurmontables issues de la consid\'eration de l'espace-temps lorentzien le plus g\'en\'eral en vue d'\'etudier la relativit\'e g\'en\'erale qui est bel et bien la th\'eorie qui explique authentiquement les lois de l'expansion de l'Univers.\\
Signalons enfin que, pour nous, le paradoxe des jumeaux de Langevin n'a aucun lien avec la r\'ealit\'e puisque le temps propre relativiste n'est qu'un param\`etre pratique utile qui n'a aucune signification r\'eelle en terme du temps absolu universel.

\section{Mati\`{e}re, \'{E}nergie, Masse et Trous noirs}

Notons pour commencer que la globalit\'e de cette \'etude est
bas\'ee sur les grands principes s\'erieusement confirm\'es de la
M\'ecanique et de la Physique, \'etablis par Newton, Einstein,
Hubble et beaucoup d'autres et qui co\"{\i}ncident dans des cas
particuliers sp\'ecifiques avec des lois codifi\'ees, mesur\'ees
et v\'erifi\'ees, \'etablies par un grand nombre de physiciens
comme Maxwell, Lagrange, Hamilton, Shrodinger, Bohr et Planck et
parfaitenment mod\'elis\'ees \`a l'aide des traveaux de Gauss,
d'Euler et de Riemann
entre autres.\\
Notre point de d\'epart ici est l'\'equation d'Einstein
simplifi\'ee et modifi\'ee suivante:
\begin {equation}\label{r14}
G_{ab}:=R_{ab}-\frac{1}{2}Rg_{ab}=T^{*}_{ab}
\end {equation}
o\`u les tenseurs sym\'etriques utilis\'es ici sont d\'efinis sur
la boule $ B_e(O,R(t)) \subset \mathbb R^3$ qu'on va
consid\'{e}rer tout au long de cette \'{e}tude (par souci de
simplicit\'{e} et de clart\'{e}) comme \'{e}tant la boule
euclidienne $B_{e}(O,t),$ ce qui revient en fait \`{a} supposer
que la vitesse de l'expansion est d\'{e}termin\'{e}e par la
vitesse de propagation des ondes \'{e}lectromagn\'{e}tiques et que
celle-ci \'{e}tait de tout temps \'{e}gale \`{a} $c=1$, bien que ceci n'est pas exact pour $t$ suffisamment petit (le cas
g\'{e}n\'{e}ral sera discut\'{e} \`{a} la fin de cet
article).\\
Ces tenseurs d\'ependent du temps $t > 0$ et d\'ecrivent des
ph\'enom\`enes physico-g\'eom\'etriques inh\'erents \`a l'Univers
physique \`a l'instant $t.$ Ainsi $T^{*}_{ab}$ d\'esigne dans
cette \'equation le tenseur global de la mati\`ere-\'energie qui
v\'erifie $\nabla^{a}T^{*}_{ab}=0,$ o\`u
$\nabla=\nabla(g_{ab}(t))$ est la
 connexion de Levi-civita associ\'ee \`a la m\'etrique
 $g_{ab}(t)$,conform\'ement \`a la relation $\nabla^{a} G_{ab}=0$
 (qui r\'esulte de la deuxi\`eme identit\'e de Bianchi).\\
 La distribution de la mati\`ere situ\'ee dans une r\'egion de l'Univers
 d\'etermine \`a tout instant $t$ une distribution de masse inertielle
$m_t(X)$
 cr\'eant un champ gravitationnel global et des champs de force donn\'es. Cette masse inertielle
 est assujettie \`a des transformations \'energ\'etiques
 \'evolutives (avec le temps) de diff\'erentes sortes:
 \'electro-magn\'etiques, thermo-nucl\'eaires et
 radio-actives. A l'aspect inertiel statique de la mati\`ere (caract\'eris\'e par la masse
 inertielle)et \`a l'aspect \'evolutif permanent, il faut ajouter l'aspect
 dynamique, c.\`a.d. la cr\'eation et la transformation de l'\'energie cin\'etique de la
 mati\`ere en mouvement. Notons aussi que le mot \'energie doit
 \^etre interpr\'et\'e ici dans un sens total et unifi\'e.\\
 De m\^eme $g_{ab}$ d\'esigne la m\'etrique physique (variable avec la position et le
 temps)qui, \`a tout instant $t$, refl\`ete la perturbation
 permanente de l'espace g\'eom\'etrique plate ($B_e(O,t),g_e$) caus\'ee
 par la mati\`ere, conform\'ement au principe de Mach repris par Einstein. Cette perturbation est traduite par la
 cr\'eation de la courbure spaciale qui est prise en compte dans
 l'\'equation (14) par l'interm\'ediaire du tenseur de Ricci
 $R_{ab}(t)$ et de la courbure scalaire non n\'egative $R(t)$. Ainsi, la
 m\'etrique $g_{ab}$ mesure en fait l'effet de la mati\`ere contenue dans
 l'espace plut\^ot que le volume g\'eom\'etrique de cet espace qui
 est mesur\'e conventionnelement \`a l'aide de la  m\'etrique euclidienne $g_{e}$.
 Notons enfin que ces tenseurs, contrairement aux
tenseurs riemanniens, peuvent admettre des singularit\'es qui
sont, comme on va le voir, li\'ees essentiellement au
ph\'enom\`ene de l'effondrement ("collapsing") de la mati\`ere,
engendrant ainsi ce qu'on appelle couramment un trou noir.
L'\'equation (14) s'\'ecrit
\begin {equation}\label{15}
Rg_{ab}=2(R_{ab}-T^{*}_{ab})=:G^{*}_{ab}.
\end {equation}
Elle contient, \`a tout instant $t$, toute la g\'eom\'etrie
physico-cosmique de l'Univers $U(t)$.\\
Supposons que $G^{*}_{ab}$ s'annule sur une boule
$B:=B(I,r)\subset B(O,t)$. Alors on a:
$$R_{ab} = 0 \mbox{ si
et seulement si } T^*_{ab} = 0 \mbox{ sur } B.$$Or, d'par\`es la
d\'efinition m\^eme du tenseur $T^*_{ab},$ la relation $T^{*}_{ab}
= 0$ est \'{e}quivalente \`{a} $g_{ab}=g_e$ sur $B$ et on a alors
$R= 0.$ Si $R \not=0$ sur un voisinage de $I$ dans $B$, alors on a
$g_{ab}=\frac{1}{R}G^{*}_{ab}=0$ sur ce voisinage et par suite
$R=0$, ce qui est absurde. De m\^eme, $R$ ne peut pas s'annuler en
un point isol\'e dans $B$ (en $I$ par exemple) ou m\^eme sur une
courbe d'int\'erieur vide dans $B$ (qui passe par $I$ par exemple)
puisqu'en dehors de ce point ou de cette courbe, on aurait $R
\not= 0$; ce qui est impossible. Donc si $G^{*}_{ab}=0$ sur $B\setminus I$, $g_{ab}$
ne peut pas \^etre un (0,2)-tenseur de classe $C^2$ ni nul ni non nul sur
$B$ avec $R\not=0$. Par contre $g_{ab}$ peut \^etre consid\'er\'ee
alors comme \'etant une distribution \`a support contenu dans $I$
(pour une raison de sym\'etrie) de la forme $k \delta _{I}g_e$
o\`u
$$\delta _{I}g_e(X(I),Y(I))=g_e(X(I),Y(I))$$ pour deux vecteurs
tangents quelconques $X(I)$ et $Y(I)$ en $I$ et $$\delta
_{I}g_e(X(P),Y(P))=0$$ pour $P\in B$, $P\not=I$ et $X(P)$ et
$Y(P)$ deux vecteurs tangents quelconques en $P$.\\
En effet, cette situation correspond en fait \`a la formation d'un
trou noir cr\'e\'e par l'effondrement concentrique compl\`{e}te
d'une agglom\'eration mat\'erielle ayant une tr\`es grande
densit\'e volumique (de sa masse inertielle) qui s'exprime par une
force de gravitation centrale trop importante. Par cons\'equent
$g_{ab}$ ne peut pas \^etre nulle et par suite $supp (g_{ab})$ est
r\'eduit au centre $I$ et on a
$$g_{ab}\simeq(V_e(B)-E)\delta_Ig_e, \hskip 0.5cm
dv_g\simeq(V_e(B)-E)\delta_I$$
$$dv_g(I)\simeq V_e(B)-E  \hskip 0.4cm \mbox {et} \hskip 0.4cm E_t(X)\simeq E\delta_I$$
o\`{u} $\delta_I$ est la masse de Dirac au point \emph{I} et
\emph{E} est l'\'{e}nergie de masse \'{e}quivalente \`{a} la masse
inertielle globale \emph{m} de l'agglom\'{e}ration, juste avant la
derni\`{e}re phase de l'effondrement. Le point $I$ est le centre
d'une gravitation centrale qui absorbe toute particule qui atteint
$B$. Ainsi le trou noir $B$ constitue, en quelque sorte, une
r\'egion d'absorption totale de la mati\`ere et des ondes
\'electromagn\'etiques (une r\'egion de ''no escape''). En fait
$B$ est la boule de Scwarzchild qui est caract\'{e}ris\'{e}e par
le fait que quasiment aucun signal ne peut pas \^{e}tre \'{e}mis
\`{a} partir d'elle. L'\'energie totale concentr\'ee en $I$ du
trou noir $B$ est approximativement $E \sim m$. En $I$, on peut
consid\'erer la densit\'e volumique de la mati\`{e}re-\'{e}nergie
et la courbure comme \'etant infinies. Autrement dit $I$ est une
singularit\'{e} de
l'espace physique $B(O,t)$.\\\\
Remarquons qu'on ne peut pas avoir non plus $g_{ab}$ de classe
$C^2$ avec $R\not=0$ et telle que l'une des valeurs propres
$\lambda _i(t,X)$, $i=1,2,3$, de $G^{*}_{ab}$ soit nulle (respt.
deux d'entre elles soient nulles) sur une boule $B(I,r)$
puisqu'alors on a: $\;\;\;\; vol (B,g_{ab})=0$ et $g_{ab}$ se
diagonalise sur une base orthonormale (par rapport \`a $g_e$) de
vecteurs propres sous la forme
$$\frac{\lambda _1(t,X)}{R}dx_1^2+ \frac{\lambda
_2(t,X)}{R}dx_2^2 \;\;\;\; \left(\mbox{resp. } \frac{\lambda
_1(t,X)}{R}dx_1^2 \right);$$ ce qui constitue un ph\'enom\`ene
physique incompatible avec le principe (intuitive) d'isotropie et
d'homog\'en\'eit\'e locales. Donc chacun de ces deux cas implique
l'annulation des trois valeurs propres sur $B$ et on aurait de
nouveau $G^{*}_{ab}=0$ sur $B$ et la m\^{e}me contradiction et on
se retrouve dans la situation d'un trou noir statique de
Schawrzchild
ou d'autres types de trous noirs.\\\\
Revenons \`a l'Univers $U(t)$ et consid\'erons une agglom\'eration
mat\'erielle dans une r\'egion connexe incluse dans $B(O,t)$. Soit
$m(t)\sim e(t)$ la masse inertielle totale de cette
agglom\'eration ou, d'une fa\c con \'equivalente, son \'energie
potentielle mat\'erielle globale. Distinguons deux sortes de
gravitation cr\'e\'ee par cette mati\`ere: la gravitation interne
ou d'interaction (responsable, avec les forces de liaisons, de la
coh\'esion ou de la non dispersion de l'agglom\'eration) d\'efinie
dans cette r\'egion et la gravitation externe d\'efinie tout
autour de cette r\'egion. Pour l'Univers \`a l'instant $t=0$ dont
la masse inertielle totale $M_0,$ convertie en \'{e}nergie totale
$E_0,$ est suppos\'ee finie (comme il est g\'en\'eralement admis
par les physiciens), la totalit\'e de la gravitation avant le Big
Bang est interne. Pour le centre $I$ d'un trou noir statique
$B(I,r)$, la gravitation interne est concentr\'ee en $I$ et la
gravitation externe est d\'efinie essentiellement sur $B$ o\`u
elle y est extr\^eme; en dehors de $B$ le trou noir cr\'ee une
gravit\'e newtonienne ordinaire. Pour un syst\`eme mat\'eriel
isol\'e (comme une galaxie avec son \'etendue gravitationnelle
significative) de masse globale $m$, la gravitation globale de ce
syst\`eme est essentiellement interne, tandis que si l'on
consid\`ere un astre quelconque de cette galaxie, alors on doit
distinguer entre sa gravitation interne et sa gravitation externe
\`a l'int\'erieur de la galaxie. Ceci reste vrai \`a tous les
\'echelons des
formations mat\'erielles. On va montrer plus loin que la gravit\'{e} interne est fortement li\'{e}e \`{a} l'\'{e}nergie de liaison et aux forces de liaison. \\
De m\^eme, il faudrait distinguer dans l'Univers dynamique entre
l'\'energie cin\'etique et l'\'energie potentielle inertielle d'un
syst\`eme mat\'eriel se d\'epla\c cant, le long d'une trajectoire $X(t)$, \`a une vitesse $v(t):=\|X'(t)\|_{g_t}\leq\|X'(t)\|_{g_e}<1$.
La premi\`ere est \'egale en effet \`a
$E_c(t)=\frac{1}{2}m_1(t)v^2(t)$ o\`u $m_1(t)$ est sa masse
inertielle r\'{e}duite \`a l'instant $t$ qui va \^{e}tre
d\'{e}finie plus loin. Pour l'Univers \`a l'instant $t=0$,
l'\'energie cin\'etique est nulle et l'\'energie potentielle
inertielle est $E_0 \sim M_0$. Pour un syst\`eme mat\'eriel
isol\'e, comme une galaxie se d\'epla\c cant \`a la vitesse $v$,
on a $E_c=\frac{1}{2}m_1(t) v^2$ et son \'energie totale est
$$E(t)=m(t) c^2 + \frac{1}{2}m_1(t) v^2(t)$$o\`{u} $m(t) =
\gamma(t) m_0 := (1 - \frac{v^2(t)}{c^2})^{-\frac{1}{2}} m_0$ a
\'{e}t\'{e} qualifi\'{e} par W. Kaufmann comme \'{e}tant la masse
apparente de la particule en mouvement. Rappelons que (d'apr\`es
le paragraphe 4) pour toute particule mobile dont la trajectoire
libre est d\'ecrite par $X(t)$ dans un rep\`ere virtuel fixe on a
$\widetilde{\Gamma}(t)=\nabla _{X'(t)}^{g_{ab}}X'(t)=0$, tandis
que $\Gamma(t)=\nabla_{X'(t)}^{g_{e}}X'(t)=X''(t)=0$ si et
seulement si $X(t)$ d\'esigne une trajectoire \`a vitesse
constante dans un
rep\`ere $g_{e}$-inertiel.\\\\
\subsection*{Distributions d'\'{e}nergie et de masse}

Supposons maintenant que la distribution de la masse inertielle de
la mati\`ere (en mouvement) dans l'Univers $B(O,t)$ est donn\'ee
par $m_t(X)$, \`a laquelle on associe la mesure $dm_t=:{\rho}_t$.
D\'esignons par $g_t=g_{ab}(t)$ la m\'etrique riemannienne sur
$B(O,t)$ (refl\'etant cette distribution ainsi que toutes les autres formes d'\'energie) et par $\mu
_t=dv_{g_t}=v_t(X)dX$ la mesure de densit\'e $v_t(X)$ par rapport
\`a la mesure de Lebesgue $dX$ sur $B(O,t)$. La masse inertielle
globale de l'Univers \`a l'instant $t$ est donn\'ee par:
$$M(t)=\int _{B_e(O,t)}\rho _t:=\int _{B_e(O,t)}m_t(X)dX$$
Par ailleurs, d\'esignons par $E(t,X)=E_t(X)$ la distribution de
l'\'energie potentielle g\'en\'eralis\'ee qui englobe, par
d\'efinition, toutes les manifestations de la mati\`{e}re et tous
ses effets (la distribution $m_t(X)$, l'\'{e}nergie pure des trous
noirs, la gravit\'e, l'electromagn\'etisme et les forces
d'interaction) et d\'esignons par $\nu_t:=E_t(X)dX$ la mesure
associ\'ee \`a cette distribution. Dans ces conditions on a: $\nu
_t=dX-\mu _t$ et cette relation exprime le fait que $\nu _t$
mesure le d\'efaut du volume physique r\'eel d'un domaine (dans
$U(t)$), contenant une distribution mat\'erielle, par rapport au
volume spacial euclidien de ce domaine suppos\'e vide de
mati\`ere. Cette relation s'\'ecrit $\mu _t=dX-\nu _t$ ou aussi
$v_t(X)=1-E_t(X)$ exprimant ainsi que $\mu _t$ mesure le volume
physique r\'eel en tenant compte de la modification des distances
euclidiennes impos\'ee par le m\'etrique $g_t$ qui elle m\^eme
refl\`ete l'existence de la mati\`ere-\'{e}nergie dans ce m\^eme
domaine. On a naturellement
$$\rho_t \leq \nu_t \leq dX$$ou, en d'autres termes
$$m_t(X) \leq E_t(X) \leq1.$$Tout ceci est bien confirm\'{e} par
notre calcul explicite des m\'{e}triques correspondant \`{a} la
gravitation uniforme et \`{a} la gravitation centrale
\'{e}tudi\'{e}es au paragraphe 4. D'autre part, ces relations
justifient la caract\'{e}risation de
la m\'{e}trique $g_{ab}$ des trous noirs \'{e}tablie ci - dessus.\\
Le principe d'\'equivalence et la loi de la conservation de
l'\'energie donnent
\begin{eqnarray*}
  E(t)&:=& \int_{B(O,t)}E_{t}(X)dX := \int _{B(O,t)\backslash {\bigcup _{\alpha \in A}}B_{\alpha}} E_t(X)dX + \sum_{\alpha \in A} e_\alpha \\
   &=&E(0)=:E_0\sim M_0
\end{eqnarray*}
o\`{u} $e_\alpha$ d\'{e}signe l'\'{e}nergie du trou noir
$B_\alpha$ pour tout $\alpha \in A$.\\Ainsi, consid\'erons, dans
un premier temps, le demi-c\^one de l'espace-temps
$$C=\{(x,y,z,t)\in \mathbb R^4;\;x^2+y^2+z^2 \leq t^2, t\geq 0\}=\bigcup _{t\geq 0} B_e(O,t) \times
\{ t \}$$et munissons le de la m\'etrique $\eta$ d\'efinie par
$$\eta(t,X)=dt^2-g_e(X)$$ induite par la m\'etrique de Minkowski
d\'efinie sur l'espace virtuel $\mathbb R^4$. Au sein de ce
demi-c\^one se produit la cr\'eation dynamique de l'espace
g\'eom\'etrico-temporel r\'eel en expansion permanente , $B(O,t)$,
formant ainsi l'espace r\'eel physique: $U(t)=(B(O,t),g_t)$
toujours en expansion dans $\mathbb R^3$.\\Notons que, sur
l'int\'erieur de $C$, $\eta$ est bien une m\'etrique riemannienne.
Consid\'erons la fonction de l'\'energie potentielle
g\'en\'eralis\'ee $E(t,X)=E_t(X)$ pour $X \in B(O,t)$ et supposons
que $E$ est continue sur $C$ et que ses d\'eriv\'ees partielles
d'ordre $\leq 2$ existent et sont continues et born\'ees sur
$C^{*}=C\backslash \{O\}$. L'\'energie globale de l'Univers \`a
l'instant $t>0$ s'\'ecrit
$$E(t)=\int _{B(O,t)}E(t,X)dX=\int _{0}^{t}dr\int _{S_e(O,r)}E(r,X)d\sigma
_r\equiv E_0$$et par suite
$$E'(t)= \int _{S(O,t)}E(t,X)d\sigma_t=:S(t)=0$$pour tout $t>0$.
Ainsi la fonction du temps $S$ n'est autre que $E_0\delta_
{\mathbb R^+}$ et on a
$$E_t|_{S(O,t)}=0 \mbox { pour } t>0.$$\\
Notons qu'\`a l'int\'erieur d'un trou noir $B(I,r)\subset B(O,t)$,
au cas o\`u il existe, on a:
$$E_t(X)=e(I)\delta _I=m(I)\delta _I,$$ o\`u $e(I)$ est l'\'energie potentielle de masse \'{e}quivalente \`a la masse inertielle $m(I)$. Ainsi $e(I)$ est une partie
de l'\'energie initiale $E_0$ de l'Univers originel qui s'est
reconcentr\'ee \`a un moment donn\'e (post\'erieur au Big Bang) au
point $I$; tandis que $E(t,X)$ = $e(I)\delta_I$ d\'esigne la
fonction de l'\'energie potentielle de masse g\'en\'eralis\'ee sur
$B(I,r)$. Notons aussi que la fonction $E_t(X)$ est nulle dans les
r\'egions de l'Univers consid\'er\'ees comme \'etant d\'epourvues
de la mati\`ere et de ses effets . Signalons enfin que, bien que
la fonction $E(t,X)$ est loin d'\^etre de classe $C^2$, on peut
quand m\^eme raisonnablement l'approcher par une telle fonction
(id\'ealisant de la sorte l'Univers) ou alors la consid\'erer,
ainsi que toutes ses d\'eriv\'ees partielles, au sens des
distributions.\\\\
\subsection*{\'{E}quation de la mati\`{e}re - \'{e}nergie}

Consid\'erons maintenant la partie $C_1$ du demi-c\^one de
l'espace-temps $C$ situ\'ee entre $t=t_1$ et $t=t_2$. On a
$\partial C_1=B(O,t_1)\bigcup B(O,t_2)\bigcup S_1 $ o\`u $S_1$ est
la fronti\`ere lat\'erale de $C_1$. Consid\'erons le champ de
force $F_{\eta}$ d\'efini sur $C^{*}$ et d\'erivant de la fonction
potentielle totale $E(t,X)$ i.e.
$$F_{\eta}(t,X):= -\nabla
^{\eta}E(t,X):= - grad_{\eta}E(t,X).$$Ainsi, si $u(t)=(t,X(t))$
est une trajectoire donn\'ee et si on a $F_{\eta}(u(t))=0$, alors
on a:
$$\eta(F_{\eta}(u(t)),u'(t))=0 \Leftrightarrow \eta(\nabla ^{\eta}E(u(t)),u'(t))=0$$$$\Leftrightarrow
dE(u(t)).u'(t)=0 \Leftrightarrow \frac{d}{dt}E(u(t)) = 0
\Leftrightarrow E(u(t))=cte.$$ Par ailleurs, la relation
$$||F_{\eta}(u(t)||_{\eta}=||\nabla ^{\eta}E(u(t))||_{\eta}=0$$
\'{e}quivaut \`{a}
$$dE(u(t)).\nabla ^{\eta}E(u(t))=0$$ et \`{a}
$$\frac{\partial E}{\partial t}(u(t))^2 =|\nabla ^{g_e}E_t(X(t))|^2=\sum _{i=1}^{3}
\frac {\partial E}{\partial x_i}(u(t))^2.$$De m\^eme on a
$$||u'(t)||_{\eta}=0 \Leftrightarrow ||(1,X'(t))||_{\eta}=0\Leftrightarrow |X'(t)|=1,$$
ce qui veut dire que la vitesse eucildienne est \'egale \`a 1 et
par suite on a $$|X(t)-X(t_0)|=t-t_0 \mbox{ pour } t \geq t_0 \geq
0$$ et la trajectoire dans $C$ se r\'eduit \`a un rayon d'un
c\^{o}ne de lumi\`{e}re.\\ D\'esignons par $d\eta$ la mesure
associ\'ee \`a $\eta$ dans $C$ et par $\Delta _{\eta}$
l'op\'erateur de Laplace-Beltrami sur $C$ ($d\eta$ est une mesure
de densit\'e $f(t,X)\geq 0$ par rapport \`a la mesure de Lebesgue
sur $C$ avec $f(t,X)=0$ sur $\partial C\setminus O$). D'apr\`es le
th\'eor\`eme de Stokes on a (en d\'{e}signant par $\overrightarrow{n}$ le
vecteur normal de $S_1$ qui permet d'appliquer le th\'eor\`eme de Stockes comme il est expliqu\'e dans $[4]$, p. $434$):
$$\int _{C_1}\Delta _{\eta} E(t,X)d\eta=\int _{C_1} div_{\eta}(\nabla^{\eta}E(t,X))d\eta$$$$=
\int _{B(O,t_2)} \eta
(\nabla^{\eta}E(t_2,X),\frac{\partial}{\partial
t})f(t_2,X)dX$$$$-\int _{B(O,t_1)} \eta
(\nabla^{\eta}E(t_1,X),\frac{\partial}{\partial t})f(t_1,X)dX+\int
_{S_1}\eta (\nabla^{\eta}E(t,X),\vec{n})d\eta$$$$= \int_{B(O,t_2)}
\frac{\partial E}{\partial t}(t_2,X)f(t_2,X)dX - \int_{B(O,t_1)}
\frac{\partial E}{\partial t} (t_1,X)f(t_1,X)dX.$$ Ainsi, on
obtient:
$$\int _{C_1}\Delta _{\eta} E(t,X)d\eta=\int _{t_1}^{t_2}dt\int _{B(O,t)}\Delta _{\eta} E(t,X)f(t,X)dX$$
$$=\int _{0}^{t_2}dt\int _{B(O,t)}\Delta _{\eta} E(t,X)f(t,X)dX-
\int _{0}^{t_1}dt\int _{B(O,t)}\Delta _{\eta} E(t,X)f(t,X)dX$$
$$=\int _{B(O,t_2)}\frac {\partial E}{\partial t}(t_2,X)f(t_2,X)dX-
\int _{B(O,t_1)}\frac {\partial E}{\partial
t}(t_1,X)f(t_1,X)dX=F(t_2)-F(t_1)$$ 
avec
$$F(t):=\int
_{B(O,t)}\frac {\partial E}{\partial t}(t,X)f(t,X)dX$$ Par suite
$$\int _{B(O,t_2)}\Delta _{\eta} E(t_2,X)f(t_2,X)dX=F'(t_2)$$et
$$\int _{B(O,t_1)}\Delta _{\eta} E(t_1,X)f(t_1,X)dX=F'(t_1)$$
Ainsi on a
$$\int _{B(O,t)}\Delta _{\eta}
E(t,X)f(t,X)dX=F'(t)$$pour tout $t>0$; relation qui s'\'ecrit
$$\int _{0}^{t}dr\int _{S(O,r)}\Delta _{\eta} E(r,X)f(r,X)d\sigma _r=
\frac{d}{dt}\int _{B(O,t)} \frac {\partial E}{\partial
t}(t,X)f(t,X)dX$$ce qui implique
$$F^{''}(t) = \int _{S(O,t)}\Delta _{\eta} E(t,X)f(t,X)d\sigma _t=
\frac{d^2}{dt^2}\int _{B(O,t)} \frac {\partial E}{\partial
t}(t,X)f(t,X)dX$$$$=\frac{d^2}{dt^2}\int _{0}^{t}dr\int
_{S(O,r)}\frac {\partial E}{\partial r}(r,X)f(r,X)d\sigma _r$$
$$=\frac{d}{dt}\int _{S(O,t)}\frac {\partial E}{\partial t}(t,X)f(t,X)d\sigma
_t=0$$pour tout $t>0$, puisque $f(t,X) \equiv 0$ sur
$S(O,t)$.\\Par cons\'equent, on a
$$F'(t) = \int
_{B(O,t)}\Delta _{\eta} E(t,X)f(t,X)dX=cte.$$ 
Or ceci implique
$$\int _{C_1}\Delta _{\eta} E(t,X)d\eta=a(t_2-t_1)$$avec $a=F'(t)$
et par suite on a, pour $t>0$ et \emph{C}(\emph{t}) = $\{$
$($\emph{x},\emph{y},\emph{z},\emph{r}$)$ $\in$ $\mathbb{R}^4$;\\
$x^2+y^2+z^2 \leq r^2, 0\leq r\leq t$ $\}$ :
$$\int _{C(t)}\Delta _{\eta} E(t,X)d\eta=at.$$Ainsi, en effectuant
un changement de variables
$$(t,X)\rightarrow (\lambda s,\lambda
X)\;\;\mbox { pour }\;\; \lambda >0 \;\;\mbox { dans } C(t),$$on
obtient
$$\int _{C(s)}\Delta _{\eta} E(\lambda s,\lambda X)\lambda ^4 d\eta=a\lambda
s$$
ou
$$\int _{C(s)}\Delta _{\eta} E(\lambda (s, X))
d\eta=\frac{a}{\lambda^3}s.$$Or, on a
$$\int _{C(s)}\Delta _{\eta} E(s,X)d\eta=as , $$d'o\`u
$$\int _{C(s)}\Delta _{\eta} E(s,X)d\eta=\lambda ^3\int _{C(s)}\Delta _{\eta} E(\lambda (s,
X))d\eta ; $$ce qui implique ($\lambda$ \'{e}tant arbitraire)
$$\int _{C(s)}\Delta _{\eta} E(s,X)d\eta=as=0 . $$Par suite, on obtient $$a=0,\;\;F(t)=cte\;\;\mbox { pour
}\;\;t>0$$et on a$$\int _{C_1}\Delta _{\eta} E(t,X)d\eta=0\;\mbox{
pour tout }\;t_1\;\mbox { et }\;t_2 \; \mbox { tels que }
0<t_1<t_2.$$
Montrons que l'on a alors
$$\Delta _{\eta}
E(t,X)=0\;\mbox{ sur }\; C^*.$$En effet, si on suppose que $\Delta
_{\eta} E(t_0,X_0)>0$ par exemple, alors on aura $\Delta _{\eta}
E(t,X)>0$ sur un voisinage $B$ de $(t_0,X_0)$ dans $\{t_0\}\times
B(O,t_0)$ et si on consid\`ere la r\'eunion $C'$ de tous les
demi-c\^ones "causaux" futurs ayant leur sommet sur $B$ et en
raisonnant de la m\^eme mani\`ere que ci-dessus sur une partie
$C'_1$ de $C'$ situ\'ee entre $t_1$ et $t_2$ avec $t_0<t_1<t_2$,
on montre que
$$\int _{C'_1}\Delta _{\eta}
E(t,X)d\eta=0,\;\mbox{ pour tout $t_1$ et $t_2$ suffisamment proches de $t_0$}
$$ (la contribution de
chaque demi-c\^{o}ne \'etant nulle puisque $E(t,X)$ peut \^etre consid\'er\'e comme \'etant constant sur la surface de chaque semi-c\^one pour tels $t_1$ et $t_2$) et par suite
$$\int
_{B}\Delta _{\eta} E_{/B}(t_0,X)f(t_0,X)dX=0$$ce qui est
contradictoire puisque $\Delta _{\eta} E(t_0,X)$ est suppos\'e
continu et positif sur \emph{B}. Par cons\'equent, on a
\begin{equation}\label{16}
\Delta _{\eta}E(t,X)=0\mbox{ sur }C^*
\end{equation}
Ainsi on a, pour tout $t>0$
$$\hskip 4cm \frac{\partial ^2}{\partial
t^2}E(t,X)-\Delta E(t,X)=0 \hskip 4cm (E^*)$$
sur $B(O,t)$ tout en ayant $E(t,X)|_{S(O,t)}=0.$\\
Ce m\^eme r\'esultat peut \^etre en fait obtenu en dehors des
trous noirs (et des autres singularit\'es) en supposant une
certaine r\'egularit\'e de la distribution $E(t,X)$ sur une
r\'egion de $C$ obtenue en excluant des petits voisinages des
trous noirs \'evoluant avec l'expansion. Si $B(I_{\alpha}(t),r_\alpha(t))$
est un trou noir inclus dans $B(O,t)$, et si l'on suppose que
$\sum _{\alpha \in A}e_{\alpha}$ est n\'egligeable par rapport \`a
$E_0$ pour $t\in[t_1,t_2]$ ou, tout au moins, constante sur
$[t_1,t_2]$, on peut appliquer le principe de la conservation de
l'\'energie sur cette r\'egion de $C$ et obtenir l'\'equation
($E^*$) sur $B(O,t)\backslash \bigcup
_{\alpha \in A}B_{\alpha}$.\\\\
Consid\'erons maitenant l'Univers dynamique r\'eel $(U(t))_{t > 0}
=(B_e(O,t),g_t)_{t > 0}$ et le demi-c\^one $C$ muni de la
m\'etrique riemannienne $h$ d\'efinie sur l'int\'erieur de $C$ (et la m\'etrique lorentzienne non n\'egative sur $C^*$) par
$$h(t,X)=dt^2-g_t(X)=dt^2-g(t,X)$$o\`u $g_t$ n'est autre que la
m\'etrique remannienne $g_{ab }$ d\'efinie sur $B_e(O,t)$ par l'\'equation
(14). En d\'esignant par $d\beta$ la mesure associ\'ee \`a $h$ sur
$C^*$ et en appliquant le th\'eor\`eme de Stokes sur une partie
quelconque $C_1$ de $C^*$ (d\'efinie de la m\^eme mani\`ere que
pr\'ec\'edemment) on peut suivre les m\^emes d\'emarches que
ci-dessus en utilisant les demi-c\^ones de lumi\`eres \`a
l'int\'erieur de $C$ dont les rayons sont des g\'eod\'esiques
adapt\'ees aux m\'etriques $g_t$ et $h$ (i.e. g\'eodesiques nulles pour $h$). Ainsi, en rempla\c cant
$d\eta$ par $d\beta$, $dX$ par $d\mu$, $\Delta_{\eta}$ par
$\Delta_{h}$ et $F_{\eta}(t,X_t)$ par $F_h(t,X_t)= -
\nabla^{h}E(t,X_t)= - grad_{h}E(t,X_t)$ et en tenant compte des
m\^emes hypoth\`eses, on obtient
$$\int _{B(O,t)}\Delta_{h}E(t,X)k(t,X)dX=0 \hskip 0.5cm \mbox { pour tout }t>0
$$avec
$k(t,X) \geq 0$ sur \emph{C} et $k(t,X)=0$ sur $\partial C\setminus O$,
$$\int _{C_1}\Delta_{h}E(t,X)d\beta=0 \hskip 0.5cm \mbox { pour }0<t_1<t_2,$$
$$\Delta_{h}E(t,X)=0 \hskip 0.5cm \mbox{ sur } C^*$$et
\begin{equation}\label{17}
\frac{\partial^2}{\partial t^2}E(t,X)-\Delta_{g_t}E(t,X)=0
\end{equation}
sur $B(O,t)$ avec $E(t,X)|_{S(O,t)}=0$ pour tout $t>0$.\\
Cette \'equation co\"{i}ncide approximativement avec l'\'equation $(E^*)$, pour $t\gg1$, parce qu'alors la m\'etrique $g_t$ pourra \^etre remplac\'ee approximativement par $g_e$ lorsqu'on tra\^ite avec des r\'esultats macroscopiques globaux.\\\\

\noindent{\bf Remarque}: Le facteur de Lorentz $\gamma=(1-v^2)^{-\frac{1}{2}}$ s'introduit tout naturellement \`a l'aide de la m\'etrique d'\'evolution $h=dt^2-g_t$ sur $C$. En effet, cette relation implique $\tau^2=t^2-d^2$, o\`u $\tau$ d\'esigne l'intervalle entre deux \'ev\`enements qui respectent le principe de causalit\'e dans $C$ et $d$ est la distance spatiale mesur\'ee \`a l'aide de $g_t$. Cette relation implique 
$$\frac{\tau^2}{t^2}=1-v^2=\frac{1}{\gamma^2}\quad\mbox{ou}\quad t=\gamma\tau.$$

\subsection*{Temps universel et temps propre}
\normalsize{Reprenons l'exp\'erience du ralentissement d'horloges mobiles (voir [7], p. 43); on a alors consid\'er\'e que la pr\'eriode $T_p$ donn\'ee par les horloges a\'eroport\'ees et celle indiqu\'ee par les horloges au sol v\'erifient la relation 
$$T=\gamma T_p\qquad\mbox{o\`u}\qquad\gamma=(1-v^2)^{-\frac{1}{2}}\simeq 1+\frac{v^2}{2}$$
puisque la vitesse relative par rapport \`a la terre $v\ll 1$ pour $c=1$.\\
Or, suivant que le vol, qui s'effectue \`a la vitesse $v$ le long d'une trajectoire circumpolaire \`a la distance $R$ de l'axe de rotation de la terre, est orient\'e vers l'ouest ou vers l'est, les vitesses d'entra\^inement \`a consid\'erer  sont respectivement 
$$v_w=v+\Omega R\qquad\mbox{and}\qquad v_e=v-\Omega R$$
o\`u $\Omega\simeq 7.3\times 10^{-5}$ rad s$^{-1}$ est la vitesse de rotation de la terre autour de l'axe des p\^oles.\\
Il est alors consid\'er\'e que la diff\'erence relative de la p\'eriode calcul\'ee pour une horloge dans une avion se d\'epla\c{c}ant vers l'ouest vaut
$$\frac{T_w-T_p}{T_p}=\gamma_w-1=\frac{(v+\Omega R)^2}{2}\simeq v(v+2\Omega R)$$
tandis que, pour l'avion se d\'epla\c{c}ant vers l'est, elle vaut 
$$\frac{T_e-T_p}{T_p}=\gamma_e-1=\frac{(v-\Omega R)^2}{2}\simeq v(v-2\Omega R).$$
Il est alors observ\'e que les d\'erives, qui sont $275$ ns pour le d\'eplacement vers l'ouest et $40$ ns pour le d\'eplacement vers l'est, sont en accord avec les pr\'evisions th\'eoriques de la relativit\'e restreinte.\\
Dans le cadre de notre mod\`ele nous obtenons des r\'esultats semblables d'une fa\c{c}on r\'eellement physique. En effet, notre unit\'e de temps du syst\`eme international est bas\'ee sur la vitesse de la lumi\`ere dans le vide, c'est \`a dire sur la m\'etrique $g_e$ (qui correspond \`a la densit\'e d'\'energie $\rho=0$, en n\'egligeant la constante cosmologique $\Lambda$) et sur la m\'etrique $-\eta=dt^2-g_e$.\\
A la surface de la terre, les m\'etriques sont $g$ et $h=dt^2-g$, qui sont diff\'erentes de celles \`a l'altitude des avions se dirigeant vers l'ouest et vers l'est que nous allons noter $g_1$ et $h_1=dt^2-g_1$.\\
La diff\'erence de ces $2\times 3$ m\'etriques est due essentiellement aux diff\'erences du champ gravitationnel et du champ magn\'etique.\\
Rappelons qu'au sein de notre mod\`ele, les m\'etriques tiennent compte de ces deux champs et elles contractent les distances:
$$\|X'(t)\|_g\leq\|X'(t)\|_{g_1}\leq\|X'(t)\|_{g_e}.$$ 
De plus, le temps universel pour nous est intrins\`equement li\'e \`a la vitesse de la lumi\`ere dans le vide $c=1$ par rapport \`a un r\'ef\'erentiel suppos\'e fixe et nous avons la relation entre le temps propre $\tau$ (qui d\'epend de la m\'etrique et de la vitesse $v$ du r\'ef\'erentiel) avec le temps universel $t$ est exprim\'ee par 
$$\frac{\tau}{t}=\frac{1}{\gamma}=(1-v^2)^{-\frac{1}{2}}.$$
Donc, dans l'exp\'erience pr\'ec\'edente, il a fallu comparer les deux temps propres $\tau$ et $\tau_1$ (relatifs \`a $g$ et $g_1$) au temps universel $t$ avant de les comparer l'un \`a l'autre; ce qui donne les m\^emes r\'esultats. Nos montres sur terre sont r\'egl\'ees d'une part sur la vitesse de la lumi\`ere dans le vide utilisant la m\'etrique $g_e$ et d'une autre part sur la m\'etrique $g$ qui est diff\'erente de $g_e$ ainsi que de $g_1$. Le rapport $\displaystyle \frac{\tau_1}{\tau}=\frac{\tau_1/t}{\tau/t}$ nous donnera des r\'esultats plus r\'ealiste physiquement pour les trois montres utilis\'ees.\\
La relativit\'e est une notion attach\'ee au temps propre $\tau$ et non pas au temps universel $t$. On n'a pas le droit de d\'efinir la notion du temps par rapport \`a une m\'etrique privil\'egi\'ee autre que $g_e$ et ni de r\'egler nos montres \`a proximit\'e de la terre, par exemple, ou \`a proximit\'e d'une autre plan\`ete ou d'un trou noir. Seul le temps pour un observateur suppos\'e fixe se trouvant dans un milieu qui puisse \^etre assimil\'e au vide absolu a un caract\`ere intrins\`eque. M\^eme le temps propre $\tau$ pour un observateur mobile se trouvant dans le vide, o\`u la m\'etrique est $g_e$, n'a rien d'intrins\`eque et ne peut pas \^etre utilis\'e pour d\'efinir n'importe quel unit\'e de temps. Seul pour un observateur fixe dans le vide, on a $\tau=t$. Pour un trou noir de centre $I$, $B(I,r)$, et d'\'energie $e$ on a $\tau=t$ pour tout r\'ef\'erentiel se d\'epla\c{c}ant dans $B\setminus I$ puisque l'on a  
$$v_{g_t}:=\|X'(t)\|_{g_t}=0,$$
tandis que pour un r\'ef\'erentiel atteignant $I$, pour $t=t_0$, on a
$$\frac{\tau}{t}=\sqrt{1-e^2}=\sqrt{1-\|X'(t)\|_{g_t}}$$
en consid\'erant que l'\'energie totale de l'Univers $E_0$ est l'unit\'e d'\'energie.\\

\subsection*{Champ de force, acc\'{e}l\'{e}ration et
g\'{e}od\'{e}siques}

Par ailleurs, posons
$$F_{g_t}(X):= - \nabla ^{g_t}E_t(X)$$et
$$F_{g_e}(X):= - \nabla ^{g_e}E_t(X)$$pour $X\in B(O,t)$ et
consid\'erons une trajectoire $X(t)$ dans l'Univers dynamique
$(U(t))_{t > 0}$. On a alors

$$\frac{d}{dt}E(u(t))=\frac{d}{dt}E(t,X(t))=\frac {\partial E}{\partial t}(t,X(t))+
\frac{d}{ds}E_t(X(s))|_{s=t}$$
$$=\frac {\partial E}{\partial t}(t,X(t))+
dE_t(X(s)).X'(s)|_{s=t}=\frac{\partial}{\partial
t}E(t,X(t))+dE_t(X(t)).X'(t)$$ainsi que
$$\eta(\nabla^{\eta}E(u(t)),u'(t))=\frac{d}{dt}E(u(t))=h(\nabla^{h}E(u(t)),u'(t))$$
$$=\frac {\partial E}{\partial t}(u(t))- g_t(\nabla ^{g_t}E_t(X(t)),X'(t))=
\frac {\partial E}{\partial t}(u(t)) - g_e(\nabla
^{g_e}E_t(X(t)),X'(t)).$$D'o\`u
$$g_t(F_{g_t}(X(t)),X'(t))=g_e(F_{g_e}(X(t)),X'(t))$$
$$=\frac{d}{dt}E(t,X(t))-\frac {\partial E}{\partial t}(t,X(t))=dE_t(X(t)).X'(t)=
\frac{d}{ds}E_t(X(s))|_{s=t};$$ce qui donne, conform\'ement \`a la
loi fondamentale g\'en\'eralis\'ee de la M\'ecanique
$$F_{g_e}(X(t))= - \nabla^{g_e}E_t(X(t))=\Gamma(t) = \nabla^{g_e} _{X'(t)} X'(t)=X''(t)$$et
$$F_{g_t}(X(t))= - \nabla^{g_t}E_t(X(t))=\nabla_{X'(t)}^{g_t}X'(t)=\widetilde{\Gamma}(t).$$
Ainsi $F_{g_e}$(\emph{X}) n'est autre que le champ global de
toutes les forces naturelles. La force agissante sur un point
mat\'{e}riel auquel on attribue la masse $m_t$ est en fait
$m_tF_{g_e}$(\emph{X}) = - $m_t \nabla^{g_e} E_t$(\emph{X}). Pour
une particule mat\'{e}rielle ponctuelle en mouvement, la force appliqu\'{e}e
est \emph{m}(\emph{X}(\emph{t}))$F_{g_e}$(\emph{X}(\emph{t})) =
\emph{m}(\emph{X}(\emph{t}))$\Gamma$(\emph{t}). En particulier si
$X(t)$ est une g\'eod\'esique pour $g_e$ avec $||X'(t)||_{g_e}=1$,
 alors $u(t)$ est une g\'eod\'esique pour $\eta$ avec $||u'(t)||_{\eta}=0$
et par suite on a: $u(t)$ est un rayon d'un c\^{o}ne de
lumi\`{e}re dans $(C, \eta)$, $E(u(t))=cte$ et
$$F_{g_e}(X(t))= - \nabla^{g_e}E_t(X(t))=\Gamma(t)=X''(t)=0.$$De
m\^eme, si $X(t)$ est une g\'eod\'esique pour $g_t$ avec
$||X'(t)||_{g_t}=1 \Leftrightarrow u(t)$ est une g\'eod\'esique
pour $h$ avec $||u'(t)||_{h}=0$ et par suite on a: $u(t)$ est un
rayon d'un c\^one de lumi\`ere, pour la m\'{e}trique $h$, dans
$C$, $E(u(t))=cte$ et
$$F_{g_t}(X(t))= - \nabla^{g_t}E_t(X(t))=\nabla_{X'(t)}^{g_t}X'(t)=\widetilde{\Gamma}(t)=0.$$
Rappelons que l'on a en fait
$$F_{g_t}(X(t))=\widetilde{\Gamma}(t)=0$$pour tout mouvement libre dans
$(U(t))_{t> 0}$; ce qui constitue une g\'en\'eralisation des deux
lois
d'inertie de Newton.\\\\

Signalons enfin que si l'on suppose l'existence virtuelle et
purement th\'eorique de l'espace $\mathbb R^3$ et du demi-c\^one
de l'espace-temps avant leurs existences r\'eelles
physico-temporelles, qui ont \'et\'e mod\'elis\'es beaucoup plus
tardivement par Euclide et Descarte d'une part et par Galil\'ee et
Newton d'autre part, alors on peut consid\'erer l'Univers originel
comme \'etant $E_0\delta_{\mathbb R^3}$ et $E_0\delta_{C}$. Par
contre, apr\`es le Big Bang, on a, dans le demi-c\^one de
l'espace-temps, $E(t,X)=0$ sur $\partial C\setminus O$ et $E(0,O)=E_0$. De
m\^eme on a $\Delta _h E(t,X)=0$ sur $ C^*$ et $\Delta _h
E=E_0\Delta _h \delta_{C}$ o\`u l'on a pos\'e, pour une fonction
de classe $C^2$ $\phi(t,X)$ \`a support compact dans $C$:
$$<\Delta _h \delta_{C},\phi(t,X)>=\lim_{t\longrightarrow 0}\Delta _h\phi(t,O).$$

\subsection*{Conclusion}
\noindent Il r\'esulte, d'apr\`es ce qui pr\'ec\`ede que notre
Univers r\'eel, physique, g\'eom\'etrique et dynamique,
mod\'elis\'e par $(U(t))_{t > 0} =(B(O,t),g_t)_{t > 0}$, est
caract\'eris\'e d'une mani\`ere \'equivalente par chacune des
notions suivantes:\\
a) La distribution mat\'erielle de (la densit\'e de) la masse
inertielle $m_t(X)$ avec les $e_\alpha(t) \delta_{I(t)}$
correspondant aux trous noirs dispers\'{e}s
dans \emph{B}(\emph{O},\emph{t}).\\
b) Le champ scalaire regularis\'e de l'\'energie potentielle
totale repr\'esent\'e par la distribution de l'\'energie
potentielle g\'en\'eralis\'ee $E_t(X)$.\\
c) La m\'etrique riemannienne r\'egularis\'ee $g_t$ sur $B(O,t)$.\\
d) Le tenseur de la mati\`ere-\'energie modifi\'e $T^{*}_{ab}$.\\
e) La mesure physique $\mu _t$ sur $B(O,t)$ ($\mu _t=v_t(X)dX$
o\`u $dX$ est la mesure de Lebesgue sur
$B(O,t)$ pour tout $t\geq 0$) mais aussi par la mesure $\nu _t=E_t(X)dX$.\\
f) Le champ vectoriel $\nabla ^h E(t,X)$ sur $C$ (o\`u
$h=dt^2-g_t$ est la m\'etrique riemannienne sur le demi-
c\^one de l'espace-temps).\\
g) Le champ total et global - $\nabla ^{g_e}E_t(X)=F_{g_e}(X) (
=\Gamma
_t=\nabla _{X'(t)}^{g_e}X'(t)=X''(t)$ pour toute particule mat\'{e}rielle en mouvement libre dans $(U(t))_{t > 0}$).\\
h) L'ensemble de toutes les g\'eod\'esiques dynamiques $X(t)$ pour
la m\'etrique $g_t$ \'evoluant avec le
temps (i.e. v\'erifiant $\widetilde{\Gamma}(t)=\nabla _{X'(t)}^{g_t}X'(t)=F_{g_t}(X(t))= - \nabla ^{g_t}E_t(X(t))=0$).\\

Notons enfin que notre mod\`ele est parfaitement cons\'equent dans
le sens o\`u, d'une part, il prouve \`a post\'erori la
l\'egitimit\'e de tous les principes m\'ecaniques et physiques
d\'ecouverts par les grands physiciens de l'humanit\'e, et d'une
autre part on a:\\
$E(t,X) = E_t(X)\equiv 0$ dans un domaine $D$ de l'Univers
dynamique \emph{C} situ\'{e} entre $t=t_1$ et $t=t_2
\Leftrightarrow \nu_t=0\Leftrightarrow T^{*}_{ab}\equiv 0$ dans
les domaines $D_t$ de $B(O,t)$ qui correspondent \`a $D
\Leftrightarrow g_t=g_e$ dans $D_t\Leftrightarrow \mu_t=dX$ sur $
D_t \Leftrightarrow h=\eta$ dans le domaine \emph{D} de $C$ $
\Leftrightarrow \Gamma_t=X''(t)\equiv 0$ pour les mouvements
libres de $D\Leftrightarrow$ Les trajectoires des mouvements
libres dans $D$ sont les g\'eod\'esiques constitu\'ees
de lignes droites.\\\\
\subsection*{Remarque} La propri\'{e}t\'{e}, signal\'{e}e pr\'{e}c\'{e}demment, de notre m\'{e}trique physique $g_t$ montre que notre espace -
temps muni de la m\'{e}trique $h_t$ = $dt^2 - g_t$ v\'{e}rifie les
trois postulats (l\'{e}g\`{e}rement modifi\'{e}s) des th\'{e}ories
m\'{e}triques de la gravitation qui stipulent
que\\

(\emph{i}) l'espace - temps est muni d'une m\'{e}trique,\\

(\emph{ii}) les trajectoires des corps en chute libre sont des
g\'{e}od\'{e}siques.\\

(\emph{iii}) Dans un r\'{e}f\'{e}rentiel local, les lois de la
physique non gravitationnelle sont celles de la physique classique
(et non pas celles que l'on \'{e}crit \`{a} l'aide de la
th\'{e}orie de la relativit\'{e} restreinte).\\\\
Ceci prouve que notre mod\`{e}le gravitationnel v\'{e}rifie le
principe d'\'{e}quivalence d'Einstein.\\\\
Signalons pour finir que, dans notre mod\`ele l'espace
g\'eom\'etrique n'existait pas autour du point de concentration de
l'\'energie originelle $E_0$ avant le Big Bang, tandis que
l'espace g\'eom\'etrique $B(I,r)$ existe autour du point de
concentration de l'\'energie du trou noir $I$ mais il est muni
d'une m\'etrique nulle en dehors de \emph{I}. Quant \`a une partie
de l'espace g\'eom\'etrique de l'Univers $B(O,t)$ qui ne contient
pas de mati\`ere ni ses effets, alors elle existe bel et bien et
elle est munie de la m\'etrique $g_e$. Le champ de gravitation y
est nul tandisqu'autour du centre $I$ d'un trou noir $B(I,r)$ le
champ gravitationnel est omnipr\'esent.
\\\\
\section{Energie, Pseudo-ondes et Fr\'equences}
Consid\'erons maintenant l'\'equation des ondes (ou l'\'{e}quation
de la mati\`{e}re - \'{e}nergie)
$$\Box E(t,X)=\frac {\partial^2E}{\partial t^2}(t,X)-\Delta
E(t,X)=0\;\;\;\;\;\;\;\;\;\;\;(E^*)$$et (en utilisant la
m\'{e}thode de s\'{e}paration des variables) d\'eterminons, pour
tout $t>0$, les solutions sur $B(O,t)$ qui v\'erifient
$E(t,X)_{|S(O,t)}=0$. Consid\'erons donc les fonctions de la forme
$$E(t,X)=f_0(t)F_0(X)\;\mbox{ pour un }\;t_0>0 \;\mbox{ fix\'e v\'erifiant }$$
$$\Box f_0(t)F_0(X)=0\;\;\;\;\;\;\;\;\;\;\;\;\;\;\;\;\;\;\;\;\;\;\;\;\;\;\;\;\;\;\;\;\;\;\;\;\;\;\;\;\;\;\;\;\;(E_0)$$pour $0<t\leq
t_0$ et $X\in B(O,t_0)$ avec $F_{{0}_{|S(O,t_0)}}=0$. Il est bien
connu que les solutions $f_0$ et $F_{0}$ de l'\'equation $(E_0)$
s'obtiennent \`a partir des fonctions propres $\varphi
_{t_0,i}(X)$ associ\'ees au probl\`eme de Dirichlet sur la boule
$B(O,t_0)$ \'{e}quip\'{e}e de la m\'{e}trique $g_e$ et que, si
$\lambda _i(t_0)$ est la suite croissante des valeurs propres de
l'op\'erateur de Laplace-Beltrami $-\Delta$ qui leur sont
associ\'ees, alors la solution correspondante de $(E_0)$,
$f_{0,i}(t)\varphi _{t_0,i}(X)$, est d\'efinie par
$$\Delta \varphi_{t_0,i}(X)=-\lambda _i(t_0)\varphi_{t_0,i}(X)$$et
$$f''_{0,i}(t)+\lambda _i(t_0)f_{0,i}(t)=0.$$
Prenons une de ces
solutions qu'on va noter$$f_0(t)\varphi_{t_0}(X);$$on a alors,
pour $0\leq t\leq t_0$
$$f''_{0}(t)+\lambda (t_0)f_{0}(t)=0.$$
Consid\'erons maintenant l'isom\'etrie qui identifie $(B(O,t_0),g_e)$ \`a $(B(O,1),t_0^2g_e)$ obtenue par l'homoth\'etie $X\rightarrow\frac{X}{t_0}$. Alors, si $\mu_0$ et $\mu(t_0)$ d\'esignent les valeurs propres ayant le m\^eme rang que $\lambda(t_0)$ associ\'ees respectivement au probl\`eme de Dirichlet sur $(B(O,1),g_e)$ et $(B(O,1),t_0^2g_e)$, on a $\lambda(t_0)=\frac{\mu_0}{t_0^2}$ $(\lambda(t_0)=t_0^2\mu(t_0),\;\nabla_{g_e}\varphi(\frac{X}{t_0})=\frac{1}{t_0}\nabla_{g_e}\varphi(X)$ et $\Delta_{t_0^2g_e}=\frac{1}{t_0^2}\Delta_{g_e}\varphi)$ et $f_0$ est la solution de l'\'equation
$$f''_{0}(t)+\frac{\mu_0}{t_0^2}f_{0}(t)=0, \;0\leq t\leq t_0\;(t_0>0).$$
La solution de cette \'equation est \'evidemment la fonction
p\'eriodique
$$
f_0(t)= f_0(0)\cos \frac{\sqrt \mu_0}{t_0}t+\frac{t_0}{\sqrt
\mu}f'_0(0)\sin \frac{\sqrt \mu_0}{t_0}t.
$$
La solution de l'\'equation $(E_0)$ correspondant \`a la valeur
propre $\lambda(t_0)=\frac{\mu_0}{t_0^2}$ est d\'efinie par
\begin{equation}\label{18}
E_{\mu_0}(t,X)=(f_0(0)\cos \frac{\sqrt \mu_0}{t_0}t+\frac{t_0}{\sqrt
\mu_0}f'_0(0)\sin \frac{\sqrt \mu_0}{t_0}t)\varphi_{t_0}(X)
\end{equation}
$\mbox{pour}\;X\in B(O,t_0),\;t_0>0\;\mbox{ et }\;0 < t\leq
t_0.$\\
De m\^eme, si $h_{\mu_0}(t)\psi_{\mu_0}(X)$ est la solution du probl\`eme de Dirichlet sur la boule unit\'e $(B(O,1),g_e)$ associ\'ee \`a la valeur propre $\mu_0$, on a, pour tout $t\leq t_0$ et $X\in B(O,t_0)$:
$$f_{\mu_0}(t)=h_{\mu_0}\left(\frac{t}{t_0}\right)\quad\mbox{et}\quad\varphi_{t_0,\mu_0}(X)=\psi_{\mu_0}\left(\frac{X}{t_0}\right).$$
En effet, l'\'equation
$$h''_{\mu_0}(t)+\mu_0h_{\mu_0}(t)=0$$
est \'equivalente \`a l'\'equation
$$f''_{\mu_0}(t)+\frac{\mu_0}{t_0^2}f_{\mu_0}(t)=0,$$
et par suite on a:
$$E_{\mu_0}(t,X)=\left(h_{\mu_0}(0)\cos\frac{\sqrt \mu_0}{t_0}t+\frac{1}{\sqrt
\mu_0}h'_{\mu_0}(0)\sin \frac{\sqrt \mu_0}{t_0}t\right)\psi_{\mu_0}\left(\frac{X}{t_0}\right).$$
$E_{\mu_0}(t,X)$ est donc une
fonction p\'eriodique de p\'eriode $T_0= 2\pi\frac{t_0}{\sqrt
\mu}$ et de fr\'equence $f(t_0)=\frac{1}{2\pi}\frac{\sqrt{\mu_0}}{t_0}$.\\
Remarquons qu'on pourrait aussi consid\'{e}rer
$E(t,X)=k_0(t)G_0(X)$ comme \'{e}tant une solution de
l'\'{e}quation (17). Une solution correspondant \`{a} une valeur
propre $\alpha(t_0)$ de l'op\'{e}rateur de Laplace-Beltarmi
$-\Delta_{g_{t_0}}$ sur la vari\'{e}t\'{e} reimannienne
$(B(O,t_0),g_{t_0})$ serait alors de la forme
$$E(t,X) = k_0(t) \theta_{t_0}(X)$$
avec $$k_0^{''}(t) + \alpha(t_0)k_0(t) = 0$$ et $$\Delta_{g_{t_0}}
\theta_{t_0}(X) = -\alpha(t_0) \theta_{t_0}(X)$$Cependant, dans ce
cas, on ne pourra pas ramener le probl\`{e}me consid\'{e}r\'{e}
\`{a} l'\'{e}tude du probl\`{e}me de Dirichlet sur l'espace
$(B(O,1),t_0^2g_1)$ \`{a} moins que l'application $X \longrightarrow
\frac{X}{t_0}$ ne soit une isom\'{e}trie de $(B(O,t_0),g_{t_0})$
sur $(B(O,1),t_0^2g_1),$ ce qui est, dans les meilleurs des cas, une
grossi\`{e}re approximation.\\\\
En fait les solutions $E_\mu$ ci-dessus ne peuvent pas \^etre
assimil\'ees \`a des solutions de notre \'equation de la
mati\`ere-\'{e}nergie dans l'Univers dynamique $(U(t))_{t > 0}$,
i.e. l'\'equation d'\'{e}volution $(E^*)$, que sur un intervalle
de temps $t_1\leq t\leq t_0$ avec $t_0>t_1>0$ et $t_1$
suffisemment relativement proche de $t_0$ de telle sorte que les
fonctions propres $\varphi _t(X)$ et les valeurs propres $\lambda
(t)$ correspondant \`a $t\in [t_1,t_0]$ puissent \^etre
consid\'er\'ees comme \'etant des bonnes approximations de
$\varphi_{t_0}(X)$ et de $\lambda (t_0)$ respectivement. De plus,
pour obtenir des bonnes approximations p\'eriodiques de
l'\'equation $(E^*)$ sur $(U(t))_{t_1\leq t \leq t_0}$, il faut
que la p\'eriode $T_0= 2\pi\frac{t_0}{\sqrt \mu}$ soit
significativement inf\'erieure \`a $t_0-t_1$; donc que $\mu$ soit
significativement sup\'erieure \`a
$4\pi^2(\frac{t_0}{t_0-t_1})^2$.\\
En donnant \`a $t_0$ des valeurs croissantes convenables $t_i$, on
obtient des solutions approch\'ees p\'eriodiques (de p\'eriodes
$T_i=2\pi\frac{t_i}{\sqrt \mu}$) de notre probl\`eme sur des
couronnes juxtapos\'ees de $B(O,t)$ pour \emph{t} = $\mbox{sup}_i$
\emph{t}$_i$.\\En rempla\c cant dans la relation (18) $t$ par
$t_0$ et en la r\'e\'ecrivant en fonction de la variable $t$ au
lieu de $t_0$, on obtient la solution $E_{\mu}$ de $(E^*)$
d\'efinie sur $B(O,t)$, pour $t>0$, par
$$E_{\mu}(t,X)=f_{\mu}(t)\varphi_{t,\mu}(X)=(f_{\mu}(0)\cos \sqrt
\mu+\frac{t}{ \sqrt{\mu}}f'_{\mu}(0)\sin \sqrt
\mu)\varphi_{\mu}(X) \;\;\;\;(18^{'})$$ 
o\`u $\varphi_{\mu}(X)=\psi_{\mu}(\frac{X}{t})$, $f_{\mu}(0)=h_{\mu}(0)$ d\'epend de $\mu$ et $f'_{\mu}(0)$
d\'epend de $\mu $ et de $t$. Cette solution peut \^etre
approch\'ee, sur des couronnes appropri\'ees $B(O,t)\backslash
B(O,t')$, par des fonctions p\'eriodiques de p\'eriodes
$T(t)=2\pi\frac{t}{\sqrt \mu}$ et de fr\'equences
$f(t)=\frac{1}{2\pi}\frac{\sqrt \mu}{t}$. $E_{\mu}$ est donc une
pseudo-onde (qu'on va appeler abusivement une onde) de
pseudo-p\'eriode $T(t)=2\pi\frac{t}{\sqrt \mu}$ et de
pseudo-fr\'equence $f(t)=\frac{1}{2\pi}\frac{\sqrt \mu}{t}$
respectivement (toutes les deux d\'ependant du temps $t$).\\
Ainsi, pour pouvoir assimiler ces solutions \`a des ondes sur un
intervalle de temps significatif, posons $t=e^{\alpha}$,
$\mu=4\pi^2 e^{2\beta}$ et remarquons que l'on a alors
$T=e^{\alpha-\beta}$ et par suite il faut prendre
$e^{\alpha-\beta}<< e^{\alpha}$. On doit donc prendre $\beta>>0$
et par suite les valeurs propres
$\mu>>0$.\\
Lorsqu'il s'agit d'une trajectoire libre $X(t)$ sur un intervalle
de temps donn\'e, i.e.
$$\widetilde{\Gamma}(t)=\nabla_{X'(t)}^{g_t}X'(t)=0\;\mbox{ pour }\; t\in I,$$
alors l'onde $E_{\mu}(t,X(t))$ correspondant \`a une valeur propre
$\mu$ v\'erifie
$$ - \nabla
^{g_t}E_{\mu}(t,X(t))=F_{g_t}(X(t)) = \widetilde{\Gamma}(t) = 0.$$
Or, lorsque la m\'{e}trique est euclidienne, le principe d'inertie
de Newton stipule que la trajectoire \emph{X}(\emph{t}) d'une
particule quelconque est lin\'{e}aire et uniforme (i.e. $X(t)$ est
une g\'{e}od\'{e}sique et ${\parallel X^{'}(t)\parallel}_{g_e}$ =
$v$ = cte) si et seulement si le champ de force qui agit sur la
particule est nul le long de la trajectoire \emph{X}(\emph{t}),
i.e.
$$ F_{g_e}(t) = - \nabla^{g_e} E(t, X(t)) = \nabla^{g_e}_{X^{'}(t)}
X^{'}(t) = X^{''}(t) = 0$$et l'\'{e}nergie de cette particule est
alors conserv\'{e}e le long de cette trajectoire i.e.
$$ E(t,X(t)) = \mbox{ cte }.$$

Dans le cas de notre m\'{e}trique physique $g_t$, ce m\^{e}me
principe se g\'{e}n\'{e}ralise de la mani\`{e}re suivante.\\

La trajectoire \emph{X}(\emph{t}) d'une particule est une
g\'{e}od\'{e}sique par rapport \`{a} la m\'{e}trique $g_t$ (avec
${\parallel X^{'}(t)\parallel}_{g_t}$ = cte) si et seulement si
$$ - \nabla^{g_t} E(t, X(t)) = F_{g_t}(t) = \nabla^{g_t}_{X^{'}(t)} X^{'}(t)=
{\widetilde{\Gamma}}(t) = 0$$et l'\'{e}nergie ponctuelle de cette
particule est alors conserv\'{e}e le long de cette
g\'{e}od\'{e}sique i.e.
$$E(t,X(t)) = \mbox{ cte }.$$
Dans notre situation, on a donc
$$ E_\mu(t,X(t)) = f_\mu(t)\varphi_{t,\mu}(X(t)) =
e(\mu)$$(o\`{u} $e$($\mu$) est une constante d\'{e}pendant de
$\mu$) et
$$\triangle E_\mu(t,X(t)) = f_\mu(t) \frac{\mu}{t^2}
\varphi_{t,\mu}(X(t)) = \frac{\mu e(\mu)}{t^2}.$$Pour les
particules originelles se propageant \`{a} la vitesse $1$ (ou $v(t)$), on a
$X(t)\in S(O,t)$ (ou $X(t)\in S(O,R(t))$) et
$E_\mu$(\emph{t},\emph{X}(\emph{t})) = 0; ce sont les ondes
\'{e}lectromagn\'{e}tiques originelles qui ont fa\c{c}onn\'{e} le
demi - c\^{o}ne de l'espace - temps. Pour une particule
mat\'{e}rielle se propageant le long d'une g\'{e}od\'{e}sique (par
rapport \`{a} $g_t$) \`{a} une vitesse \emph{v} = ${\parallel
X^{'}(t)\parallel}_{g_t} < $1, on a bien
$$E_\mu(t,X(t)) = e_0(\mu) > 0.$$
Signalons que l'\'{e}nergie d'un photon autre que les photons
originels est \'{e}galement positive.
\\
\subsection*{L'\'{e}nergie de Planck - Einstein}

En adaptant maintenant le principe ondulatoire de Planck-Einstein
\`a notre situation on doit avoir, pour tout point mat\'eriel ou
immat\'eriel $X$ se d\'epl\c cant librement dans $(U(t))_{t > 0}$:
$$E_{\mu}(t,X(t))=h_{\mu}(t)f_{\mu}(t)=h_{\mu}(t)\frac{1}{2\pi}
{\frac{\sqrt \mu}{t}}=\overline{h}_{\mu}(t)
{\frac{\sqrt\mu}{t}}$$
o\`u $f_\mu$(\emph{t}) d\'{e}signe ici la
fr\'{e}quence et $\overline{h}_{\mu}(t)$ remplace en quelque sorte
la constante de Planck. Or cette relation implique que l'on a
$$\overline{h}_{\mu}(t)=tc(\mu)$$
o\`u $c(\mu)=\frac{e(\mu)}{\sqrt{\mu}}$ est une
constante ne d\'ependant que de $\mu$; d'o\`u
$$E_{\mu}(t,X(t))=c(\mu)\sqrt {\mu} = e(\mu).\;\;\;\;\;\;\;\;\;\;\;\;\;\;(e)$$
\textbf{Remarques:} $1^\circ$) Signalons quand m\^{e}me que, comme
on va le voir dans le paragraphe suivant, cette constante
d\'{e}pend en fait du temps lorsqu'il s'agit d'un tr\`{e}s long
intervalle d'\'{e}volution. Cette d\'{e}pendance est d\^{u}e au
refroidissement perp\'{e}tuel de l'Univers cosmique qui n'est
perceptible qu'\`{a} une grande \'{e}chelle temporelle. De plus, on va montrer au paragraphe $12$ que, si $\rho(t)$ d\'esigne la densit\'e moyenne de l'\'energie dans l'Univers et $f(t)$ d\'esigne la fr\'equence moyenne de la mati\`ere-\'energie du cosmos, alors $E(t,X(t))=:e(t)=h(t)f(t)=\rho(t)\propto\frac{1}{t^3}$ et par cons\'equent $f(t)\propto\frac{1}{t^4}$.\\
$2^\circ$) Contrairement \`{a} la constante de Planck (qui est
g\'{e}n\'{e}ralement suppos\'{e}e constante), notre constante
${\overline{h}}_{\mu}(t)$ est proportionnelle au temps $t$.\\\\

Par ailleurs, consid\'{e}rons une agglom\'{e}ration mat\'{e}rielle
occupant un domaine $D_t$ de \emph{B}(\emph{O},\emph{t}). Celui-ci
est subdivis\'{e} en une r\'{e}union de sous-domaines $D_{t,n}$
sur lesquels sont d\'{e}finies des distributions
\'{e}nerg\'{e}tiques $E_n(X_t)$ qui co\"{\i}nc-\\ident sur chaque
domaine occup\'{e} par une particule mat\'{e}rielle fondamentale
avec une distribution mat\'{e}rielle constante $m_n(X_t)$ =: $m_n$
(les particules fondamentales seront classifi\'{e}es au paragraphe
8). Les distributions \'{e}nerg\'{e}tiques $E_n(X_t)$,
d\'{e}finies sur les autres sous-domaines, seront \'{e}galement
suppos\'{e}es constantes. On a donc
$$D_t=\bigcup_{1 \leq n \leq N}D_{t,n}$$ avec $$ E_n(X_t) = e_n \sim m_n \mbox { pour } X_t \in
D_{t,n}.$$ L'\'{e}nergie du domaine $D_t$ sera donc
$$E(D_t)=\sum _{n}vol(D_{t,n})m_n=:\sum
_{n}V_n(t)m_n,$$ o\`{u} \emph{vol} ($\emph{D$_{t,n}$}$)
d\'{e}signe ici le volume euclidien du sous-domaine
\emph{D$_{t,n}$}.
 \\ Evidemment, m\^eme si
$D_t$ constitue un syst\`eme isol\'e tel que \emph{vol}($D_t$)
reste constant, alors les $V_n(t)$ \'evoluent avec le temps. Ceci
est d\^u \`a un grand nombre de ph\'enom\`enes dynamiques
\'evolutifs: transformations mati\`ere$-$\'energie pure,
radiations de tout genre, d\'esint\'egration, collision, fission,
fusion et interactions
chimiques, nucl\'eaires et thermiques...\\\\
Dans l'Univers dynamique, il faut tenir compte de l'\'energie
cin\'etique de la mati\`ere en mouvement. Or on ne peut parler de
masse inertielle, d'\'energie potentielle, d'\'energie cin\'etique
ou de quantit\'e de mouvement (i.e $m v$) que d'un point
mat\'eriel ou d'un domaine mat\'eriel circulant \`a des vitesses
inf\'erieures \`a 1. L'\'energie cin\'etique d'une particule
mat\'erielle effectuant la trajectoire $X(t)$ est en effet
$\frac{1}{2}m_1(X(t))X'(t)^2$ o\`{u} $m_1(X(t)) = \gamma_1(X(t))
m_0$, $m_0$ \'{e}tant la masse au repos de la particule et
$\gamma_1(X(t))$ \'{e}tant un facteur qui d\'{e}cro\^{i}t de 1
\`{a} 0 lorsque la vitesse de la particule cro\^{i}t de 0 \`{a} 1,
exprimant ainsi la perte de masse subie par la particule par les
radiations produites sous l'effet de l'acc\'{e}l\'{e}ration. Ce
facteur peut \^{e}tre d\'{e}termin\'{e} th\'{e}oriquement ou
exp\'{e}rimentalement pour les diff\'{e}rents types de particules.
La masse inertielle ou l'\'energie potentielle de masse d'un
domaine mat\'eriel $D_t$ \`a l'instant $t$ est donn\'ee par
$$\rho_t(D_t)=\int_{D_t}m_1(X_t)dX_t.$$ Son
\'energie cin\'etique est donn\'ee par
$$\frac{1}{2}\int_{D_t}m_1(X_t)v(X_t)^2dX_t.$$ Lorsqu'il s'agit de
domaines vides ( ne contenant pas de mati\`ere) et travers\'es par
des radiations comme les ondes \'electro-magn\'etiques (rayons
lumineux visibles ou invisibles, rayons X, rayons $\gamma$) on ne
peut parler ni de masse inertielle ni d'\'energie potentielle, de
quantit\'e de mouvement $m v$ ou d'\'energie cin\'etique
$\frac{1}{2}m v^2$. On ne peut parler que de l'\'energie de l'onde
propag\'ee, du faisceau d'ondes et du photon. N\'{e}anmoins, on
peut d\'{e}finir (dans le cadre de notre mod\`{e}le) le vecteur
pulsation $\overrightarrow{p}$ associ\'{e} \`{a} une onde
\'{e}lectromagn\'{e}tique (ou \`{a} un photon) par
$\overrightarrow{p(t)}=\frac{1}{c^2}E(t,X(t))\overrightarrow{X'(t)}$
qu'on pourrait appeler le
$^{\mbox{\guillemotleft}}$momentum$^{\mbox{\guillemotright}}$ du
photon et on a alors $p(t)c =E(t)$ ou $p c =E$ en accord avec la
th\'eorie de la relativit\'e puisqu'ici $|| X^{'}(t)||_{g_e} =c$.
Ainsi l'\'energie cin\'etique d'un point mat\'eriel $X$, tel que
$m_0(X)=m_0$ et effectuant une trajectoire $X(t)$, est
$\frac{1}{2}m_1(X(t))X'(t)^2$ et son \'energie totale est
$m(t)c^2+\frac{1}{2}m_1(t)X'(t)^2$ tant que sa vitesse est $<c$ avec $m(t)=\gamma(t)m_0$. Lorsqu'on tra\^ite l'\'equation $(17)$ (avec la m\'etrique $g_t$ \`a la place de la m\'etrique $g_e$) on doit remplacer la vitesse euclidienne par $\|X'(t)\|_{g_t}$ et alors nous obtenons $\|X'(t)\|_{g_t}\simeq c$ et $p(t)c\simeq E(t)$ pour les ondes \'electromagn\'etiques sauf \`a l'int\'erieur d'un trou noir $B\setminus I$ o\`u l'on a $\|X'(t)\|_{g_t}=0$. \\
Pour une particule ponctuelle mat\'{e}rielle, ayant une masse au
repos $m_0$, la quantit\'{e} $m_1(t) = \gamma_1(t) m_0$ va
\^{e}tre appel\'{e}e la masse r\'{e}duite et $m(t) = \gamma (t)
m_0 = (1 - \frac{v^2(t)}{c^2})^{-\frac{1}{2}} m_0$ est la masse
apparente de la particule en mouvement. N\'eamoins lorsqu'il
s'agit d'un trou noir \emph{B} d'\'energie de masse initiale
\emph{E} circulant \`{a} une vitesse quelconque $v,$ on peut
consid\'erer que son
\'energie totale est alors $\frac{E}{c^2}(c^2+\frac{1}{2}v^2)$ d'apr\`{e}s la fameuse relation d'Einstein $E = mc^2$.\\
Cependant, la distribution $E_t(X)$ est enti\`{e}rement
d\'etermin\'ee par la distribution $m_t(X)$ en y ajoutant les
\'{e}nergies des trous noirs. Il en est ainsi pour le champ de
forces global $F_{g_e}(X)=\nabla^{g_e}E_t(X)$ et de
l'acc\'el\'eration $\Gamma(X(t))=X''(t)$ pour tout mouvement
libre. Donc la m\'etrique $g_t$, qui est li\'ee intrins\`equement
\`a la distribution $E_t(X)$ et qui v\'erifie
$\widetilde{\Gamma}(X(t))=\nabla _{X'(t)}^{g_t}X'(t)=0$ pour tout
mouvement libre, tient compte de toutes les manifestations de la
mati\`ere y compris des champs \'electromagn\'etiques, de toute
sorte d'interactions et des forces de liaison qui en r\'esultent
et non pas seulement des
forces gravitationnelles.\\
Notons que l'\'energie cin\'etique d'un syst\`eme n'est pas
n\'ecessairement conserv\'ee ni globalement ni localement comme le
montre, par exemple, la transformation d'une partie de l'\'energie
cin\'etique d'un syst\`eme en chaleur lors d'une collision. Par
contre, le principe de la conservation globale de la quantit\'e de
mouvement ou "momentum" d'un syst\`eme isol\'e (comme l'Univers
tout entier) est valide. On a donc, \`a tout instant $t$:
$$\int _{B(O,t)} \frac{1}{c^2}E_t(X_t)\overrightarrow{v}(X_t)dX_t = 0.$$
En particulier, le centre de gravit\'e de
l'Univers est fixe. Cette relation s'\'ecrit, pour
$B$(\emph{O},\emph{t}) = $\bigcup_{n}D_n$(\emph{t}) o\`{u} chaque
$D_n(t)$ est caract\'{e}ris\'{e} par sa densit\'{e} mat\'erielle $m_n:=\frac{e_n}{c^2}$,
$$\sum_{n}\int _{D_n(t)}m_n \overrightarrow{v} (X_t)dX_t = 0.$$Lorsque, pour un domaine
isol\'e $D(t)=\bigcup_{n}D_{n}(t)$, on attribue au centre de
gravit\'e $G_n(t)$ de chaque $D_n(t)$ le vecteur vitesse
r\'{e}sultant $\overrightarrow{v_n}(t)$, on obtient
$$\sum_nV(D_n(t))\frac{E_{t,n}}{c^2}(G_n(t))\overrightarrow{v_n}(t) = \overrightarrow{a}$$ou
$$\sum_nV_n(t)m_n\overrightarrow{v_n}(t) = \overrightarrow{a}.$$
Lorsqu'il s'agit d'une particule de masse $m$ soumise \`a des
forces ext\'erieures constantes, occupant un domaine
$D=\bigcup_nD_n(t)$ et circulant \`a la vitesse
$\overrightarrow{v}(t)$, on a:
$$\sum_nV_n(t)m_n \overrightarrow{v_n}(t) = m \overrightarrow{v}(t).$$Pour un atome, par exemple, de masse
$m$ et de vitesse $\overrightarrow{v}$ ayant $k_1$ \'electrons et
$k_2$ quarks occupant respectivement les volumes $V_{1,i}$ et
$V_{2,j}$ et ayant des vitesses respectives
$\overrightarrow{v}_{1,i}$ et $\overrightarrow{v}_{2,j}$, on a
$$\sum _{i=1}^{k_1}V_{1,i}m_{1,i} \overrightarrow{v}_{1,i}(t)+
\sum _{j=1}^{k_2}V_{2,j}m_{2,j} \overrightarrow{v}_{2,j}(t) = m
\overrightarrow{v}.$$ Le syst\`{e}me solaire ob\'{e}\"{\i}t
quasiment \`{a} ce m\^{e}me sch\'{e}ma.\\
En effet, supposons que le syst\`{e}me constitu\'{e} du soleil et
de \emph{N} plan\`{e}tes est isol\'{e} (ce qui n'est pas le cas
puisque le syst\`{e}me solaire appartient \`{a} la voie
lact\'{e}e) et d\'{e}signons par $\overrightarrow{v}$ le vecteur
vitesse absolu (i.e. par rapport \`{a} un rep\`{e}re virtuel fixe)
du centre de gravit\'{e} de ce syst\`{e}me.\\De m\^{e}me
d\'{e}signons par $\overrightarrow{v}_i$, pour \emph{i} =
1,...,\emph{N}, le vecteur vitesse absolu du centre de gravit\'{e}
de la $i^{\emph{\`{e}me}}$ plan\`{e}te et par
$\overrightarrow{v}_0$ celui qui est associ\'{e} au soleil. Toutes
les trajectoires c\'{e}lestes sont des g\'{e}od\'{e}siques par
rapport \`{a} la m\'{e}trique cosmologique $g_t$, i.e.
$\nabla_{\overrightarrow{v}_i(t)}^{g_t} \overrightarrow{v}_i (t)$
= 0 pour tout \emph{i} = 0,1,...,\emph{N}, et on a
${||\overrightarrow{v}_i||}_{g_t} = v_i$ o\`{u} $v_i$ est une
constante.\\Ainsi, on a
$$\sum_{i=0}^{N}m_i \overrightarrow{v}_i = m_0
\overrightarrow{v}_0 + \sum_{i=1}^{N} m_i \overrightarrow{v}_i =
(\sum_{i=0}^{N} m_i) \overrightarrow{v} = m_0 \overrightarrow{v} +
(\sum_{i=1}^{N} m_i)\overrightarrow{v}.$$Or, pour tout \emph{i} =
1,...,\emph{N}, on a $$\overrightarrow{v}_i = \overrightarrow{v} +
\overrightarrow{u}_i$$ o\`{u} $\overrightarrow{u}_i$ est le
vecteur vitesse relatif par rapport au soleil; d'o\`{u}
$$m_0 \overrightarrow{v_0} + \sum_{i=1}^{N} m_i
(\overrightarrow{v}+\overrightarrow{u_i}) = m_0 \overrightarrow{v}
+ (\sum_{i=1}^{N}m_i) \overrightarrow{v}$$ Le centre de
gravti\'{e} du syst\`{e}me \'{e}tant quasiment le m\^{e}me que
celui du soleil, on a $\overrightarrow{v}_0 \simeq
\overrightarrow{v}$ et par suite on obtient $$\sum_{i=1}^{N}m_i
\overrightarrow{u}_i = 0.$$ Notons aussi, qu'\`a part l'Univers
tout entier, nul autre syst\`eme n'est durablement isol\'e (y
compris les galaxies, les trous noirs et \'evidemment les
syst\`emes \`a l'\'echelle plan\'etaire). N\'eamoins, c'est la
distribution (essentiellement locale) $E_t(X)$ qui r\'egit le
mouvement libre dans tous les syst\`emes locaux et microlocaux.\\
Ainsi au niveau d'un atome, par exemple, le mouvement libre des
\'electrons dans leurs orbites respectifs, ou plut\^{o}t la
trajectoire d'un point mat\'eriel de chaque \'electron, est r\'egi
par l'acc\'el\'eration $\Gamma(t)=-\nabla^{g_e}E_t(X(t))$ et
v\'erifie $\widetilde{\Gamma}(t)=\nabla _{X'(t)}^{g_t}X'(t)=0.$
Toutefois, \`a la suite d'un apport \'energ\'etique ext\'erieur
(thermique ou \'electro-magn\'etique par exemple), l'\'electron
subit des transformations \'energ\'etiques comme le changement de
son niveau d'\'energie (changement d'orbite) ou m\^{e}me la
s\'eparation de l'atome originel. De m\^{e}me signalons que dans
le cas contraire, l'\'emission de
photons assure le principe de la conservation de l'\'energie.\\\\
Ceci reste valable pour un noyau qui est, de plus, soumis aux
forces d'interaction nucl\'eaire qui, \`a l'aide d'une stimulation
externe ou d'un processus naturel, conduit \`a des transformations
mat\'erio-\'energ\'etiques comme la d\'esint\'egration, la
fission, la fusion, l'excitation et les radiations et \`a des
r\'eactions chimiques conduisant toutes \`a des transferts ou
\`{a} des lib\'erations d'\'energie ob\'e\"{\i}ssant au principe
de la conservation de l'\'energie, au principe du meilleur
\'equilibre \'energ\'etique possible et au principe m\'ecanique de
la moindre
action, mais aussi au principe d'exclusion de Pauli.\\

Signalons enfin que chaque solution
\emph{E}(\emph{t},\emph{X}(\emph{t})) de l'\'{e}quation ($E^*$) se
propageant suivant $X(t)$ est de la forme $E
_{\mu_0}(X(t))=f_{\mu_0}(t)\varphi_{t,\mu_0}(X(t))$ et non pas
\'egale \`a une combinaison lin\'eaire de telles solutions, comme
on peut le v\'erifier \`{a} l'aide du ph\'enom\`ene de la
diffraction de la lumi\`ere. Un rayon lumineux uniphotonique ne
peut pas se diffracter en plusieurs rayons caract\'{e}ris\'{e}s
par plusieurs photons diff\'erents de celui du
rayon incident.\\
La relation $f_{\mu}(t)=\frac{1}{2\pi}\frac{\sqrt \mu}{t}$ montre
que la fr\'equence ondulatoire de la mati\`ere-\'energie est une
fonction qui cro\^it avec $\mu$ et d\'ecro\^it avec $t$. Donc,
juste apr\`es le Big Bang (pour $t$ assez petit), toutes les
propagations ont un caract\`ere ondulatoire d'autant plus
prononc\'e que lorsque $\mu$ est assez grand. Pour $t$ assez
grand, seuls les $\mu$ d'ordre tr\`es grand donnent lieu \`a un
caract\`ere ondulatoire bien perceptible. Cependant, on a
$v(t)=\lambda(t)f_{\mu}(t)$ o\`u $\lambda(t)$ est la longueur
d'onde et $v(t)<1$. Or pour $t<<1$, on a $f_{\mu}>>1$ pour tout
$\mu$ et par suite on a $\lambda<<1$. On peut donc concevoir
intuitivement que lorsque $t=0$ (c.\`a.d. avant le Big Bang), on a $\lambda=0$ et il n'y a pas
de propagation ni mat\'erielle ($v<1$) ni immat\'erielle ($v=1$).
Par ailleurs, lorsque $v_e(t)$ est la vitesse euclidienne de propagation d'une
onde \`a partir d'un temps infinit\'esimal $t\sim0$, on a
$$v_e(t)=\frac{1}{2\pi}\frac{\sqrt{\lambda^2(t)\mu}}{t}$$et par
suite, pour $t\gg0$ o\`u $v_e\sim 1$, on a
$$\frac{\lambda^2(t)\mu}{t^2}\leq 4\pi^2.$$
Ainsi, pour les ondes de vitesse $v_e\sim1$ (lumi\`ere, rayons X,
rayons $\gamma$), on a $\lambda^2(t)\mu=4\pi^2t^2$ et
$\displaystyle\lambda(t)=\frac{2\pi t}{\sqrt \mu}$. N\'eamoins, pour les premiers originels photons on a $E(t,X(t))=h_\mu(t)f_\mu(t)=0$, $v(t)\neq0$ et $\displaystyle\lim_{t\rightarrow0}\lambda(t)=0$ en plus que $\displaystyle\lim_{t\rightarrow0}h_{\mu}(t)=0$ et $\displaystyle\lim_{t\rightarrow0}f_{\mu}(t)=+\infty$.\\
Pour revenir \`a notre point de d\'epart, ce sont ces ondes l\`a
qui cr\'eent, \`a partir du Big Bang, l'espace g\'eom\'etrique
dont l'expansion se produit \`a une vitesse croissante $v(t)$ qui est actuellement tr\`es proche de $1$ et qui tend
d\'{e}finitivement vers 1. L'Univers mat\'eriel corpusculaire
s'\'elargit \`a une vitesse inf\'erieure \`a 1 et dont l'
"acc\'el\'eration", assujettie \`a la possibilit\'e de la
perception mat\'erielle, reste \`a d\'eterminer d'une fa\c con de
plus en plus pr\'ecise et doit
th\'eoriquement tendre \`{a} \^etre nulle.\\\\

\textbf{Remarque:} Dans notre mod\`ele, on a repr\'esent\'e toute
\'energie concentr\'ee en un point par $e(I)\delta_I$ ou par
$m(I)\delta_I$. Si on avait suppos\'e, d'une facon \'equivalente,
que la masse-\'energie originelle $M_0\sim E_0$ est concentr\'ee sous forme d'une boule
de mati\`ere-\'energie ayant un volume extr\^{e}mement r\'{e}duit et une
densit\'e de masse-\'energie tr\`es importante (du genre d'\'{e}toile \`{a}
neutron ultradense), alors ceci n'aurait pas, du point de vue
math\'ematique, affect\'e notre mod\'elisation \`a cause de
l'\'etendue pharamineuse de l'espace g\'eom\'etrique virtuel.
Ainsi, ce fait nous permet d'\'{e}viter d'avoir recours
\`{a} la notion d'infinit\'{e} (densit\'{e} infinie, courbure infinie, pression infinie et temp\'{e}rature infinie).\\\\

\section{Repercussions sur la Physique moderne}
\subsection*{Temp\'{e}rature et pression}

Le grand absent de notre \'{e}tude jusqu'\`{a} pr\'{e}sent est le
facteur temp\'{e}rature. Pourtant, la temp\'{e}rature est une
caract\'{e}ristique inh\'{e}rente \`{a} l'expansion: L'Univers est
en expansion et en refroidissement permanents. Par ailleurs, la
temp\'{e}rature est indissociable de toutes les formes
d'\'{e}nergie: La chaleur, les radiations (via le spectre
thermal), l'\'{e}lectricit\'{e}, la fourniture de l'\'{e}nergie
interne des galaxies (via la fusion et les fluctuations de la
pression interne) et surtout, faut-il le rappeler, la
temp\'{e}rature caract\'{e}rise l'\'{e}nergie cin\'{e}tique
moyenne des mol\'{e}cules dans les \'{e}tats d'\'{e}quilibre
thermaux:\\
$$<E_{c}> = \frac{3}{2}kT$$
Ainsi, on peut dire bri\`{e}vement que la temp\'{e}rature
intervient dans tous les \'{e}tats d'\'{e}quilibre
\'{e}nerg\'{e}tique des syst\`{e}mes et les fa\c{c}onne.\\
Commen\c{c}ons donc par pr\'{e}ciser que la relation obtenue aux
paragraphes pr\'{e}c\'{e}dents pour caract\'{e}riser les
trajectoires $X(t)$ consid\'{e}r\'{e}es comme libres (i.e.
$\nabla^{g_t}_{X'(t)}X'(t)$ = 0), \`{a} la fois pour les
particules mat\'{e}rielles ponctuelles et pour les radiations (les
photons), et qui s'\'{e}crit\\
$$E_\mu(t,X(t)) = e(\mu)$$
n'est en fait valable que sur un petit intervalle de temps o\`{u}
l'on peut consid\'{e}rer la temp\'{e}rature comme \'{e}tant
constante. En effet, bien que la m\'{e}trique $g_t$ tient compte
implicitement (par d\'{e}finition) de la temp\'{e}rature ambiante
$T(t),$ alors il faut reconna\^{i}tre que la relation
pr\'{e}c\'{e}dente devrait s'\'{e}crire sous la forme\\
$$E_\mu(t,T(t),X(t)) = e(\mu,T(t))$$
\`{A} la d\'{e}pendance de l'\'{e}nergie \emph{E$_\mu(t,X(t))$} =
\emph{h$_\mu(t)f_\mu(t)$ de $\mu$}, il faut ajouter
n\'{e}cessairement sa d\'{e}pendance de $T(t)$ \`{a} travers la
d\'{e}pendance de la m\'{e}trique elle-m\^{e}me de $T(t).$ D'autre
part, l'\'{e}nergie $E(t,X(t)),$ v\'{e}hicul\'{e}e par les
radiations qui nous arrivent \`{a} partir des temps et des
distances lointains (et originels), est att\'{e}nu\'{e}e, non
seulement \`{a} cause des collisions, mais aussi \`{a} cause du
refroidissement de l'Univers ondulatoire. L'affaiblissement des
fr\'{e}quences $f_\mu(t)$ (i.e. le rallongement des longueurs
d'onde $\lambda_\mu(t)$)est contrebalanc\'{e} par l'accroissement
des $h_\mu(t).$ D'autre part, il est \'{e}vident que si les ondes
se propagent suivant les trajectoires $X(t)$ avec une \'{e}nergie
$E(t,X(t)) = \mbox{cte},$ alors on
ne pourrait en aucun cas avoir, pour tout $t\gg0$, \\
$$\int _{B(O,t)}E_t(X_t)\,dX_t \;=\; \mbox{cte}.$$
Si on adapte le mod\`{e}le du gaz parfait \`{a} l'Univers tout
entier, on peut supposer que l'on ait en permanence \\
$$P(t)V(t) = K(t)T(t);$$\\
ce qui implique, pour $t>0$ suffisamment grand\\
$$P(t)t^3 = K^{'}(t)T(t).$$\\
Notons qu'au voisinage de 0 la situation pourrait ne pas \^{e}tre
la m\^{e}me, puisque, pour $t$ assez petit, le volume de
$B_e(O,R(t))$ pourrait ne pas \^{e}tre proportionnel \`{a} $t^3.$
Ceci est d\^{u} au fait que la vitesse de propagation des
radiations v\'{e}rifiant $\nabla^{g_t}_{X'(t)}X'(t)$ = 0 pourrait
\^{e}tre originellement plus petite que 1 (vitesse de la
lumi\`{e}re dans le vide). En effet, la m\'{e}trique $g_t$
contracte les distances d'une mani\`{e}re tr\`{e}s substancielle
au d\'{e}but de l'expansion \`{a} cause de l'importance des forces
gravitationnelles et des autres forces d'interaction pr\`{e}s de
l'origine et de l'\'{e}normit\'{e} de la densit\'{e}
\'{e}nerg\'{e}tique et des intensit\'{e}s de la temp\'{e}rature et
de la pression.\\
Signalons que, pour \'{e}viter le recours \`{a} la notion d'une
\'{e}nergie gigantesque concentr\'{e}e en un point \`{a} l'origine
du temps $t=0$ avec une pression et une temp\'{e}rature toutes les
deux infinies, on pourrait concevoir que notre \'{e}tude peut
\^{e}tre refaite en consid\'{e}rant l'Univers quasi-originel \`{a}
partir du moment o\`{u} il est r\'{e}duit \`{a} une boule de rayon
euclidien $r_0$ suffisamment petit et \'{e}tudier l'expansion
\`{a} partir d'un temps $t_0>0$ assez petit. L'Univers sera
alors consid\'{e}r\'{e}, \`{a} l'instant $t_0,$ comme \'{e}tant
une boule au sein de laquelle les radiations sont
caract\'{e}ris\'{e}es par une pression et une temp\'{e}rature
extr\^{e}mement grandes (tout en \'{e}tant finies). Cette
situation \'{e}voluera avec le d\'{e}roulement du temps vers un
\'{e}tat qualifi\'{e} de soupe de quarks et de leptons avant la
formation des hadrons suivie de celle des nucl\'{e}ons, des atomes
et des galaxies marquant le passage d'un \'{e}tat \`{a} dominance
radiationnelle \`{a} l'\'{e}tat \`{a} dominance mat\'{e}rielle. La
temp\'{e}rature a bien \'{e}videmment jou\'{e} (avec la pression)
un r\^{o}le essentiel au cours de cette \'{e}volution qui a
conduit \`{a} la situation actuelle caract\'{e}ris\'{e}e par une
temp\'{e}rature moyenne approximative de 2,74 K. Pourtant, un
grand nombre de moyens techniques, exp\'{e}rimentaux et
th\'{e}oriques nous permettent d'aller de plus en plus loin dans
nos investigations afin de d\'{e}couvrir de plus en plus
profond\'{e}ment les \'{e}tats originels de notre Univers et des
lois qui r\'{e}gissent
son \'{e}volution.\\
Montrons, sur quelques exemples, que l'on peut retrouver un
certain nombre de r\'{e}sultats confirm\'{e}s en Physique moderne
en se basant sur notre mod\`{e}le et sans avoir recours ni \`{a}
la partie erronn\'{e}e du deuxi\`{e}me postulat de la
relativit\'{e} restreinte (i.e. l'ind\'{e}pendance de la vitesse
des ondes \'{e}lectromagn\'{e}tiques vis \`{a} vis de la vitesse
du
r\'{e}f\'{e}rentiel inertiel) ni au principe d'incertitude.\\

\subsection*{Remarques sur quelques formules relativistes}

Dans la suite de ce paragraphe, on va souligner des remarques et
\'{e}tablir des propri\'{e}t\'{e}s et r\'{e}sultats bas\'{e}s sur
la r\'{e}futation de l'interpr\'{e}tation erron\'{e}e du
deuxi\`{e}me postulat de la relativit\'{e} restreinte et la notion
relativiste de l'espace - temps ainsi que sur la canonicit\'{e}
des \'{e}quations de Maxwell et celle du principe de la constance
de la vitesse de la lumi\`{e}re obtenue en utilisant les
d\'{e}rivations $\frac{d_1}{dt}$ et $\frac{d_*}{dt}$ qui tiennent
compte du mouvement du rep\`{e}re
inertiel (voir paragraphes 2, 3 et 4).\\

\'{E}tant donn\'{e} que la vitesse est, par nature, une variable
continue sur $]$0,1$]$, nous avons opt\'{e} pour ne pas faire une
distinction nette, selon leur vitesse, entre particules
relativistes et non relativistes, concernant soit leur \'{e}nergie
soit leur "momentum". \`{A} partir de quelle vitesse
devrions-nous utiliser les formules relativistes:\\\\
$$p=\frac{mv}{\sqrt{1-\frac{v^2}{c^2}}}=:\gamma\emph{mv}, \hskip
0.15cm \emph{E}=\sqrt{p^2c^2+(mc^2)^2}, \hskip 0.15cm
\emph{E}=\gamma\emph{m}c^2 \hskip 0.15cm \mbox{et} \hskip 0.15cm
\emph{v}=\frac{pc^2}{E} ?$$\\\\
Ces formules ne peuvent co\"{i}cinder avec les formules classiques
donnant l'\'{e}nergie d'une particule pour n'importe
quelle vitesse non nulle. En effet\\
$$\emph{E} = \emph{m}c^2+\frac{1}{2}\emph{mv$^2$} =
\gamma\emph{m}c^2 \Longleftrightarrow \gamma = 1+\frac{v^2}{2c^2}
\Longleftrightarrow \frac{1}{1-\frac{v^2}{c^2}} =
1+\frac{v^2}{c^2}+\frac{v^4}{4c^4}$$
$$\Longleftrightarrow 1 = 1 - \frac{v^4}{c^4}+\frac{v^4}{4c^4} -
\frac{v^6}{4c^6} \Longleftrightarrow 0 = - 3\frac{v^4}{4c^4} -
\frac{v^6}{4c^6}$$

De m\^{e}me, on a, dans le cadre classique, pour
\emph{v}$\ll$1\\\\
$$\emph{p = mv} , \hskip 0.15cm \emph{E$_c$} =
\frac{1}{2}\emph{mv$^2$} = \frac{1}{2}\emph{pv} \hskip 0.15cm et
\hskip 0.15cm \emph{E} = \emph{m}c^2 + \frac{1}{2}\emph{mv$^2$},$$
tandis que, dans le cadre de la relativit\'{e}, on a pour les
photons:
$$p c= \frac{hc}{\lambda} = h f = E$$
et pour les particules mat\'{e}rielles on a $v = \frac{pc^2}{E}.$
Pour ces derni\`{e}res, les deux notions ne peuvent pas
co\"{\i}ncider que pour $v = 0.$ En effet, si $v>0,$ alors la
relation $$ v = \frac{pc^2}{E} = \frac{m
vc^2}{E}\qquad\mbox{implique}\qquad E = mc^2,$$ ce qui est
contradictoire
($E=mc^2$ $\Longrightarrow$ $v=0$).\\

\textbf{Remarque:} Signalons bri\`{e}vement que tous les
r\'{e}sultats et formules \'{e}tablis en utilisant la partie
erronn\'{e}e du deuxi\`{e}me postulat peuvent \^{e}tre \'{e}tablis
plus pr\'{e}cis\'{e}ment d'une mani\`{e}re coh\'{e}rente.
N\'{e}anmoins, l'utilisation des formules relativistiques conduit
\`{a} des r\'{e}sultats approximatifs
tr\`{e}s utiles.\\

\subsection*{La fameuse relation \textbf{$E = mc^2$}}

Dans le cadre de notre mod\`{e}le, on a pour les points
mat\'{e}riels et immat\'{e}riels:
$$E_\mu\left(t,T(t),X(t)\right) =
h_\mu\left(t,T(t)\right)f_\mu\left(t,T(t)\right) = e(\mu,T(t))$$
et on a, pour toute particule mat\'{e}rielle fondamentale de masse
$m(t)$ occupant un domaine $D_t$ de volume $V(D_t)\;=:\;V(t)$
\`{a} l'instant $t$ et pour toute vitesse $v(t)\leq v_e(t)<1$:
$$E(t)=
\int_{D_t}E_\mu(t,T(t),X(t)) d\,X_t = \int_{D_t}h_\mu(t,T(t))
f_\mu(t,T(t))dX_t$$
$$ \hskip 4.4cm = h_\mu\left(t,T(t)\right)f_\mu(t,T(t))V(t)=e(\mu,T(t))V(t)$$

$$p(t)= \int_{D_t}m(X_t)v(X_t)
d\,X_t=m(t)v(t)$$

$$E_c(t)=\frac{1}{2}\int_{D_t}m_1(X_t)v^2(X_t)d\,X_t
= \frac{1}{2}m_1(t)v^2(t)$$ et
$$E(t)=m(t)c^2
+\frac{1}{2}m_1(t)v^2(t)=m(t)(c^2 +
\frac{1}{2}\frac{m_1(t)}{m(t)}v^2(t))=\rho(t)V(t)(c^2 +
\frac{1}{2}\frac{m_1(t)}{m(t)}v^2(t),$$
o\`u $\rho(t)$ d\'esigne la densit\'e de masse de la particule mat\'erielle. D'o\`{u}
$$h_\mu(t,T(t))f_\mu(t,T(t))V(t) = \rho (t) V(t) (c^2 +
\frac{1}{2}\frac{m_1(t)}{m(t)}v^2(t))$$et par suite
$$E(t,X(t))=h(t)f(t)=\rho(t)(c^2 +
\frac{1}{2}\frac{m_1(t)}{m(t)}v^2(t))$$ o\`{u} l'on a
d\'{e}sign\'{e} $E_\mu(t,T(t),X(t))$ par $E(t,X(t)),$
$h_\mu(t,T(t))$ par $h(t)$ et
$f_\mu(t,T(t))$ par $f(t).$\\
Pour $v=0,$ on obtient $E(t_0,X(t_0))=\rho(t_0)c^2,$ ce qui
implique que l'\'{e}nergie ponctuelle d'une mati\`{e}re au repos
est \'{e}gale \`{a} sa densit\'{e} volumique d'\'{e}nergie
retrouvant ainsi la relation
$$\emph{E$_0$} = \emph{m$_0$}c^2 \hskip 0.15cm (i.e. \hskip 0.15cm \emph{E} =
\emph{mc$^2$}).$$ 
L'\'{e}nergie totale de la particule mat\'erielle, \`{a}
l'instant $t,$ est donc
$$E(t)\;=\;m(t)(c^2 + \frac{1}{2}\frac{m_1(t)}{m(t)}v^2(t))$$
 o\`{u} $m(t)$ n'est autre que la masse initiale (au repos)
$m_0\;=\;m(0)$ multipli\'{e}e par un facteur d\'{e}pendant de la
vitesse et du temps : $m(t)\;=\;\gamma(t)m_0.$\\
Ce facteur a \'{e}t\'{e} d\'{e}termin\'{e} exp\'{e}rimentalement
(plusieurs ann\'{e}es avant la th\'{e}orie de la relativit\'{e}
restreinte) par le physicien W. Kaufmann qui a qualifi\'{e}
l'expression $\gamma m$ de masse apparente. Le facteur $\gamma(t)$
n'est autre que le facteur de Lorentz
$$\gamma(t) = \gamma = \frac{1}{\sqrt{1-\beta^2}} =
(1-\frac{v^2(t)}{c^2})^{-\frac{1}{2}}$$pour $\beta =
\frac{v(t)}{c}$.\\

Ainsi, en \'{e}crivant $m(t)=\gamma(t)m_0$ et $m_1(t) =
\gamma_1(t) m_0$, on obtient
\begin{equation} \label {r19}
E(t)=\gamma(t)m_0\left(c^2 + \frac{1}{2}
\frac{\gamma_1(t)}{\gamma(t)}v^2 \right) \simeq \gamma(t) m_0 c^2,
\end{equation}
puisque pour les petites vitesses comme pour les grandes vitesses
le terme $\frac{1}{2} \frac{\gamma_1(t)}{\gamma(t)} v^2$ est
n\'{e}gligeable \`{a} c\^{o}t\'{e} de $c^2$, et
\begin{equation} \label {r20}
p(t)=\gamma(t)m_0v
\end{equation}
et pour $v(t) \neq 0$, on a
$$\frac{E(t)}{p(t)} = \frac{c^2 + \frac{1}{2}\frac{\gamma_1(t)}{\gamma(t)}v^2}{v} \simeq \frac{c^2}{v}.$$
Ainsi, pour $v(t)\equiv0,$ on obtient $p(t)\equiv0$ et $E(t)
\equiv \gamma(t)m_0 c^2,$ ce qui, avec $$E(t)\equiv m_0
c^2,\;\;\mbox{redonne}\;\; \gamma(t)\equiv\gamma(0)=1 =
\gamma_0.$$
 De plus, pour $v\sim c,$ on a $E(t) \simeq p(t) c$ et pour $v\sim0,$ on a $E(t) \simeq m_0 c^2$ et
$p(t)\sim0.$\\
Enfin, pour $v\ll 1,$ on a $\gamma\simeq 1$, $m(t)\simeq m_0$, $\gamma_1\simeq 1$, $m_1\simeq m_0$ et $E(t)\simeq m_0c^2+\frac{1}{2}m_0v^2$.\\

\noindent En d\'{e}rivant la relation approximative (19) et la relation (20)
par rapport \`{a}
$t,$ on obtient\\
\begin{equation} \label {r21}
E^{'}(t)= m_0\gamma^{'}(t) c^2
\end{equation}
\begin{equation} \label{r22}
p^{'}(t)=m_0\gamma^{'}(t)v + m_0\gamma(t)v^{'}
\end{equation}
et en les d\'{e}rivant par rapport \`{a} la vitesse, on obtient\\\
\begin{equation} \label {r23}
\frac{dE}{dv} = m_0\frac{d\gamma}{dv} c^2
\end{equation}
\begin{equation} \label {r24}
\frac{dp}{dv}=m_0\frac{d\gamma}{dv}v + m_0\gamma(t)
\end{equation}\\\\

L'\'{e}quation approximative (19) et l'\'equation (20) sont bien
cons\'{e}quentes puisque (21) et (22) impliquent
$$ p'(t) = \frac{E'(t)}{c^2} v + m_0 \gamma(t) v'$$
et par suite
$$ \frac{dp}{dv}v' = \frac{1}{c^2} \frac{dE}{dv}v'v + m_0
\gamma(t) v'$$ et les \'{e}quations (23) et (24) impliquent
$$\frac{dp}{dv} = \frac{1}{c^2} \frac{dE}{dv}v + m_0
\gamma(t)$$qui est \'{e}quivalente \`{a} celle du dessus.\\
La relation approximative (19) et la relation (20) ne sont autres que les
fameuses relations d'Einstein pour l'\'{e}nergie et le
"momentum".\\\\
\textbf{Remarque:} La diff\'{e}rentation des relations exactes
(19) et (20) peut donner des indications sur le facteur
$\gamma_1$.\\\\
Notons qu'une particule mat\'{e}rielle de masse $m(t)$ ne peut pas
atteindre la vitesse $v = 1$ tout en gardant une masse finale
$m_f> 0$, puisque $$ \lim_{v \rightarrow 1} {\gamma(v) =
+\infty}$$L'\'{e}nergie n\'{e}cessaire qu'il lui faut fournir pour
atteindre une telle vitesse serait infinie.
\\

\subsection*{La quantit\'{e} de mouvement, l'\'{e}nergie et la
masse} 
Dans le cadre de notre mod\`{e}le, on a privil\'{e}gi\'{e}
(comme l'a fait originellement Einstein) la notion de la masse au
repos $m_0$ de chaque particule. Cependant, on a adopt\'{e}, pour
une particule ayant une vitesse significative, la notion de la
masse d\'{e}pendant de la vitesse et du temps sous la forme
$$ m(t) = \gamma (t)\hskip 0.02cm m_0$$
o\`{u} $\gamma(t) = (1+\frac{v^2(t)}{c^2})^{-\frac{1}{2}}$. On a
aussi adopt\'{e} l'expression
$$ E_c (t) = \frac{1}{2}\hskip 0.02cm m_1(t) \hskip 0.02cm v^2(t) = \frac{1}{2} \hskip 0.02cm \gamma_1 (t) \hskip 0.02cm
m_0\hskip 0.02cm v^2(t)$$pour l'\'{e}nergie cin\'{e}tique et
l'expression
$$\overrightarrow{p(t)} = m(t) \hskip 0.02cm \overrightarrow{v(t)} = \gamma (t) \hskip 0.02cm m_0 \hskip 0.02cm \overrightarrow{v(t)}$$
pour la quantit\'e de mouvement ("momentum") d'une particule en mouvement.\\
La quantit\'{e} de mouvement d'une particule est classiquement
d\'{e}finie comme \'{e}tant
$$\overrightarrow{p} \hskip 0.05cm = \hskip 0.05cm m\hskip 0.02cm \overrightarrow{v} \hskip 2cm \mbox{ avec }
\hskip 1cm \frac{d\overrightarrow{p}}{dt} \hskip 0.05cm = \hskip
0.05cm \overrightarrow{F}$$ o\`{u} $\overrightarrow{F}$ = $m
\overrightarrow{\Gamma}$ est la force qui agit sur la particule,
tandis que la d\'{e}finition classique de l'\'{e}nergie
cin\'{e}tique est
$$ E_c \hskip 0.05cm = \hskip 0.05cm \frac{1}{2}\hskip 0.02cm m \hskip 0.02cmv^2.$$
Einstein a bien montr\'{e} que ces deux d\'{e}finitions sont
\'{e}rronn\'{e}es pour des particules \`{a} grande vitesse. En
effet, un exemple simple ([2], ,p.112) montre que la
d\'{e}finition classique de l'\'{e}nergie cin\'{e}tique,
conduisant \`{a} la relation $v$ = $\sqrt{\frac{2E_c}{m}},$
contredit la loi fondamentale de la relativit\'{e} sp\'{e}ciale
(et de la physique en g\'{e}n\'{e}ral) qui stipule que la vitesse
d'une particule
mat\'{e}rielle ne peut pas exc\'{e}der $c$ = 1.\\\\

Par contre, le deuxi\`{e}me exemple ([2], p.113) qui a \'{e}t\'{e}
utilis\'{e} pour montrer la non conservation de la quantit\'{e} de
mouvement lors d'une collision de deux particules \emph{A} et
\emph{B} de m\^{e}me masse \emph{m} et ayant des vitesses
$\overrightarrow{v}$ et -$\overrightarrow{v}$ dans un rep\`{e}re
r\'{e}f\'{e}rentiel donn\'{e} $S^{'}$, ne permet pas de tirer les
cons\'{e}quences \'{e}tablies en utilisant les relations
relativistes du changement de rep\`{e}re. Pour nous, la
quantit\'{e} du mouvement est bien conserv\'{e}e lorsqu'on utilise
le rep\`{e}re \emph{S} pour lequel la particule \emph{B} est au
repos. Ceci est clairement montr\'{e} par la figure $11.$\\\\
D'un autre c\^{o}t\'{e}, il est clair ([2],p.112) que les formules
classiques $p = mv$ et $F = \frac{dp}{dt}$ conduisent \`{a} une
contradiction avec le principe fondamentale qui stipule que la
vitesse d'un objet de masse non nulle ne peut pas exc\'{e}der
celle de la
lumi\`{e}re.\\
De m\^{e}me la relation $F=\frac{dp}{dt}$, pour $F \neq 0$,
conduit, dans le cadre de notre mod\`{e}le, \`{a} une
contradiction de la m\^{e}me nature qu'auparavant. En effet,
consid\'{e}rons (\`{a} titre d'exemple) un \'{e}lectron de masse
au repos $m_0$ qui est acc\'{e}l\'{e}r\'{e} dans un champ
\'{e}lectrique $E$ telle que la force \'{e}lectrique $F$
exerc\'{e}e sur l'\'{e}lectron soit une constante non nulle.
Lorsqu'on \'{e}crit
$$ F \hskip 0.2cm =\hskip 0.2cm m(t)\hskip 0.2cm \Gamma(t)\hskip 0.2cm = \hskip 0.2cm \gamma(t) \hskip 0.1cm m_0 \hskip 0.1cm \Gamma(t)$$
et
$$ p \hskip 0.2cm = \hskip 0.2cm m(t) \hskip 0.1cm  v(t) \hskip 0.2cm = \hskip 0.2cm \gamma(t)\hskip 0.1cm  m_0 \hskip 0.1cm v(t)$$
pour $m(t)\neq0,$ propri\'et\'e qui se traduit par $\gamma(t)\neq
0$ et $v(t)<1,$ alors
$$ F \hskip 0.04cm =\hskip 0.04cm  \frac{dp}{dt}\hskip 0.02cm  \Leftrightarrow \hskip 0.02cm \gamma(t)\hskip 0.04cm m_0 \hskip 0.04cm \Gamma(t) \hskip 0.04cm=
\hskip 0.04cm \frac{d}{dt}\hskip 0.04cm (\gamma(t)\hskip 0.04cm
m_0\hskip 0.04cm v(t)) \hskip 0.02cm \Leftrightarrow \hskip 0.02cm
\frac{F}{m_0} \hskip 0.02cm =\hskip 0.02cm \frac{d}{dt}\hskip
0.02cm(\gamma(t)\hskip 0.02cm v(t)).$$\emph{F} \'{e}tant une
constante, on aura alors
$$ \gamma(t) v(t) = \frac{F}{m_0} t + C = \frac{F}{m_0}t +
\gamma(\tau) v(\tau) - \frac{F}{m_0} \tau$$pour $\tau >$ 0.\\Or,
ceci implique
$$\frac{d}{dt}(\gamma(t) v(t)) = \gamma^{'}(t) v(t) + \gamma(t)
\Gamma(t) = \frac{F}{m_0} = \gamma(t) \Gamma(t)$$et par suite, on
obtient
$$ \gamma^{'} (t) = 0 \hskip 2cm \mbox{ (puisque } v(t) \neq 0
\mbox{ ) }$$ et
$$ \gamma (t) = \gamma = \mbox{ cte } \hskip 1cm \mbox{ et }
\hskip 1cm m(t) = \gamma m_0 = \mbox{ cte  ,}$$ce qui est
impossible puisqu'une particule acc\'{e}l\'{e}r\'{e}e ne peut pas
avoir une masse constante.\\
De m\^{e}me, la relation pr\'{e}c\'{e}dente implique

$$ v(t) =
\frac{F}{m_0 \gamma} t + \frac{C}{\gamma}$$
$$\hskip 2.2cm = \frac{F}{m} t + \frac{C}{\gamma} = \Gamma t +
\frac{C}{\gamma}$$ avec $\Gamma$ constante strictement positive,
ce qui est impossible \'{e}galement puisque la vitesse d'une
particule de masse non nulle ne peut pas exc\'{e}der 1.\\
Ainsi la relation $\frac{dp}{dt}$ = \emph{F} ne peut pas \^{e}tre
approximativement vraie que pour des vitesses minimes par rapport
\`{a} la vitesse de la lumi\`{e}re lorsque $\gamma^{'}$ (\emph{t})
$\sim $ 0, $\gamma$ (\emph{t}) $\sim $ 1 et \emph{m}(\emph{t})
$\sim m_{0}$ \`{a} condition que \emph{m}(\emph{t}) soit
diff\'{e}rent de 0. Dans cette situation, on peut \'{e}crire
$$ \frac{dp}{dt} = \frac{d}{dt} m_0 \hskip 0.05cm v(t) = m_0
\hskip 0.05cm \Gamma(t) = F(t)$$ et pour \emph{v} = cte, on a
$$ \frac{dp}{dt} = \frac{d}{dt} m_0 \hskip 0.01cm v = 0.$$
Rappelons que, dans le cadre de notre mod\`{e}le, on a
$$p(t) \hskip 0.09cm = \hskip 0.09cm  m(t)\hskip 0.05cm v(t) \hskip 0.09cm =
\hskip 0.09cm \gamma(t) \hskip 0.05cm m_0 \hskip 0.05cm v(t)$$
$$E_c(t) \hskip 0.09cm = \hskip 0.09cm \frac{1}{2} m_1(t)\hskip 0.05cm v^2(t) \hskip 0.09cm =
\hskip 0.09cm \frac{1}{2} \gamma_1(t) \hskip 0.05cm m_0 \hskip
0.05cm v^2(t)$$
$$E(t) \hskip 0.09cm = \hskip 0.09cm  m(t)\hskip 0.05cm (c^2 + \frac{1}{2}\frac{m_1(t)}{m(t)}v^2(t)) \hskip 0.09cm =
\hskip 0.09cm \gamma(t) \hskip 0.05cm m_0 \hskip 0.05cm (c^2 +
\frac{1}{2}\frac{\gamma_1(t)}{\gamma(t)}v^2(t))$$pour
\emph{m}(\emph{t}) $\neq$ 0 et
$v(t)\leq v_e(t)<1$.\\
Ces formules sont bien conformes avec les deux lois fondamentales
de conservation (de l'\'{e}nergie et de la quantit\'{e} de
mouvement).\\
En r\'{e}alit\'{e}, le principe de la conservation de la
quantit\'{e} de mouvement est clairement exprim\'{e} dans le cadre
de notre mod\`{e}le o\`{u} l'on a
\begin{eqnarray*}
 \nabla^{g_t}_{X^{'}(t)} \overrightarrow{p(t)}&=& \nabla^{g_t}_{X^{'}(t)} \hskip 0.05cm \gamma(t) m_0 \hskip 0.05cm X^{'}(t) \hskip 0.1cm  = \hskip
0.1cm m_0 \nabla^{g_t}_{X^{'}(t)} 
\gamma(t)  X^{'}(t)\\
&=& m_0 (\gamma(t)
\nabla^{g_t}_{X^{'}(t)} X^{'}(t) + X^{'}(t).
\gamma(t) X^{'}(t))\\
 &=& m_0 (\gamma(t) \nabla^{g_t}_{X^{'}(t)} X^{'}(t) +
\gamma^{'}(X(t))X^{'}(t))=\overrightarrow{F}.
\end{eqnarray*}
Par cons\'{e}quent, si
\emph{X}(\emph{t}) est une g\'{e}od\'{e}sique, on obtient
$$ \widetilde{\Gamma}(t) \hskip 0.1cm = \hskip 0.1cm
\nabla^{g_t}_{X^{'}(t)}\hskip 0.05cm X^{'}(t) \hskip 0.1cm =
\hskip 0.1cm 0$$ et
$$ \| X^{'}(t) \|_{g_t} \hskip 0.1cm = \hskip 0.1cm v \hskip 0.1cm
= \hskip 0.1cm \mbox{cte};$$ce qui donne $\gamma$(\emph{t}) = cte
et $\gamma^{'}(t)=0,$ et par suite on obtient
$$ \overrightarrow{F}=\nabla^{g_t}_{X^{'}(t)} \hskip 0.04cm \overrightarrow{p(t)}
\hskip 0.09cm = \hskip 0.09cm 0.$$ Cette quantit\'{e} est
r\'{e}ellement nulle pour toute trajectoire libre
\emph{X}(\emph{t}) i.e. pour toute g\'{e}od\'{e}sique par rapport
\`{a} la m\'{e}trique physique qui tient compte de toutes les
forces naturelles agissant sur la particule. Par cons\'{e}quent,
on peut affirmer que la relation
$$ \nabla^{g_t}_{X^{'}(t)} \overrightarrow{p(t)} = \overrightarrow{F}$$est plus cons\'{e}quente que la relation classique
$\frac{d\overrightarrow{p}}{dt} = \overrightarrow{F}$ comme elle confirme l'\'equivalence: $\overrightarrow{F}(X(t))=0\Longleftrightarrow X(t)$ est une g\'eod\'esique pour $g_t$.\\\\
Finalement notons que, dans le cadre de la relativit\'{e}, la
relation $\overrightarrow{F} = \frac{d\overrightarrow{p}}{dt}$
conduit \`{a} la relation ([2], (4.104))
$$\overrightarrow{\Gamma} = \frac{d\overrightarrow{v}}{dt} =
\frac{ \overrightarrow{F}-
\overrightarrow{\beta}(\overrightarrow{F}.\overrightarrow{\beta})}{m
\gamma}$$ qui montre qu'\`{a} grande vitesse
l'acc\'{e}l\'{e}ration n'est pas parall\`{e}le \`{a} la force
tandis que le champ de force $F_{g_t}$ le long d'une trajectoire
$X(t)$ est donn\'{e} dans le cadre de notre mod\`{e}le par
$$ F_{g_t} (X(t)) = - \nabla^{g_t} E_t(X(t)) =
\widetilde{\Gamma}(T).$$\\

\subsection*{Remarques sur la th\'{e}orie quantique}

En ce qui concerne le principe d'incertitude, il nous a sembl\'{e}
illogique de se baser sur des exp\'{e}riences comme celle o\`{u}
des particules arrivent sur un \'{e}cran \`{a} travers deux
petites fentes l\'{e}g\`{e}rement espac\'{e}es l'une de l'autre
pour en conclure que le fait de savoir \`{a} travers desquelles
des deux fentes sont pass\'{e}es les particules, pourrait
alt\'{e}rer r\'{e}ellement le ph\'{e}nom\`{e}ne physique. Il est
vrai que le moyen utilis\'{e} pour conna\^{i}tre l'origine
pourrait alt\'{e}rer le r\'{e}sultat en influen\c{c}ant
\'{e}nerg\'{e}tiquement les parcours, mais ceci est un
ph\'{e}nom\`{e}ne conjoncturel et technique qui ne nous autorise
pas \`{a} en d\'{e}duire que notre pure connaissance pourrait
changer les r\'{e}sultats d'une telle exp\'{e}rie-\\nce qui sont
d\'{e}termin\'{e}s objectivement par les conditions physiques
elles-m\^{e}me. Par ailleurs, notre capacit\'{e}, qu'elle soit
th\'{e}orique ou pratique, de d\'{e}couvrir une loi de la nature
ne changera en rien l'objectivit\'{e} de cette loi. La longue
histoire de la d\'{e}couverte dans tous les domaines prouve
l'objectivit\'{e} de ces lois ind\'{e}pendamment de notre
capacit\'{e} conjoncturelle th\'{e}orique, approximative ou
exp\'{e}rimentale (technique) de les d\'{e}couvrir. L'utilisation
des niveaux \'{e}nerg\'{e}tiques grandissant des particules et des
longueurs d'onde de plus en plus petites, par exemple, a permis de
r\'{e}aliser des progr\`{e}s inimaginables dans le domaine de la
compr\'{e}hension de notre Univers physique \`{a} travers les
temps ainsi que celle de la mati\`{e}re et de sa structure, en
particulier celle des nucl\'{e}ons, quarks et autres leptons et
hadrons, mais aussi du raffinement de nos connaissances sur les
quantisations au niveau de l'\'{e}nergie et des moments angulaires
et intrins\`{e}ques aussi bien atomiques que nucl\'{e}aires. Les
fonctions de Schr\"{o}dinger et la statistique quantique ont
permis de jeter des lumi\`{e}res d\'{e}terminantes sur notre
Univers en fournissant des m\'{e}thodes efficaces pour
l'interpr\'{e}tation des r\'{e}sultats obtenus par
exp\'{e}rimentation et en conduisant \`{a} des approximations
puissantes au niveau de la mod\'{e}lisation des ph\'{e}nom\`{e}nes
naturels. Ces derniers comportent par leur nature et d'une
fa\c{c}on inh\'{e}rente des incertitudes dues \`{a} la multitude
des facteurs (\'{e}nerg\'{e}tiques et dynamiques essentiellement
\'{e}volutifs) qui gouvernent les niveaux \'{e}nerg\'{e}tiques,
les trajectoires,
les interactions...\\\\
Ces ph\'{e}nom\`{e}nes sont loin d'\^{e}tre r\'{e}guliers
(diff\'{e}rentiables) mais ils sont continus. Un \'{e}lectron qui,
par exemple, change son orbite (en \'{e}volution continue et
permanente) pour un niveau plus haut (en absorbant un photon) ou
pour un niveau plus bas (en c\'{e}dant un photon), passe une
infime fraction de seconde pour effectuer ce passage assurant
\`{a} la fois la continuit\'{e} de sa trajectoire et la loi de la
conservation de l'\'{e}nergie au cours de ce passage.\\
Par ailleurs, notons que la fonction $\psi$ solution de
l'\'{e}quation de Schr\"{o}dinger
$$-\frac{\overline{h}^2}{2m}\frac{d^2\psi}{dx^2} +
V(x)\psi(x) = E\psi(x)$$par exemple, qui donne la probabilit\'{e}
de trouver la particule par unit\'{e} de distance \emph{x} \`{a}
l'aide de la distribution
$$\frac{dP}{dx} = |\psi(x)|^2$$
a \'{e}t\'{e} d\'{e}termin\'{e}e d'une fa\c{c}on exp\'{e}rimentale
et pr\'{e}dective. Cette fonction est d'une nature
compl\`{e}tement diff\'{e}rente de notre fonction $\psi(X).$ Cette
derni\`{e}re provient de la r\'{e}solution de l'\'{e}quation
d'\'{e}nergie $(E^{*})$ dont les solutions sont de la forme
$$E(t,X) = g(t)\psi(\frac{X}{t})$$
caract\'{e}risant ainsi, \`{a} la fois, l'\'{e}nergie et la
fr\'{e}quence d'un point mat\'{e}riel ou immat\'{e}riel et de
toute particule mat\'{e}rielle ponctuelle \`{a} un instant
donn\'{e}. De m\^{e}me, il ne faut pas confondre ni l'une ni
l'autre avec la trajectoire $X(t)$ de cette particule. Ainsi, pour
un pendule simple ou un oscillateur harmonique (quantique), par
exemple, la fr\'{e}quence d'oscillation est une chose et celle des
points mat\'{e}riels est une autre. Lorsque, par exemple, le
pendule se trouve en \'{e}quilibre vertical stable, la
fr\'{e}quence du point mat\'{e}riel est d\'{e}termin\'{e}e par son
\'{e}nergie caract\'{e}ristique $E(t,X(t)) = h(t)f(t),$ son
\'{e}nergie potentielle et son \'{e}nergie cin\'{e}tique sont
nulles et son \'{e}nergie de masse est \emph{E} = \emph{mc$^2$} =
\emph{m}. Cependant, on ne peut pas parler ni de sa p\'{e}riode
\emph{T} ni de sa fr\'{e}quence \emph{f} = $\frac{v}{\lambda}$. On
ne peut donc pas supposer qu'elle a une fr\'{e}quence non nulle et
d\'{e}terminer, \`{a} l'aide du principe d'incertitude, son
\'{e}nergie minimale de base (ground state energy) qui serait non
nulle (aussi petite qu'elle soit) contrairement aux principes
newtoniennes de la Physique classique. Ceci reste valable
\'{e}galement en ce qui concerne une balle de tennis ou une
miniscule bille, par exemple, se trouvant au repos dans une caisse
(suppos\'{e}e \'{e}galement au repos). De m\^{e}me, le niveau
minimum d'\'{e}nergie (bel et bien quantis\'{e}) de l'\'{e}lectron
dans un atome d'hydrog\`{e}ne correspond \`{a} l'orbite
associ\'{e}e au rayon de Bohr
$$\emph{r} = \emph{a$_0$} = \frac{4\pi\varepsilon_0\overline{h}^2}{me^2}$$
qui est lui-m\^{e}me d\'{e}termin\'{e} par l'\'{e}nergie minimale
$$\emph{E$_m$} = (\frac{1}{2}\emph{mv$^2$} -
\frac{e^2}{4\pi\varepsilon_0r})_\emph{m} =
(\frac{\overline{h}^2}{2\pi r^{2}} -
\frac{e^2}{4\pi\varepsilon_0r})_\emph{m}$$o\`{u} $\overline{h}$
est ici la constante de Planck classique.\\
Ceci n'a rien \`{a} avoir avec le principe d'incertitude mais il
est plut\^{o}t d\^{u} au fait que l'\'{e}nergie totale minimale
que peut avoir un \'{e}lectron au sein d'un atome d'hydrog\`{e}ne
est finie et caract\'{e}ris\'{e}e (en moyenne) par les deux
constantes
$a_0$ et $\overline{h}$.\\\\
Pr\'{e}cisons enfin que le principe
d'incertitude, sous ses deux formes: \\
\emph{$\Delta$x$\Delta$p$_x$} $ \geq $ $\frac{\overline{h}}{2}$ et
\emph{$\Delta$E$\Delta$t} $ \geq $ $\frac{\overline{h}}{2}$, n'est
en r\'{e}alit\'{e} qu'une cons\'{e}quence l\'egitime de
l'utilisation des fonctions de Schr\"{o}dinger:
$$\psi(x) =
\frac{1}{\sqrt{2\pi}}\int_{-\infty}^{+\infty} g(k)e^{-ikx} dk$$ et
$$g(k) = \frac{1}{\sqrt{2\pi}}\int_{-\infty}^{+\infty} \psi(x)e^{ikx}
dx$$ afin de d\'{e}terminer la probabilit\'{e} de localiser une
particule donn\'{e}e soumise \`{a} des contraintes donn\'{e}es
dans une position donn\'{e}e. Ce principe est, en effet, obtenu
\`{a} partir des distributions
$$\frac{dP}{dx} =
|\psi(\emph{x})|^2,\hskip 0.5cm \frac{dP}{dk} = |g(\emph{k})|^2$$
et de leurs d\'{e}viations standards $\sigma$$_x$ et $\sigma$$_k$
en utilisant la relation de De Broglie \emph{p} =
$\frac{h}{\lambda}$. Par cons\'{e}quent, le principe d'incertitude
obtenu ainsi confirme simplement que cette approche
particuli\`{e}re et l'utilisation de cette m\'{e}thode
particuli\`{e}re comportent d'une fa\c{c}on inh\'{e}rente
l'incertitude ainsi quanitfi\'{e}e. Mais ceci ne veut pas dire que
la position \emph{x} de la particule et son moment \emph{p$_x$}
\`{a} un instant donn\'{e} ne sont pas bien d\'{e}finis ou qu'ils
ne peuvent pas \^{e}tre d\'{e}termin\'{e}s \`{a} une meilleure
pr\'{e}cision par un proc\'{e}d\'{e} th\'{e}orique ou
exp\'{e}rimental plus performant. En effet, nul ne peut affirmer
qu'on ne pourra pas un jour mettre au point un moyen technique ou
th\'{e}orique pouvant \^{e}tre utilis\'{e} pour mesurer une
largeur de fente ou un ordre de grandeur d'une particule beaucoup
plus petits que ceux atteints actuellement \`{a} l'aide de la
diffusion des particules en cours aujourd'hui. On pourrait peut
\^{e}tre utiliser des rayons $\gamma$ ultra-\'energ\'etiques dont
les longueurs d'onde sont beaucoup plus petites et inventer un
proc\'{e}d\'{e} interm\'{e}diaire qui rend les effets de tels
rayons accessibles \`{a} notre sensibilit\'{e} ou notre
compr\'{e}hension. On peut esp\'{e}rer la m\^{e}me chose en ce qui
concerne la mise au point de proc\'{e}d\'{e}s nouveaux pour
mesurer la position \emph{x} et la composante $p_{x}$ de la
quantit\'{e} de mouvement \emph{p} d'une particule donn\'{e}e qui
am\'{e}lioreront les incertitudes $\Delta x$ et $\Delta p_{x}$
ainsi que celui de leur produit $\Delta x \Delta p_{x}$ qui est
limit\'{e} actuellement par
$\frac{\overline{h}}{2}$ avec les proc\'{e}d\'{e}s actuels.\\\\
\'{E}videmment, on peut utiliser les \'{e}quations de
Schr\"{o}dinger et la Statistique quantique en tant qu'une
approche conduisant \`{a} des approximations qualitatives et
quantitatives efficaces (avec, tout naturellement, une marge
d'incertitude) des ph\'{e}nom\`{e}nes physiques \'{e}tudi\'{e}s
dans les cas o\`{u} les mesures approximatives effectives
s'av\`{e}rent difficiles
\`{a} effectuer. Ces probl\`{e}mes vont \^{e}tre discut\'{e}s d'une fa\c{c}on plus pr\'{e}cise dans les paragraphes suivants.\\\\

Notons que, dans le cadre de notre mod\`{e}le, les notions de
longueur d'onde et de fr\'{e}quence sont des caract\'{e}ristiques
d'un point mat\'{e}riel ou immat\'{e}riel et non pas d'une
particule mat\'{e}rielle m\^{e}me si cette derni\`{e}re pourrait
\^{e}tre consid\'{e}r\'{e}e physiquement comme \'{e}tant
ponctuelle comme un \'{e}lectron par exemple. C'est dans ce
contexte pr\'{e}cis qu'il faut interpr\'{e}ter le caract\`{e}re
ondulatoire de la mati\`{e}re. Nous consid\'{e}rons donc le fait
d'attribuer une longueur d'onde $\lambda = \frac{h}{p}$ et une
fr\'{e}quence $f = \frac{v}{\lambda}$ \`{a} une particule
mat\'{e}rielle ponctuelle circulant \`{a} une vitesse \emph{v}
comme \'{e}tant une m\'{e}thode pratique pour effectuer des
calculs approch\'{e}s ne correspondant pas \`{a} une vraie
trajectoire p\'{e}riodique et elles ne peuvent pas \^{e}tre
utilis\'{e}es pour calculer l'\'{e}nergie exacte d'une telle
particule \`{a} l'aide de la relation
$$E = \sqrt{(pc)^2+(mc^2)^2} = \sqrt{p^2+m^2} =
\sqrt{\frac{h^2}{\lambda^2}+m^2}= \sqrt{\frac{h^2f^2}{v^2}},$$par
exemple, qu'il s'agisse d'une particule relativiste ou non
relativiste (voir paragraphe 8).\\
De m\^{e}me, lorsqu'on utilise la relation 
$$p= \gamma m
v=\frac{h}{\lambda}=\frac{hf}{v}$$ 
pour une particule
mat\'{e}rielle ponctuelle non relativiste, alors on obtient
$$ E_T\;:=\;hf\;=\;\gamma mv^2$$
ce qui implique (conform\'{e}ment aux formules relativistes)
$$\gamma m\;=\;\gamma mv^2$$
et par suite  $v^2=1,$ ce qui est absurde. Quant aux "momentum" $p$ d'un photon, on a bien: $p=E=hf=\frac{h}{\lambda}$ et par suite, on a $\lambda=\frac{h}{p}.$\\\\

De plus, l'utilisation de l'expression de la quantit\'{e} de
mouvement relativiste $p=\frac{h}{\lambda}$ pour les particules
mat\'{e}rielles conduit, dans le cas de l'\'{e}lectron de l'atome
d'hydrog\`{e}ne, \`{a} une contradiction flagrante. En effet, on a
alors ([2],p.139):
$$ \frac{<E_c>}{<E_p>} = -
\frac{<\frac{mv^2}{2}>}{ke^2<\frac{1}{r}>} = -
\frac{<\frac{mv^2}{2}>}{ke^2 \frac{mv}{\overline{h}}}$$o\`{u} l'on
a utilis\'{e} la formule
$$<\frac{1}{r}> \simeq \frac{2\pi}{\lambda} =
\frac{p}{\overline{h}}$$et l'expression non relativiste de $E_c$,
qui est valable pour les niveaux d'\'{e}nergie de l'atome en
question.\\Ainsi, on obtient
$$ - \frac{1}{2} = - \frac{v \overline{h}}{2ke^2} = - \frac{v
\overline{h}}{2 \alpha \overline{h} c} = - \frac{1}{2 \alpha c}
v;$$ce qui est contradictoire puisque $v$ d\'{e}croit lorsque $r$
croit.\\\\
D'un autre c\^{o}t\'{e}, la comparaison de notre expression de
l'\'{e}nergie ondulatoire $E(t) = h_\mu(t)f_\mu(t)$ \`{a} celle de
De Broglie-Planck-Einstein qu'on va \'{e}crire\\ $E = h_Pf_D$
(o\`{u} $h_P$ est la constante de Plank et $f_D$ est la
fr\'{e}quence de De Broglie) donne $h_\mu(t)f_\mu(t) = h_Pf_D$ et
parsuite
$$h_P = \frac{h_\mu(t)f_\mu(t)}{f_D(\mu,t)} =
\frac{e(\mu,t)}{f_D(\mu,t)}$$ et$$ f_D(\mu,t) =
\frac{1}{h_P}h_\mu(t)f_\mu(t) = \frac{1}{h_P}e(\mu,t).$$
\\
\textbf{Remarque: } Une \'{e}tude suppl\'{e}mentaire plus
syst\'{e}matique des limites de la th\'{e}orie quantique sera
effectu\'{e}e dans le paragraphe suivant.\\

\subsection*{R\'{e}percussions sur quelques autres notions}

Signalons qu'\`{a} la lumi\`{e}re de notre mod\`{e}le, on pourra
r\'{e}examiner et pr\'{e}ciser un grand nombre de notions et de
facteurs jouant un r\^{o}le important en Physique moderne (la loi
de Hubble, la d\'{e}pendance du spectre des radiations des
puissances de la temp\'{e}rature, le probl\`{e}me de la
r\'{e}unification des forces fondamentales...) en \'{e}vitant
l'utilisation de l'aspect erronn\'{e} du deuxi\`{e}me postulat de
la relativit\'{e} restreinte et du principe d'incertitude et en
confinant les statistiques quantiques dans leur juste r\^{o}le et
leur juste et importante port\'{e}e et enfin en r\'{e}tablissant
la d\'{e}pendance du temps (ou de la distance) et de la
temp\'{e}rature de certaines notions et constantes. Ainsi, le
d\'{e}calage vers le rouge et vers le bleu (le "redshift" et le
"blueshift"), par exemple, sont d\^{u}s \`{a} la d\'{e}pendance de
la longueur d'onde du temps et de la distance et non pas \`{a} la
vitesse de la source. La vitesse de la source a pour seul effet
d'\'{e}loigner ou de rapprocher plus ou moins vite
l'\'{e}metteur de l'analyseur - r\'{e}cepteur. La d\'ependance du facteur du``redshift" $z$ de la distance explique bien les grandes valeurs observ\'ees de $z$ $(z>2)$ qui semblent \^etre contradictoire \`a la th\'eorie classique de la relativit\'e g\'en\'erale. Ces valeurs l\`a prouvent uniquement l'existence de pulsars tr\`es distants de nous, par exemple.\\
De m\^{e}me, on peut montrer facilement, dans le cadre de notre
mod\`{e}le, que la vitesse relative de deux galaxies isotropiques
quelconques, s'\'{e}loignant dans la direction de l'expansion est
proportionnelle \`{a} la distance les s\'{e}parant. Ceci nous
permet d'introduire le facteur $R(t)$ caract\'{e}risant
l'expansion et suivre les travaux de Hubble en posant $r =
r_0R(t)$ et
$$H(t) = \frac{\frac{dR}{dt}}{R} \mbox{ avec }
H_0  = \left(\frac{dR}{dt}\right)_{t = t_0}$$ pour obtenir la loi
de
Hubble\\
$$v = \frac{dr}{dt} = H_0r$$
Ainsi, en d\'{e}signant par $m$ la masse totale d'une galaxie
situ\'ee sur une sph\`ere de rayon $R(t)$ suffisemment grand et
par $M$ la masse totale de la boule de rayon $R(t),$ on peut
obtenir les relations classiques (J.W.Rohlf, p.$552$):
$$E_c = \frac{1}{2}mr_0^2\left(\frac{dR}{dt}\right)^2$$
$$V = -\frac{4 \pi m G r_0^2 R^2 \rho}{3}$$
o\`{u} $V$ est l'\'energie potentielle de la galaxie et $\rho$ est
la densit\'{e} moyenne de la masse dans
la boule.\\
Ensuite, suivant Einstein, on introduit le param\`{e}tre de
courbure $K(t)$ qui est, dans le cadre de notre mod\`{e}le,
intrins\`{e}quement li\'{e} \`{a} la m\'{e}trique $g_t$ qui
refl\`{e}te, elle-m\^{e}me, la distribution \'{e}nerg\'{e}tique de
l'Univers. En appliquant le principe de la conservation de
l'\'{e}nergie, on obtient alors l'\'{e}quation de Friedmann\\
$$\left(\frac{dR}{dt}\right)^2 = \frac{8 \pi \rho G R^2}{3} - K$$
Bien que cette \'{e}quation pourrait donner des renseignements sur
l'\'{e}volution de l'Univers, signalons que notre mod\`{e}le est
diff\'{e}rent de celui d'Einstein-de Sitter puisqu'il est bas\'{e}
sur une autre conception de l'espace et du temps d'une part et
puisque le param\`{e}tre de courbure \emph{K} qui appara\^{i}t ici
d\'{e}pend du temps et ne peut pas \^{e}tre nul d'une autre part.
Signalons aussi que notre mod\`{e}le n'est pas conforme au premier
postulat du principe cosmologique qui stipule que l'Univers
para\^{i}t (exactement) le m\^{e}me quelque soit la position
galactique de l'observateur tout en \'{e}tant conforme \`{a} la
deuxi\`{e}me affirmation de ce principe qui stipule que (dans le
sens pr\'{e}cis\'{e} plus haut) la vitesse relative des galaxies
est proportionnelle \`{a} leur distance. Cependant, la
r\'{e}adaptation de la th\'{e}orie de la relativit\'{e}
g\'{e}n\'{e}rale,
des lois de Hubble et des \'{e}quations de Friedmann - Einstein \`{a} notre mod\`{e}le sera effectu\'{e}e au cours des paragraphes 10, 11 et 12.\\\\
La r\'{e}examination approfondie de ces notions et postulats et
des r\'esultats qui en d\'ecoulent n\'{e}cessite \'{e}videmment un
travail collectif laborieux et
assidu.\\\\
Tout ce qui pr\'{e}c\`{e}de nous pousse \`{a} croire que la
Physique est une science exacte, mais cette science ne pourra nous
\^{e}tre r\'{e}v\'{e}l\'{e}e que progressivement, et souvent d'une
mani\`{e}re approximative, en mariant l'exp\'{e}rimentation \`{a}
la th\'{e}orie. Ceci est essentiellement d\^{u} \`{a} la
complexit\'{e} des ph\'{e}nom\`{e}nes naturels (bien que les lois
de la Physique sont essentiellement simples) d'une part et aux
limites impos\'{e}es par nos moyens techniques et pratiques d'une
autre part. Cependant, tout ce qui reste dans le domaine de la
Physique (et non pas dans celui de la M\'{e}thaphysique) est
r\'{e}gi par un certain nombre de principes et de lois dont la
majorit\'{e} a \'{e}t\'{e} d\'{e}couverte par des voies
exp\'{e}rimentales et th\'{e}oriques. La voie th\'{e}orique
utilise essentiellement les Math\'{e}matiques qui, bien qu'elles
soient purement intellectuelles et th\'{e}oriques, sont
initi\'{e}es et
dynamis\'{e}es par la Physique et la Technologie.\\\\
En conclusion, les Math\'{e}matiques et la Physique sont
indissociables comme le sont la th\'{e}orie (intellectuelle) et la
pratique (exp\'{e}rimentale et utilitaire); d'o\`{u} la
n\'{e}cessit\'{e} de toutes les sciences, auxquelles il faut
ajouter l'imagination et la Philosophie pour aller encore plus
loin dans l'\'{e}pop\'{e}e de la
d\'{e}couverte scientifique dans tous les domaines.\\\\
{\bf {Remarque.}} D'autres r\'{e}percussions fondamentales sur
diff\'{e}rentes branches de la physique moderne seront
d\'{e}velopp\'{e}es dans les paragraphes suivants.\\

\section{The limits of Quantum theory}
 In this section we aim to specify the proper domain of the Quantum
theory efficiency. We will show that the wave Quantum Mechanics
(based on Schr\"odinger's equations) and quantum Statistics
essentially constitute experimental and approximate tools that
result in a probabilistic and predictive approach for explaining
physical phenomena. Consequently, they can not constitute a proper
theoretical framework for instituting any intrinsic or canonical
physical law. The De Broglie wavelength, which is a canonical
feature of electromagnetism, is simply a practical object that is
useful only for approximately studying the pointlike material
particles behavior. Moreover the uncertainty principle is only a
legitimate consequence of the Schr\"odinger probabilistic process
and can not be considered as a universal principle. More
fundamentally, we consider that the legitimacy of the wave quantum
Mechanics (which is derived from classical Mechanics), is based on
its ability to provide, in the macroscopic cases, approximate
results that coincide with those given by Newton, Lagrange and
Hamilton's Mechanics. Classical Mechanics institutes laws for
idealized physical situations (when sufficient data are known),
whereas quantum Mechanics predicts and explains experimental
observed results; the legitimacy of the latter is insured by the
Bohr correspondence principle. Indeed, we show that quantum
Statistics uniquely relies on the very physical characteristics of
both the realized experiment and the involved particles (such as
distances, symmetries, masses, charges, momenta
and spins) as well as on mathematical Logics.\\
Quantum Mechanics can and must be used in microscopic subatomic
phenomena when our present means can not result in a theoretical
formulation.

\subsection{The De Broglie Wavelength}
\normalsize{The early $20^{th}$ century was marked by three
fundamental discoveries: the photon by Einstein, the Planck
constant and the Bohr model for the hydrogen atom. The Quantum
period has begun. The quantized nature of light as well as of
energy levels was clearly proved. Contemporaneously, many
experiments and facts have shown that matter has also some wavy
nature. The success of some new notions and the partial success of
some others led De Broglie to translate the notion of wavelength
from electromagnetism to matter particles and bodies. He then
defined the wavelength of a particle as
$$\lambda\;=\;\frac{h}{p}$$
where $p$ is the relativistic momentum of the particle and $h$ is
the Planck's constant. This was a practical and useful
approximation for analyzing the energy, momentum and speed of
particles. Nevertheless, this notion, joined to relativistic
formulas, on one hand, and to the Bohr model, on the other hand,
leads to some obvious contradictions. The wavy nature of matter
has to be explained
more generally and more precisely.\\

\noindent\textbf{Remark $1.$}$\;\;$ In the previous sections, we
have proved that some interpretations of the special relativity
second postulate are false. We also proved that the frequency $f$
is a characteristic feature of a (material or immaterial) point
that is extended only to fundamental particles (quarks and
leptons). Integration of the relation $E=hf$ over the domain
occupied by a fundamental particle gives the famous relation
$$E_0\;=\;m_0c^2\;:=\;m_0$$
for at rest matter and
$$E(t)\;=\;\gamma(t)m_0 (c^2+\frac{1}{2} \frac{\gamma_1(t)}{\gamma(t)} v^2(t)) \simeq \gamma(t) m_0 c^2$$
for a material particle into movement, where $\gamma_1(t)$
decreases from 1 to 0 when the speed increases from 0 to 1 and
$\gamma(t)$ is the Lorentz factor; $\gamma(t) m_0$ is qualified by
W. Kaufmann as being the apparent mass and we call
$\gamma_1(t) m_0$ the reduced mass of the particle into movement.\\

So, when we attribute a De Broglie wavelength
$\lambda=\frac{h}{p}$ and a frequency $f=\frac{v}{\lambda}$ to a
pointlike material particle, they do not correspond to a real
periodic movement (or trajectory) and can not be used for making
exact calculation of relativistic or non relativistic particle
energy by means of formulas such as
$$p\;=\;\gamma mv\quad,\quad E\;=\;\sqrt{p^2+m^2}\;=\;\gamma
m\quad\mbox{and}\quad v\;=\;\frac{p}{E},$$
where we have taken $c=1$
for simplicity.\\
For non relativistic particles, we obtain in this
way:
$$p\;=\;\frac{h}{\lambda}\;=\;\frac{hf}{v}\;=\;\frac{E}{v}\;=\;\frac{\gamma
m}{v}$$ which yields
$$\gamma mv\;=\;\frac{\gamma m}{v}$$
and then $v^2=1,$ which is absurd.\\

Moreover the Bohr model shows clearly that, for energy levels
$E_n$ with $n$ sufficiently large, we have $f_{orb}\simeq f_{rad}$
(whereas, for lower $n,$ we have $f_{orb}\neq f_{rad}$). So, when
we consider two consecutive high levels $E_1$ and $E_2$
corresponding to frequencies $f_1$ and $f_2,$ wavelengths
$\lambda_1$ and $\lambda_2$ and speeds $v_1$ and $v_2,$ we have
$\left(\mbox{using the relation}
\;\;E=\sqrt{p^2+m^2}=\sqrt{\frac{h^2}{\lambda^2}+m^2}=\sqrt{\frac{h^2f^2}{v^2}+m^2}\right),$
\begin{eqnarray*}
  f_2 &=& \frac{\Delta E}{h}\;=\;\frac{E_2-E_1}{h}\;=\;\frac{\sqrt{p_2^2+m^2}-\sqrt{p_1^2+m^2}}{h}\\
   &=&
   \frac{\sqrt{\frac{h^2}{\lambda_2^2}+m^2}-\sqrt{\frac{h^2}{\lambda_1^2}+m^2}}{h}\;=\;\sqrt{\frac{1}{\lambda_2^2}+\frac{m^2}{h^2}}\;-\;\sqrt{\frac{1}{\lambda_1^2}+\frac{m^2}{h^2}}.
\end{eqnarray*}
Consequently, we obtain
$$\frac{v_2}{\lambda_2}\;=\;\frac{1}{\lambda_2}\sqrt{1+m^2\frac{\lambda_2^2}{h^2}}\;-\;\frac{1}{\lambda_1}\sqrt{1+m^2\frac{\lambda_1^2}{h^2}}$$
and
\begin{eqnarray*}
  v_2 &=& \sqrt{1+\frac{m^2}{p_2^2}}\;-\;\frac{\lambda_2}{\lambda_1}\sqrt{1+\frac{m^2}{p_1^2}}\\
   &=& \sqrt{1+\frac{m^2}{p_2^2}}\;-\;\frac{p_1}{p_2}\sqrt{1+\frac{m^2}{p_1^2}} \\
   &=& \sqrt{1+\frac{m^2}{p_2^2}}\;-\;\sqrt{\frac{p_1^2}{p_2^2}+\frac{m^2}{p_2^2}}\\
   &=& \sqrt{1+\frac{m^2}{p_2^2}}\;-\;\sqrt{\frac{v_1^2}{v_2^2}+\frac{m^2}{p_2^2}} \\
   &=&\sqrt{1+\frac{m^2}{p_2^2}}\;-\;\sqrt{\left(\frac{n+1}{n}\right)^2+\frac{m^2}{p_2^2}}\;<\;0
\end{eqnarray*}
which is absurd.\\
We obtain a similar contradiction when we use $f_1=\frac{\Delta
E}{h}.$
\subsubsection*{The particle in a box case}
We consider now a small ball (or particle) in a fixed box of
length $L.$ When we are looking for the ground state energy by
using the De Broglie wavelength notion $\lambda=\frac{h}{p}$ and
the Schr\"odinger equation
$$-\frac{\overline{h}^2}{2m}\frac{d^2\psi}{dx^2}\;=\;E\psi,$$
we arbitrarily exclude the case where the particle speed is $v_0=0.$\\
Now, the introduction of the speed notion implies necessarily the
introduction of the time progress notion. Let then
$E_0=\frac{h^2}{8mL^2}$ be the quantum ground state energy that
corresponds to the speed $v_0\neq 0.$ If $|v_0|=a_0$ (a positive
constant), we obtain $|p_0|=ma_0$ when using classical Physics
formulas and $|p_0|=\gamma_0 ma_0$ when using relativistic ones,
where $\gamma_0=\frac{1}{\sqrt{1-v_0^2}}.$ But $<p_0>=0$ implies
$p_0(x)=\pm|p_0|$ and $v_0(x)=\pm a_0$ which constitute a
physically and mathematically inconceivable phenomena
($v_{0}$(\emph{x}) can not pass from $-a_{0}$ to +$a_{0}$
instantaneously). So $|p_0|$ is time dependent which is
contradictory as (according to generally accepted notions)
$$|p_0|\;=\;\frac{h}{\lambda_0}\;=\;\frac{h}{2L}\;=\;\mbox{const}.$$
Therefore, we have either $v_0=0,$ which implies $E_0=p_0=0$ (in
accordance with the classical physics minimal energy) and
$\lambda_0$ has no a real existence, or all quantities $v_0,p_0,E_0$
and (if we put $\lambda_0=\frac{h}{p_0}$) $\lambda_0$ depend on
time. In that case $L$ obviously depends on time, which is absurd
unless the legitimacy of the theoretically exact measurements'
existence is thoroughly questioned.

\subsubsection*{The pendulum and quantum harmonic oscillator case}
When we consider a pendulum, we assume either $v_1=0,$ which
corresponds to the stable vertical equilibrium state and implies
$E_1=p_1=0$ in accordance with classical Newtonian Physics, or
$v_1\neq0.$ In that case the ground state energy within the wave
quantum Mechanics framework is
$$E_1\;=\;\frac{\overline{h}\omega_1}{2}\;=\;\frac{hf_1}{2},$$
which is a non vanishing constant, since the very physical nature
of the pendulum notion imposes the attribution of a frequency
$f_1=\frac{1}{T_1}$ to the theoretically periodic movement of the
pendulum.\\
Now, when we incorrectly identify the De Broglie wavelength
$\lambda_1=\frac{h}{p_1}$ with the wavelength
$\lambda_1=\frac{v_1}{f_1}$ of the periodic movement (where $p_1$
and $v_1$ are respectively the mean scalar momentum and speed), we
obtain
$$E_1\;=\;\frac{hf_1}{2}\;=\;\frac{1}{2}\frac{h}{\lambda_1}\lambda_1
f_1\;=\;\frac{1}{2}p_1v_1$$ which is the corresponding mean
kinetic energy. The same results are valid for a quantum harmonic
oscillator. Furthermore, we obtain for the first excited state
$E_2=3E_1$ ([2], (7.121)). Now, it is physically and
mathematically undeniable that mean speed, mean momentum and mean
kinetic energy depend continuously on the initial displacement of
both pendulum and harmonic oscillator. But displacement is a
continuous variable; therefore the energy levels can not be
quantized by means of Schr\"odinger's equation.

\subsection{The uncertainty principle}
We begin this subsection by noticing that it seems strongly
illogical that, after some experiments such as the one where
particles hit a screen through small slits slightly spaced, we
conclude that the fact of knowing the origin of the particles
hitting the screen could really alter the physical phenomena. It
is true that the means used in order to know this origin can alter
the results by modifying the particles momenta and trajectories
but this is only a technical and circumstantial phenomena that
does not allow us to conclude that our pure knowledge can
transform the physical results that are determined objectively by
the real physical conditions. Beside of that, our theoretical or
practical capacity to discover any law of Nature does not
influence the objective reality of this law. The very long history
of discoveries in all domains shows the objectivity of natural
laws independently of our circumstantial (theoretical,
approximate, experimental or technical) capacity to discover them.
The use of progressively increasing energetic levels of particles
(i.e. increasing «frequencies» or decreasing «wavelengths»), for
instance, has permitted the realization of important progress in
the understanding of our physical universe through the ages as
well as the understanding of matter (nucleons, quarks, hadrons,
leptons) structure and also the refinement of our knowledge on the
quantization of both energy levels inside atoms and angular (as
well as intrinsic) atomic and
nuclear momentum.\\
The Schr\"odinger function and quantum Statistics have also
permitted a jump in our comprehension of the universe by giving
effective methods for discovering and interpreting all results
that are obtained from experimentation and led to powerful
approximations
for natural phenomena rules.\\
However, these phenomena contain essentially a part of
uncertainties due to the multitude of energetic and dynamic
evolutive factors which govern all aspects of matter-energy:
energy levels,
trajectories, interactions, transformations...\\
Although these phenomena are far from being regular
(differentiable), they are continuous. An electron, for instance,
that changes its orbit (which is permanently evolving) for a
higher energy level (when absorbing a photon) or for a lower
energy level (together with photon emission), spends an
infinitesimal time fraction before achieving its final state.
During this transition, continuity of trajectory and energy
conservation are both insured since we have
$$E_0\;\equiv\;E_e(t)\;\pm\;k(t)E_p\;\equiv\;E_e\;\pm\;E_p$$
where $E_0$ is the initial electron energy, $E_e$ its final
energy, $E_p=hf$ is the photon energy and $k(t)$ is a continuous
function that increases from $0$ to $1.$ Indeed, since photon is
fundamentally a quantum object with a fixed wavelength, its
existence is essentially related to time and distance. Its
formation (and its absorption) takes an infinitesimal fraction of
time and needs an infinitesimal extent of distance; moreover it
can not exist in a static state (i.e. independently of motion).
Then, formation and existence of photon need time, distance,
motion and speed notions. Its absorption and emission are
necessarily related to time and energy change notions.\\\\
In the following, we will give some arguments aiming to show that
the uncertainty principle, that can be written as $\Delta x\Delta
p_x\geq \frac{\overline{h}}{2}$ and $\Delta E\Delta t\geq
\frac{\overline{h}}{2},$ can not be of a canonical and universal
nature. It is only a legitimate consequence of the use of the
Schr\"odinger's equations
$$\psi(x)\;=\;\frac{1}{\sqrt{2\pi}}\int_{-\infty}^{+\infty}g(k)e^{-ikx}d\,k$$
and
$$g(k)\;=\;\frac{1}{\sqrt{2\pi}}\int_{-\infty}^{+\infty}\psi(x)e^{ikx}d\,x$$
in order to find out the probability of localizing a given
particle, under some constraints, in a given position and
determining its momentum. The uncertainty principle is stated
after using the probability distributions
$$\frac{dP}{dx}\;=\;|\psi(x)|^2\qquad\mbox{and}\qquad\frac{dP}{dk}\;=\;|g(k)|^2$$
and their standard deviations $\sigma_x$ and $\sigma_k$ as well as
the De Broglie relation $p=\frac{h}{\lambda}.$ Therefore, this
principle states simply that this particular approach and the use
of this particular method hold within themselves the uncertainty
so
quantized.\\\\
Moreover, the very definition of $\sigma_x$ and $\sigma_k$ only
means that there is a large probability for the position $x$ to be
within a distance less than $\sigma_x$ to the mean value $\langle
x \rangle$ and for the component $p_x$ of the momentum $p$ to be
within an interval less than $\sigma_k$ about the mean value
$\langle p_x \rangle$. However, it is obvious that there is lesser
probability of finding $x$ within a distance lesser than
$\sigma_x$ to $\langle x \rangle$ and a non negligible probability
for $x$ to be at a distance larger than $\sigma_x$ to $\langle x
\rangle$. Similarly, we can assert the same properties for $p_x$ and
$\sigma_k$. So, $\sigma_x$ and $\sigma_k$ only determine a
probabilistic estimation and they can not institute a sharp
limiting for the uncertainty of both position and momentum and of
their product.\\
The same reasoning can be produced when commenting on the
Heisenberg relation $\Delta A \Delta B \geq \frac{1}{2} | \langle
M \rangle |$ for any two observables $A$ and $B$ where $M$ is
defined by $\widehat{M} = -i [\widehat{A},\widehat{B}]$
([3],11.021) and particularly for $x$, $p_x$ and $-i
[\widehat{x},\widehat{p_x}] = \overline{h}.$\\\\
Nevertheless, this does not mean that the position $x$ of the
particle or its momentum $p_x,$ at a given time, can not be well
defined or can not be determined with more precision when more
physical information are specified by more efficient theoretical
or
experimental processes.\\

\subsection*{ Scaling problem}

More generally, the problem of determining the position, the
trajectory and other characteristics (such as momentum and energy,
for instance) of subatomic particles, which move with very large
speed, was one basic problem in the heart of the fundation of
quantum theory.\\
Indeed, in spite of our fantastic technical progress, we are,
until now, incapable of visualizing or perceiving these minuscule
particles and their movement and even of distinguishing between
them. The time and distance scales that suit our perception
actually are infinitely large regarding their infinitely small
world. Our centimeters and grams and our seconds are really
gigantic and inappropriate for analyzing this microworld (or
rather this nano or femtoworld).\\In spite of using the most
sophisticated means, the electron motion around the nucleus
appears for us as a foggy scene because of the infinitely small
size of the electron orbit and the infinitely large speed of the
electron. Not only we are incapable of determining its trajectory,
but we are still at the stage of contenting ourself with
determining the probability of finding it at such and such region
of the minute space around the nucleus.\\\\As for quarks, the
ultrasophisticated means and the ultraclever methods are necessary
in order to get some scanty information concerning their existence
and their characteristics which are ultra-fluctuating and even
ultra-ephemeral. However, all that does not prevent us from
conceding that the electron, for example, has, at every fixed
time, a precise position and that it has a well defined speed and
trajectory during an infinitesimal fraction of nanosecond in spite
of all evolutions it may undergo. \\\\
In order to convince ourselves that this nanoworld respects
mechanical and physical laws during infinitesimal time interval,
we can imagine that a minicreature (or a nanocreature) that is as
intelligent as us but infinitely more sensitive than us regarding
the infinitesimal distances and time-intervals making them (when
living inside the nanoworld of atoms) capable of discerning
(without using sophisticated technical means that would alter
physical characteristics) between infinitesimal particles and
noting the fractions of nanodistances between them as well as the
fractions of nanoseconds separating two minute events and finally
of perceiving the tiny transformations and fluctuations that occur
within infinitesimal space and time. Moreover, we have to imagine
that these intelligent creatures possess the means and the good
will of communicating us their observations along infinitesimal
time - intervals after registrating and schemetizing them and
above all after enlarging and rescaling them in order to make us
capable of reading the slightest details concerning positions at
very precise time and trajectories (during infinitesimal time
intervals) of the nanoparticles of this nanoworld. This has to be
done in such a manner that, for instance, the foggy scene of the
electron motion transforms for us into interlacing lines. All that
we need is to
enlarge the distances and to slow down the motions.\\\\

In other respects, we can say, for instance, that the ground state
energy of the hydrogen atom in the Bohr model is determined by the
finiteness of the electron energy and has nothing to do with the
uncertainty principle. Indeed, when an electron moves from an
energy level corresponding to a $V_1$ potential energy to another
level corresponding to a $V_0$ potential energy then it releases a
photon $\gamma$ with $E(\gamma)$ energy. If $m_i(t)$, $m^{'}_i(t)$
and $v_i(t)$ denote respectively the electron's apparent mass, the
electron's reduced mass and the electron's speed that correspond
to the $V_i$ levels, for $i=0,1,$ then we must have
$$m_1(t)c^2+\frac{1}{2}m^{'}_1(t)v_1^2(t)+V_1-E(\gamma)=m_0(t)c^2+\frac{1}{2}m^{'}_0(t)v_0^2(t)+V_0,$$
which yields
$$\Delta
V=V_1-V_0=m_0(t)c^2+\frac{1}{2}m^{'}_0(t)v_0^2(t)-m_1(t)c^2-\frac{1}{2}m^{'}_1(t)v_1^2(t)+E(\gamma).$$
This shows that $\Delta V$ and consequently $V_0$ are finite.\\
Likewise, we can state that, when we use our proper expression for
the kinetic energy $E_k$ of the Bohr atom electron (for instance),
the energy and the kinetic energy can not exceed the absolute
value of the potential energy for arbitrary $r$ because then we
would have
$$\frac{ke^2}{r} \leq \gamma(t)m_0 (c^2+\frac{v^2}{2}) <
\gamma(t)m_0(c^2+\frac{c^2}{2})$$which is impossible for
sufficiently
small $r$.\\
Therefore, there exists a finite minimal potential energy
corresponding to a finite minimal energy level for the electron
inside the hydrogen atom. This level is, as experiments
show, $V_0\simeq -13,6$ ev.\\
Besides, we notice that the inverse process to the above one takes
place after an electromagnetic or a thermal energy absorption
which leads the atom to an excited state and can even lead the
electron to a pure "separation" from its original atom and even
(occasionally) with a large kinetic energy. In that case we have
(using obvious notations)
$$m_0(t)c^2+\frac{1}{2}m^{'}_0(t)v_0^2(t)+\Delta E+V_0=m_e
c^2+\frac{1}{2}m^{'}_ev^2.$$ We mention that, for a non uniform
movement, the mass $m_0(t)$ is variable because of the radiation
phenomenon that comes with such a movement.\\
Theoretical and experimental measurements of the hydrogen atom
ground state energy show that this energy is characterized by the
planck constant $\overline{h}$ and the Bohr radius
$$r\;=\;a_0\;=\;\frac{4\pi\varepsilon_0\overline{h}^2}{me^2}$$
where $m$ is the electron mass corresponding to this energy level.
The ground state energy actually is determined by the minimal
value
$$E_m\;=\;\left(\frac{1}{2}mv^2-\frac{e^2}{4\pi
\varepsilon_0r}\right)_m\;=\;\left(\frac{\overline{h}^2}{2\pi
r^2}-\frac{e^2}{4\pi \varepsilon_0r}\right)_m.$$
\subsection{Classical versus quantum Mechanics}
It is well known that classical Mechanics and Physics are based on
some principles and laws that derive from a theoretical
formulation essentially obtained from idealizing real physical
systems and phenomena. This does not mean that observation and
experiments are less important than theoretical formulations of
classical Physics since these formulations stem from those
observations and then are adopted and improved after many
confrontations, inspections and verifications. Quantum Mechanics
consists of several predictive rules that derive from a huge
number of experiments and ends up by founding the powerful
probabilistic quantum Statistics. Some rules and results become
postulates, principles or laws because none has
observed exceptions that contradict them.\\

For our part, we maintain that the wave quantum theory structure,
which leans upon Schr\"odinger equation, is essentially
established with the (declared or undeclared) aim to be unified
with the Lagrangian and Hamiltonian Mechanics by the intermediate
of the Hamilton-Jacobi equation:
$$\mathcal{H}\left(q_j,\frac{\partial S}{\partial q_j},t\right)+\frac{\partial S}{\partial
t}=0$$ where $S$ is the Hamilton's principal function. This
equation reduces, in the well known particular case, where the
Hamiltonian is written as
$$\mathcal{H}=\frac{1}{2m}p^2+V(r,t),\quad\mbox{with}\quad p=\nabla
S\quad\mbox{and}\quad\mathcal{H}=-\frac{\partial S}{\partial t},$$
to
$$\frac{1}{2m}\left|\nabla S\right|^2+V(r,t)+\frac{\partial S}{\partial
t}=0$$ that is
$$\mathcal{H}=\frac{1}{2m}\left|\nabla S\right|^2+V(r,t).$$
Thus, the wave quantum theory is based, on one hand, upon the
notion of Schr\"odinger's wave functions (having the general form
of $\Psi(r,t)=A_0(r,t)\exp\left(i\sigma(r,t)\right)$), stationary
waves, plane, quasiplane and packet waves and, on the other hand,
upon the following eikonal equation (which is obtained when
putting $S=\overline{h}\sigma$):
$$\frac{\overline{h}^2}{2m}\left|\nabla \sigma\right|^2+V(r,t)+\overline{h}\frac{\partial \sigma}{\partial
t}=0$$ and finally upon the Schr\"{o}dinger's equation:
$$-\frac{\overline{h}^2}{2m}\nabla ^2\Psi+V(r,t)\Psi=i\overline{h}\frac{\partial \Psi}{\partial t}.$$
This latter equation is written, for a time-independent potential
$V$ and for $\Psi(r,t)\;=\;\psi(r)\exp(-i\omega t),$ as the
classical time-independent Schr\"{o}dinger's equation
$$-\frac{\overline{h}^2}{2m}\nabla ^2\psi+V\psi\;=\;\overline{h}\omega\psi\;=\;E\psi.$$
Then starts the mechanism that relates wave quantum Mechanics to
Hermitian operators (associated with Observables) and to
expectation values by means of relations such as
$$\hat{p}=-i\overline{h}\nabla\qquad,\qquad\hat{\mathcal{H}}\left(\hat{q}_j,\hat{p}_j,t\right)\Psi=i\overline{h}\frac{\partial \Psi}{\partial
t},$$
$$\hat{\mathcal{H}}\left(\hat{q}_j,\hat{p}_j\right)\psi=E\psi\qquad\mbox{for}\qquad\mathcal{H}=E=\overline{h}\omega$$
and (as a particular case)
$$\hat{\mathcal{H}}=-\frac{\overline{h}^2}{2m}\nabla ^2\;+\;V(r,t),$$
as well as the relations
$$<r>\;=\;\int \psi^*\hat{r}\psi\,d\tau$$
and
$$<p>\;=\;\int \psi^*\hat{p}\psi\,d\tau.$$
Moreover, when $\psi$ is represented with the Hamiltonian
eigenfunctions (i.e. $\hat{\mathcal{H}}\psi_n\;=\;E_n\psi_n$ for
$\psi\;=\;\sum\alpha_n\psi_n$), we get
$$<\mathcal{H}>\;=\;<E>\;=\;\sum|\alpha_n|^2E_n.$$
All this is accompanied by the uncertainty principle and extended
by
the Heisenberg matrix quantum theory.\\
It is very convenient to write down here the following quotation
of
$[3]$ that illuminates the preceding with a specific example:\\
"Attention is now directed to wave Mechanics and the immediate
objective is to derive the fundamentals of this branch of quantum
theory in a way that takes inspiration from one of Schr\"odinger's
lines of thought. As a specific example, from which broader
conclusions may be readily deduced, consider an electron moving in
a prescribed field characterized by a scalar potential
$\varphi(r,t)$ and at most a negligible vector potential $A(r,t).$
The wave which, according to experimental evidence, is in some way
associated with this electron is called the wave function and is
denoted $\Psi(r,t).$ The program of derivation begin by assuming
properties for the $\Psi-$wave such that, in a classical
situation, a packet of these waves moves according to the laws of
Newtonian mechanics and thereby "explains" the motion of the
electron. This is the spirit of the correspondence principle since
it expects as a first requirement that the new Mechanics should
predict,in a classical context, behavior appropriate to that
context. The hypotheses involved in this program are by no means
gratuitous but are suggested by Hamilton-Jacobi theory and by De
Broglie's results. Once the fundamental properties of the
$\Psi-$wave have been determined in this way, it is an easy matter
to derive the linear wave equation which $\Psi$ must obey. This
equation stands at the apex of wave mechanics; from it an enormous
number of deductions, some within the domain of classical
Mechanics but most going far beyond that domain, can be made. It
is of course, in the agreement between such deductions and the
results of experimentation that the ultimate justification of the
theory
lies".\\

The successful reconciliation between both theories has gone
beyond the status of a justification process and has led to a
hurried and non justified conclusion asserting that there exists,
in fact, a unique Mechanics which is "naturally" the quantum
Mechanics having two branches that are the wave and the matrix
quantum Mechanics; the latter, initiated by Heisenberg, is
considered as more general than the former. Moreover it is
declared that classical Mechanics is a particular case of the
quantum one and it has  to be limited to macroscopic situations.
For our part, we think that there is actually a unique theoretical
Mechanics based upon well approved mechanical and physical laws,
even though there are other ones to be discovered, checked and
improved. Many fundamental laws have been established by Newton,
Lagrange, Hamilton, Maxwell and his predecessors, Einstein, Planck
and Bohr beside of a large number of physicists and mathematicians
such as Gauss, Euler, Riemann, Fourier, Laplace, Hilbert,
Schr\"{o}dinger and
many others.\\
We have to admit that this Mechanics is not presently completely
adapted for studying infinitesimal phenomena and therefore it must
be superseded by quantum Mechanics as an efficient means for
studying microscopic phenomena such as the dynamic behavior, the
energy and the structure of particles. These phenomena are
presently beyond the reach of our measurement means and tools and
of our analyzing capacity. Until further decisive technological
and theoretical progress, the analysis of these phenomena needs
the predictive and probabilistic methods of the quantum Statistics
guided by the quantum theory of Schr\"{o}dinger, Heisenberg, Born,
Fermi, Dirac, Pauli and many others. This theory was in fact
inaugurated by Einstein, Planck and Bohr who have definitely
proved the quantum nature of waves and energy levels beside of the
quantization of electrical charges. The efficiency of these
methods are fortunately increased by numerical methods progress
and the presently huge capacity of
empirical data treatment.\\
However, we can state that, although some natural phenomena are
quantized, there are a lot of others that are not. Electrical
charges and energy levels inside the atom, for instance, are
quantized. Electromagnetic waves are constituted with integer
numbers of photons but wavelengths, speed, masses and energies,
for instance, are continuous variables evolving (themselves)
continuously with the variable that essentially gives the
continuity meaning: the
time.\\

\noindent\textbf{Remark $2.$}$\;\;$ In the previous sections, we
have established that the material and immaterial point energy is
given by $E(t)=h(t)f(t)$ where $h(t)$ and $f(t)$ depend on time
and $E(t)$ depends also on time by the intermediate of the
temperature and environment. Frequency, wavelength and energy are
then continuous mathematical objects.
\subsubsection*{Schr\"odinger probability density and classical
probability} The general Schr\"odinger equation, where
$\Psi(r,t)=A_0(r,t)\exp i\sigma(r,t),$ implies the following
equation
$$\nabla.\left(A_0^2\overline{h}\frac{\nabla\sigma}{m}\right)\;+\;\frac{\partial
A_0^2}{\partial t}\;=\;0.$$ Comparison of this equation with the
continuity equation of a substance of density $\rho$ having a
current density $\emph{\textbf{J}}=\rho \emph{\textbf{v}}$ (i.e. $\nabla\cdotp(\rho\emph{\textbf{v}})+\frac{\partial\rho}{\partial t}=0$) has led
Born to identify $\Psi^*\Psi=A_0^2$ to an imaginary substance density
$\rho.$ Then, he interpreted $\Psi^*\Psi$ as being the probability
of localizing the particle having $\Psi$ as its wave function.
Namely, the probability of finding the particle at time $t$ in a
given volume element $d\,\tau$ at position \textbf{r} is
$$dP(\textbf{r},t)\;=\;\Psi^*\Psi(\textbf{r},t)d\,\tau.$$
Since $\Psi^*\Psi$ is interpreted as a probability density, it
must obey the normalization condition
$$\int_{\mathbb{R}^3}\Psi^*\Psi d\,\tau\;=\;1.$$
However, this fundamental notion joined to another fundamental one
in Quantum theory which is the quantum measuring apparatus leads
to a paradox which is clearly explained in the following quotation
of
$[3]:$\\
"Such an apparatus does not detect that a particular system is in
a certain final state, rather it places the system in its final
state and does so with a probability that depends upon the degree
to which the final state was involved in the composition of
the initial state $!$\\
The basic paradox of quantum Mechanics exhibits itself here with
unusual clarity; a distribution of measurement results is
generated obeying a known calculus of probabilities without any
apparent internal mechanism to explain how such a distribution
comes into being. Many physicists accept this at face value,
reasoning that the ultimate theory of the universe will probably
contain elements which are incomprehensible in terms abstracted
from macroscopic experience; hence, if Quantum theory is the
ultimate theory, it is not surprising that a paradox of the type
just described should be incorporated in its makeup. Others, not
satisfied with such a state of affairs, incline toward «hidden
variable» theories. On this view point, the pre-measurement
systems of such apparatus, although quantum mechanically
indistinguishable, are actually
distinguishable in some yet more fundamental ways".\\

\noindent\textbf{Remark $3.$}$\;\;$ In the next section, we will
give a general classification of fundamental particles. Using
Dirac operator, we show that there are originally two types of
electrons that have two opposite "spins". This classification
gives a coherent explanation of the Stern-Gerlach experiment
results
which conversely give an argument that sustains it.\\

\noindent Apart from this paradox and this discussion, let us
consider, as an example, the classical case of a particle in a
box. If $\psi_n$ denotes, for large $n,$ the stationary solution
of the Schr\"{o}dinger equation, then the probability distribution
$\frac{dP}{dx}=\left|\psi_n(x)\right|^2$ can be compared to the
classical probability which is in that case equal to
$\frac{1}{L}.$ The reconciliation between these two notions
increases with increasing $n$ (c.f. $[2]$) and ends up by a sort
of justification of the Bohr correspondence principle.
Nevertheless, stationary solutions are generally considered as
being highly improbable and essentially ephemeral and the utmost
probable solutions are constituted with finite or infinite linear
combination of such solutions. For our part, we think that only
the limit cases (i.e. infinite linear combination of stationary
solutions) reveal the real physical probability of finding the
particle at a given position and
this probability is the classical one.\\
Likewise, we consider that, for the harmonic oscillator, only the
limit cases (taking parity into account) have genuine real value
and they clearly give good approximate results as (using here and
below
the notations of $[3]$):\\
$$<x>=0\;\qquad,\;\qquad<F>=0\;\qquad \mbox {and}\;\qquad<E>=\frac{1}{2}KA_0^2.$$
These results are naturally obtained, within idealized conditions,
from the well established laws of classical Mechanics and Physics.
\subsubsection*{Wave packet and Born statistical interpretation}
It is generally admitted that a wave mechanical packet represents
the center of mass of a system of particles rather than a single
particle since, in that case, there may be no particle present at
the site of the packet. This point of view which excludes the
identification of a packet wave with a particle is called the Born
statistical interpretation. However, when we attribute to a wave
packet a definite centroid to which we associate the expectation
values $<r>$ for the position and $<p>$ for the average momentum
of all individual momenta of the packet wave components, we
obtain, according to Ehrenfest's theorem that $<p>$ is equal to
the particle mass times the velocity of the centroid, and both
$<r>$ and $<p>$ obey the laws of classical Mechanics. Contrary to
the discussion about centroid of probability, hidden variables,
multiple worlds or the real existence of particle entities, we
maintain that what precedes gives only a new justification to the
legitimacy of using wave Quantum approach when studying dynamical
phenomena where classical Mechanics formulations are unreachable.
For us Ehrenfest's theorem states that statistical wave Quantum
approach is, as well as the idealizing classical Physics one, just
an approximate description of
the real physical phenomenon.\\
\subsubsection*{Relationship between wave functions and
trajectories} It is clear that a wave function $\Psi (r,t) =
A(r,t) e^{i \sigma (r,t)}$ associated with a particle (such as an
electron moving around a nucleus) that satisfies a
Schr\"{o}dinger's equation is specified by its eikonal function
$\sigma$ and its normalized amplitude $A$. The eikonal $\sigma$
which satisfies the eikonal equation determines the Hamilton's
principal function $S$ which (theoretically) determines the exact
trajectory of (the center of mass of) the particle $(q_j(t))_j$.
The particle trajectory can not be clearly perceived or specified
with our present means. All we can perceive is its gross location
at some fraction of time without discerning the particular line
that is described by it because of the too many loops that are
carried out by (the center of mass of) the particle during any
fraction of time. So the role of the amplitude of the wave
function $\Psi$ is to indicate the probability of finding the
particle in a given region within the clouded region formed by the
very swift particle into movement. Therefore $S$ is associated
with the classical Newton-Lagrange-Hamilton Mechanics whereas
$\Psi$ is associated with the quantum wave Mechanics and $\sigma$
is the connection between them.\\
Now, when we are dealing with two particles into movement, for
instance, there are two wave functions $\Psi_1$ and $\Psi_2$, two
Hamilton's principal functions $S_1$ and $S_2$ and possibly a wave
function $\Psi$ associated with the system formed with both
particles and the Hamilton's principal function $S$ associated
with the center of mass of the system. If the two particles are
distinguishable there are two trajectories and two probabilities
and as usual the probability of finding each of them inside two
pre-indicated regions is the product of the two probabilities. If
the two particles are indistinguishable bosons, then $\Psi$ is
symmetrical and the two trajectories can be arbitrarily close to
each other and they form a dense cloud which is more dense than
the cloud formed by two indistinguishable fermions in virtue of
the Pauli exclusion principle. This fact may explain the smaller
probability of finding the (indifferently located) two fermions in
a given region than that of finding two (indifferently located)
bosons in a comparable region. Each of these probabilities is
obtained by adding the amplitudes before squaring the resulting
amplitude. It is normal that the new probabilities are related to
the trajectory of each pair of particles as well as to the
properties
of each of them.\\\\
Finally, we can state that there is no antagonism between the
results of quantum and classical Mechanics. The former deals only
with microscopic physical situations (that roughly involve
moderately small "wavelengths") where classical Physics is
presently not efficient enough. Both quantum and classical
Mechanics are applicable in macroscopic physical situations,
roughly characterized by very small "wavelengths". In that case,
if classical Physics which involves the Newtonian, Lagrangian and
Hamiltonian confirmed idealizing laws is not easy to use, we can
use quantum Mechanics which involves the wave packet approximate
notion (with the imprecise De Broglie wavelength notion and the
quantum Statistic approach).\\

Let us consider, for instance, the case of two rectangular
barriers, one with relatively abrupt inclines and the other with a
relatively gradual inclines for the potential levels (c.f. [3],
p.170). The incident microscopic particle (or the packet wave) has
a relatively large wavelength for the former barrier and a
relatively small one for the second barrier. Classical Mechanics
and quantum Mechanics both give a very little probability for the
occurrence of tunneling phenomenon for the second barrier. For the
first barrier type (if a sharp incline could really exist),
classical Mechanics gives a null probability for both macroscopic
and microscopic particles. When such a tunneling does exist, we
can explain it, and other similar phenomena, by noticing that
kinetic and mass energies transform easier into potential energy
for abrupt potential energy inclines than for gradual ones,
provided high energies are involved. This also means that such
transformations are easier in smaller fraction of time. This
proves, by the way, that if we have better knowledge of the
particle physical situations, we can refine our probabilistic
expectations. The uncertainty principle corresponds to the case
where minimal physical conditions are known
about particles and systems.\\

Conversely, determinism in Mechanics is achieved when all physical
conditions are exactly known and practically realized. Initial
conditions then imply, as Laplace asserted, a unique solution that
extends wherever and whenever all conditions are known and
satisfied. If, for example, we consider a ball that is in an
actual stable equilibrium on a punctual vertex of a cone, then it
must (in ideal conditions) stay indefinitely in that state. The
solution is unique. If it is not really in a stable equilibrium
(as it is probably the case), then it rolls downward along a cone
ray in a given direction. In the idealized former case, only a
given (yet infinitesimal) applied force can make it roll down in a
given direction. This force can be determined, a posteriori,
according to Newton's laws. Then, we can not state in any case
that a well determined initial conditions for a physical system
can result, unexpectadly, on several solutions. The real problems
is the possibility of defining entirely and exactly the initial
conditions in order to predict the solution yet
in an ideal surrounding situation.\\

\subsection*{Conclusion}
We can now summarize the preceeding study by stating that:
\begin{description}
    \item[$\bullet$] A particle is never reduced to a single point.
    \item[$\bullet$] Any particle has at any time $t$ a centroid.
    \item[$\bullet$] Even for a pointlike particle the centroid can not have a
    definite geometrical position inside the particle during any
    small time interval since a particle is permanently evolving due
    to internal and external interactions and energy
    transformations.
    \item[$\bullet$] If we use the Hamilton-Jacobi equation and a specific
    Schr\"odinger function $\Psi$ that determines an eikonal
    function $\sigma$ which is proportional to a Hamilton's principal function $S,$ we can determine $\sigma$ as the solution of the equation
    $$\frac{\overline{h}^2}{2m}\left|\nabla \sigma\right|^2+V(r,t)+\overline{h}\frac{\partial \sigma}{\partial
t}=0.$$ $S$ can then be theoretically determined. Therefore, if we
assume that the particle centroid is fixed relatively to the
particle and if the initial conditions are well defined as well as
all surrounded conditions, then $S$ can be used to exactly
determine the trajectory $q=(q_j)_j$ and the momentum $p=(p_j)_j.$
    \item[$\bullet$] Since such specifications are quasi impossible, we have to
    content ourselves with using the probability density
    $$dP=\Psi^*\Psi d\,\tau$$and then our knowledge of both position
    and momentum are limited by the uncertainty principle.
    \item[$\bullet$] Nevertheless, if we could have some specific information
    about the particle and some surrounding physical conditions, we
    can hope to set down some constraints on the centroid and the momentum. Then,
    when we take a given point as approximate centroid, we can
    determine a fictitious trajectory for this point and deduce,
    using some estimates, that the trajectory of the real centroid is
    within a space tube about the fictitious trajectory during a
    reasonable time interval. If this theoretically possible
    situation is realized in practice, then (using in a similar way some estimates for the momentum) we can obtain a smaller
    uncertainty than the limit given by the uncertainty principle.
    \item[$\bullet$] Finally, when we use a wave packet for a particle (in
    macroscopic cases or in the short wave limit and the Bohr's
    correspondence principle case), we may have approximate values $<r>$
    for the position and $<p>$ for the momentum of the particle but
    then, it is sometimes possible to use the idealizing classical Physics for
    getting better approximate values.
     \item[$\bullet$] {\bf About Schr\"odinger equation:} We have already noticed that the trajectory $X(t)$ of an electron inside atoms is a geodesic with respect to the
ambiant metric $g_t$ (i.e. $\nabla^{g_t}_{X'(t)}X'(t)=0$) as long as it does not change its orbit where it has a well determined energy and approximatly a
constant speed $v$. In other respects, if $V(r,t)$ is the potential energy of the electron, then the Hamiltonian is given by $\mathcal{H}=\frac1{2m}p^2+V(r,t)=T+V$.
Letting $S(r,t)$ be the Hamilton's principal function then we have approximatly:
$$\overrightarrow{p}=\nabla S\qquad\mbox{and}\qquad \mathcal{H}=-\frac{\partial S}{\partial t}$$
and so the Hamilton-Jacobi equation can be written as
$$\frac1{2m}|\nabla S|^2+V(r,t)+\frac{\partial S}{\partial t}=0.$$
Let $\displaystyle \Psi(r,t)=A_0(r,t)e^{i\sigma}(r,t)$ be the Schr\"odinger's wave function, i.e. the average function of the wave packet associated to the electron. As
explained in Section 6.03 of [3] it is possible, under classical conditions and the non relativistic cases, to characterize the wave function $\Psi$ by its eikonal function $\sigma$ and, as it is proven in [3], the Schr\"odinger equation (6.021), i. e.
$$-\frac{h}{2m}\nabla^2\Psi+V\Psi=i\overline{h}\frac{\partial\Psi}{\partial T}$$\\
for $\Psi$ is equivalent to the eikonal equation (6.015), i. e.
$$\frac{\overline{h}^2}{2m}|\nabla \sigma|^2+V(r,t)+\overline{h}\frac{\partial \sigma}{\partial t}=0.$$
which is nothing but the Hamilton- Jacobi equation when using $S=\overline{h}\sigma$ where $\overline{h}$ is the Planck constant,
which can be considered as being constant along the orbit of the electron before changing its energy level. Therefore, we have just proved that,
under classical conditions (i.e. in the short wave length limit), the Shr\"{o}dinger equation is nothing but the above Hamilton-Jacobi
equation which can be written as $$\frac1{2m}|\nabla S|^2+V(r,t)=-\frac{\partial S}{\partial t}=\mathcal{H}=T+V(r,t)$$
which gives
$$\frac1{2m}|\nabla S|^2=T=\frac12mv^2,$$
that is, the trivial identity $$\frac1{2m}|\nabla S|^2=m^2v^2=p^2.$$
So the Schr\"odinger equation is an approximate result of the Lagrange-Hamilton-Jacobi Mechanichs and since the latter gives the Newtonian Mechanics
and the Heisenberg matricial Mechanics is equivalent to the wave quantum one, the quantum Mechanics is only a practical approximate result of the
unique theoretical Newton-Lagrange-Hamilton Mechanics.
\end{description}}

\subsection{Remarks on the quantum Statistics foundation} The aim
of this section is to show that only physical characteristics of
an interference problem (particle types, momenta, distances,
symmetries) determine the general quantum Statistics schemes.

\subsubsection*{The two slits problem scheme}
Assume that a large number $n$ of identical particles having a given momentum $p$ are directed perpendicularly toward two parallel slits $S_1$ and $S_2$ extremely close to each other and having both the same infinitely small width. We further assume that we get, on a screen located behind the slits, only two possible outcomes $E_1$ and $E_2$ having respectively $n_1$ and $n_2$ events such as $n=n_1+n_2.$ Finally, we assume that, between the $n_1$ particles reaching $E_1,$ $r_1$ particles originate from the slit $S_1$ and $s_1$ particles originate from the slit $S_2$ and that between the $n_2$ particles reaching $E_2,$ $r_2$ particles originate from $S_1$ and $s_2$ particles originate from $S_2.$\\
We then have
$$n_1=r_1+s_1\quad\hbox{and}\quad n_2=r_2+s_2.$$
Let $A_1$ and $A_2$ be two complex numbers such as
$$|A_1|^2=\frac{n_1}{n_1+n_2}\quad\hbox{and}\quad |A_2|^2=\frac{n_2}{n_1+n_2}$$
and $\alpha\in[0,2\pi]$ such that
$$|A_1|=\cos\alpha=\sqrt{\frac{n_1}{n_1+n_2}}=\sqrt{\frac{n_1}{n}}$$
and
$$|A_2|=\sin\alpha=\sqrt{\frac{n_2}{n_1+n_2}}=\sqrt{\frac{n_2}{n}}.$$
We then have
$$A_1=|A_1|e^{i\theta_1}=\cos\alpha e^{i\theta_1}$$
and
$$A_2=|A_2|e^{i\theta_2}=\sin\alpha e^{i\theta_2}.$$
The probability that an $E_1$ event (resp. $E_2$ event) originates
from $S_1$ is
$$|B_1|^2=\frac{r_1}{r_1+s_1}=\frac{r_1}{n_1}\quad(\hbox{resp.}\;|B_2|^2=\frac{r_2}{r_2+s_2}=\frac{r_2}{n_2})$$
and the probability that an $E_1$ event (resp. $E_2$ event)
originates from $S_2$ is
$$|C_1|^2=\frac{s_1}{r_1+s_1}=\frac{s_1}{n_1}\quad(\hbox{resp.}\;|C_2|^2=\frac{s_2}{r_2+s_2}=\frac{s_2}{n_2})$$
for $B_1, B_2, C_1, C_2 \in \mathbb{C}$.\\
We then have
\begin{eqnarray*}
 |B_1|&=&\sqrt{\frac{r_1}{n_1}}=\cos\beta\quad\quad |B_2|=\sqrt{\frac{r_2}{n_2}}=\cos\gamma\\
  |C_1|&=&\sqrt{\frac{s_1}{n_1}}=\sin\beta\quad\quad |C_2|=\sqrt{\frac{s_2}{n_2}}=\sin\gamma
\end{eqnarray*}
and
\begin{eqnarray*}
B_1&=&|B_1|e^{i\beta_1}=\cos\beta e^{i\beta_1}\quad\quad B_2=|B_2|e^{i\beta_2}=\cos\gamma e^{i\beta_2}\\
C_1&=&|C_1|e^{i\gamma_1}=\sin\beta e^{i\gamma_1}\quad\quad
C_2=|C_2|e^{i\gamma_2}=\sin\gamma e^{i\gamma_2}.
\end{eqnarray*}
Under these conditions, the relations
$$\left\{\begin{array}{c}
|A_1|^2=|B_1+C_1|^2\\
|A_2|^2=|B_2+C_2|^2
\end{array}
\right.
$$
are equivalent to the system
$$\left\{\begin{array}{c}
\cos^2\alpha=\cos^2\beta+\sin^2\beta+2\cos\beta\sin\beta\cos(\beta_1-\gamma_1)\\
\sin^2\alpha=\cos^2\gamma+\sin^2\gamma+2\cos\gamma\sin\gamma\cos(\beta_2-\gamma_2)
\end{array}
\right.
$$
or also to the system
$$\left\{\begin{array}{c}
\displaystyle\frac{n_1}{n}=1+2\sqrt{\frac{r_1s_1}{n_1^2}}\cos(\beta_1-\gamma_1)\\
\displaystyle\frac{n_2}{n}=1+2\sqrt{\frac{r_2s_2}{n_2^2}}\cos(\beta_2-\gamma_2).
\end{array}
\right.
$$
If we assume a perfect symmetry of the physical system, we can
state that equal number of particles passes through $S_1$ and $S_2$ which gives
$$r_1+r_2=s_1+s_2=\frac{n}{2}=:m$$
and that $r_1=s_2$ and $r_2=s_1$ which yield
$$r_1+s_1=r_2+s_2=m=n_1=n_2$$
and the above system is reduced to
$$\left\{\begin{array}{c}
\displaystyle\frac{m}{2m}=1+\frac{2}{m}\sqrt{r_1s_1}\cos(\beta_1-\gamma_1)\\
{}\\
\displaystyle\frac{m}{2m}=1+\frac{2}{m}\sqrt{r_2s_2}\cos(\beta_2-\gamma_2)
\end{array}
\right.
$$
and, putting $a=\cos(\beta_1-\gamma_1)$ and
$b=\cos(\beta_2-\gamma_2),$ to
$$\left\{\begin{array}{c}
\displaystyle\frac{m}{2}=m+2a\sqrt{r_1s_1}\\
{}\\
\displaystyle\frac{m}{2}=m+2b\sqrt{r_2s_2}.
\end{array}
\right.
$$
Adding these two equations, we get
$$2a\sqrt{r_1s_1}+2b\sqrt{r_2s_2}=-m$$
or
$$2b\sqrt{(m-r_1)(m-s_1)}=-2a\sqrt{r_1s_1}-m,$$
which gives
$$4b^2(m-r_1)(m-s_1)=4a^2r_1s_1+m^2+4a\sqrt{r_1s_1}m$$
that is
$$(4b^2-1)m^2-4[(r_1+s_1)b^2+a\sqrt{r_1s_1}]m+4(b^2-a^2)r_1s_1=0.$$
As $m$ is assumed to be an arbitrary large number, we obtain
$$\left\{\begin{array}{c}
\displaystyle 4b^2-1=0\\
\displaystyle b^2(r_1+s_1)+a\sqrt{r_1s_1}=0\\
\displaystyle b^2=a^2
\end{array}
\right.
$$
and then
$$\left\{\begin{array}{c}
\displaystyle b=\pm\frac{1}{2}\\
\displaystyle a=\pm b\\
\displaystyle b^2(r_1+s_1)+a\sqrt{r_1s_1}=0.
\end{array}
\right.
$$
These relations imply successively
\begin{eqnarray*}
  a &=& -\frac{r_1+s_1}{\sqrt{r_1s_1}}b^2= -\frac{r_1+s_1}{\sqrt{r_1s_1}}a^2,\\
  1 &=& -\frac{r_1+s_1}{\sqrt{r_1s_1}}a,
\end{eqnarray*}
$$a=-\frac{1}{2}\quad \hbox{and}\quad \frac{r_1+s_1}{\sqrt{r_1s_1}}=2$$
and finally $r_1=s_1$ which implies
$$r_1 = s_1 = r_2=s_2=\frac{m}{2}\quad n_1=n_2=m.$$
Reciprocally, $n_1=n_2=m,$ with $r_1+s_1=r_2+s_2=m,$ is the only solution to the considered system, what is perfectly legitimate as we have considered a perfect symmetrical system.\\
We notice that, the solution of this problem can not be given by
$$\left\{\begin{array}{c}
\displaystyle|A_1|^2=|B_1|^2+|C_1|^2\\
\displaystyle|A_2|^2=|B_2|^2+|C_2|^2
\end{array}
\right.
$$
or equivalently by
$$\left\{\begin{array}{c}
\displaystyle\cos^2\alpha=\cos^2\beta+\cos^2\gamma\\
\displaystyle\sin^2\alpha=\sin^2\beta+\sin^2\gamma
\end{array}
\right.
$$
since this latter leads to
$$1=1+1.$$
We can sum up the preceding by noting that the above physical
problem reduces to the determination of $6$ unknowns:
$$|A_1|^2,\;|A_2|^2,\;|B_1|^2,\;|B_2|^2,\;|C_1|^2,\;|C_2|^2,$$
knowing $5$ equations:
\begin{eqnarray*}
   &{}&|A_1|^2+|A_2|^2=|B_1|^2+|B_2|^2=|C_1|^2+|C_2|^2=1, \\
   &{}& |A_1|^2=|B_1+C_1|^2,\quad |A_2|^2=|B_2+C_2|^2. 
\end{eqnarray*}
This system is equivalent to a system of six unknowns, $n_1$, $n_2$, $r_1$, $r_2$, $s_1$ and $s_2$, that satisfy the three relations:
$$n_1+n_2=n\qquad r_1+s_1=n_1\qquad r_2+s_2=n_2.$$
The symmetry of the physical problem reduces the unknowns by $2$ and the arbitrariness of $n$ permits to uniquely resolve the system.\\

\noindent We assume now that there are, as previously, two slits $S_1$ and
$S_2$ and that only three possible outcomes $E_1,$ $E_2$ and
$E_3.$ By processing as before, we notice that $9$ variables are
associated to this problem, which are (using obvious notations)
the following:
\begin{eqnarray*}
&{}&|A_1|^2,\;|A_2|^2,\;|A_3|^2,\;|B_1|^2,\;|B_2|^2,\;|B_3|^2,\;|C_1|^2,\;|C_2|^2,\;|C_3|^2.
\end{eqnarray*}
The equations relating them are only $6$:\\
\noindent$\bullet$\;$|A_1|^2+|A_2|^2+|A_3|^2=|B_1|^2+|B_2|^2+|B_3|^2=|C_1|^2+|C_2|^2+|C_3|^2=1$.\\
\noindent$\bullet$\;$|A_1|^2=|B_1+C_1|^2,$ $|A_2|^2=|B_2+C_2|^2,$ $|A_3|^2=|B_3+C_3|^2$.\\
This system is equivalent to a system of $9$ unknowns, $n_1$, $n_2$, $n_3$, $r_1$, $r_2$, $r_3$, $s_1$, $s_2$ and $s_3$, that satisfy the $4$ relations:
$$n_1+n_2+n_3=n\qquad r_1+s_1=n_1\qquad r_2+s_2=n_2\qquad r_3+s_3=n_3.$$
The symmetry of the physical problem reduces the unknowns by $3$ ($n_1=n_3$, $r_1=r_3$, $s_1=s_3$) and the arbitrariness of $n$ permits to uniquely resolve the system.\\
Thus, pushing this reasoning to the very end, we can show that the basis of the Probability and Statistics associated with the above problem is determined by its physical characteristics. It is also shown that if the occurrence of an event $E_i$, to which is associated the complex number $A_i$, can be realized by two different ways then the probability of this occurrence, which is the amplitude of $A_i$ squared, is obtained by first adding the complex numbers that are associated to both probabilities and then squaring the amplitude of the sum.\\ 

\subsubsection*{Quantum Statistics versus classical Statistics}
Let us  consider two particles $P_1$ and $P_2$ that can only occupy two independent physical states $a_1$ and $a_2.$\\
Several questions can be formulated concerning the occupation distribution and the answer to these questions fundamentally depends on the physical characteristics of these particles.\\

\noindent$1^\circ)\;$ If the two particles are distinguishable and each of them can occupy with the same probability each of both states, we can assert that the probability that $P_1$ be in $a_1$ and $P_2$ be in $a_2$ is equal to the probability that $P_1$ be in $a_2$ and $P_2$ be in $a_1$ which is equal to the probability that $P_1$ and $P_2$ be both in $a_1$ or in $a_2.$ All these probabilities are then equal to $\frac{1}{2}\times\frac{1}{2}=\frac{1}{4}.$\\

\noindent$2^\circ)\;$ Under the same conditions as previously, we can assert that if we know that particle $P_1$ is in state $a_1,$ for example, then the probability that $P_2$ be in $a_1$ is equal to that of $P_2$ be in $a_2$ and both probabilities are equal to $\frac{1}{2}.$\\

\noindent$3^\circ)\;$  Always under the same conditions, we can also assert that the probability that both particles be in the same state $a_1$ (or $a_2$) is $\frac{1}{4}$ and the probability that $P_1$ be in $a_1$ and $P_2$ in $a_2$ (or $P_1$ be in $a_2$ and $P_2$ in $a_1$) is $\frac{1}{4}.$ Finally we can assert that the probabilities that both particles be in the same state ($a_1$ or $a_2$) and that these two particles be in different states ($a_1$ and $a_2$ or $a_2$ and $a_1$) are both equal to $\frac{1}{2}.$\\

We now suppose that particles $P_1$ and $P_2$ are indistinguishable and that the question is:\\
\noindent What is the probability that both particles be in the same state (without specifying which of the two states)? The answer then depends on the physical nature of the particles.\\

\noindent$4^\circ)\;$ If both particles are bosons (i.e. of integer spin or also having symmetrical wave function), then experimentation shows that the probability that both particles be in the same state ($a_1$ or $a_2$) is twice the probability of being in two different states $(P_1\;\hbox{in}\;a_1\;\hbox{and}\;P_2\;\hbox{in}\;a_2\;\hbox{or}\;P_1\;\hbox{in}\;a_2\;\hbox{and}\;P_2\;\hbox{in}\;a_1).$ Thus the probability of each of the two first cases is $\frac{1}{3}$ and that of each of the two last cases is $\frac{1}{6}.$ We can then state that, in accordance with Bose-Einstein Statistics, the probability that both particles be in the same state is $\frac{2}{3}$ and the probability of being in two different states is $\frac{1}{3}.$\\

\noindent$5^\circ)\;$ If both particles are fermions (i.e. of fractional spin or also having an antisymmetrical wave function) then experimentation shows that (in accordance with the Pauli exclusion principle) the probability that both particles be in the same state is null and the probability that $P_1$ be in $a_1$ and $P_2$ in $a_2$ is the same as the probability that $P_1$ be in $a_2$ and $P_2$ in $a_1$ which is $\frac{1}{2}.$ So, the probability that these two particles  be in two different states is $1.$\\

We consider now the triple experiment of collisions between particles $^4$He and $^3$He ([2],p.$340$) where we study the probability of the right angle scattering of these types of particles; the first is a boson and the second is a fermion.\\

\noindent$1^\circ)\;$ For the $^4$He$\;$-$\;$$^3$He scattering, we
have two distinguishable particles and we naturally ask for
determining the probability that the particle $^4$He be scattered
upward and the particle $^3$He be scattered downward and vice
versa. These two probabilities are equal and then we can as well
ask for determining the global probability $P_{34}$ of the right
angle scattering. This problem must be resolved with classical
Statistics. If $P_{34}$ is the probability of right angle
scattering that is observed after a large number of scattering
experiments, we can conclude that the number of events that the
particle $^4$He is scattered upward or downward is the same (this
is due to the physical symmetry of the collision:$\;$An equal
number of $^4$He particles comes slightly above or slightly under
the collision axis) and we have
$$P_{34}=a^2+a^2.$$
If we associate the amplitude $A$ to each of these probabilities
and the amplitude $B$ to the global probability of the right angle
scattering, we obtain
$$P_{34}=|B|^2=|A|^2+|A|^2=2|A|^2.$$
In other respects, for identical (indistinguishable) particles, the natural question is:\\
\noindent What is the probability of a right angle scattering of these particles independently of knowing the origin (from the right or the left) of those that are scattered upward and those that are scattered downward?\\

\noindent$2^\circ)\;$ When we consider the $^3$He - $^3$He
collision, experiments show that the probability of a right angle
scattering is null and if the amplitude $A$ is associated with the
probability of an hypothetical upward right angle scattering of
$^3$He particles coming from the right or from the left and the
amplitude $B$ with the global probability of a right angle
scattering, we have
$$P_{33}=|B|^2=|A-A|^2=0.$$
\noindent$3^\circ)\;$ Conversely, when we consider the $^4$He -
$^4$He collision, we can attribute the positive number $a^2$ to
the probability of the upward right angle scattering for $^4$He
particles coming from the right (or from the left) and we can
conclude that the global probability of a right angle scattering
is
$$P_{44}=a^2+a^2+a^2+a^2=4a^2.$$
If we associate the amplitude $A$ to the probability of an upward
right angle scattering of particles coming from the right and from
the left, and the amplitude $B$ to the probability of the global
right angle scattering, we obtain
$$P_{44}=|B|^2=|A+A|^2=4|A|^2.$$
Consequently, the physical reality of the experiment has determined that the probability of a right angle scattering of the $^4$He-$^4$He collision is twice the $^4$He-$^3$He collision and this is independent of the fact of knowing or no the number of the particles that have followed any one of the possible trajectories. The fact of knowing such details can not obviously alter the answer to any question of the type:\\
\noindent What is the probability of a right angle scattering for two beams of particles having well defined physical characteristics (momentum, spin, charge, mass)?  Only these characteristics hold the answer.\\

Concerning the above collisions, we notice that the charges, the
masses and the spin have made the difference. The Pauli exclusion
principle prevents the two particles $^3$He to get sufficiently
closer to each other in order to cause a right angle scattering.
The masses' inequality of the two particles $^3$He and $^4$He
disadvantages them to get sufficiently closer to each other in order to
cause such a scattering as it would be the case for two identical particles ${}^4$He.
\subsubsection*{Interference Bragg condition}
Let us consider the light diffraction experiment through a slit of
width $d$ $([2],\;\hbox{p.}5)$. The destructive interference is
achieved for
$$n\lambda=d \sin \theta_n$$
where $\lambda$ is the wavelength of the used light and $\theta_n$
is given by
$$\sin \theta_n=\frac{\Delta L}{d/2}=\frac{2\Delta L}{d}=n\frac{\lambda}{d}.$$
If we use the De Broglie relation $p=\frac{h}{\lambda},$ we can
write it as
$$n\frac{h}{p}=d \sin\theta_n$$
or
$$|\overrightarrow{p}.\overrightarrow{d}|=nh.$$
The destructive (or constructive) interference condition is then
traduced by a momentum quantization condition on the light's
photon. The momentum $p$ of the photon is inversely proportional
to the real wavelength of the light's ray. When we consider the
Bragg scattering of $X$ rays through a given crystal
$([2],\;\hbox{p.}142),$ we recover the same condition for a
constructive interference, that is
$$n\lambda=2d\sin\theta_n$$
where $\lambda$ is the wavelength of the used $X$ ray, $d$ is the distance between two adjacent layers of the crystal and $\theta_n$ is the angle that makes the ray with the plane of the crystal.\\
Again this condition can be written as
$$|\overrightarrow{p}.\overrightarrow{d}|=n\frac{h}{2}$$
which is a sort of a quantization on the momentum $p=\frac{h}{\lambda}$ of the used photon.\\\\
We consider now the Davisson-Germer experiment. The obtained
condition on the scattered electrons' maxima is exactly the same
as the previous one, namely
$$n\lambda=2d\sin\theta_n$$
where $\lambda$ denotes here the De Broglie «wavelength» attributed to the electron:$\;\lambda:=\frac{h}{p}.$\\
We have already seen that this practical and useful notion is derived from the fundamental notion that characterizes the used electrons (as well as all other particles) namely the momentum $p.$\\
The «constructive interference» intrinsic condition, which is
traduced here by the reflected electrons' maxima in some directions
is in fact the quantization relation
$$|\overrightarrow{p}.\overrightarrow{d}|=n\frac{h}{2}.$$
A similar interpretation can be furnished concerning the Thomson-Reid experiment. This phenomenon contributes to consider that electrons possess a wavy nature similar to electromagnetic waves. This is true in a certain sense but we do not have to deduce that the interference phenomenon here is identical to that of the electromagnetic waves and that every particle possesses a real wavelength identical to that of the electromagnetic wave. The only two fundamental common points between particles and waves (or more exactly photons) is the momentum and its quantization which is associated with the experiment physical characteristics. Recall that material particles possess other characteristics (mass, charge, spin) that photon does not possess.\\

The interference problem during a scattering from crystal is
related, beside of the physical nature of the electron, to the
atomic structure of the crystal and to the layout of the energy
bands within the crystal and to the Fermi gap of the material as
it is shown by the fact of recovering the Bragg condition when
analyzing the wave numbers
$$k=\pm\frac{2\pi}{\lambda}=\pm\frac{n\pi}{a}$$
where $a=n\frac{\lambda}{2}$ characterizes the gaps between the crystal energy bands ([2], p.373). This clearly shows that the electrons scattering (or their reflection similar to the electromagnetic wave reflection) is advantaged for some angles that are determined by a given momentum of the electrons, the energy bands of the crystal (the relation $a=n\frac{\lambda}{2}$ is, in fact, $p=n\frac{h}{2a}$)  and the Fermi gap that characterizes the material taking into account that all three factors are readily quantized.\\

Finally, we notice one of the numerous contradictions to which
leads the formula $\lambda=\frac{h}{p}$ when stated for material
particles. Indeed, when we attribute to the electron inside the
hydrogen atom (in accordance with the Bohr model) the wavelength
$\lambda$ that satisfies
$$\lambda\simeq 2\pi<r>,$$
we get ([2],p.139)
$$<\frac{1}{r}>\simeq\frac{2\pi}{\lambda}=\frac{p}{\overline{h}}.$$
Thus, using the formula
$$E_k=\frac{1}{2}mv^2$$
we obtain
\begin{eqnarray*}
  -\frac{1}{2} &=& \frac{<E_k>}{<V>}=-\frac{\frac{1}{2}m<v^2>}{ke^2<\frac{1}{r}>}\simeq- \frac{\frac{1}{2}m<v^2>}{ke^2<\frac{m<v>}{\overline{h}}>} \\
   &=& -\frac{<v>\overline{h}}{2ke^2}=-\frac{<v>\overline{h}}{2\alpha\overline{h}c}=-\frac{1}{2\alpha c}<v>
\end{eqnarray*}
which gives
$$\frac{1}{\alpha c}<v>\simeq 1$$
or
$$<v> \simeq \alpha c.$$
This approximate relation is obtained independently of the electron mass (in both cases: constant or depending on the speed) and independently of the energy levels, the momentum and the mean radius $<r>.$\\
But, we know that $\alpha\simeq\frac{1}{137}$ is quasi-constant for the significative energy scale of the hydrogen atom. Nevertheless, the relation $<v>=\alpha c$ is correct only for the ground state of the hydrogen atom.\\\\

\section{Mati\`{e}re, antimati\`{e}re et forces fondamentales}

Consid\'{e}rons l'Univers dynamique $U(t)$ en tant qu'espace
riemannien\\
$(B_e(O,t),g_t)$ o\`{u} $g_t$ est la m\'{e}trique physique \`{a}
l'instant $t > 0$. Consid\'{e}rons ensuite l'op\'{e}rateur de
Laplace-Beltrami $-\Delta$ sur $(B_{e}(O,t),g_e).$ Si $E(t,X)$ est
la distribution de la mati\`{e}re-\'{e}nergie de l'Univers
r\'{e}el \`{a} l'instant $t,$ alors elle v\'{e}rifie
l'\'{e}quation de la mati\`{e}re-\'{e}nergie suivante:
$$\Box E(t,X) = \frac{\partial^2}{\partial t^2}E(t,X) -
\Delta E(t,X) = 0 \hskip0.6cm \mbox{pour}\hskip 0.3cm X \in B(O,t)
\hskip 1.5cm (E^*)$$avec$$E(t,X)|_{S(O,t)} = 0 \mbox { pour tout }
t
> 0.$$ Une solution de cette \'{e}quation $E_\mu(t,X)$ s'\'{e}crit
sous la forme
$$E_\mu(t,X) = g_\mu(t)\psi(\frac{X}{t}) \mbox { pour \emph{X}
$\in$ $B_e$(\emph{O},\emph{t});}$$ $\mu$ \'{e}tant une valeur
propre quelconque associ\'{e}e au probl\`{e}me de Dirichlet sur la
boule unit\'{e} $B_e$(\emph{O},1) (relatif au laplacien -$\Delta$)
et $\psi$ \'{e}tant la fonction propre qui lui est associ\'{e}e.\\
Soit \emph{D} l'op\'{e}rateur de Dirac d\'{e}fini par la structure
spinorielle de la vari\'{e}t\'{e} riemannienne
($B_e$(\emph{O},\emph{t}),$g_e$). Les champs spinoriels sont, dans
ce contexte pr\'{e}cis, les sections
$$\Phi:B_e(O,t) \longrightarrow B_e(O,t)\times \Sigma_3$$
o\`{u} $\Sigma_3$ $\simeq$ $\mathbb{C}^{2^{[\frac{3}{2}]}}$ =
$\mathbb{C}^2$.\\
Ces champs spinoriels sont alors identifiables \`{a} des
fonctions:
\begin{eqnarray*}
\Phi:&B_e(O,t)& \longrightarrow \mathbb{C}^2\simeq \mathbb{R}^4\\
& X& \longrightarrow (\Phi_1(X),\Phi_2(X))\begin{array}{l}
{}\\
 =(\varphi_1(X)+i\varphi_2(X),\varphi_3(X)+i\varphi_4(X))\\
\simeq(\varphi_1(X),\varphi_2(X),\varphi_3(X),\varphi_4(X))\\
\end{array}
\end{eqnarray*}
Par ailleurs, on
a dans ce cas\\
\begin{description}
    \item[i)] $\;D^2 := D\circ D = -\left(%
\begin{array}{cc}
  \Delta & 0 \\
  0 & \Delta \\
\end{array}%
\right)\simeq -\left(%
\begin{array}{cccc}
  \Delta & 0 & 0 & 0 \\
  0 & \Delta & 0 & 0 \\
  0 & 0 & \Delta & 0 \\
  0 & 0 & 0 & \Delta \\
\end{array}%
\right)$
    \item[ii)] $\;D$ est un op\'{e}rateur elliptique formellement
auto-adjoint d'ordre 1.\\
\end{description}

Par cons\'{e}quent, l'ensemble des solutions de l'\'{e}quation
$$\hspace{1.5cm}\frac{\partial^2}{\partial t^2}\overrightarrow{E}(t,X) - D \overrightarrow{E}(t,X) = 0\;\; \mbox { avec }\;\; \overrightarrow{E}(t,X)|_{S_e(O,1)} = 0\hspace{1.5cm}(D)$$
d\'etermine un espace hilbertien admettant une base hilbertienne
de
fonctions propres $\Phi^p=(\varphi_1^p,\varphi_2^p,\varphi_3^p,\varphi_4^p),$ pour $p\in\mathbb{Z},$ de vecteurs propres associ\'es au probl\`eme de Dirichlet d\'efini en utilisant l'op\'erateur de Dirac \`a la place de l'op\'erateur de Laplace-Beltrami sur la boule unit\'e $B_e(O,1).$\\
De plus, pour une valeur propre simple $\mu_n>0$ de $-\Delta$,
$\lambda_n$ = $\sqrt{\mu_n}$ est une valeur propre de $D$ \`a
laquelle est associ\'{e} le vecteur propre
$$\Phi^n=(\varphi_1^n,\varphi_2^n,\varphi_3^n,\varphi^n_4)$$
et on a
$$D\;\Phi^n=\lambda_n\Phi^n$$
et
$$D\circ D\;\Phi^n=-\Delta\;\Phi^n=\mu_n\Phi^n.$$
D'o\`u
\begin{eqnarray*}
D\circ D (\varphi_1^n, \varphi_2^n\varphi_3^n,\varphi_4^n)&=& -(\Delta\varphi_1^n, \Delta\varphi_2^n,\Delta\varphi_3^n,\Delta\varphi_4^n)\\
&=&\mu_n(\varphi_1^n, \varphi_2^n,\varphi_3^n,\varphi_4^n);
\end{eqnarray*}
ce qui implique que $\varphi_i^n,$ pour $i=1,2,3,4,$ est une
fonction propre associ\'{e}e \`a la valeur propre $\mu_n$ relatif
au probl\`{e}me de Dirichlet classique sur $B_e(O,1)$ et on a:
$$\varphi_2^n=a_n\varphi_1^n\quad\varphi_3^n=b_n\varphi_1^n\quad\varphi_4^n=c_n\varphi_1^n.$$
Remarquons, qu'en consid\'{e}rant l'op\'{e}rateur de
Laplace-Beltrami $\displaystyle{-\Delta_{g_t}}$ sur
\emph{B}(\emph{O},$t$), on ne pourra pas avoir une relation simple
entre les valeurs propres relatives \`{a} $-\Delta_{g_{t}}$ et
celles relatives \`{a} l'op\'{e}rateur de Dirac $D_{g_{t}}$ que
dans le cas o\`{u} la courbure scalaire associ\'{e}e \`{a} $g_{t}$
est constante, ce qui pourrait \^{e}tre le cas tout au d\'{e}but
de l'\'{e}volution
de l'Univers.\\
Fixons maintenant trois valeurs propres simples $\mu_1,\mu_2$ et
$\mu_3$ du probl\`{e}me de Dirichlet sur la boule unit\'{e}
$B_e$(\emph{O},1), relatif \`{a} l'op\'{e}rateur de
Laplace-Beltrami $-\Delta$ et posons $\lambda_i=\sqrt{\mu_i}$ pour
$i=1,2,3.$ Signalons ensuite que nous pensons que la
mod\'{e}lisation math\'{e}matique qui va suivre et qui vise \`{a}
classifier les diff\'{e}rents types de la mati\`{e}re et de
l'antimati\`{e}re, ne d\'{e}pend pas du choix des $\mu_i.$ Ce
choix correspond, \`{a} notre avis, \`{a} l'instauration d'une
\'{e}chelle de mesure concernant toutes les notions fondamentales
li\'{e}es \`{a} la mati\`{e}re-\'{e}nergie, le temps et les
distances. Nous supposons alors que notre choix correspond au
syst\`{e}me international de mesure (S.I.) qui conduit \`{a}
toutes les constantes universelles classiques
de la Physique.\\
Dans cette optique, nous r\'{e}examinons, \`{a} titre d'exemple,
les relations suivantes (successivement \'{e}tablies dans les
paragraphes pr\'{e}c\'{e}dents) concernant l'\'{e}nergie
$E$($t,X$($t$)):
$$E_\mu(t,X(t)) = h_\mu(t)f_\mu(t) = c(\mu) \sqrt{\mu} = e(\mu),$$
$$E_\mu(t,X(t),T(t)) = h_\mu(t,T(t))f_\mu(t,T(t)) = c(\mu,t) \sqrt{\mu} = e(\mu,t)$$
et
$$h_P = \frac{e(\mu,t)}{f_D(\mu,t)} =
\frac{E_\mu(t,X(t))}{f_D(\mu,t)}$$ o\`{u} $h_P$ est la constante
de
Planck et $f_D$($\mu,t$) est la fr\'{e}quence de De Broglie.\\
Avec notre choix (fix\'{e}) de $\mu$, on peut \'{e}crire la
deuxi\`{e}me relation sous la forme
$$E(t) := E(t,X(t)) = h(t)f(t)$$o\`{u} \emph{h}, \emph{f} et \emph{E}
d\'{e}pendent naturellement du temps. La derni\`{e}re relation
s'\'{e}crit
$$h_P(t) = \frac{E(t,X(t))}{f_D(t)}$$qui, \`{a} son tour, s'\'{e}crit
$$E(t,X(t)) = h_P(t) f_D (t)$$qui n'est autre que la relation
classique
$$E = h_P f_D$$avec la seule diff\'{e}rence que $E$, $h_P$ et $f_D$ d\'{e}pendent ici du temps $t$ par le biais de la
temp\'{e}rature $T(t).$ Les variations de cette d\'{e}pendance est
n\'{e}gligeable
\`{a} l'\'{e}chelle cosmique et temporelle actuelle.\\\\
Notons aussi que, selon notre mod\`{e}le, $h(t)$ et $f(t)$
changent, \`{a} temp\'{e}rature constante, avec la distance (ou le
temps) mais pas leur produit. Par contre $h(t), f(t)$ et
$h(t)f(t)= E(t)$ changent avec la temp\'{e}rature. De m\^{e}me
$f_{D}$ change avec l'\'{e}nergie et par suite avec la
temp\'{e}rature. Nous pensons de plus que ce que l'on mesure dans
la plupart des exp\'{e}riences sont les grandeurs $f(t)$ et
$\lambda(t)$ et non pas $f_{D}$
et $\lambda_{D}$.\\

\subsection*{Classification de la mati\`{e}re, l'antimati\`{e}re
et l'\'{e}nergie} Consid\'erons maintenant le sous-espace propre
$E_{\lambda_1},$ engendr\'e par
$$\Phi_1=(\varphi_1,a_1\varphi_1,b_1\varphi_1,c_1\varphi_1)=:(\psi_1,\psi_2,\psi_3,\psi_4).$$
Posons ensuite
\begin{eqnarray*}
\overline{\Phi}_1=(\overline{\varphi}_1,a_1\overline{\varphi}_1,b_1\overline{\varphi}_1,c_1\overline{\varphi}_1)&:=&(-\varphi_1,-a_1\varphi_1,-b_1\varphi_1,-c_1\varphi_1)\\
&=:&(\overline{\psi}_1,\overline{\psi}_2,\overline{\psi}_3,\overline{\psi}_4)
\end{eqnarray*}
et consid\'erons les vecteurs de $\mathbb{R}^8$
$$\Gamma_1= (\psi_1, \psi_2,
\psi_3, \psi_4, \overline{\psi}_1, \overline{\psi}_2,
\overline{\psi}_3, \overline{\psi}_4)$$
 et
$$\Gamma_2 = (\overline{\psi}_1, \overline{\psi}_2, \overline{\psi}_3,\overline{ \psi}_4,
\psi_1, \psi_2, \psi_3, \psi_4).$$ On a alors
\begin{eqnarray*}
D\times D\;\Gamma_1&=&\lambda_1\Gamma_1,\\
D\times D\;\Gamma_2&=&\lambda_1\Gamma_2
\end{eqnarray*}
avec
$$\Delta \psi_i=\mu_1\psi_i,\qquad\Delta \overline{\psi}_i=\mu_1\overline{\psi}_i.$$
Rempla\c cons dans $\Gamma_1,$ $\psi_1$ par $e^{-}_{1/2},$
$\psi_2$ par $\nu_e,$ $\psi_3$ par $u_{1/2},$ $\psi_4$ par
$d_{1/2},$ $\overline{\psi}_1$ par $e^+_{-1/2},$
$\overline{\psi}_2$ par $\overline{\nu}_e,$ $\overline{\psi}_3$
par $\overline{u}_{-1/2}$ et $\overline{\psi}_4$ par
$\overline{d}_{-1/2}$ et dans $\Gamma_2,$ $\psi_1$ par
$e^{-}_{-1/2},$ $\psi_2$ par $\nu_e,$ $\psi_3$ par $u_{-1/2},$
$\psi_4$ par $d_{-1/2},$ $\overline{\psi}_1$ par $e^+_{1/2},$
$\overline{\psi}_2$ par $\overline{\nu}_e,$ $\overline{\psi}_3$
par $\overline{u}_{1/2}$ et $\overline{\psi}_4$ par
$\overline{d}_{1/2}.$ En r\'eordonnant les composantes de
$\Gamma_1$ et $\Gamma_2$ dans $\mathbb{R}^8,$ on obtient les
vecteurs d'\'energie pure
$$\Gamma_1=\left(  e_{1/2}^-, e^+_{-1/2}, \nu_e, \overline{\nu}_e, u_{1/2}, \overline{u} _{-1/2},  d_{1/2}, \overline{d}_{-1/2}\right)$$
$$\Gamma_2=\left(  e_{1/2}^+, e^-_{-1/2}, \overline{\nu}_e, \nu_e, \overline{u}_{1/2}, u _{-1/2}, \overline{d}_{1/2}, d_{-1/2}\right)$$
qui repr\'esentent chacun une polarisation diff\'erente de la m\^eme onde \'electrom-\\agn\'etique ou du m\^eme photon.\\
Ainsi, on a associ\'e \`a la solution du probl\`eme de Dirichlet $\psi_1$ l'\'electron $e^-_{1/2}$ dans $\Gamma_1$ et $e^-_{-1/2}$ dans $\Gamma_2,$ \`a la solution $\psi_2$ le neutrinos $\nu_e,$ \`a la solution $\psi_3$ le quark $u_{1/2}$ dans $\Gamma_1$ et $u_{-1/2}$ dans $\Gamma_2$ et \`a la solution $\psi_4$ le quark $d_{1/2}$ dans $\Gamma_1$ et  $d_{-1/2}$ dans $\Gamma_2$ et enfin \`a chaque solution $\overline{\psi}_i$ l'antiparticule de la particule fondamentale associ\'ee \`a $\psi_i$ avec un spin oppos\'e \`a celui de la particule.\\
Nous pensons enfin que, dans $\Gamma_1$ (resp. $\Gamma_2$), les couples $(e^-_{1/2},e^+_{-1/2})$ (resp. $(e^+_{1/2},e^-_{-1/2})$) \'etant dans la premi\`ere case, chacun des couples impliquant les neutrinos-antineutrinos et les quarks-antiquarks peut se situer dans chacune des trois autres cases de $\mathbb{R}^8\simeq\mathbb{R}^2\times\mathbb{R}^2\times  \mathbb{R}^2\times\mathbb{R}^2.$\\
Ceci \'evoque, en effet, la sym\'etrie de couleurs dans le
mod\`ele standard et pourrait expliquer l'existence de chacun des
quarks et des antiquarks en trois variantes distinctes.
L'existence de trois couleurs attribu\'ees \`a chaque saveur de
quark est bien confirm\'ee par le taux de formation de hadrons
lors des exp\'eriences d'annihilations \'electron-positron. Ceci
permet \'{e}galement d'expliquer l'existence de plusieurs types de gluons avec des couleurs mix\'{e}es.\\
En proc\'edant de la m\^eme mani\`ere en consid\'erant, \`a la
place de $\mu_1$ et $\lambda_1$ les valeurs propres $\mu_2$ et
$\lambda_2$ et ensuite les valeurs propres $\mu_3$ et $\lambda_3,$
on obtient les vecteurs d\'energie pure
\begin{eqnarray*}
\Gamma'_1&=&\left(\mu^-_{1/2},\mu^+_{-1/2},\nu_\mu, \overline{\nu}_\mu,c_{1/2},\overline{c}_{-1/2},s_{1/2},\overline{s}_{-1/2}\right)\\
  \Gamma'_2&=&\left(\mu^+_{1/2},\mu^-_{-1/2},\overline{\nu}_\mu,\nu_\mu,\overline{c}_{1/2},c_{-1/2},\overline{s}_{1/2}, s_{-1/2}\right)\\
  \Gamma''_1&=&\left(\tau^-_{1/2},\tau^+_{-1/2},\nu_\tau, \overline{\nu}_\tau, b_{1/2},\overline{b}_{-1/2}, t_{1/2},\overline{t}_{-1/2}\right)\\
  \Gamma''_2&=&\left(\tau^+_{1/2},\tau^-_{-1/2},\overline{\nu}_\tau,\nu_\tau,\overline{b}_{1/2}, b_{-1/2}, \overline{t}_{1/2},t_{-1/2}\right).
    \end{eqnarray*}
Ainsi, on trouve (dans le cadre de notre mod\`ele) que les
particules fondamentales sont au nombre de $24,$ douze particules:
$$\;e^-,\;\mu^-,\;\tau^-,\;\nu_e,\;\nu_\mu,\;\nu_\tau,\;u,\;d,\;s,\;c,\;b\;\mbox{et}\; t$$
 ainsi que leurs douze antiparticules. Chacune des trois permi\`eres particules (ainsi que chaque antiparticule) existe en deux variantes qui correspondent \`a deux spins oppos\'es. Chacune des trois suivantes existent uniquement avec un spin n\'egatif $-1/2$ (des particules gauches). Les six saveurs de quarks existent avec deux spins oppos\'es chacun mais aussi avec trois couleurs diff\'erentes.\\
Chacune de ces particules et antiparticules (\`a l'exception probablement des neutrinos et des antineutrinos) est form\'ee d'une distribution $E(t,X)$ sur un domaine $D_t$ formant ainsi une masse mat\'erielle (ou antimat\'erielle) donn\'ee $m(t)$ avec une densit\'e $\rho(t)$ donn\'ee.\\
Ainsi, les particules fondamentales (conformement \`a notre mod\`ele) sont les particules issues des solutions $\Phi^n$ du probl\`eme de Dirichlet (\emph{D}) relatif \`a l'op\'erateur de Dirac sur $B_e(O,1)$ qui d\'eterminent chacune $4$ solutions du probl\`eme de Dirichlet relatif \`a l'op\'erateur de Laplace-Beltrami sur $B_e(O,1).$ Ces solutions d\'eterminent, d'une part, des vecteurs \'energie $\Gamma$ associ\'es au trois g\'en\'erations de leptons-quarks et, d'une autre part, d\'eterminent des solutions de l'\'equation de la mati\`ere-\'energie $(E^*)$ donnant naissance \`a des distributions $E(t,X)$ sur des domaines $D_t$ formant de la sorte les particules (et antiparticules) fondamentales ayant des masse-\'energies donn\'ees \'evoluant avec le temps, la temp\'erature et toute sorte d'interaction.\\
Toutes ces particules (\`{a} l'exception des neutrinos) sont
soumises \`a des interactions entres elles-m\^eme et avec des
photons. Toutes peuvent s'annihiler avec leurs antiparticules pour
cr\'eer des photons et des gluons.
Inversement, les photons et les gluons peuvent donner naissance \`{a} des paires de particules - antiparticules.\\
Seuls les \'electrons, les neutrinos et le quark $u$ sont
absolument stables contre toute d\'esint\'egration. Les cinq
quarks les plus massifs (par ordre d\'ecroissant:
$\;t,b,c,s\;\mbox{et}\; d$) peuvent se transformer par
d\'esint\'egration naturelle et par des interactions faibles ou
\'electromagn\'etique pour donner naissance \`a des quarks moins
massifs. Le quark le moins massif $u$ a besoin d'un apport
\'energ\'etique pour se tranformer en un autre quark. Cet apport
peut \^{e}tre fourni lors d'int\'{e}ractions avec des
\'{e}lectrons ou des antineutrinos (par exemple). Le quark
\emph{u} peut aussi se transformer lors d'une transformation
nucl\'{e}aire s'effectuant avec une
augmentation de l'\'{e}nergie de liaison. Les familles leptoniques $\mu$ et $\tau$ donnent toujours naissances \`a des \'electrons (ou des positrons) et \`a des neutrinos (ou des antineutrinos).\\
Les neutrinos sont cr\'e\'es au cours d'op\'erations
d'annihilations, de d\'esint\'egrat-\\ions et d'interactions
faibles entre particules de tout genre. Ils sont absolument
stables et \'{e}lectriquement neutres et constitueraient des
particules mat\'{e}rielles, \`a moins qu'ils ne soient une forme
d\'energie pure, ce qui est pour nous plus probable. Toutefois,
les masses infinit\'esimales hypoth\'{e}tiques de ces particules
ne sont pas encore d\'{e}termin\'{e}es d'une mani\`{e}re
pr\'{e}cise. Notons que les seules particules mat\'{e}rielles
\'{e}l\'{e}mentaires stables (i.e. de dur\'{e}e de vie
illimit\'{e}e) sont l'\'{e}lectron (charg\'{e} n\'{e}gativement),
le quark \emph{u} (charg\'{e} positivement) et les trois neutrinos
(\'{e}lectriquement neutres). A ces particules
\'{e}l\'{e}mentaires stables, il faut ajouter la seule particule
mat\'{e}rielle composite consid\'{e}r\'{e}e jusqu'\`{a}
pr\'{e}sent comme \'{e}tant absolument stable contre toute sorte
de d\'{e}sint\'{e}gration (ou plus pr\'{e}cis\'{e}ment de dur\'ee
de vie sup\'{e}rieure \`{a} $10^{32}$ ans) qui est le proton qui a
une charge \'{e}lectrique oppos\'{e}e \`{a} celle de
l'\'{e}lectron. Les particules immat\'{e}rielles (de masse nulle)
d'\'{e}nergie pure qui sont les photons (et les gluons) sont
\'{e}galement stables. Elles rec\`{e}lent en elles-m\^{e}mes la
pr\'{e}existence de la mati\`{e}re et la capacit\'{e} de
cr\'{e}ation des diff\'{e}rents types de mati\`{e}re et
d'antimati\`{e}re et la capacit\'{e} d'interaction avec toute
sorte de mati\`{e}re et de se transformer en toute sorte
d'\'{e}nergie.\\

\subsection*{ Les forces naturelles}

En ce qui concerne les forces naturelles, il r\'{e}sulte de notre
\'{e}tude globale qu'il existe essentiellement deux forces
fondamentales inh\'{e}rentes \`{a} la cr\'{e}ation, la formation
et l'\'{e}volution de l'Univers. La premi\`{e}re est la force
\'{e}lectromagn\'{e}tique trouvant son origine d\`{e}s la
formation de deux sortes de mati\`{e}re (et d'antimati\`{e}re);
l'une charg\'{e}e positivement et l'autre n\'{e}gativement.
L'addition arithm\'{e}tique des charges de l'une et de l'autre (de
l'\'{e}lectron et des six quarks) conduit \`{a} l'\'{e}galit\'{e}
$$ 1 + \frac{1}{3} + \frac{1}{3} +\frac{1}{3} = \frac{2}{3} +
\frac{2}{3} + \frac{2}{3}$$L'attraction entre deux particules
charg\'{e}es diff\'{e}remment rappelle l'unit\'{e} originelle de
la mati\`{e}re-\'{e}nergie. La r\'{e}pulsion entre les particules
pareillement charg\'{e}es rappelle l'expansion, le mouvement et la
dispersion originels de la mati\`{e}re. Cette force a une
port\'{e}e illimit\'{e}e tout en \'{e}tant proportionnelle aux
produits des charges et \`{a} l'inverse du carr\'{e} de la
distance. Elle s'exerce \`{a} tous les \'{e}chelons de la
mati\`{e}re (et de l'antimati\`{e}re) charg\'{e}e : quark-quark,
quark-lepton, lepton-lepton et plus particuli\`{e}rement :
\'{e}lectron-\'{e}lectron, \'{e}lectron-proton, proton-proton
\`{a} l'int\'{e}rieur de l'atome et entre les atomes et plus
g\'{e}n\'{e}ralement entre toutes les agglom\'{e}rations de
mati\`{e}re charg\'{e}e. La force \'{e}lectromagn\'{e}tique
r\'{e}sulte des champs \'{e}lectriques et
\'{e}lectromagn\'{e}tiques qui existent naturellement autour de
toute mati\`{e}re charg\'{e}e statique ou en mouvement. Elle n'a
pas besoin d'\^{e}tre v\'{e}hicul\'{e}e par un interm\'{e}diaire
quelconque. Les ondes \'{e}lectromagn\'{e}tiques ne sont que des
radiations \'{e}mises par des charges en acc\'{e}l\'{e}ration ou
par des \'{e}lectrons changeant leur niveau \'{e}nerg\'{e}tique
(pour ne parler que des radiations atomiques et non pas des
radiations nucl\'{e}aires) d'une fa\c{c}on ind\'{e}pendante des
r\'{e}cepteurs possibles de ces radiations. L'\'{e}change
d'\'{e}nergie v\'{e}hicul\'{e} par les radiations
\'{e}lectromagn\'{e}tiques (\'{e}change d'une grande quantit\'{e}
de photons) a pour seul effet la modification du champ
\'{e}lectromagn\'{e}tique ambiant modifiant ainsi le champ de
force ambiant qui n'est autre que le champ de gradient de la
distribution de la mati\`{e}re-\'{e}nergie
\emph{E}(\emph{t},\emph{X}). Cette distribution est
refl\'{e}t\'{e}e, comme on l'a d\'{e}j\`{a} vu, par la
m\'{e}trique r\'{e}elle physique qui, \`{a} son tour,
caract\'{e}rise la partie
ambiante de l'Univers physique (de ces particules et ces agglom\'{e}rations) qui est en permanente \'{e}volution.\\

La deuxi\`{e}me force fondamentale est la gravit\'{e}
g\'{e}n\'{e}raliz\'{e}e qui est bas\'{e}e sur la tendance des
particules \`{a} s'attirer et de la mati\`{e}re \`{a} se
contracter pour retrouver son \'{e}tat quasi-originel ultra
condens\'{e} comme les \'{e}toiles \`{a} neutrons et m\^{e}me pour
se retrouver \`{a} l'\'{e}tat de trou noir. Cette tendance est
contrecarr\'{e}e par la r\'{e}pulsion \'{e}lectromagn\'{e}tique et
l'exclusion fermionique d'une part et par l'\'{e}nergie
cin\'{e}tique li\'{e}e au mouvement initi\'{e} par le Big Bang
d'une autre part. La force gravitationnelle s'exerce d'une
mani\`{e}re tout \`{a} fait naturelle \`{a} tous les \'{e}chelons
de la mati\`{e}re: quark-quark, nucl\'{e}on-nucl\'{e}on,
noyau-\'{e}lectron, atome-atome, astre-m\'{e}t\'{e}orite,
astre-astre, galaxie-galaxie etc. Elle est proportionnelle au
produit des masses et inversement proportionnelle au carr\'{e} de
la distance. Elle a une port\'{e}e illimit\'{e}e et n'a pas besoin
ni de graviton ni de gluon pour la v\'{e}hiculer. Une masse
donn\'{e}e est entour\'{e}e par un champ gravitationnel qui attire
proportionnellement toute autre masse situ\'{e}e aux alentours.
C'est \'{e}galement le cas des trous noirs qui constituent
probablement une grande
partie de l'\'{e}quivalent de la masse de l'Univers.\\
Bien que les deux forces fondamentales s'exercent
simultan\'{e}ment et ne s'excluent pas l'une l'autre (les deux
champs existent bel et bien autour d'une mati\`{e}re charg\'{e}e),
l'\'{e}tude des lois qui gouvernent ces deux forces montre,
qu'\`{a} l'\'{e}chelle non cosmologique, la force
\'{e}lectromagn\'{e}tique est pr\'{e}pond\'{e}rante sur la force
gravitationnelle \`{a} moins qu'il ne s'agisse de deux
agglom\'{e}rations mat\'{e}rielles de masses assez importantes et
de faibles charges \'{e}lectriques (comme une pomme situ\'{e}e
\`{a} quelques m\`{e}tres de la surface de la terre). Cependant,
il faut noter que les formules (statiques) caract\'{e}risant ces
deux forces, notamment $F = k\frac{q_1q_2}{r^2}$ et $F =
G\frac{m_1m_2}{r^2}$, sont inad\'{e}quates \`{a} une \'{e}chelle
subatomique. En effet, en ce qui concerne la force
gravitationnelle, il faut tenir compte du mouvement important
(vibrationnel et autre) des sous particules (tr\`es
\'energ\'etiques et situ\'ees \`a des distances infimes les unes
des autres) et introduire des corrections (g\'{e}n\'{e}ralement
qualifi\'{e}es de relativistes) qui tiennent compte des vitesses,
des acc\'{e}l\'{e}rations, des fr\'{e}quences, des moments et de
l'\'{e}volution des masses. Quant \`{a} la force
\'{e}lectromagn\'{e}tique, alors les corrections qui doivent
\^{e}tre effectu\'{e}es \`{a} la loi de Coulomb sont
d\'{e}terminantes. \`{A} l'int\'{e}rieur des atomes et des
nucl\'{e}ons, c'est n\'{e}cessairement la loi de Lorentz $F$ =
$q$($E$ + $v \wedge B$) qui doit \^{e}tre appliqu\'{e}e. C'est le
champ magn\'{e}tique global (orbital et intrins\`{e}que) $B$ qui
effectue les corrections indispensables aux forces
\'{e}lectromagn\'{e}tiques r\'{e}pulsives et attractives au sein
de l'atome, du noyau et des nucl\'{e}ons en tenant compte de
l'important fait que des charges en mouvement m\^{e}me si elles
sont de m\^{e}me nature (positive ou n\'{e}gative) cr\'{e}ent des
champs de force pouvant \^{e}tre attractives ou r\'{e}pulsives
suivant les directions du mouvement; ces deux effets pouvant
\^{e}tre extr\^{e}mement forts lorsque la vitesse $v$ est tr\`{e}s
grande ou "relativiste". L'association de ces champs de force
(incluant l'effet d'\'{e}cran) \`{a} celui de la force de
gravitation (corrig\'{e}e par les effets dynamiques et leurs
cons\'{e}quences multi-\'{e}nerg\'{e}tiques) conduit au champ de
force interne \`{a} chaque \'{e}chelon sans oublier les
forces externes et les interactions internes et externes.\\

Concernant les facteurs classiques $\alpha$, $\alpha_g$, $\alpha_s$,
et $\alpha_w$, seul le facteur
\\ \'el\'ectromagn\'etique $\alpha$ est naturellement d\'efini
par(c.f.[2])
$$ \alpha = \frac{2\pi k e^2}{E_{photon}\lambda_{photon}}=\frac{k e^2}{\overline{h}c}.$$
Il est \'evident que la d\'efinition de $\alpha_g$ par
$$\alpha_g = \frac{G m_p m_p}{1240}$$
et
$$\alpha_g(E)=\frac{G(\frac{E}{c^2})^2}{\overline{h}c} $$
est moins naturelle. Il en est de m\^eme pour les d\'efinitions
$$\alpha_w(E)\simeq \left(\frac{kg^2}{\overline{h}c}\right)\left(\frac{E}{m_w c^2}\right)^2\simeq\alpha\left(\frac{E}{m_w
c^2}\right)^2$$ et $$\alpha_s(E)\simeq
\frac{12\pi}{(33-2m_f)\ln(\frac{E^2}{\Lambda^2})}$$ pour $g\simeq e$
et $\Lambda = 0,2$ GeV.

\noindent Lorsqu'on admet la l\'egitimit\'e de ces d\'efinitions, en
tenant compte de la d\'ependance de ces derniers facteurs de
l'\'energie et de la distance, on obtient, au lieu du sh\'ema de la
figure $(15)$ (c.f. [6]), le sh\'ema de la figure $(15')$ pour le
facteur unifi\'e $\alpha$ qui d\'epend fortement de la distance
consid\'er\'ee.

\subsection*{Interactions fortes et faibles}

Consid\'{e}rons, \`{a} titre d'exemple, un neutron constitu\'{e}
d'un \'{e}tat de liaison (bound state) de deux quarks charg\'{e}s
n\'{e}gativement et d'un troisi\`{e}me charg\'{e} positivement,
symbolis\'{e} par $udd$. Deux causes fondamentales poussent ces
trois quarks \`{a} former un \'{e}tat de liaison, muni d'une
certaine coh\'{e}sion au sein de cette particule. Ce sont
l'unit\'{e} originelle de la mati\`{e}re et leurs charges
fractionnelles. Un quark individuel n'existe plus dans la nature,
tr\`{e}s peu de temps apr\`{e}s le Big Bang, ce qui induit l'effet
de confinement. Il en est de m\^{e}me pour une particule de charge
\'{e}lectrique fractionnelle. Le champ de force int\'{e}rieur qui
emp\^{e}che la dispersion de ces trois quarks ou la s\'{e}paration
de l'un d'entre eux r\'{e}sulte du champ de gravit\'{e} tripolaire
corrig\'{e}e et du champ \'{e}lectromagn\'{e}tique total; c'est ce
champ global qui est responsable de la coh\'{e}sion interne du
neutron. L'\'{e}nergie de liaison (binding energy) est \'{e}gale
\`{a}
$$m_u c^2 + 2 m_d c^2 - m_n c^2$$
Cet \'{e}tat de liaison n'est pas statique, comme pour tous les
hadrons, et elle n'est pas stable, comme pour tous les hadrons
\`{a} l'exception du proton. En effet, il y a, au sein du neutron,
un \'{e}change permanent de particules d'\'{e}nergie pures (i.e.
de masse nulle), appel\'{e}es gluons, d'une mani\`{e}re analogue
\`{a} l'\'{e}change permanent de photons au sein d'un atome. De
plus, il y a une interaction \'{e}lectromagn\'{e}tique (et
gravitationnelle) entre chacun de ces trois quarks avec les quarks
avoisinants contenus dans les nucl\'{e}ons voisins du m\^{e}me
noyau, de la m\^{e}me mani\`{e}re qu'il y a une interaction
\'{e}lectromagn\'{e}tique entre protons et \'{e}lectrons de deux
atomes voisins au sein d'une mol\'{e}cule. \`{A} tout ceci
s'ajoute l'attraction \'{e}lectromagn\'{e}tique entre tous les
nucl\'{e}ons au sein du noyau, doubl\'{e}e des forces
d'interaction \'{e}lectromagn\'{e}tique entre le noyau et les
\'{e}lectrons et entre toutes les particules de l'atome, sans
oublier les interactions \'{e}lectromagn\'{e}tiques entre l'atome
et l'ext\'{e}rieur. Les interactions \'{e}lectromagn\'{e}tiques
peuvent conduire \`{a} la d\'{e}sint\'{e}gration (qualifi\'{e}e de
faible) du neutron qui se transforme en proton par le
proc\'{e}d\'{e} du $\beta^--$d\'{e}sint\'{e}gration
($\beta^--$decay), symbolis\'{e} par $n \longrightarrow p + e^- +
\overline{\nu_e}$, qui s'effectue en fait par l'\'{e}mission, par
le quark \emph{d}, d'une particule "virtuelle" (qui est, \`{a}
notre avis bien r\'{e}elle) not\'{e}e $W^{\ast^-}$ qui donne
naissance \`{a} un \'{e}lectron accompagn\'{e} d'un antineutrinos
selon le sch\'{e}ma
$$ W^{\ast^-} \longrightarrow e^- + \overline{\nu_e}$$
Soulignons quand m\^{e}me que dans un grand nombre de processus
d'interactions et de d\'{e}sint\'{e}grations (dites faibles) trois
particules not\'{e}es $W^+, W^-$ et $Z^0$ existent bel et bien; ce
sont des bosons mat\'{e}riels ayant des \'{e}nergies de masses
bien d\'{e}termin\'{e}es (bien qu'on leur attribue des valeurs
approximatives essentiellement \`{a} cause de leur nature
dynamique et \'{e}volutive) et ayant une tr\`{e}s courte dur\'{e}e
de vie. Signalons, \`{a} titre d'exemple, quelques sch\'{e}mas
repr\'{e}sentatifs de d\'{e}sint\'{e}grations et d'interactions
(fig.7) (appel\'{e}s diagrammes de Feynman) faisant intervenir des
quarks, des leptons et des neutrinos impliquant ces trois
particules agissant en tant qu'interm\'{e}diaires d'interactions
dites faibles (cf. [2]).\\\\

La description pr\'{e}c\'{e}dente de l'\'{e}tat de liaison
constituant le neutron reste valable lorsqu'on consid\`{e}re un
proton form\'{e} d'un \'{e}tat de liaison entre deux quarks $u$
(charg\'{e}s positivement) et un quark $d$ (charg\'{e}
n\'{e}gativement) au sein du noyau d'un atome donn\'{e}. L'effet
de r\'{e}pulsion \'{e}lectrique entre deux quarks charg\'{e}s
positivement au sein d'un proton ou de deux protons au sein du
noyau de cet atome est contrebalanc\'{e} par les forces
attractives cr\'{e}\'{e}es par le mouvement, les champs
\'{e}lectromagn\'{e}tiques (surtout les champs magn\'{e}tiques) et
les champs de gravitation conduisant ainsi \`{a} un \'{e}quilibre
\'{e}nerg\'{e}tique \`{a} tout instant (bien qu'\'{e}voluant en
permanence) entre \'{e}nergie de masse, \'{e}nergies potentielles
et \'{e}nergie cin\'{e}tique. L'\'{e}nergie de liaison au sein du
proton est \'{e}gale \`{a}
$$2 m_u c^2 + m_d c^2 - m_p c^2$$
et au sein du noyau \`{a}
$$\Sigma m_p c^2 + \Sigma m_n c^2 - m_N c^2$$
Notons que, pour les noyaux l\'{e}gers, la stabilit\'{e} du noyau
contre la d\'{e}sint\'{e}-\\gration "forte" comme, par exemple, la
fission naturelle et la d\'{e}sint\'{e}gration de type $\alpha$
($\alpha$-decay), n\'{e}cessite l'\'{e}galit\'{e} approximative
entre les deux nombres \emph{Z} et \emph{A}-\emph{Z} de l'atome,
tandis que la stabilit\'{e} des noyaux lourds n\'{e}cessite un
nombre de neutrons nettement sup\'{e}rieur \`{a} celui des
protons. L'effet de r\'{e}puls-\\ion pour un tel noyau est assez
important lorsque \emph{Z} est proche de \emph{A}-\emph{Z} et ne
peut pas \^{e}tre durablement contrebalanc\'{e} par la composante
attractive de l'effet \'{e}lectromagn\'{e}tique et par l'effet
attractif de la gravit\'{e}. Les $\beta$-d\'{e}sint\'{e}grations
dites "faibles" et l'\'{e}mission des rayons \emph{X} sont
toujours possibles naturellement et artificiellement (au moyen des
interactions faibles et des radiations \'{e}lectromagn\'{e}tiques)
sauf pour le proton qui est, comme on l'a d\'{e}j\`{a}
signal\'{e}, suppos\'{e} absolument stable. Notons, l\`{a} aussi,
que les interactions dites fortes sont effectivement
v\'{e}hicul\'{e}es par l'interm\'{e}diaire des gluons qui sont des
bosons de masse nulle \'{e}chang\'{e}s (\'{e}mis et absorb\'{e}s)
\`{a} courte distance au sein du nucl\'{e}on et du noyau,
contrairement aux d\'{e}sint\'{e}grations de type $\alpha$,
$\beta$ et $\gamma$ qui sont d'une nature diff\'{e}rente. Faut-il
pour autant consid\'{e}rer les gluons comme \'{e}tant
fondamentalement diff\'{e}rents des photons
\'{e}lectromagn\'{e}tiques?\\
Notre r\'{e}ponse est Non; qu'ils soient virtuels ou r\'{e}els,
intervenant dans des interactions dites fortes comme celle
repr\'{e}sent\'{e}e par (fig.8) ou dans des op\'{e}rations du
genre $g \longleftrightarrow q + \overline{q}$, ils ne sont pas
essentiellement diff\'{e}rents des photons. Ces derniers se
pr\^{e}tent, d'une fa\c{c}on similaire, \`{a} des \'{e}changes,
des interactions et des annihilations. Il suffit alors d'invoquer
les annihilations du genre $\gamma \longleftrightarrow e^+ + e^-$
et les annihilations avec des productions en paires comme celles
repr\'{e}sent\'{e}es par (fig.9), en les rapprochant du
ph\'{e}nom\`{e}ne du jet lors d'une collision entre \'{e}lectrons
et positrons, repr\'{e}sent\'{e}e par: $e^+ + e^- \longrightarrow
q + \overline{q} + g$, par exemple. La diff\'{e}rence entre ces
deux types de particules d'\'{e}nergie pure r\'{e}side dans le
fait que les gluons se pr\`{e}tent uniquement \`{a} des
\'{e}changes d'une port\'{e}e et d'une dur\'{e}e extr\^{e}mement
courtes tandis que les photons sont \'{e}chang\'{e}s \`{a} toute
port\'{e}e mais aussi \'{e}mis dans des conditions diverses
ind\'{e}pendemment de tout
receveur potentiel.\\
Par cons\'{e}quent, nous ne pensons pas qu'il y a une
troisi\`{e}me force fondamentale, ind\'{e}pendante des deux forces
fondamentales d\'{e}crites auparavant, qui serait bas\'{e}e sur
des charges dites fortes (ou couleurs) inh\'{e}rentes aux quarks
(et antiquarks) et qui serait v\'{e}hicul\'{e}e par
l'interm\'{e}diaire des gluons (comme les photons, les gluons sont
des messagers d'interactions non pas de charges). N\'{e}anmoins,
chacun des six quarks existe bel et bien avec trois couleurs
diff\'{e}rentes de la m\^{e}me fa\c{c}on que certaines particules
existent avec deux spins oppos\'{e}s. Les trois couleurs
constituent en fait trois variantes d'une m\^{e}me particule sans
que l'on puisse d\'{e}tecter (jusqu'\`{a} pr\'{e}sent) aucune
diff\'{e}rence physique entre elles contrairement \`{a} la
diff\'{e}rence physique notable entre deux variantes d'une
m\^{e}me particule poss\'{e}dant deux spins oppos\'{e}s. Notre
point de vue est d'ailleurs appuy\'{e} par le fait qu'il n'existe
pas de particule \'{e}l\'{e}mentaire assez stable (tel un
nucl\'{e}on par exemple) constitu\'{e}e de quarks portant des
charges de m\^{e}me signe ni de noyau, par exemple, form\'{e}
uniquement de deux protons ou de deux neutrons: La conjonction de
la gravit\'{e} et des forces \'{e}lectromagn\'{e}tiques ne le
permettant pas, alors les ''charges fortes'' ou les ''forces
fortes'' n'auraient-elles pas
jou\'{e} un r\^{o}le favorisant?\\\\
Ainsi, nous pouvons affirmer que, bien que les forces et charges
dites fortes n'existent pas r\'{e}ellement, les interactions et
liaisons fortes existent bel et bien. Les liaisons fortes,
organiquement li\'{e}es aux deux forces fondamentales, contribuent
avec elles (dans le respect de toutes les r\`{e}gles
\'{e}nerg\'{e}tiques) \`{a} former tous les \'{e}tats de liaison
(bound state) entre quarks au sein des hadrons et entre hadrons au
sein des noyaux. La stabilit\'e de ces \'etats de lisaison est
directement li\'ee \`a la stabilit\'e de l'\'equilibre
\'energ\'etique qui se cr\'ee plus ou moins temporairement entre
les \'energies potentielles (\'electromagn\'etique et
gravitationnelle), les \'energie cin\'etiques vibrationnelle et
rotationnelle (thermodynamiques) et les \'energies de masse de
toutes les composantes. Les interactions et les liaisons fortes
sont essentiellement li\'{e}es \`{a} la transformation de la
mati\`{e}re en \'{e}nergie et r\'{e}ciproquement. Ces
transformations sont v\'{e}hicul\'{e}es par l'interm\'{e}diaire
des gluons \'{e}chang\'{e}s entre quarks favorisant plusieurs
mutations \'{e}nerg\'{e}tiques transformant ainsi leur nature et
celle des hadrons qui les contiennent. On peut avoir un changement
de masse, d'\'{e}nergie de liaison (binding energy) ou de charge
\'{e}lectrique (accompagn\'{e} d'\'{e}mission d'\'{e}lectron ou de
positron, par exemple) conduisant \`{a} toutes les transformations
nucl\'{e}aires possibles. Ceci explique d'ailleurs la tr\`{e}s
courte port\'{e}e ainsi que la tr\`{e}s courte dur\'{e}e de vie
des gluons, contrairement aux photons qui
sont \'{e}mis sans condition pr\'{e}alable d'\'{e}change.\\\\

De m\^{e}me, toutes les interactions faibles faisant intervenir
les bosons $W^+$, $W^-$ et $Z^0$ (dont on a cit\'{e} quelques
exemples) et les d\'{e}sint\'{e}grations faibles (\'{e}tudi\'{e}es
en parall\`{e}le avec des interactions du genre
$$e^+ + e^- \longrightarrow \psi' \longrightarrow \pi^+ + \pi^- +
\psi \mbox { avec } \psi \longrightarrow e^+ + e^- ,$$ou du genre
$$e^+ + n \longrightarrow \overline{\nu_e} + p \hskip 1cm , \hskip 1cm e^- + p
\longrightarrow \nu_e + n ,$$
$$p + p \longrightarrow d + e^+ + \nu_e \hskip 1cm \mbox { et } \hskip 1cm \gamma + d
\longrightarrow n + p)$$ne d\'{e}c\`{e}lent pas l'existence d'une
autre force fondamentale (qualifi\'{e}e de faible) qui serait
bas\'{e}e sur des charges (dites faibles) port\'{e}es par les
hadrons et les leptons et qui seraient v\'{e}hicul\'{e}es par les
trois bosons mentionn\'{e}s plus haut. Ces trois particules ont une tr\`es courte dur\'ee de vie, une port\'ee extr\`emement courte et servent uniquement en tant qu'interm\'ediaires des interactions dites faibles entre les quarks et les leptons \`a tr\`es courtes distances.\\\\
 Les interactions
\'{e}lectromagn\'{e}tiques, les lois de conservation et la loi
d'exclusion de Pauli sont suffisantes pour expliquer tous ces
ph\'{e}nom\`{e}nes ainsi que d'autres comme celui de la formation
du deuteron \emph{d} \`{a} partir d'un neutron et d'un proton et
l'inexistence  d'un \'{e}tat de liaison de deux protons ni de deux
neutrons, par exemple. Elles sont tout autant suffisantes pour
expliquer \'{e}galement le cycle (solaire)
$$p + p \longrightarrow d + e^+ + \nu_e$$
$$p + d \longrightarrow ^3He + \gamma$$
$$^3He + ^3He \longrightarrow p + p + \alpha$$
ou l'interaction $$\hskip 1cm n + ^3He \longrightarrow \alpha +
\gamma,$$ \`{a}
titre d'exemple.\\
\textbf{Remarque:} D'autres discussions sur les \'{e}tats de
liaison et les forces fondamentales seront fournies dans les
paragraphes 11 et 13.\\\\

\subsection*{\'{E}nergie de liaison et Mati\`{e}re-\'{E}nergie}

L'\'{e}nergie globale de l'Univers $E_0$ \'{e}tait concentr\'{e}e,
avant le Big Bang, en un point singulier de l'espace-temps (i.e.
le sommet du demi-c\^{o}ne de l'espace-temps) o\`{u} l'on peut
consid\'{e}rer que toute l'\'{e}nergie \'{e}tait une sorte
d'\'{e}nergie de liaison. L'apparition, apr\`{e}s le Big Bang, de
la mati\`{e}re sous forme de hadrons avec leur \'{e}nergie de
liaison interne et plus particuli\`{e}rement des neutrons et
protons formant (plus tard) des noyaux avec leur propre
\'{e}nergie de liaison n'\'{e}tait pas la seule forme de
transformation de l'\'{e}nergie initiale $E_0$. En effet, il faut
ajouter \`{a} l'\'{e}nergie de masse de ces particules et
agglom\'{e}rtions de particules et \`{a} leur \'{e}nergie de
liaison inh\'{e}rente \`{a} leur formation l'\'{e}nergie
cin\'{e}tique de la mati\`{e}re en mouvement et leur \'{e}nergie
potentielle d'interaction qui sont essentiellement les
\'{e}nergies potentielles gravitationnelle et
\'{e}lectromagn\'{e}tique. Ces deux \'{e}nergies potentielles sont
en fait deux formes d'\'{e}nergie de liaison. Dans un atome la
quantit\'{e}
$$ (m_N + \Sigma m_i - m_a) c^2 ,$$o\`{u} $m_N$, $\Sigma m_i$ et
$m_a$ sont respectivement les masses du noyau, des \'{e}lectrons
et de l'atome, est l'\'{e}nergie de liaison de l'atome (i.e. la
diff\'{e}rence de masse ($m_N + \Sigma m_i$)-$m_a$ transform\'{e}e
en \'{e}nergie) est \'{e}troitement li\'{e}e \`{a} l'\'{e}nergie
potentielle \'{e}lectromagn\'{e}tique \`{a} l'int\'{e}rieur de
l'atome (l'\'{e}nergie potentielle gravitationnelle \`{a}
l'int\'{e}rieur de l'atome est n\'{e}gligeable). Le passage d'un
\'{e}lectron d'un niveau potentiel \`{a} un autre est
n\'{e}cesairement accompagn\'{e} d'un changement de l'\'{e}nergie
de liaison \`{a} l'int\'{e}rieur de l'atome. Le passage de
l'\'{e}lectron d'un atome d'hydrog\`{e}ne du niveau fondamental
\`{a} un niveau sup\'{e}rieur s'exprime par une \'{e}nergie
potentielle \'{e}lectromagn\'{e}tique attractive moins importante
et une \'{e}nergie de liaison plus faible de sorte que
$$ e^{'}_l := (m^{'}_p + m^{'}_e - m^{'}_H)c^2 < (m_p + m_e - m_H)c^2 =:
e_l$$ Signalons que la masse $m^{'}_H$ de l'atome et la masse
$m^{'}_e$ de l'\'{e}lectron \`{a} l'int\'{e}rieur de l'atome ont
augment\'{e}, la vitesse et l'\'{e}nergie cin\'{e}tique de
l'\'{e}lectron ayant diminu\'{e}. L'absorption par l'\'{e}lectron
d'un photon d'\'{e}nergie $e_p$ augmente l'\'{e}nergie potentielle
n\'{e}gative de l'\'{e}lectron et augmente l\'{e}g\`{e}rement sa
masse en diminuant sa vitesse et son \'{e}nergie cn\'{e}tique.
L'\'{e}nergie totale de l'\'{e}lectron (potentielle +
cin\'{e}tique) est quand m\^{e}me en hausse. Dautre part,
l'\'{e}nergie globale de l'atome a augment\'{e} de $e_p$. Sa masse
globale a augment\'{e} de $\frac{e_p}{c^2}$. L'\'{e}lectron ayant
absorb\'{e} l'\'{e}nergie du photon $e_p$, l'augmentation de la
masse de l'atome provient de la transformation de la
diff\'{e}rence d'\'{e}nergie de liaison $e_l - e_l^{'}$ en masse:
$$ \frac{e_l - e^{'}_l}{c^2} = m^{'}_H - m_H = \frac{e_p}{c^2}$$
L'\'{e}nergie potentielle gravitationnelle s'interpr\`{e}te de la
m\^{e}me mani\`{e}re en terme d'\'{e}nergie de liaison li\'{e}e
\`{a} la trasnformation de masse en \'{e}nergie et vice versa.
Pour une plan\`{e}te ayant une orbite stable autour d'un p\^{o}le
de gravitation (disons une \'{e}toile), il y a un \'{e}quilibre
stable entre les \'{e}nergies de masse, les \'{e}nergies
cin\'{e}tiques, l'\'{e}nergie potentielle gravitationnelle et
l'\'{e}nergie de liaison en n\'{e}gligeant les \'{e}nergies
radiationnelles thermique et gravitationnelle. Lorsque la
plan\`{e}te d\'{e}crit une orbite se rapprochant continuellement
du p\^{o}le, l'\'{e}nergie potentielle gravitationnelle devient de
plus en plus n\'{e}gative, la vitesse orbitale augmente
continuellement ainsi que l'\'{e}nergie cin\'{e}tique tandis que
l'\'{e}nergie de liaison augmente et la masse de la plan\`{e}te
diminue ainsi que la masse du syst\`{e}me form\'{e} de la source
de gravitation et de la plan\`{e}te. Une partie de la
diff\'{e}rence de masse se transforme en \'{e}nergie de liaison du
syst\`{e}me caus\'{e}e par le champ gravitationnel devenant de
plus en plus intense; l'autre partie se transforme en \'{e}nergie
cin\'{e}tique.\\
Dans le cas extr\^{e}me d'une plan\`{e}te absorb\'{e}e par un trou
noir, l'\'{e}nergie de masse se transforme, \`{a} la fin, presque
enti\`{e}rement en \'{e}nergie de liaison sous forme d'une
augmentation de l'\'{e}nergie du trou noir.\\
Un ph\'{e}nom\`{e}ne semblable se produit lors de la r\'{e}duction
continuellement progressive des orbites d'un syst\`{e}me binaire;
l'\'{e}nergie potentielle gravitationnelle n\'{e}gative devient de
plus en plus importante, l'\'{e}nergie de liaison augmente,
l'\'{e}nergie cin\'{e}tique
aussi, alors que la masse totale du syst\`{e}me diminue.\\\\
Notons que le passage de l'\'{e}lectron de l'atome d'hydrog\`{e}ne
d'une orbite \`{a} une autre plus pr\`{e}s du noyau pr\'{e}sente
le m\^{e}me sch\'{e}ma sachant que l'\'{e}nergie potentielle
\'{e}lectromagn\'{e}tique joue le r\^{o}le de l'\'{e}nergie
potentielle gravitationnelle et l'\'{e}quilibre
\'{e}nerg\'{e}tique (momentan\'{e}) est de nouveau assur\'{e} par
l'interm\'{e}diaire du nouveau mouvement orbital apr\`{e}s
l'\'{e}mission d'un photon ayant une \'{e}nergie bien
d\'{e}termin\'{e}e: L'\'{e}nergie totale de l'\'{e}lectron
diminue, son \'{e}nergie potentielle \'{e}lectromagn\'{e}tique
diminue, sa vitesse et son \'{e}nergie cin\'{e}tique augmentent,
sa masse diminue, l'\'{e}nergie et la masse de l'atome diminuent
et l'\'{e}nergie de liaison augmente. La m\^{e}me chose se produit
au sein des noyaux o\`{u} l'\'{e}nergie de liaison entre les
nucl\'{e}ons se cr\'{e}e au d\'{e}triment de la diminution de la
somme des masses individuelles. Dans ce cas les \'{e}nergies
potentielles \'{e}lectromagn\'{e}tique et gravitationnelle sont
caus\'{e}es par les constituants charg\'{e}s des nucl\'{e}ons,
i.e. les quarks, bien que chaque nucl\'{e}on constitue un \'{e}tat
de liaison entre trois quarks de parfums, de spins ou de couleurs
diff\'{e}rents.\\\\
En r\'{e}sum\'{e}, l'\'{e}nergie initiale $E_0$ de l'Univers se
trouve apr\`{e}s le Big Bang sous forme d'\'{e}nergie de masse,
d'\'{e}nergie cin\'{e}tique et d'\'{e}nergie de liaison ou
d'interaction \`{a} caract\`{e}re gravitationnel ou
\'{e}lectromagn\'{e}tique sans oublier les \'{e}nergies pures (qui
peuvent \^{e}tre consid\'{e}r\'{e}es comme une sorte d'\'{e}nergie
de liaison) des trous noirs et l'\'energie pure des photons. L'\'{e}nergie de masse de la
mati\`{e}re visible n'\'{e}tant qu'une petite partie de
l'\'{e}nergie de masse totale de l'Univers, nous pensons que ce
que l'on appelle mati\`{e}re noire ou \'{e}nergie noire est
constitu\'{e}e de trous noirs (dont l'\'{e}nergie a le
caract\`{e}re d'une \'{e}nergie de masse gravitationnelle),
d'\'{e}toiles \`{a} neutrons et de naines brunes ou autres
\'{e}toiles invisibles associ\'{e}es \`{a} un grand nombre de
syst\`{e}mes binaires et enfin de toute forme de mati\`{e}re
ordinaire invisible comme les plan\`{e}tes. Quant \`{a} la
classification de l'\'{e}nergie de fond associ\'{e}e aux
neutrinos, elle reste assez \'{e}nigmatique. Les neutrinos n'ont
probablement pas de masse et bien qu'ils n'ont pas d'interactions
\'{e}lectromagn\'{e}tiques, ils contribuent d'une fa\c{c}on
significative \`{a} l'\'{e}nergie radiationnelle de l'Univers.\\\\

\subsection*{ Description sommaire de l'Univers}

Remarquons pour finir que notre mod\`{e}le global est tout \`{a}
fait compatible avec la description classique des diff\'{e}rents
stades de l'\'{e}volution de l'Univers, d'une part, et de la
mati\`{e}re, de
l'antimati\`{e}re et de l'\'{e}nergie, d'une autre part.\\\\
1. Tout au d\'{e}but de l'expansion, l'Univers de taille
infiniment petite ($t \ll$ 1) \'{e}tait domin\'{e} par les
radiations ultra-\'{e}nerg\'{e}tiques, dans un \'{e}tat
d'\'{e}quilibre thermal parfait, ayant des fr\'{e}quences
infiniment grandes (i.e. de longueurs d'ondes infiniment petites)
donnant lieu \`{a} une densit\'{e} de radiation infiniment grande
sous une pression et \`{a} une temp\'{e}rature infiniment grandes
\'{e}galement;
toutes les trois d\'{e}croissant tr\`{e}s rapidement.\\\\
2. Ensuite commence le stade qualifi\'{e} de soupe de quarks et de
leptons (sans doute avec leurs antiparticules) qui sont
diff\'{e}rement charg\'{e}s, suivi de la formation des protons,
des neutrons et (sans doute) des neutrinos ainsi que de leurs
antiparticules. Cette formation est devenue possible avec
l'att\'{e}nuation relative de l'\'{e}nergie gigantesque des
radiations originelles et de la pression et temp\'{e}rature originelles.\\\\
3. Les r\`{e}gles de conservation et d'exclusion favorisent
certaines interactions \'{e}lectromagn\'{e}tiques au d\'{e}pens
d'autres conduisant \`{a} la baisse de formation des neutrinos
(neutrinos freeze) et \`{a} la disparition progressive de
l'antimati\`{e}re (comme les positrons) au profit de
l'augmentation du rapport du nombre des protons sur celui des
neutrons. Tout ceci est gouvern\'{e} par des \'{e}quilibres
\'{e}nerg\'{e}tiques impliquant l'effet de stabilit\'{e} et de
dur\'{e}e de vie.\\\\
4. Survient ensuite la formation des noyaux l\'{e}gers stables et
des autres formations mat\'{e}rielles plus ou moins stables et des
atomes, accompagnant la tendance \`{a} la diminution de la
densit\'{e} des radiations au profit de la densit\'{e}
mat\'{e}rielle rendant ainsi possible la formation de toutes les
agglom\'{e}rations mat\'{e}trielles \`{a} partir des atomes et
mol\'{e}cules
jusqu'aux galaxies.\\\\
\`{A} partir de ce stade qualifi\'{e} de "photons freeze", la
tendance vers une pr\'{e}dominance de la densit\'{e}
mat\'{e}rielle sur celle des radiations bien que la densit\'{e} g\'{e}n\'{e}rale, la pression
g\'{e}n\'{e}rale et la temp\'{e}rature cosmique ne font que
baisser \`{a} cause de l'expansion, toujours en cours, qui
s'effectue \`{a} une vitesse avoisinant 1 et tendant vers 1.\\\\

En fin de compte, signalons (pour r\'{e}sumer) qu'\`{a} la
lumi\`{e}re de ce qui pr\'{e}c\`{e}de, on peut supposer que l'on
ait, pour tout petit intervalle de temps, la formule
$$v(t) = \lambda(t) f(t) = \frac{1}{2 \pi}
\frac{\sqrt{\mu}}{t} \lambda(t)$$o\`{u} \emph{v}(\emph{t}),
$\lambda$(\emph{t}) et \emph{f}(\emph{t}) = $\frac{1}{2 \pi}
\frac{\sqrt{\mu}}{t}$ sont respectivement la vitesse de
l'expansion, la longueur d'onde et la fr\'{e}quence des ondes
responsables de l'expansion (cr\'{e}ant l'espace
g\'{e}om\'{e}trique de notre Univers r\'{e}el). On peut concevoir
alors que l'on ait
$$\lambda(t) = \frac{2 \pi}{\sqrt{\mu}} \hskip 0.1cm t \hskip
0.1cm
\frac{a(t)}{b(t)} \mbox { et } v(t) = \frac{a(t)}{b(t)}$$avec\\
(i) \emph{a}(\emph{t}) est une fonction strictement croissante
v\'{e}rifiant $$\lim_{t\rightarrow 0^+} a(t) = 0 \mbox { et
}\lim_{t\rightarrow +\infty} a(t) = +
\infty.$$\\
(ii) \emph{b}(\emph{t}) est une fonction strictement croissante
v\'{e}rifiant $$\lim_{t\rightarrow 0^+} b(t) = b > 0 \mbox { et
}\lim_{t\rightarrow +\infty} b(t) = +\infty.$$\\
(iii)\emph{v}(\emph{t}) = $\frac{a(t)}{b(t)}$ est strictement
croissante avec
$$\lim_{t\rightarrow 0^+} \frac{a(t)}{b(t)} = 0 \mbox { et }
\lim_{t\rightarrow +\infty} \frac{a(t)}{b(t)} = 1.$$ (De telles
fonctions existent \'{e}videmment; on peut prendre \`{a} titre
d'exemple,
$$v(t) = \frac{\ln(1+t)}{\ln(1+\alpha+t)}\mbox { avec
}\ln(1+\alpha) = b).$$\\
La d\'{e}termination de \emph{a}(\emph{t}), \emph{b}(\emph{t}) et
\emph{b} n\'{e}cessite \'{e}videmment des mesures aussi
pr\'{e}cises que possible en utilisant des moyens techniques aussi
bien vari\'{e}s que sophistiqu\'{e}s comme des t\'{e}lescopes
ultrapuissants et des acc\'{e}l\'{e}rateurs nucl\'{e}aires
engageant des \'{e}nergies gigantesques afin d'aller le plus loin
possible dans le temps et dans l'espace pour acqu\'{e}rir une
meilleure compr\'{e}hension de l'Univers et de la
mati\`{e}re-\'{e}nergie originels. Notons que la grandeur du
nombre \emph{b} de l'exemple pr\'{e}c\'{e}dent, \`{a} titre
d'exemple, est d\'{e}cisive: Pour \emph{b} petit ou infiniment
petit, l'\^{a}ge de l'Univers est proche de celui qui est
g\'{e}n\'{e}ralement avanc\'{e} actuellement et qui va \^etre
d\'etermin\'e approximativement au paragraphe $11;$ tandis que,
pour \emph{b} grand ou infiniment grand, alors l'\^{a}ge de notre
Univers est beaucoup plus grand que l'on ne pense et que son
\'{e}volution jusqu'au d\'{e}but du stade actuellement scrutable a
pris beaucoup de temps. Dans ce dernier cas, le temps $T_0$
(exprim\'{e} en seconde) au bout duquel l'Univers a atteint la
taille $B_e$(\emph{O},1) (o\`{u} 1 repr\'{e}sente ici
3$\times$10$^8$m), qui correspondrait \`{a} une vitesse
d'expansion significative qui sera plus tard assez proche de 1,
est assez grand et pourrait \^{e}tre m\^{e}me tr\`{e}s grand ($T_0
\gg$ 1). Par suite, si on suppose que la taille actuelle de
l'Univers est approximativement \emph{t} $\times$
(3$\times$10$^8$m), alors l'\^{a}ge effectif de l'Univers \`{a}
partir du temps \emph{t} = 0 serait assez proche de \emph{t} +
$T_0$. Si, par contre, on suppose que le temps qui nous s\'{e}pare
du Big Bang est \emph{t}, alors la taille de l'Univers serait
actuellement proche de \emph{t} - $T_0$.\\

Cette \'{e}ventualit\'{e} est appuy\'{e}e par la validit\'{e} de
la th\'{e}orie de la relativit\'{e} g\'{e}n\'{e}rale en ce qui
concerne l'interaction du champ gravitationnel et des ondes
\'{e}lectromagn\'{e}tiques. L'influence de la force
gravitationnelle sur les ondes a \'{e}t\'{e} confirm\'{e}e par
l'observation de plusieurs ph\'{e}nom\`{e}nes naturels et par
plusieurs exp\'{e}riences dont celle effectu\'{e}e par Pound et
Rebka. Cette influence est d'ailleurs reconfirm\'{e}e par notre
mod\`{e}le puisque le champ gravitationnel ''courbe'' les
g\'{e}od\'{e}siques et contracte les distances. Si $X(t)$
d\'{e}crit la trajectoire d'une onde \'{e}lectromagn\'{e}tique
dans un rep\`{e}re virtuel fixe, alors on a
$\nabla_{X^{'}(t)}^{g_t} X^{'}(t)$ = 0 et, en g\'{e}n\'{e}ral,
$\|X^{'}(t)\|_{g_e}$ = \emph{c} = 1 conform\'{e}ment au premier
postulat de la relativit\'{e} restreinte d'Einstein tandis que
l'on a $\|X^{'}(t)\|_{g_t} < \|X^{'}(t)\|_{g_e}$ au sein d'un
champ gravitationnel cons\'{e}quent. Ainsi, bien que
l'interpr\'{e}tation de l'exp\'{e}rience de Pound-Rebka est, selon
notre mod\`{e}le, diff\'{e}rente de celle accept\'{e}e
g\'{e}n\'{e}ralement, elle montre bien l'existence de l'action de
la force gravitationnelle sur le photon. En effet: D'apr\`{e}s
Pound et Rebka, si la gravitation ne cause pas un blueshift
$\Delta_{1} E$ = \emph{gL} lorsque le photon $\gamma$ se dirige
vers la terre et si la temp\'{e}rature est r\'{e}ellement
constante et le vide est absolu, alors on obtiendrait une
r\'{e}sonance optimale pour un \'{e}metteur fixe. Le fait qu'une
telle r\'{e}sonance est obtenue en variant la position de
l'\'{e}metteur implique, selon eux, qu'il est n\'{e}cessaire de
provoquer un redshift $\Delta_{0} E$ = -$\beta E$, avec $\beta =
\frac{v}{c}$ et le mouvement doit s'op\'{e}rer vers le haut.
Lorsque le photon est dirig\'{e} vers le haut, on devrait
op\'{e}rer un mouvement de l'\'{e}metteur toujours vers le haut
pour provoquer un blueshift $\Delta_{0}E$ = $\beta E$ qui
compenserait le redshift $\Delta_{1} E$ =
-\emph{gL}.\\
Notre interpr\'{e}tation co\"{\i}ncide en partie avec celle de
Pound et Rebka. S'il n'y avait pas un blueshift (ou un redshift)
gravitationnel $\Delta_{1} E$ = \emph{gL} et si la temp\'{e}rature
est absolument constante et le vide absolu est parfaitement
respect\'{e}, alors l'\'{e}nergie du photon ne changerait pas et
le ph\'{e}nom\`{e}ne de r\'{e}sonance optimale se produirait sans
varier la position de l'\'{e}metteur. L'\'{e}nergie \emph{E} du
photon serait, d'apr\`{e}s notre mod\`{e}le, constante et elle est
donn\'{e}e, pour n'importe quelle distance (dans ces m\^{e}mes
conditions), par
$$E = h(t) f(t)$$bien que \emph{h}(\emph{t}) et \emph{f}(\emph{t}) soient
variables avec le temps. Cette quantit\'{e} est \'{e}gale aussi
\`{a} $h_{P}f_{D}$ comme on l'a d\'{e}j\`{a} signal\'{e}
auparavent. Par suite, \emph{E} et $f_{D}$ seraient constantes et
Pound et Rebka n'auraient pas eu besoin de faire mouvoir la
source. Par contre, si les deux conditions ci-dessus ne sont pas
respect\'{e}es, alors, m\^{e}me si le blueshift (ou le redshift)
gravitationnel n'existait pas, on aurait besoin de provoquer un
d\'{e}calage appropri\'{e} dans les deux directions. \'{E}tant
donn\'{e} que tout laisse \`{a} croire que $\Delta_{1} E$ =
$\pm$\emph{gL} existe, on a bel et bien besoin de provoquer un
$\Delta_{0} E$ ad\'{e}quat. \`{A} position fixe, ceci pourrait
\^{e}tre fourni, en cas de propagation vers la terre par les
fluctuations de temp\'{e}rature et par le d\'{e}faut du vide
absolu, mais, dans le sens contraire, ceci ne peut pas se produire
puisque les deux $\Delta_{i} E$ (\emph{i} = 0,1) seraient de
m\^{e}me signe. D'o\`{u} la n\'{e}cessit\'{e} d'effectuer une
correction $\Delta_{0} E$ (qui serait n\'{e}cessairement du type
blueshift dans le cas o\`{u} le sens de propagation est vers le
haut) qui, d'apr\`{e}s notre mod\`{e}le, ne pouvait s'effectuer
qu'\`{a} l'aide de fluctuations de vitesse (ou
acc\'{e}l\'{e}rations) ou de variations de distance. Cette
derni\`{e}re implique une variation de l'effet de temp\'{e}rature
et du d\'{e}faut du vide absolu.
\subsection*{Mod\'elisation g\'en\'erale de l'Univers}
Remarquons aussi que notre \'{e}tude aurait pu \^{e}tre
enti\`{e}rement refaite en prenant comme point de d\'{e}part
l'Univers \emph{U}($t_0$) = ($B_e$(\emph{O},$R$($t_0$)),$g_{t_0}$)
pour $t_0$ quelconque pourvu que l'on ait des renseignements
suffisants sur \emph{U}($t_0$) et que l'on puisse remonter le
temps jusqu'\`{a} cet instant privil\'{e}gi\'{e} tout en ayant des
connaissances satisfaisantes sur les m\'{e}canismes du recul et de
la progression de l'Univers dans le temps. Consid\'{e}rons, par
exemple, l'Univers \`{a} un instant $t >$ 0 comme \'{e}tant
r\'{e}duit \`{a} la boule $B_{e}(O,R(t))$ munie de la m\'{e}trique
physique $g_t(X)$ pour $|X| \leq R(t)$ qui est d\'{e}termin\'{e}e
par la distribution de l'\'{e}nergie g\'{e}n\'{e}ralis\'{e}e
$E_{t}(X)  = E(t,X).$ L'\'{e}quation de la mati\`{e}re-\'{e}nergie
$(E^{*})$ qui est alors v\'{e}rifi\'{e}e sur $B_{e}(O,R(t))$
devient
$$\square E(t,X) = \frac{\partial^2 E}{\partial t^2}(t,X) - \Delta
E(t,X) = 0$$
avec
$$E(t,X)_{|S_e(O,R(t))} = 0.$$En ramenant la
r\'{e}solution de ce probl\`{e}me \`{a} celle du probl\`{e}me de
Dirichlet sur la boule unit\'{e} $B_{e}(O,1)$ et en choisissant
une valeur propre particuli\`{e}re $\mu$, on obtient la solution
pseudo-p\'{e}riodique
$$E_\mu(t,X) = (f_\mu(0)\cos \sqrt{\mu} \frac{t}{R(t)} +
\frac{R(t)}{\sqrt{\mu}}f^{'}_\mu(0)\sin \sqrt{\mu}
\frac{t}{R(t)})\psi_\mu (\frac{X}{R(t)})$$de pseudo-p\'{e}riode
$T_\mu$(t) = $2 \pi \frac{R(t)}{\sqrt{\mu}}$ et de
pseudo-fr\'{e}quence $f_\mu$(t) = $\frac{1}{2 \pi}
\frac{\sqrt{\mu}}{R(t)}$.\\
Cette fonction v\'{e}rifie, pour une trajectoire
g\'{e}od\'{e}sique (relativement \`{a} $g_{t}$) quelconque $X(t)$
la relation
$$E(t) := E_\mu(t,X(t)) = h_\mu(t) f_\mu(t) =: h(t)f(t) = h_P
f_D$$o\`{u} $h_{P}$ est la constante de Planck et $f_{D}$ est la
fr\'{e}quence de De Broglie. Toutes les formules d\'{e}j\`{a}
\'{e}tablies dans des contextes simplifi\'{e}s peuvent \^{e}tre
adapt\'{e}es \`{a} ce contexte g\'{e}n\'{e}ral o\`{u} l'on
consid\`{e}re l'Univers dynamique $U(t)$ comme \'{e}tant
mod\'{e}lis\'{e}, \`{a} tout instant $t,$ par l'espace riemannien
($B_{e}(O,R(t)),g_{t}$) o\`{u} $R(t)$ est le rayon effectif de
l'Univers et $g_{t}$ est la m\'{e}trique physique r\'{e}elle qui
est d\'{e}termin\'{e}e par la distribution de la mati\`ere-\'energie, le
temps et la temp\'{e}rature (ou la pression). Notons, qu'en
d\'{e}finitve, $R(t)$ est \'{e}gal \`{a} tout instant $ t>0$ \`{a}
$$R(t) = R(t_0) + k(t)(t - t_0)$$pour un $t_{0}$ quelconque avec
$k(t) \sim1$ pour $t \gg 1,$ $k(t)$ est croissante et $$\lim_{t
\rightarrow +\infty}\emph{k(t)} = 1.$$Tout ceci nous conduit \`{a}
penser que le demi-c\^{o}ne de l'espace et du temps a plut\^{o}t
la forme esquiss\'{e}e \`{a} la figure $10,$ \`{a} moins que la
vitesse de propagation des ondes \'{e}lectromagn\'{e}tiques et
celle de l'expansion de l'espace ont \'{e}t\'{e} depuis toujours
\'{e}gales \`{a} $1,$ auquel cas le demi-c\^{o}ne de l'espace et
du
temps serait bel et bien repr\'{e}sent\'{e} par la figure $3.$ N\'eamoins, nous pensons que le premier cas est logiquement beaucoup plus probable. L'\'equation de la mati\`ere-\'energie s'\'ecrit alors sous la forme
$$\frac{1}{v^2(t)} \frac{\partial^2 E}{\partial t^2}(t,X) - \Delta
E(t,X) = 0$$
o\`u $v(t)$ est originellement extr\^emement petit $(v(t)\ll1)$ pour un intervalle de temps inconnu avec $v(t)$ est perp\'etuellement croissante devenant, plus tard, tr\`es proche de $1$ avec $\lim\limits_{t\rightarrow+\infty}v(t)=1$. Ceci est d\^u au gigantesque champ de gravitation originel \`a l'\'epoque o\`u l'\'energie globale de l'Univers \'etait concentr\'ee en un point singulier \`a l'instar d'un gigantesque trou noir. \\\\

Signalons enfin que notre d\'{e}marche progressive a \'{e}t\'{e}
d\'{e}lib\'{e}r\'{e}ment (et subjectivement) choisie et maintenue
pour \'{e}viter les complications qui pourraient r\'{e}sulter de
la multitude des facteurs entrant en jeu dans la construction de
ce mod\`{e}le et pour y arriver le plus simplement et le plus
clairement possible tout en ayant conscience qu'il y a
\'{e}norm\'{e}ment de points \`{a} d\'{e}tailler, \`{a}
pr\'{e}ciser, \`{a} \'{e}lucider, \`{a} ajouter et \`{a}
d\'{e}couvrir.\\\\

\section{La relativit\'{e} g\'{e}n\'{e}rale revue et
simplifi\'{e}e}

Dans ce paragraphe, on va montrer que notre Univers dynamique est
globalement d\'{e}crit par la th\'{e}orie de la relativit\'{e}
g\'{e}n\'{e}rale r\'{e}adapt\'{e}e \`{a} notre mod\`{e}le. Une
fois cette r\'{e}adaptation effectu\'{e}e, un grand nombre de
probl\`{e}mes relevant de la Physique th\'{e}orique et de la
Cosmolgie (entre autres) seront pos\'{e}s sans \'{e}quivoque et
pourront \^{e}tre r\'{e}solus d'une fa\c{c}on plus pr\'{e}cise,
plus claire et plus simple.

\subsection*{Pr\'eliminaires}
Notre \'{e}tude sera ici essentiellement bas\'{e}e sur les
r\'{e}sultats \'{e}tablis dans les paragraphes pr\'{e}c\'{e}dents
d'une part et sur ceux expos\'{e}s dans le grand classique de la
relativit\'{e} g\'{e}n\'{e}rale de R.Wald: "General Relativity".\\

Rappelons que, dans le cadre de notre mod\'{e}lisation, l'aspect
dynamique de notre Univers (qui est en expansion permanente \`{a}
une vitesse avoisinant la vitesse de la lumi\`{e}re dans le vide pour $t\gg1$)
est caract\'{e}ris\'{e} par le demi - c\^{o}ne de l'espace -
temps:
$$C = \{ (x, y, z, t); x^2 + y^2 + z^2 \leq t^2, t > 0 \} =
\bigcup_{t > 0} B(O,t)\times \{t\}$$muni de la m\'{e}trique
riemannienne
$$h = dt^2 - g_t.$$
Ici $g_t$ est la m\'{e}trique riemannienne r\'{e}gularis\'{e}e
d\'{e}finie, \`{a} tout instant \emph{t}, sur l'Univers
assimil\'{e} \`{a} la boule \emph{B}(\emph{O},\emph{t}) qui
refl\`{e}te, \`{a} l'instant \emph{t}, la consistence physique de
l'Univers. Cette consistence est d\'{e}crite enti\`{e}rement par
la distribution $E_t(X)$ de l'\'{e}nergie g\'{e}n\'{e}ralis\'{e}e,
sur \emph{B}(\emph{O},\emph{t}), qui englobe la distribution
mat\'{e}rielle $m_t(X)$ ainsi que toutes les manifestations de la
mati\`{e}re - \'{e}nergie (gravit\'{e}, \'{e}lectromagn\'{e}tisme,
ph\'{e}nom\`{e}nes thermodynamiques...).\\
L'Univers dynamique sera donc repr\'{e}sent\'{e}, \`{a} chaque
instant $t_0$, par l'hypersurface intersection de ce demi -
c\^{o}ne de $\mathbb{R}^4$ avec l'hyperplan de $\mathbb{R}^4$
d'\'{e}quation \emph{t} = $t_0$; cet hypersurface sera not\'{e}
$\Sigma_{t_0}$ et
\emph{C} sera not\'{e} \emph{M}.\\
Ainsi, il est clair que dans le cadre de notre mod\`{e}le,
$\Sigma_{t_0}$ est une surface de Cauchy compacte et que notre
vari\'{e}t\'{e} de l'espace - temps (\emph{M},$h_{ab}$) est
stablement causale, globalement hyperbolique et asymptotiquement
plate. Le champ de vecteur $(\frac{\partial}{\partial t})^a$ sur
\emph{M} est orthogonal \`{a} toutes les hypersurfaces $\Sigma_t$.
Ces propri\'{e}t\'{e}s vont simplifier \'{e}norm\'{e}ment les
bases de la th\'{e}orie de la relativit\'{e} g\'{e}n\'{e}rale
ainsi que son utilisation pour expliquer l'Univers dynamique.\\

\subsection*{\'{E}quations tensorielles d'Einstein - Formulation lagrangienne}

Notre point de d\'{e}part est l'\'{e}quation
$$\hskip 2cm ^{(3)}G_{ab}(t) := \hskip 0.1cm ^{(3)}R_{ab}(t) - \frac{1}{2} \hskip 0.2cm ^{(3)}R \hskip 0.2cm ^{(3)}g_{ab}(t) =: \hskip 0.1cm^{(3)}T_{ab}^{*}(t) \hskip 2cm (25) $$
d\'{e}finie sur (\emph{B}(\emph{O},\emph{t}),$g_t$), o\`{u}
$^{(3)} R_{ab}$ et $^{(3)}R$ d\'{e}signent respectivement la
courbure de Ricci et la courbure scalaire associ\'{e}es \`{a} la
m\'{e}trique physique $g_t$. Le tenseur
$^{(3)}T_{ab}^{*}$(\emph{t}) est le tenseur de la mati\`{e}re -
\'{e}nergie d\'{e}crivant \`{a} tout instant $t > 0$ la
consistence physique de l'Univers li\'{e}e \`{a} l'existence
m\^{e}me de la mati\`{e}re - \'{e}nergie en \'{e}volution
permanente remplissant l'espace g\'{e}om\'{e}trique vide
\emph{B}(\emph{O},\emph{t}). Conform\'{e}ment \`{a} notre
d\'{e}finition, la relation $^{(3)}T_{ab}^{*}$(\emph{t}) $\equiv$
0 sur un domaine $D \subset B$(\emph{O},\emph{t}) est
\'{e}quivalente au fait que le domaine \emph{D} est absolument
vide (i.e. \emph{D} est quasiment \`{a} l'abri de toutes les
manifestations et effets de la mati\`{e}re - \'{e}nergie) et par
suite on a alors
$$g_{ab}(t) \equiv g_e \;\;\;\;\;\; \mbox { sur }\;  D. $$
Ensuite, on consid\`{e}re, sur (\emph{M}, $h_{ab}$) l'\'{e}quation
d'Einstein (l\'{e}g\`{e}rement modifi\'{e}e):
$$\hskip 2cm  ^{(4)} G_{ab} := \hskip 0.1cm ^{(4)} R_{ab} - \frac{1}{2} \hskip 0.2cm
^{(4)}R\hskip 0.1cm h_{ab} = \hskip 0.1cm ^{(4)} T_{ab}^{*} \hskip
4.8cm (26)$$o\`{u} $^{(4)}R_{ab}$ et $^{(4)}R$ d\'{e}signent
respectivement les courbures associ\'{e}s \`{a} $h_{ab}$ sur
\emph{M} et $^{(4)}T_{ab}^{*}$ est (\`{a} une constante pr\`{e}s)
le tenseur de stress-\'{e}nergie g\'{e}n\'{e}ralis\'{e}
d'Einstein. Ainsi, avec ces notations on a:
$$^{(3)}T_{ab}^{*} = 0 \Rightarrow g_t = g_e \Rightarrow h = dt^2
- g_e = \eta_{ab}$$o\`{u} $\eta_{ab}$ est, \`{a} un signe
pr\`{e}s, la m\'{e}trique plate de Minkowski sur le demi -
c\^{o}ne
\emph{C} et on a alors $^{(4)}T_{ab}^{*}$ = 0.\\
Notons que notre vari\'{e}t\'{e} d'espace - temps \emph{M}
\'{e}volue avec le temps. \`{A} tout instant $t_0$, on a
$$M = C(t_0) = \{ (x, y, z, t); x^2 + y^2 + z^2 \leq t^2, 0 < t
\leq t_0 \}. $$Dans la suite on va utiliser intensivement les
r\'{e}sultats mentionn\'{e}s dans ([4]) par souci d'une tr\`{e}s
appr\'{e}ciable \'{e}conomie. Ainsi, pour plus de facilit\'{e} et
de clart\'{e}, on va modifier l\'{e}g\`{e}rement nos notations
pour se conformer \`{a} celles utilis\'{e}es dans cette m\^{e}me
r\'{e}f\'{e}rence. Par cons\'{e}quent, notre m\'{e}trique sur
l'espace-temps en \'{e}volution \emph{M} = \emph{C}(\emph{t}) sera
not\'{e}e $ ^{(4)} g_{ab}$. Elle d\'{e}signera une m\'{e}trique de
signature lorentzienne et on a
$$^{(4)}g_{ab} = - dt^2 + \hskip 0.1cm ^{(4)}h_{ab}$$
o\`{u} $^{(4)}h_{ab}$ d\'{e}signe ici la m\'{e}trique induite par
$^{(4)} g_{ab}$ sur $\Sigma_t$ de telle sorte que notre
m\'{e}trique physique sur l'Univers \emph{B}(\emph{O},\emph{t}),
pr\'{e}c\'{e}demment not\'{e}e $g_t$, s'identifiera \`{a} la
m\'{e}trique $^{(3)}h_{ab}$ obtenue par la restriction de
$^{(4)}h_{ab}$ sur $\Sigma_t$.\\

On est maintenant pr\^{e}t \`{a} obtenir les formulations
lagrangienne et hamiltonienne ad\'{e}quates adapt\'{e}es au
nouveau contexte dans lequel on traite la th\'{e}orie de la
realtivit\'{e} g\'{e}n\'{e}rale qui correspondra r\'{e}ellement
\`{a} la loi g\'{e}n\'{e}rale de notre Univers dynamique.
\'{E}videmment cette loi correspond en fait \`{a} une
id\'{e}alisation de notre Univers r\'{e}el par l'interm\'{e}diaire
de la r\'{e}gularisation de la m\'{e}trique $g_t$ sur
\emph{B}(\emph{O},\emph{t}). En effet, la m\'{e}trique r\'{e}elle
est loin d'\^{e}tre de classe $C^2$ \`{a} cause des
singularit\'{e}s se r\'{e}duisant essentiellement aux trous noirs.
Compte tenu de ces adaptations, nos deux \'{e}quations (26) et
(25) s'\'{e}crivent:
$$\hskip 3cm ^{(4)} R_{ab} - \frac{1}{2} \hskip 0.2cm ^{(4)}R \hskip 0.2cm
^{(4)} g_{ab} = \hskip 0.1cm ^{(4)} T_{ab}^{*} \hskip 5cm (E)$$et
$$\hskip 3cm ^{(3)} R_{ab} - \frac{1}{2} \hskip 0.2cm ^{(3)}R
\hskip 0.2cm ^{(3)} h_{ab} = \hskip 0.1cm ^{(3)} T_{ab}^{*} \hskip
5cm (E^*)$$avec
$$^{(3)} T_{ab}^{*} = 0 \Leftrightarrow \hskip 0.1cm ^{(3)} h_{ab}
= \hskip 0.1cm ^{(3)} g_e \Leftrightarrow \hskip 0.1cm ^{(4)}
g_{ab} = \eta_{ab}$$o\`{u} $\eta_{ab}$ est la m\'{e}trique de
Minkowski.\\
Ainsi, \emph{B}(\emph{O},\emph{t}) est l'espace virtuel vide dans
lequel vit notre Univers physique r\'{e}el et \emph{C}(\emph{t})
est l'espace virtuel dans lequel \'{e}volue l'Univers dynamique
conform\'{e}ment \`{a} la th\'{e}orie de l'expansion de
l'Univers.\\
En utilisant une base orthonorm\'ee pour la m\'etrique
riemannienne $g_{ab},$ l'\'{e}quation (\emph{E}) implique
$$ ^{(4)} R = - \hskip 0.2cm
^{(4)} T^{*}:=-^{(4)}T^{*a}_a$$et l'\'{e}quation ($E^{*}$)
implique (en utilisant une base orthonorm\'{e}e par rapport \`{a}
$^{(3)} h_{ab}$ compl\'{e}t\'{e}e par le champ de vecteur
$(\frac{\partial}{\partial t})^a = (1,0,0,0))$
$$ ^{(3)} R = -2 \hskip 0.2cm ^{(3)}
T^{*}:=-2^{(3)}T^{*a}_a.$$ Ainsi la forme volume $^{(3)}
\varepsilon_{abc}$ =: $^{(3)} \varepsilon$ associ\'{e}e \`{a}
$^{(3)} h_{ab}$ n'est autre que $\sqrt{h} e_{abc}$ o\`{u}
$e_{abc}$ =: $^{(3)}e$ est la forme volume canonique (euclidien)
de $\mathbb{R}^3$; la forme volume $\varepsilon_{abcd}$ =:
$^{(4)}\varepsilon$ associ\'{e}e \`{a} $^{(4)} g_{ab}$ n'est autre
que $\sqrt{-g} e_{abcd}$ = $\sqrt{h} e_{abcd}$, o\`{u} $e_{abcd}$
=: $^{(4)} e$ est la forme volume canonique de $\mathbb{R}^4$
(ici, $g$ et $h$ sont respectivement les d\'eterminants des
matrices associ\'ees \`a $g_{ab}$ et $h_{ab}$ lorsqu'elles sont
\'ecrites \`a l'aide des bases canoniques de $\mathbb{R}^4$ et
$\mathbb{R}^3),$ et on a
$$\hskip 3cm ^{(3)} \varepsilon =
i_{\frac{\partial}{\partial t}} \hskip 0.2cm ^{(4)} \varepsilon
\hskip 3cm \mbox{ (produit int\'{e}rieur) }$$ puisque, dans le
cadre de notre mod\`{e}le, le champ de vecteur unitaire orthogonal
aux hypersurfaces $\Sigma_t$ n'est autre que $\overrightarrow{n}$
= $(\frac{\partial}{\partial t}) ^ a$. Signalons aussi que la
fonction caract\'{e}risant le flot du temps n'est autre que la
quatri\`{e}me coordonn\'{e}e \emph{t} puisque, dans notre
mod\`{e}le, la notion relativiste du temps propre n'a
pas d'existence.\\

Ainsi, suivant R.Wald, l'action de Hilbert associ\'{e}e \`{a}
l'\'{e}quation d'Einstein dans le vide (o\`{u} $g_{ab}(t)$
d\'{e}signe ici la m\'{e}trique d'Einstein dans le vide sur
l'espace-temps $M=C(t)$):
$$\hskip 4cm ^{(4)}R_{ab} - \frac{1}{2} \hskip 0.2cm ^{(4)}R\hskip 0.3cm
^{(4)}g_{ab} = 0 \hskip 4cm (E_0)$$sera donn\'{e}e par
$$S_G\hskip 0.1cm[ g^{ab} ] = \int_{M} {\cal L}_G  \hskip 0.2cm ^{(4)}
e$$o\`{u}
$${\cal L}_G = \sqrt{-g} \hskip 0.2cm ^{(4)}R =
\sqrt{h} \hskip 0.2cm ^{(4)}R.$$On a alors (pour une famille \`a
un param\`etre $(g_{ab})_\lambda$ ([4], E.1.18)):
$$\frac{dS_G}{d\lambda} = \int \frac{d{\cal L}_G}{d\lambda} \hskip 0.2cm ^{(4)} e =
\int \nabla^a \upsilon _a \sqrt{-g}\hskip 0.2cm ^{(4)}e + \int ( R
_{ab} - \frac {1}{2} R \hskip 0.1cm g_{ab}) \delta g^{ab}
\sqrt{-g} \hskip 0.2cm ^{(4)} e$$ o\`u $\delta
g_{ab}=\frac{d(g_{ab})_\lambda}{d\lambda}|_{\lambda=0}$ et
$\upsilon_a=\nabla^b(\delta g_{ab})-g^{cd}\nabla_a(\delta
g_{cd});$ et, en n\'{e}gligeant le premier terme du second membre
comme \'{e}tant l'int\'{e}grale par rapport \`{a} $^{(4)}
\varepsilon$ d'une divergence, on a (E.1.19):
$$ \frac{\delta S_G}{\delta g^{ab}} = \sqrt{-g}( R_{ab} - \frac{1}{2} R \hskip 0.1cm g_{ab}).$$
Ainsi, on a
$$ \frac{\delta S_G}{\delta g^{ab}} =0\quad\Longleftrightarrow\quad (E_0).$$
Mais, en tenant compte de la contribution du terme provenant de la
fronti\`{e}re, l'action r\'{e}elle devrait \^{e}tre modifi\'{e}e
pour devenir
$$ S^{'}_G \hskip 0.1cm = \hskip 0.1cm S_G + 2 \int_{\dot{U}}
K.$$Ici, $\dot{U}$ est la fronti\`{e}re de la partie du demi -
c\^{o}ne de l'espace - temps comprise entre les deux hypersurfaces
de Cauchy $\Sigma_{t_1}$ et $\Sigma_{t_2}$ avec $ 0
< t_1 < t_2.$\\
Le second terme est l'int\'{e}grale, sur la fronti\`{e}re de
\emph{U}, de la courbure scalaire induite par la m\'{e}trique
$g_{ab}$ sur les trois hypersurfaces $\Sigma_{t_1}$,
$\Sigma_{t_2}$ et celle constitu\'{e}e par la fronti\`{e}re
lat\'{e}rale. Sur cette derni\`{e}re la courbure est nulle et
l'action $S^{'}_G$ s'\'{e}crit
$$S^{'}_G = S_G + 2 \int_{\Sigma_{t_2}} K \hskip 0.2cm ^{(3)}
\varepsilon - 2 \int _{\Sigma_{t_1}} K \hskip 0.2cm ^{(3)}
\varepsilon.$$Rappelons que \emph{K} est ici (E.1.39)
$$K = K^a_{\hskip 0.1cm a} = h^a_{\hskip 0.1cm b} \nabla_a
(\frac{\partial}{\partial t})^b$$o\`{u} $$K_{ab} = \frac{1}{2}
{\cal L}_{\frac{\partial}{\partial t}} \hskip 0.1cm g_{ab} =
\frac{1}{2} {\cal L}_{\frac{\partial}{\partial t}} \hskip 0.1cm
h_{ab} = \frac{1}{2} {\dot{h}}_{ab}$$est la courbure
extrins\`{e}que des hypersurfaces $\Sigma_t$ (c.f. E.2.30). Ceci
implique que la courbure scalaire extrins\`{e}que sur $\Sigma_t$
n'est autre que
$$ \hskip 3cm K = \frac{1}{2}
\dot{h} = \frac{1}{2} \dot{g} \hskip 1.5cm (K(t) =
\frac{1}{2}\dot{h}(t) = \frac{1}{2}\dot{g}(t))$$ $\dot{h}$ et
$\dot{g}$ \'{e}tant la trace commune de $^{(4)} \dot{h}_{ab}$ =
$^{(4)} \dot{g}_{ab}$ et de $^{(3)}
\dot{h}_{ab}$.\\
\subsection*{Formulation hamiltonienne}

Passons maintenant \`{a} la formulation Hamiltonienne associ\'{e}e
\`{a} l'\'{e}quation ($E_0$). En utilisant les notations de [4],
on constate que dans le cadre de notre mod\`{e}le, on a:
$$ \hskip 3cm N = 1 \hskip 3cm \mbox{ (La fonction lapse est 1) }
$$et
$$ \hskip 3.6cm N^a = 0 \hskip 2.5cm \mbox{ (Il n'y a pas de "shift
vector") }$$ainsi que ((E.2.26), (E.2.27), (E.2.28) et (E.2.29))
$$ R = 2 \left(G_{ab} \left(\frac{\partial}{\partial t}\right)^a \left(\frac{\partial}{\partial
t}\right)^b - R_{ab} \left(\frac{\partial}{\partial t}\right)^a
\left(\frac{\partial}{\partial t}\right)^b\right),$$
$$ G_{ab} \left(\frac{\partial}{\partial t}\right)^a \left(\frac{\partial}{\partial
t}\right)^b = \frac{1}{2} ( ^{(3)} R - K_{ab} K^{ab} + K^2),$$
$$ R _{ab} \left(\frac{\partial}{\partial t}\right)^a
\hskip 0.1cm \left(\frac{\partial}{\partial t}\right)^b = K^2 - K_{ac}\hskip
0.1cm K^{ac}-\nabla_a\left(\left(\frac{\partial}{\partial t}\right)^a\nabla_c\left(\frac{\partial}{\partial t}\right)^c\right)+\nabla_c\left(\left(\frac{\partial}{\partial t}\right)^a\nabla_a\left(\frac{\partial}{\partial t}\right)^c\right)$$ 
et
$${\cal L}_G = \sqrt{h} (\hskip 0.1cm ^{(3)} R + K_{ab}\hskip 0.1cm
K^{ab} - K^2)$$avec toujours:
$$K_{ab} = \frac{1}{2} \dot{h}_{ab}\hskip 0.2cm.$$En d\'{e}finissant le
"momentum" canoniquement conjugu\'{e} \`{a} $h_{ab}$ par
((E.2.31))
$$ \Pi^{ab} = \frac{\partial {\cal L}_G}{\partial \dot{h}_{ab}} =
\sqrt{h} ( K^{ab} - K \hskip 0.1cm h^{ab})$$et l'espace de
configuration comme \'{e}tant l'ensemble des m\'{e}triques
riemanniennes asymptotiquement plates $h_{ab}$ sur $\Sigma_t$, on
d\'{e}finit la densit\'{e} hamiltonienne associ\'{e}e \`{a}
l'action gravitationnelle $S_G$ (la diff\'{e}rence des deux autres
termes figurant dans l'expression de $S'_G$ pouvant \'{e}tre
n\'{e}glig\'{e}e pour $t_2$ tr\`{e}s proche de $t_1$), par
((E.2.32))
$${\cal H}_G = \Pi^{ab} \hskip 0.1cm \dot{h}_{ab} - {\cal L}_G$$
$$ \hskip 4cm = \sqrt{h} ( - \hskip 0.1cm ^{(3)} R + \frac{1}{h}
\Pi^{ab} \hskip 0.1cm \Pi_{ab} - \frac{1}{2h} \Pi^2)$$o\`{u}
$$\Pi = \Pi^a_a \hskip 0.2cm.$$Le hamiltonien sera la fonction
d\'{e}finie, pour tout $\Sigma_t$, par
$$ H (g_{ab}, \Pi^{ab}) = \int_{\Sigma_t} {\cal H}_G \hskip 0.2cm
^{(3)} e.$$La formulation hamiltonienne r\'{e}sultant d'une
variation de $h_{ab}$ avec $\delta h_{ab}$ = 0 sur les $\Sigma_t$
est \'{e}quivalente \`{a} ($E_0$). Les $h_{ab}$ \'{e}tant
asymptotiquement plates ($h_{ab} \cong g_e$ sur un voisinage de
$S(O,t)$ pour $ t \gg 1$), les solutions de ($E_0$) sont les
solutions du syst\`{e}me hamiltonien, libre de toute contrainte,
suivant:
$$\dot{h}_{ab} = \frac{\delta H_G}{\delta \Pi^{ab}} =
\frac{2}{\sqrt{h}} (\Pi_{ab} - \frac{1}{2} \Pi \hskip 0.1cm
h_{ab})$$
$$\dot{\Pi}^{ab} = -\frac{\delta H_G}{\delta h_{ab}} = - \sqrt{h}
(\hskip 0.1cm ^{(3)} R^{ab} - \frac{1}{2} \hskip 0.1cm ^{(3)} R
\hskip 0.1cm h^{ab})$$
$$\hskip 3cm + \frac{1}{2} \frac{1}{\sqrt{h}} h^{ab} (\Pi_{cd} \hskip 0.1cm \Pi^{cd} - \frac{1}{2}\Pi^2)$$
$$\hskip 3cm - \frac{2}{\sqrt{h}} (\Pi^{ac} \hskip 0.1cm \Pi_c^b -
\frac{1}{2} \Pi \hskip 0.1cm \Pi^{ab}).$$

Ce syst\`{e}me se r\'{e}duit, dans le cadre de notre mod\`{e}le,
\`{a} douze \'{e}quations \`{a} douze inconnues ind\'{e}pendantes. Une
solution $h_{ab}$ de ce syst\`{e}me n'est autre que notre
m\'{e}trique $g_t$, d\'{e}finie sur
\emph{B}(\emph{O},\emph{t}),dans le cas (virtuel) o\`{u} le
tenseur $^{(3)} G^{*}_{ab} = \hskip 0.1cm ^{(3)} T_{ab}^{*}$
d\'{e}crit uniquement l'effet de la gravit\'{e}.\\
\subsection*{Remarques}
$1^\circ$) Les deux contraintes (E.2.33) et (E.2.34) i.e. 
$${}^{(3)}R+h^{-1}\Pi^{ab}\Pi_{ab}-\frac{1}{2}h^{-1}\Pi^2=0$$ 
et 
$$D_a(h^{-\frac{1}{2}}\Pi^{ab})=0,$$
qui
proviennent de la variation de $H_G$ par rapport \`{a} $N$ et
$N_a$ (la contrainte (E.2.34) pouvant \^{e}tre contourn\'{e}e en
utilisant les super-espaces de wheeler) n'ont pas d'existence dans
le cadre de notre mod\`{e}le, comme dans notre mod\`ele il existe un seul canonique ``slicing'' du temps de l'espace-temps. Elles constituent en fait
l'h\'{e}ritage de la relativit\'{e} sp\'{e}ciale au sein de celle
de la relativit\'{e} g\'{e}n\'{e}rale standard. D'ailleurs
m\^{e}me pour cette derni\`{e}re, $N$ et $N_a$ ne constituent pas
des vraies variables dynamiques (voir [4]).\\
Les contraintes (E.2.33) et (E.2.34) sont en fait \'{e}quivalentes
aux deux \'{e}quations (10.2.28) et (10.2.30) de [4] suivantes:
$$ D_b K^b_a - D_a K ^b_b = 0$$
et
$$ ^{(3)} R + (K_a^a)^2 - K_{ab} K ^{ab} = 0$$
appel\'{e}es contraintes des valeurs initiales de la
relativit\'{e} g\'{e}n\'{e}rale.\\
Ces m\^{e}mes contraintes peuvent aussi \^{e}tre exprim\'{e}es,
\`{a} l'aide de la deuxi\`{e}me forme fondamentale (not\'{e}e
\'{e}galement $K_{ij}$ dans [5]) de l'hypersurface $S$ de type
espace de la vari\'{e}t\'{e} lorentzienne $M$ (ayant une courbure
de Ricci nulle), de la mani\`{e}re suivante (c.f. (9.7) et (9.8)
de [5])
$$ R_S- K_{jk} K^{jk} + K_j^j K^k_k = 0$$
et
$$ K^k_{j;k} - 3H_{;j} = 0$$
Notons que la relation (10.3) de [5] i.e. $\dot{g}_{jk}$ =
-2$\lambda K_{jk}$, o\`{u} $\lambda$ est le coefficient de $
-dt^2$ dans la m\'{e}trique lorentzienne (10.1) de [5], qui se
r\'{e}duit, dans le cadre de notre mod\`{e}le, \`{a} 1, montre
(dans ce cadre pr\'{e}cis) que la courbure extrins\`{e}que de $S$
dans $M$ n'est autre que la seconde forme fondamentale sur
$S$.\\\\
Soit maintenant $\displaystyle M_\infty=\lim_{t\rightarrow+\infty}C(t)=C\subset\mathbb{R}^4$; comme $\displaystyle\lim_{t\rightarrow+\infty}\rho(t)=0$, la m\'etrique lorentzienne non n\'egative de $M_\infty$ devient $h=dt^2-g_e=-\eta$, qui est une m\'etrique riemannienne sur l'int\'erieur de $M_\infty$. Si $\Sigma_t\simeq B(O,t)$ est muni de $g_t$, le tenseur de courbure extrins\`eque de $\Sigma_t$ dans $M_\infty$ est donn\'e (c.f. \cite[5.5]{8}) par le th\'eor\`eme de Gauss
$$R(x,y,u,v)=\ell(x,u) \ell(y,v)-\ell(x,v)\ell(y,u)$$ 
puisque l'on a ici $\tilde{R}(x,y,u,v)=0$, et la courbure scalaire extrins\`eque est donn\'ee par
$${}^{(3)}R_{\Sigma_t}=\sum_{i,j=1}^{3}R(e_i,e_j,e_i,e_j)=\sum_{i,j=1}^{3}(\ell(e_i,e_i)\ell(e_j,e_j)-\ell(e_i,e_j)^2)$$
o\`u $e_i$, pour $i=1,2,3$ est une base orthonorm\'ee (pour $g_t$) de $T_m\Sigma_t$ et $\ell$ est la deuxi\`eme forme fondamentale de $\Sigma_t$. Cette relation n'est autre que la premi\`ere contrainte ci-dessus 
$$R_s=K_{jk}K^{jk}-K_j^j K^k_k;$$
la diff\'erence de signe \'etant le r\'esultat de la diff\'erence de signature de notre m\'etrique $h_t=dt^2-g_t$.\\
De m\^eme l'\'equation de Gauss-Codazzi equation ([8], 5.8.e)) devient, dans le cadre de notre mod\`ele,
$$\sum_{i,j=1}^{3}\tilde{R}(e_i,e_j,e_i,\frac{\partial}{\partial t})=\sum_{i,j=1}^{3}(D\ell(e_j,e_i,e_i)-D\ell(e_i,e_j,e_i))$$
qui n'est autre que
\begin{eqnarray*}
0&=&\sum_{i,j=1}^{3}(D_{e_j}\ell(e_i,e_i)-D_{e_i}\ell(e_j,e_i))\\
&=&\sum_{i,j=1}^{3}(D_{e_j}K_{ii}-D_{e_i}K_{ji})\\
&=&\frac{1}{2}\sum_{i,j=1}^{3}(D_{e_j}\dot{g}_{ii}-D_{e_i}\dot{g}_{ji})
\end{eqnarray*}
qui est toujours v\'erifi\'ee.\\
Si maintenant la m\'etrique $g_t$ est repr\'esent\'ee par la matrice 
$$\left(\begin{array}{lll}
g_{11}&g_{12}&g_{13}\\
g_{21}&g_{22}&g_{23}\\
g_{31}&g_{32}&g_{33}\\
\end{array}
\right)
$$
alors on a
$$K_{ij}=\frac{1}{2}\left(\begin{array}{lll}
\dot{g}_{11}&\dot{g}_{12}&\dot{g}_{13}\\
\dot{g}_{21}&\dot{g}_{22}&\dot{g}_{23}\\
\dot{g}_{31}&\dot{g}_{32}&\dot{g}_{33}\\
\end{array}
\right)
$$
et
\begin{eqnarray*}
{}^{(3)}R&=&K_{jk}K^{jk}-K_j^jK_k^k\\
&=&\frac{1}{4}(\dot{g}^2_{11}+\dot{g}^2_{22}+\dot{g}^2_{33})+\frac{1}{2}(\dot{g}^2_{12}+\dot{g}^2_{13}+\dot{g}^2_{23})-\frac{1}{4}(\dot{g}_{11}+\dot{g}_{22}+\dot{g}_{33})^2\\
&=&\frac{1}{2}(\dot{g}^2_{12}+\dot{g}^2_{13}+\dot{g}^2_{23}-\dot{g}_{11}\dot{g}_{22}-\dot{g}_{11}\dot{g}_{33}-\dot{g}_{22}\dot{g}_{33}).
\end{eqnarray*}

\noindent $2^{\circ}$) Pour int\'{e}grer les autres effets de la
distribution de la mati\`{e}re - \'{e}nergie dans l'Univers dans
une formulation lagrangienne, on doit consid\'{e}rer une
densit\'{e} lagrangienne ${\cal L}$ donn\'{e}e par
$$ {\cal L} = {\cal L}_G + {\cal L}_M.$$
Cette densit\'{e} d\'{e}termine une action
$$S = S_G + S_M$$
dont l'extr\'{e}misation \'{e}quivaut \`{a} la r\'{e}solution de
l'\'{e}quation (\emph{E}). Dans le cas particulier d'un couplage
du champ gravitationnel avec un champ scalaire de Klein-Gordon,
${\cal L}$, ${\cal L}_G$, ${\cal L}_{KG}$, $T_{ab}^{KG}$ et
$S_{KG}$ sont explicitement donn\'{e}s et reli\'{e}s entre eux
\`{a} l'aide des relations (E.1.22) et (E.1.24) - (E.1.26) de [4].
Pour le couplage Einstein - Maxwell, on peut se r\'{e}f\'{e}rer
aux relations (E.1.23) - (E.1.26).\\
Dans un cadre plus g\'{e}n\'{e}ral o\`{u} l'on consid\`{e}re les
\'{e}quations qui gouvernent l'interaction d'un champ
\'{e}lectromagn\'{e}tique avec une poussi\`{e}re de mati\`{e}re
charg\'{e}e model\'{e}e \`{a} l'aide d'une substance charg\'{e}e
continue au sein d'un champ gravitationnel, on est amen\'{e} \`{a}
consid\'{e}rer le lagrangien
$$ L = L_1+L_2+L_3$$
avec ($S$ \'etant la courbure scalaire)
\begin{eqnarray*}
 &L_1& = - \frac{1}{8 \pi} < \cal{F}, \cal{F} > + < A, J >\\
 &L_2& = \frac{1}{2} \mu < u,u >\\
 &L_3& = \alpha S
\end{eqnarray*}
o\`{u} l'on a utilis\'{e} les notations de [5] en imposant les
conditions (1.6), (1.7) et (1.10) de [5]. Les variations par
rapport \`{a} la m\'{e}trique $g_{jk}$ donnent:
$$\hskip 3cm \delta \int (L_1+L_2) dV = \frac{1}{2} \int T^{jk} (\delta
g_{jk}) dV \hskip 3cm (1.15)$$ avec
$$ \hskip 2cm T^{jk} = \mu u^j u^k + \frac{1}{4 \pi} ( {\cal{F}} ^j_l
{\cal{F}} ^{kl} - \frac{1}{4} g^{jk} {\cal{F}} ^{il} {\cal{F}}
_{il}) \hskip 3cm (1.24)$$ et
$$ \hskip 2cm \delta \int S dV = - \int G^{jk} (\delta g_{jk}) dV \hskip 5cm
(1.26)$$ 
ce qui donne l'\'{e}quation tensorielle standard
d'Einstein dans le cas consid\'{e}r\'{e}.\\\\
Par ailleurs, notre m\'{e}trique physique $g_t$ sur l'Univers
$B(O,t)$ et notre m\'{e}trique dynamique $h = dt^2 - g_t$ sur $C^*
= C \setminus O$ sont les solutions respectives de $(E^*)$ et
$(E)$ o\`{u} les tenseurs $^{(3)} T_{ab}^{*}$ et $^{(4)}
T_{ab}^{*}$ qui y figurent refl\`{e}tent toutes les formes de la
mati\`{e}re - \'{e}nergie (masse et champ gravitationnel locaux,
masse, trous noirs et champ gravitationnel cosmiques, champ
\'{e}lectromagn\'{e}tique local, radiations neutrinos et pression
cosmiques). Les mouvements libres dans $B(O,t)$ et $C^{*}$ se
confondent avec des g\'{e}od\'{e}siques pour $g_t$ et $h$.\\
Les lagrangiens et tenseurs consid\'{e}r\'{e}s en relativit\'{e}
g\'{e}n\'{e}rale classique constituent (dans des cas particuliers)
de bonnes approximations pour les lagrangiens et les tenseurs
$^{(4)} T^{*}_{ab}$
utilis\'{e} dans le cadre de notre mod\`{e}le.\\

\subsection*{R\'{e}sum\'{e}}
L'Univers physique (id\'{e}alis\'{e})
est caract\'{e}ris\'{e}, d'une fa\c{c}on \'{e}quivalente, par:\\\\
1. La distribution $E_t(X)$ de la mati\`{e}re - \'{e}nergie, pour
(\emph{t}, \emph{X}) v\'{e}rifiant $t >$ 0 et \emph{X} $\in$
\emph{B}(\emph{O},\emph{t}).\\\\
2. Le tenseur $^{(3)} T_{ab}^{*}$ sur \emph{B}(\emph{O},\emph{t})
pour \emph{t} $>$ 0.\\\\
3. La m\'{e}trique riemannienne $g_t$ d\'{e}finie, pour tout
\emph{t} $>$ 0, sur \emph{B}(\emph{O}, \emph{t}) et v\'{e}rifiant
$$^{(3)} R_{ab} - \frac{1}{2} \hskip 0.1cm ^{(3)}R \hskip 0.1cm g_t = \hskip
0.1cm ^{(3)}T_{ab}^* . $$\\
4. La distribution \emph{E}(\emph{t}, \emph{X}) d\'{e}finie sur
\emph{C} par \emph{E}(\emph{t},
\emph{X}) = $E_t$(\emph{X}) pour $t > $0 et $X \in \Sigma_t$.\\\\
5. La m\'{e}trique $g_{ab}$ d\'{e}finie sur \emph{C} par
$$ g_{ab} = - dt^2 + \hskip 0.1cm ^{(4)} h_{ab}$$o\`{u}
$$^{(4)}h_{ab} ( \frac{\partial}{\partial t},
\frac{\partial}{\partial t}) = 0 \hskip 1cm , \hskip 0.5cm  ^{(4)}
h_{ab} (\frac{\partial}{\partial t}, \frac{\partial}{\partial
x_i}) = 0 \hskip 0.5cm \mbox{ pour } i = 1,2,3 $$et
$$^{(3)} h_{ab} = g_t \hskip 1.5cm
\mbox{ sur } \Sigma_t = \emph{B}(O,\emph{t}) \times \{ t \}.$$\\
6. Le tenseur g\'{e}n\'{e}ralis\'{e} de mass - \'{e}nergie
d'Einstein $^{(4)} T^{*}_{ab}$ v\'{e}rifiant
$$^{(4)} R_{ab} - \frac{1}{2} \hskip 0.1cm ^{(4)} R \hskip 0.1cm
^{(4)}g_{ab} =  ^{(4)} T^{*}_{ab}.$$\\
7. La densit\'{e} lagrangienne
$${\cal L} = {\cal L}_G + {\cal L}_M$$o\`{u}$${\cal L}_G = \sqrt{-g} \hskip 0.2cm ^{(4)}R$$
et ${\cal L}_M$ est la densit\'{e} lagrangienne associ\'{e}e \`{a}
l'ensemble des champs autre que le champ gravitationnel de sorte
que l'extr\'{e}misation de l'action
$$S^{'}_G = S_G + 2 \int_{\Sigma_t} K - 2
\int_{\Sigma_{t_0}}K$$par rapport aux variations v\'{e}rifiant
$\delta g_{ab}$ = 0 ou $\delta h_{ab}$ = 0 sur les $\Sigma_t$
donne les solutions des \'{e}quations (\emph{E}) et ($E^{*}$).\\\\
8. La densit\'{e} hamiltonienne ${\cal H}$ et le hamiltonien
\emph{H} = $\int_{\Sigma_t} {\cal H} $ d\'{e}finis \`{a} partir du
lagrangien ${\cal L}$ par
$${\cal H} = \Pi^{ab} \hskip 0.1cm
h_{ab} - {\cal L}$$o\`{u}$$\Pi^{ab} = \frac{\partial {\cal
L}}{\partial \dot{h}_{ab}};$$ $h_{ab}$ est alors la solution du
syst\`{e}me hamiltonien sans contraintes:
$$\dot{h}_{ab} = \frac{\delta H}{\delta \Pi^{ab}}$$
$$\dot{\Pi}^{ab} = - \frac{\delta H}{\delta h_{ab}}\hskip 0.1cm.$$
\\
\subsection*{Aspects g\'{e}n\'{e}raux de la solution}

Ainsi, on obtient, dans le cadre de notre mod\`{e}le, une solution
id\'{e}alis\'{e}e d\'{e}terministe. Notre espace - temps sera,
pour tout $t >$ 0, (\emph{C}(\emph{t}), $^{(4)} g_{ab}$)
constitu\'{e} par le d\'{e}veloppement de Cauchy maximal
associ\'{e} aux conditions initiales
$$^{(3)} h_{ab} (t_0) \hskip
1cm \mbox{ et } \hskip 1cm \frac{1}{2} \hskip0.1cm ^{(3)}
\dot{h}_{ab}(t_0)$$d\'{e}finies sur une surface de Cauchy
quelconque $\Sigma_{t_0}$ de \emph{C}(\emph{t}) avec $t_0 < t$ de
sorte que la m\'{e}trique riemannienne $^{(3)}h_{ab}(t_0)$ sur
$\Sigma_{t_0}$ s'identifie \`{a} la m\'{e}trique physique (des
paragraphes pr\'{e}c\'{e}dents) $g_{t_0}$ sur
\emph{B}(\emph{O},$t_0$) et que le tenseur $\frac{1}{2} \hskip
0.1cm ^{(3)} \dot{h}_{ab} (t_0)$ = $\frac{1}{2}$
($\dot{g_{t_0}}$)$_{ab}$ ne soit autre que le tenseur de courbure
extrins\`{e}que de l'hypersurface $\Sigma_{t_0}$ dans
l'espace - temps (\emph{C}(\emph{t}), $^{(4)} g_{ab}$).\\
Cette solution pourrait \^{e}tre assez proche de la vraie
m\'{e}trique r\'{e}elle sur un intervalle de temps d'autant plus
\'{e}tendu que l'approximation et la r\'{e}gularisation des
conditions initiales $^{(3)} h_{ab}$($t_0$) et $K_{ab}(t_{0})$ sur
$\Sigma_{t_0}$ soient proches de la r\'{e}alit\'{e} physique de
notre Univers \`{a} l'instant $t_0$. Une r\'{e}actualisation
permanente, bas\'{e}e sur l'\'{e}largissement de la banque de
donn\'{e}es disponibles et de son affinement, est
n\'{e}cessaire.\\
Notre mod\`{e}le contient le juste nombre de degr\'{e} de
libert\'{e} et ne comporte pas la notion de libert\'e de jauge ("gauge freedom").\\
En effet, la nature physique de l'Univers le force (gr\^{a}ce
\`{a} la constance de la vitesse de propagation
\'{e}lectromagn\'{e}tique et \`{a} son caract\`{e}re isotropique)
\`{a} exister, \`{a} tout instant \emph{t}, sous la forme d'une
boule de rayon euclidien \emph{t} et \`{a} \'{e}voluer sous la
forme du demi - c\^{o}ne d'espace - temps \emph{C}. Ceci oblige
tout diff\'{e}omorphisme-jauge $\psi$ \`{a} transformer le demi -
c\^{o}ne \emph{C}(\emph{t}) en un demi - c\^{o}ne
\emph{C}($t^{'}$) et \`{a} \^{e}tre de la forme $\psi$ =
(\emph{t}, $\varphi_{t}$) o\`{u} $\varphi_{t}$ est un
diff\'{e}omorphisme de \emph{B}(\emph{O},\emph{t}) sur
\emph{B}(\emph{O},$t^{'}$) v\'{e}rifiant $g_t$ = $\varphi^{*}_{t}
g_{t^{'}}$. Ceci revient tout simplement \`{a} un changement
d'\'{e}chelle purement conventionnel, \`{a} moins que $\psi$ ne
soit une transformation isom\'{e}trique de \emph{C} dans
$\mathbb{R}^4$ et $\varphi$ ne soit une transformation
isom\'{e}trique de \emph{B}(\emph{O},\emph{t}) dans
$\mathbb{R}^3$; ce qui constitue en fait des
diff\'{e}omorphismes-jauges triviaux i.e. v\'{e}rifiant
$$\psi^{*} g_{ab} = g_{ab}
\hskip 1cm \mbox{ et } \hskip 1cm \varphi^{*}_{t} g_t = g_t.$$

\subsection*{Remarques}
Conform\'{e}ment \`{a} notre mod\`{e}le, on
peut affirmer les propri\'{e}t\'{e}s suivantes:\\\\
1. La m\'{e}trique physique r\'{e}elle ne peut pas \^{e}tre
globalement obtenue par le proc\'{e}d\'{e} de lin\'{e}arisation:
La perturbation de la m\'{e}trique de Minkowski dans \emph{C} et
de la m\'{e}trique euclidienne dans \emph{B}(\emph{O},\emph{t})
d\^{u}es aux effets de la mati\`{e}re - \'{e}nergie sont loin
d'\^{e}tre "petites" surtout autour des trous noirs.\\\\
2. Notre Univers n'est \'{e}videmment ni homog\`{e}ne ni
isotropique. Il existe bel et bien une foliation
($\Sigma_t$)$_{t>0}$ de l'espace - temps, mais, pour \emph{p} et
\emph{q} $\in \Sigma_{t}$ arbitraires, il ne peut pas exister une
isom\'{e}trie de $\Sigma_{t}$ transformant \emph{p} en \emph{q}.
De m\^{e}me, il n'existe pas forc\'{e}ment une isom\'{e}trie de
\emph{C} laissant \emph{p} fixe et transformant un vecteur spacial
unitaire en \emph{p} en un autre vecteur ayant ces m\^{e}mes
propri\'{e}t\'{e}s. Par cons\'{e}quent, notre mod\`{e}le est
totalement diff\'{e}rent des deux cas du mod\`{e}le de
Robertson-Walker correpondant \`{a} \emph{K} = $\pm$ 1 bien qu'il
est, en cas d'extr\^{e}me id\'{e}alisation, tr\`{e}s proche du cas
correspondant \`{a} \emph{K} = 0 de ce m\^{e}me mod\`{e}le. En
effet, la m\'{e}trique correspondant \`{a} ce dernier cas se
r\'{e}duit, dans le cadre de notre mod\`{e}le, \`{a}
$$ ^{(4)} g = -dt^2 + a^2(t)(dx^2 + dy^2 + dz^2)$$o\`{u}$\hskip 0.4cm$ $dx^2 + dy^2 +
dz^2$ est la restriction de la m\'{e}trique euclidienne de
$\mathbb{R}^3$ \`{a} la boule
\emph{B}(\emph{O},\emph{R}(\emph{t})) qui est, faut-il le rappeler, en expansion permanente.\\
Si $g_t=a^2(t)g_e$ et $h_t=dt^2-g_t$, alors 
$$\dot{g}_t=\left(\begin{array}{lll}
2a\dot{a}&\;\;0&\;\;0\\
\;\;0&2a\dot{a}&\;\;0\\
\;\;0&\;\;0&2a\dot{a}\\
\end{array}
\right)
$$
et
$${}^{(3)}R=\frac{1}{2}(4a^2\dot{a}^2+4a^2\dot{a}^2+4a^2\dot{a}^2)=6a^2\dot{a}^2.$$

\noindent Par ailleurs, les \'{e}quations d'\'{e}volution pour la Cosmologie
homog\`{e}ne isotropique s'\'{e}crivent en supposant $c = 1$ et $G
= 1$ (c.f.(5.2.14) et (5.2.15) de [4])
$$ 3 \frac{\dot{a}^2}{a^2} = 8 \pi \rho - \frac{3K}{a^2}$$et
$$ 3 \frac{\ddot{a}}{a} = -4 \pi (\rho + 3P),$$o\`{u} $K$ est le param\`{e}tre de courbure
(qui est ici, contrairement \`{a} notre mod\`{e}le, une
constante), $\rho$ est la densit\'{e} moyenne de la mati\`{e}re
dans l'Univers et \emph{P} est la pression moyenne associ\'{e}e
aux radiations thermiques remplissant l'Univers qui constituent
ensemble le tenseur de stress - \'{e}nergie d'Einstein. Ceci
implique (5.2.18)
$$\dot{\rho} + 3 (\rho + P) \frac{\dot{a}}{a} = 0$$qui, pour le
mod\`{e}le poussi\`{e}re standard (\emph{P} = 0) conduit \`{a}
$$\rho_m a^3 = cte,$$
(o\`{u} $\rho_m$ est la densit\'{e} moyenne de la mati\`{e}re dans
le cadre de ce mod\`{e}le) et, pour le mod\`{e}le radiation
standard (\emph{P} = $\frac{\rho}{3}$), conduit \`{a}
$$\rho_r a^4 = cte$$
(o\`{u} $\rho_r$ est la densit\'{e} moyenne des radiations dans le
cadre de ce dernier mod\`{e}le).\\
Ainsi, on a, pour le mod\`{e}le poussi\`{e}re standard (5.2.21):
$${\dot{a}}^2 - \frac{C}{a} = 0 \;\;\;\;\;\; \mbox{ avec } \;\;\;\;C =
\frac{8}{3} \pi \rho a^3$$et, pour le mod\`{e}le radiation
standard , on a (5.2.22):
$${\dot{a}}^2 - \frac{C'}{a} = 0 \;\;\;\;\; \mbox{ avec } \;\;\;\;
C' = \frac{8}{3} \pi \rho a^4.$$Dans le premier cas, qui
correspondrait \`{a} notre Univers actuel, on a ([4],Table 5.1)
$$a(t) = (\frac{9C}{4})^{\frac{1}{3}} \hskip 0.2cm
t^{\frac{2}{3}}$$et dans le second, qui correpondrait au tout
d\'{e}but de la formation de l'Univers juste apr\`{e}s le Big
Bang, on a ((5.4.1) et (5.4.2)):
$$a(t) = (4C^{'})^{\frac{1}{4}} \hskip 0.2cm t^{\frac{1}{2}}$$
et
$$\rho_r(t) = \frac{3}{32 \pi G t^2}\hskip 0.1cm.$$
Par suite, en utilisant la M\'{e}canique quantique statistique, on
trouve que, pour \emph{t} infiniment petit, la temp\'{e}rature
\emph{T} de l'Univers est proportionnelle \`{a} $\rho_r^{\frac{1}{4}}$ et \`{a} $\frac{1}{a}.$\\\\
D'un autre c\^{o}t\'{e}, on a, conform\'{e}ment \`{a} notre
mod\`{e}le,apr\`{e}s une simplification grossi\`{e}re (dans le
cadre d'une Cosmologie homog\`{e}ne et isotropique de l'Univers
naissant):
$$g_t = a^2(t) (dx^2 + dy^2 + dz^2)$$sur
\emph{B}(\emph{O},\emph{R}(\emph{t})),ce qui implique
$$\int _{B(O,R(t))} dv_{g_t} = \int_{B(O,R(t))} a^3(t)dv_{g_e} =
\int_{B(O,R(t))} dX - E = \frac{4 \pi R^3(t)}{3} - E$$ o\`{u}
\emph{E} est l'\'{e}nergie totale de l'Univers. Ceci nous permet,
\`{a} titre d'exemple, de calculer $C^{'}$ en fonction de \emph{E}
et \emph{R}(\emph{t}) pour $t \ll $1 et $C$ en fonction de \emph{E} et
de \emph{t} lorsqu'on suppose
\emph{R}(\emph{t}) $\simeq t$ pour $t\gg1$.\\
De m\^{e}me, on pourrait aussi avoir des renseignements sur le
facteur (i.e. la constante d\'{e}pendante du temps) de Hubble
\emph{H}(\emph{t}) = $\frac{\dot{a}}{a}$.\\\\
3. L'Univers n'est pas sph\'{e}riquement sym\'{e}trique ni
axisym\'{e}trique. Par cons\'{e}-\\quent, la solution de
Schwarzschild n'est qu'une id\'{e}alisation de l'Univers le
r\'{e}duisant \`{a} un champ gravitationnel r\'{e}sultant d'un
noyau mat\'{e}riel sph\'{e}rique
et statique.\\\\
4. L'Univers n'est pas stationnaire. Le champ de vecteur de la
translation du temps ($\frac{\partial}{\partial t}$)$^{a}$ n'est
pas un champ de vecteur de Killing; bien qu'il peut \^{e}tre
approximativement consid\'{e}r\'{e} ainsi dans toutes les
r\'{e}gions de l'espace-temps qui correspondent aux r\'{e}gions de
$\Sigma_t \simeq B(O,t)$ qui peuvent \^{e}tre consid\'{e}r\'{e}es
durablement comme \'{e}tant \`{a} l'abri de la mati\`{e}re et de
ses effets (i.e. l\`{a} o\`{u} $h_{ab}=g_t \simeq g_e$. L'Univers
n'est pas statique non plus bien que la famille $\sum_t$ est
orthogonale \`{a} ($\frac{\partial}{\partial t}$)$^{a}$ et que
l'on a
$$g_{ab} = -dt^2 + h_{ab}$$
N\'{e}anmoins, un grand nombre de r\'{e}sultats obtenus \`{a}
partir de tels hypoth\`{e}ses, r\'{e}ductions et id\'{e}alisations
restent, qualitativement et quantitativement, plus ou moins
valables que ce soit localement ou globalement. Ceci est
particuli\`{e}rement vrai pour les r\'{e}sultats obtenus \`{a}
l'aide de ce qui est qualifi\'{e} comme limite newtonienne ou
Cosmologie homog\`{e}ne ainsi que leurs cons\'{e}quences sur
certains aspects concernant l'\'{e}volution de l'Univers et sa
structure causale. Il en est de m\^{e}me pour certaines
cons\'{e}quences de l'extension de Kruskal de la solution de
Schwarzschild concernant les trous noirs stationnaires dans le
vide et les trous noirs charg\'{e}s de Kerr, associ\'{e}s \`{a}
l'\'{e}quation d'Einstein - Maxwell, ainsi que les
propri\'{e}t\'{e}s, qualifi\'{e}es de thermodynamiques, des trous
noirs.\\\\
5. \textbf{(La constante cosmologique)} Lorsqu'on \'{e}crit
l'\'{e}quation d'Einstein en y incorporant une constante
cosmologique non nulle, on obtient
$$R_{ab} - \frac{1}{2} R g_{ab} = 8 \pi T_{ab} - \Lambda
g_{ab}.$$L'\'{e}quation d'Einstein dans le vide devient
$$R_{ab} - \frac{1}{2} Rg_{ab} = - \Lambda g_{ab}$$qui, par
contraction, donne
$$R = 4 \Lambda$$ce qui montre que la courbure de $g_{ab}$ dans le
vide d'Einstein (caract\'{e}ris\'{e} par $\rho=0$ et $T_{ab}=0$)
est non nulle.\\
Or, dans le cadre de notre mod\`{e}le, l'identit\'{e} $T^*_{ab}=0$
implique $T_{ab}=0$ mais la r\'{e}ciproque n'est pas vraie. Cette
comparaison montre que notre condition $T^*_{ab}=0$ (et alors
$^{(3)}T_{ab} =0$ et $g_t = g_e$) est une condition tr\`{e}s
restrictive et id\'{e}aliste. Elle correspond \`{a} une r\'{e}gion
de vide absolu dans $B(O,t)$ dont l'existence est tr\`{e}s
improbable. La constante $\Lambda$ ne fait que refl\'{e}ter, dans
le contexte de notre cosmologie, l'influence de la mati\`{e}re
cosmique et de la gravit\'{e} et des radiations cosmiques dans les
r\'{e}gions d\'{e}pourvues de mati\`{e}re et qui ne sont pas sous
l'influence directe d'un champ gravitationnel ou
\'{e}lectromagn\'{e}tique, i.e. les r\'{e}gions de $B(O,t)$
caract\'{e}ris\'{e}es par
$$ ^{(3)} T^*_{ab} = \Lambda g_e.$$Notre \'{e}quation tensorielle du
champ y sera alors
$$ ^{(3)} R_{ab} - \frac{1}{2} {}^{(3)}R g_t = \Lambda g_e,$$
$\Lambda$ d\'{e}pendant probablement de la r\'{e}gion en question
et du temps.\\
Par cons\'{e}quent, pour notre m\'{e}trique $g_t$ (ainsi que pour
la m\'{e}trique $h=dt^2-g_t$), on peut difficilement avoir $R_{ab}
- \frac{1}{2} R g_t = 0$ dans une r\'{e}gion (intergalactique)
quelconque de l'Univers. Ainsi la courbure scalaire $R$ et la
courbure de Ricci $R_{ab}$ ne peuvent s'annuler rigoureusement
dans une telle r\'{e}gion bien que $\Lambda$ y est n\'ecessairement extr\`emement petite.\\\\

Pourtant, un travail gigantesque et assidu reste \`{a} faire pour
traiter ces grands sujets. On peut parler aussi \`{a} propos des
modifications qu'il faut effectuer en ce qui concerne le
mod\`{e}le cosmologique d'Einstein - de Sitter - Friedmann pour
l'adapter \`{a} notre mod\'{e}lisation. Tout ceci n\'{e}cessite
l'abandon de quelques postulats et principes erron\'{e}s et non
justifi\'{e}s et d'adapter le tout \`{a} la r\'{e}alit\'{e} de
l'expansion dynamique de notre Univers qui, bien qu'il est de plus
en plus immense, il est perp\'{e}tuellement fini. Ajoutons aussi
que notre mod\`{e}le dynamique ouvre la voie \`{a}
l'\'{e}tablissement d'une th\'{e}orie quantique de la
relativit\'{e} g\'{e}n\'{e}rale (utilisant la m\'ethode de quantisation canonique) en la lib\'{e}rant des contraintes
inutiles ((E.2.33) et (E.2.34) de [2]) pr\'{e}c\'{e}demment impos\'{e}es sur le syst\`{e}me
hamiltonien d\'{e}crivant l'\'{e}volution dynamique de l'Univers
en m\^{e}me temps que des contraintes impos\'{e}es aux conditions
initiales, conduisant ainsi \`{a} bien poser le probl\`{e}me de la
relativit\'{e} g\'{e}n\'{e}rale en terme d'une formulation avec
conditions initiales clairement et simplement d\'{e}finie.

\section{Introduction \`{a} une Cosmologie revis\'{e}e}
\subsection*{\'{E}quation de Friedmann}

Dans ce paragraphe, on va adapter la cosmologie homog\`{e}ne et
isotropique d'Einstein - Friedmann - Hubble - de Sitter \`{a}
notre mod\`{e}le pour pr\'{e}ciser quelques r\'{e}sultats
approximatifs qui ont \'{e}t\'{e} \'{e}tablis mi -
th\'{e}oriquement, mi - exp\'{e}rimentalement sans \^{e}tre
rigoureusement d\'{e}montr\'{e}s.\\

L'Univers physique r\'{e}el \'{e}tant assimil\'{e}, \`{a} tout
instant $t \gg $0, \`{a} l'espace riemannien\emph{U}(\emph{t}) =
(\emph{B}(\emph{O},\emph{ct}),$g_t$) (voir [1]), on va adopter le
mod\`{e}le cosmologique macroscopique global r\'{e}duisant
l'Univers actuel \`{a} une poussi\`{e}re de galaxies
distribu\'{e}e d'une fa\c{c}on homog\`{e}ne isotropique dans la
boule $B(O,ct),$ bien que ceci ne soit pas tout \`a fait exact. On
va donc suivre les travaux de Hubble et de Friedmann (voir [2]) en
posant
$$ r = r_0 R(t)$$o\`{u} \emph{R}(\emph{t}) d\'{e}signe ici le
param\`{e}tre d'expansion. Cette \'{e}quation s'\'{e}crit, dans le
cadre de notre mod\`{e}le:
$$c t = c t_0 R(t)$$o\`{u} \emph{c} est la vitesse de la
lumi\`{e}re dans le vide. Ceci implique
$$R(t) = \frac{t}{t_0}.$$
On a alors
$$\frac{dR}{dt} = \frac{1}{t_0}$$et
$$H := \frac{\frac{dR}{dt}}{R} = \frac{1}{t}.$$
Ce r\'esultat est tout \`a fait conforme \`a notre r\'eadaptation de la cosmologie macroscopique homog\`ene et isotropique \`a notre mod\`ele qui donne, d'une part que le rayon de l'Univers est pour $t\gg1$, \'equivalent \`a $ct$ et, d'autre part, que $g_t=a^2(t)(dx^2+dy^2+dz^2)$. En effet, le calcul du volume euclidien de $U(t)=(B(O,ct),g_t)$ implique $a=ct$ et $H=\dfrac{\dot{a}}{a}=\dfrac{1}{t}$.\\
Ceci conduit \`{a} l'\'{e}quation de Friedmann
$$\hskip 4.5cm {\left(\frac{dR}{dt}\right)}^2 = \frac{ 8 \pi G \rho R^2}{3 c^2} - K(t)
\;\;\;\;\;\;\;\;\;\;\;\;\;\;\;\; \hskip 2cm (27)$$o\`{u}
\emph{K}(\emph{t}) d\'{e}signe le param\`{e}tre de courbure
(d\'{e}pendant du temps) associ\'{e} \`{a} la courbure de l'espace
caus\'{e}e par la distribution mat\'{e}rielle dans l'espace et
caract\'{e}ris\'{e}e par la m\'{e}trique riemannienne $g_t$ et $\rho=\rho(t)$ est la densit\'e moyenne de la mati\`ere-\'energie associ\'ee \`a la distribution de la mati\`ere-\'energie $E_t(X)$ dans l'Univers \`a l'instant $t\gg0$, i.e.
$$\rho(t)=\int_{B(I,r_0)}E_t(X)\,dX$$
o\`u $B(I,r_0)$ est une boule moyenne de $B(O,t)$ de volume euclidien \'egale \`a l'unit\'e de volume. Ceci
est possible gr\^{a}ce \`{a} notre hypoth\`{e}se
d'homog\'{e}n\'{e}it\'{e} et d'isotropie qui est valable dans la
mesure o\`{u} il s'agit d'\'{e}tablir des r\'{e}sultats
macroscopiques g\'{e}n\'{e}raux.\\\\
\textbf{Remarque:} Dans notre mod\`{e}le, le param\`{e}tre de
courbure qui figure dans l'\'{e}quation de Friedmann ([2], 19.58)
$$ \left(\frac{dR}{dt}\right)^2 = \frac{8 \pi G \rho R^2}{3 c^2} - K$$
d\'{e}pend du temps (ainsi que la constante gravitationnelle $G$).
Le fait de consid\'{e}rer $K$ comme \'{e}tant une constante
absolue (i.e. ind\'{e}pendente du temps), comme Friedmann et la
plupart des cosmologistes l'on fait, conduit \`{a} des
contradictions flagrantes. En effet, pour \'{e}tablir cette
\'{e}quation, Friedmann a consid\'{e}r\'{e} une boule de l'Univers
de rayon $r=: r_0 R(t)$ de masse $M$ et une galaxie situ\'{e}e sur
la sph\`{e}re correspondante de masse $m$ pour aboutir \`{a} la
relation ([2]. 19.56) qui d\'etermine l'\'energie totale de la galaxie:
$$ E = - \frac{K m r_0^2}{2}.$$
Or, la constante $K$ introduite ici d\'{e}pend n\'{e}cessairement
de l'instant arbritrairement choisie $t_0$ (i.e. $K = K(t_0)$).\\
Sinon, le fait de privil\'{e}ger un autre instant $t_1 \neq t_0$,
auquel il correspond un autre rayon $r_1 \neq r_0$, et en
proc\'{e}dant de la m\^{e}me mani\`{e}re, on arrive \`{a} la
relation
$$ E = - \frac{K m r_1^2}{2}$$qui, au cas o\`{u} $K$ est une
constante absolue, contredit le principe de la conservation de
l'\'{e}nergie (de la galaxie).\\
Par ailleurs, le fait de consid\'{e}rer $K$ et $G$ comme \'{e}tant
des constantes absolues conduit \`{a} une contradiction entre
l'\'{e}quation de Friedmann ci-dessus et le deuxi\`{e}me principe
cosmologique qui stipule que la vitesse relative des galaxies est
proportionnelle \`{a} la distance relative les s\'{e}parant. En
effet $\frac{dR}{dt}$ \'{e}tant proportionnel \`{a} $\frac{dr}{dt}
= v_r$ et $\rho$ \'{e}tant proportionnel \`{a} $\frac{1}{r^3}$ qui
est proportionnel \`{a} $\frac{1}{R^3}$, alors cette \'{e}quation
implique que $\frac{dR}{dt}$ d\'{e}cro\^{i}t en fonction de $R$
comme $\frac{\alpha}{\sqrt{R}}$ et par suite que $v_r$
d\'{e}cro\^{i}t comme $\frac{\beta}{\sqrt{r}}$.\\ 
De m\^{e}me, les
relations (19.66) et (19.67) de [2] i.e.
$$ \frac{dR}{dt} = \sqrt{\frac{8\pi G\rho_c}{3c^2}}
R^{-\frac{1}{2}}$$o\`{u} $\rho_c$ est la densit\'{e} critique
(19.65) et
$$ R = (\frac{3}{2})^ {\frac{2}{3}} (\frac{8\pi G
\rho_c}{3c^2})^{\frac{1}{3}} t^{\frac{2}{3}}$$ montre clairement
la d\'{e}faillance du mod\`{e}le d'Einstein - de Sitter -
Friedmann. En effet, ces relations impliquent
$$ \frac{dr}{dt} \propto \frac{dR}{dt} \propto
\frac{1}{R^{\frac{1}{2}}} \propto \frac{1}{t^{\frac{1}{3}}}$$qui
montre que $r$ cro\^{i}t avec le temps tandis que $v_r =
\frac{dr}{dt}$ d\'{e}cro\^{i}t.\\\\
Ainsi, l'\'{e}quation (27) s'\'{e}crit:
$$\frac{1}{t_0^2} = \frac{ 8 \pi G}{3} \frac{E}{\frac{4\pi c^3
t^3}{3}} \times \frac{R^2}{c^2} - K$$
$$\hskip 1.5cm = \frac{2 G E}{c^5 t^3}R^2 - K = \frac{2 G E}{c^5
t\hskip 0.1cm t_0^2} - K.$$On en d\'{e}duit
$$ 1 = \frac{2 G E}{c^5 t} - K t_0^2$$
D'o\`{u}
$$ \frac{2G E}{c^5 \hskip0.1cm t} = 1 + K t_0^2$$
et
$$\hskip 3cm E = \frac{c^5 t (1 + Kt_0^2)}{2 G} \equiv \frac{c^5 (t_0 + K_0
t_0^3)}{2 G_0} = \frac{c^5 C_0}{2 G_0} \hskip 2.4cm (28) $$o\`{u}
$G_0$ est la constante gravitationnelle calcul\'{e}e \`{a}
l'instant $t_0$ et
$$ \hskip 3cm C_0 := t_0 + K_0 t_0^3 = \frac{2G_0
E}{c^5}. \hskip 3cm (29)$$ De plus on peut \'{e}crire dans un
intervalle de temps significatif autour de $t_0$
$$t (1 + K t_0^2) \equiv C_0$$
En d\'{e}rivant l'expression \emph{t}(1 + \emph{K}$t_0^2$), on
obtient
$$1 + K t_0^2 + t_0^2 t K^{'} = 0$$
D'o\`{u}
$$(K + K^{'} t)t_0^2 = -1$$
ou
$$K + K^{'} t = -\frac{1}{t_0^2}\hskip 0.1cm,$$
ce qui donne
$$ \;\;\;(K t)^{'} = -\frac{1}{t_0^2}$$
et
$$ \;\;\;\;\;\;\;\;\;\;\;\; K t = -\frac{t}{t_0^2} + b$$
avec
$$ \;\;\;\;\;\;\;\;\;\;\;\;\;\;\;\;\; b = K_0 t_0 + \frac{1}{t_0}.$$
On a donc
$$ \;\;\;\;\;\;\;\;\;\;\;\;\;\;\;\;\;\;\; K t = -\frac{t}{t_0^2} + K_0 t_0 + \frac{1}{t_0}$$
et
$$\hskip 5cm \;\;\;\;\;\;\;\;\; K = -\frac{1}{t_0^2} + \frac{1+K_0 t_0^2}{t_0 t}. \hskip
4cm (30)$$

\subsection*{L'\'{e}quation de densit\'{e}}

 \'{E}crivons maintenant la relation
$$\rho(t) = \rho_m(t) + \rho_r(t)$$
o\`{u} $\rho_m$(\emph{t}) d\'{e}signe la densit\'{e} moyenne de
l'\'{e}nergie de masse et $\rho_r$(\emph{t}) celle de
l'\'{e}nergie radiationnelle. Or, notre mod\`{e}le implique
$$ \rho(t) = \frac{E}{vol (B(O,ct))} = \frac{E}{\frac{4}{3} \pi
(ct)^3} = \frac{3E}{4 \pi c^3 t^3}$$et la Cosmologie homog\`{e}ne
conduit \`{a} l'expression
$$ \rho_m(t) = \frac{c^2}{6 \pi G t^2}$$
(c.f. [4], Table 5.1 avec $C=\frac{8\pi\rho_m a^3}{3c^2}$ (p. 101)).\\
Cette valeur pourrait \^{e}tre consid\'{e}r\'{e}e comme \'{e}tant
une valeur approximative pour $\rho_m(t)$ dans le cadre de notre
mod\`{e}le.\\
L'\'{e}quation de la densit\'{e} s'\'{e}crit donc, pour $t \gg1$ et pour une constante donn\'ee $a_0$:
$$\frac{3 E}{4 \pi c^3 t^3} = \frac{c^2}{6 \pi G t^2} +
\frac{a_0}{t^3}$$ qui s'\'{e}crit
$$ \frac {3 E}{4 \pi c^3} = \frac{c^2}{6 \pi G} t + a_0$$
avec
$$ a_0= -\frac{c^2}{6 \pi G} t + \frac{3 E}{4 \pi c^3}.$$
Alors, pour $t\gg 1,$ on a
$$\hspace{1.5cm}\rho_r (t) = \frac{a_0}{t^3} = -\frac{c^2}{6 \pi G} \frac{1}{t^2} +\frac{3E}{4\pi c^3}\frac{1}{t^3}= -
\rho_m(t)+\frac{3E}{4\pi c^3}\frac{1}{t^3}.\hspace{2.5cm}( \rho)$$
R\'{e}\'{e}crivons maintenant l'\'{e}galit\'{e} des densit\'{e}s
($\rho$) pour \emph{t} = $t_0 \gg$ 1, on obtient
$$ \frac{3 E}{4 \pi c^3 t_0^3} = \frac{c^2}{6 \pi G_0 t_0^2} +
\frac{a_0}{t_0^3}$$ ce qui donne
$$ \frac{3 c^5 (t_0 + K_0 t_0^3)}{4 \pi c^3 t_0^3 \times 2 G_0} =
\frac{c^2}{6 \pi G_0 t_0^2} + \frac{a_0}{t_0^3}$$ ou
$$\hskip 4cm \frac{3 c^2}{8 \pi G_0} (K_0 + \frac{1}{t_0^2}) = \frac{c^2}{6
\pi G_0 t_0^2} + \frac{a_0}{t_0^3} \hskip 3.1cm (31) $$ Posons
ensuite
$$ K_0 t_0^3 + t_0 = \frac{4}{9} t_0 + b_0 \hskip 0.1cm ,$$
la relation (31) s'\'{e}crit alors
$$\frac{3 c^2}{8 \pi G_0} (\frac{4}{9 t_0^2} + \frac{b_0}{t_0^3})
= \frac{c^2}{6 \pi G_0 t_0^2} + \frac{a_0}{t_0^3}.$$D'o\`{u}
$$\frac{3 c^2}{8 \pi G_0} \frac{b_0}{t_0^3} = \frac{a_0}{t_0^3} \hskip 0.1cm ,$$
ce qui implique
$$ b_0 = \frac{ 8 \pi G_0 a_0}{3 c^2} \hskip 0.1cm ,$$
$$ C_0 = K_0 t_0^3 + t_0 = \frac{4}{9} t_0 + \frac{8 \pi G_0
a_0}{3 c^2}$$ et
$$ \hskip 4cm K_0 = -\frac{5}{9 t_0^2} + \frac{8 \pi G_0 a_0}{3 c^2 t_0^3} \hskip 5.4cm (32) $$
De m\^{e}me, on a (en utilisant $(28)$)
$$ \hskip 4cm \frac{2 G_0 E}{c^5} = C_0 = \frac{4}{9} t_0 + \frac{8 \pi G_0
a_0}{3 c^2} \hskip 4cm (33) $$et par suite, on retrouve la
relation (d\'{e}j\`{a} \'{e}tablie)
$$ a_0 = (\frac{2 G_0 E}{c^5} - \frac{4}{9} t_0) \times \frac{3
c^2}{8 \pi G_0}$$ou
$$ \hskip 4.5cm a_0 = \frac{3 E}{4 \pi c^3} - \frac{c^2 t_0}{6 \pi
G_0}. \hskip 5.3cm (34)$$
\\
\subsection*{\'{E}nergie, \^{a}ge et \'{e}tendue de l'Univers}

En prenant pour $t_0$ l'instant pr\'{e}sent, on obtient
$$\rho_r (t_0) = \frac{a_0}{t_0^3} = \frac{3 E}{4 \pi c^3 t_0^3} -
\frac{c^2}{6 \pi G_0 t_0^2}$$
$$ \hskip 4cm = -0.4 \times {10}^6 eV/m^3 = - 6.4 \times
{10}^{-14} J/m^3,$$comme il r\'{e}sulterait de la loi de Stefan -
Boltzmann et de l'estimation g\'{e}n\'{e}ralement admise de la
contribution des neutrinos \`{a} la densit\'{e} de l'Univers \`{a}
l'instant pr\'{e}sent.\\\\
La deuxi\`{e}me \'{e}quation de Friedmann - Einstein
$$ \frac{\frac{d^2R}{dt^2}}{R} = - \frac{4 \pi G}{3 c^2} (\rho +
3P)$$s'\'{e}crit, tout simplement, dans le cadre de notre
mod\`{e}le
$$ \rho + 3 P = 0$$
o\`{u} \emph{P} est la pression moyenne et $\rho$ est la
densit\'{e} moyenne de l'\'{e}nergie dans l'Univers. Or
$$ \rho = \rho_m + \rho_r = \rho_m + 3 P $$et par suite, on a
$$ \rho_m + \rho_r + \rho_r = 0.$$D'o\`{u}
$$\rho_m = -2 \rho_r \hskip 1cm \mbox{ et } \hskip 1cm \rho =
-\rho_r,$$ce qui est conforme avec l'\'{e}quation ($\rho$) qui se r\'eduit, apr\`es substitution, \`a $\rho=\frac{3E}{4\pi c^3t^3}=-\rho_r$.\\
Ceci montre clairement que, lors de l'\'{e}tude de la
relativit\'{e} g\'{e}n\'{e}rale, il faut consid\'{e}rer la
densit\'{e} radiationnelle moyenne et la pression moyenne au sein
de l'Univers comme \'{e}tant n\'{e}gatives; ce qui est normal
puisque leur effet est antigravitationnel. \\Ainsi, on a
$$ \rho_0 = -\frac{a_0}{t_0^3} = 6.4 \times {10}^{-14} J/m^3.$$
Par cons\'{e}quent la relation (33) implique
$$ E = \frac{c^5}{2G_0} (\frac{4}{9} t_0 - \frac{8 \pi G_0
t_0^3}{3 c^2} \times 6.4 \times {10}^{-14})$$et
$$ \frac {E}{\rho_0} = \frac{4 \pi c^3 t_0^3}{3} = \frac{c^5}{2
G_0 \times 6.4 \times {10}^{-14}} (\frac{4}{9} t_0 - \frac{8 \pi
G_0 t_0^3 \times 6.4 \times{10}^{-14}}{3 c^2})$$ D'o\`{u}
$$ \frac{4 \pi c^3 t_0^3}{3} = \frac{2 c^5 t_0}{9 G_0 \times 6.4
\times {10}^{-14}} - \frac{8 \pi G_0 c^3 t_0^3}{6 G_0} \hskip
0.1cm ,$$ce qui donne
$$ \frac {8 \pi}{3} t_0^2 = \frac{2 c^2 \times {10}^{14}}{9 G_0
\times 6.4}$$et
$$t_0^2 = \frac{3}{8 \pi} \times \frac{2 c^2 \times
{10}^{14}}{9 G_0 \times 6.4}\;\;\;\;\;\;\;\;\;\; $$
$$ \hskip 1.2cm = \frac{3 \times 2 \times 9 \times {10}^{16} \times
{10}^{14}}{8 \pi \times 9 \times 6.67 \times {10}^{-11} \times
6.4} $$
$$ \simeq 5.595 \times {10}^{38}\;\;\;\;\;\;\;\;\;\;\;\;\;$$et par suite
$$ t_0 \simeq 2.365 \times {10}^{19}. $$Ce r\'{e}sultat aurait pu
\^{e}tre obtenu directement en utilisant uniquement la
deuxi\`{e}me \'{e}quation de Friedmann. Celle - ci donne en effet
$$ \rho_m + \rho_r + \rho_r = 0$$
qui conduit \`{a}
$$ \rho_m = -2 \rho_r$$
ou
$$ \rho_m = \frac{c^2}{6 \pi G_0 t_0^2} = 12.8 \times {10}^{-14}
J/m^3$$et par suite
$$ t_0^2 = \frac {c^2}{6 \pi G \times 12.8 \times {10}^{-14}}\;\;\;\;\;\;$$
$$ \hskip 4cm = \frac{9 \times {10}^{16} \times {10}^{14}}{6
\times 3.14 \times 6.67 \times {10}^{-11} \times
12.8}\;\;\;\;\;\;\;\;\;\;\;\;\;\;\;\;\;\;\;\;\;$$
$$ \simeq
5.595 \times {10}^{38}\;\;\;\;\;\;\;\;\;\;\;\;\;\;\;$$ et
$$t_0 \simeq 2.365 \times {10}^{19} s$$
Ainsi le rayon de l'Univers est actuellement
$$ r_0 = c t_0 \simeq 7.1 \times {10}^{27} m.$$
L'\'{e}nergie totale de l'Univers est
\begin{eqnarray*}
 E &=& \frac{4 \pi c^3 t_0^3 }{3} \rho_0 = \frac{2 c^5 t_0}{9 G_0}
- \frac{4 \pi t_0^3 c^3}{3} \times 6.4 \times {10}^{-14}\\
&=& 9.57 \times {10}^{70} J ( = 19.147 \times {10}^{70} - 9.57
\times {10}^{70} )J.
\end{eqnarray*}
La valeur obtenue ci-dessus pour $\rho_m$ est conforme avec la
valeur de $\rho_m$ obtenue \`a partir de l'\'equation de densit\'e
$(\rho)$ qui donne (en rempla\c{c}ant $\rho_r(t_0)$ par $-\frac{\rho_m(t_0)}{2}$)
$$\rho_m(t_0)=\frac{3E}{2\pi c^3}\frac{1}{t_0^3}.$$
 Le param\`{e}tre de Hubble est
$$ H_0 = 4.228 \times {10}^{-20}$$
La densit\'{e} radiationnelle est actuellement
$$ \rho_r = - 0.4 \times {10}^6 eV/m^3$$
La densit\'{e} actuelle de l'\'{e}nergie mat\'{e}rielle est
$$ \rho_m = 0.8 \times {10}^6 eV/m^3$$
La masse mat\'{e}rielle (incluant les trous noirs) est
$$ M = \frac{\rho_m}{c^2} \times \frac{4 \pi c^3 t_0^3}{3} =
\rho_m \times \frac{4 \pi c t_0^3}{3}$$
$$  = 2.126 \times {10}^{54} Kg.\;\;\;\;\;\;\;\;\;\;$$La masse
\'{e}quivalente \`{a} l'\'{e}nergie totale est
$$ M_e := \frac{E}{c^2} = 1.063 \times {10}^{54} Kg$$


\subsection*{Champ de vision d'un observateur}
Soit $I$ un observateur (\`a l'instant $t=t_0$) de l'Univers $B(O,t_0)$ situ\'e \`a une distance $d$ de $O$ (fig. $14$). Cet observateur re\c{c}oit \`a l'instant $t_0$ tous les signaux \'emis \`a l'instant $t_0-s$ par les objets situ\'es \`a l'intersection de la sph\`ere $S(I,s)$ et de l'Univers \`a l'instant $t_0-s$, i.e. sur $S(I,s)\cap B(O,t_0-s)$. Le champ de vision de l'observateur $I$ est donc
$$V=\bigcup_{0<s\leq\frac{d+t_0}{2}}\left(S(I,s)\cap B(O,t_0-s)\right)$$
qui n'est autre que
$$B\left(I,\frac{t_0-d}{2}\right)\bigcup\left(\bigcup_{\frac{t_0-d}{2}\leq s\leq\frac{t_0+d}{2}}\left(S(I,s)\cap B(O,t_0-s)\right)\right)$$
comme il est indiqu\'e sur la figure $14$.

\subsection*{Comparaison avec le mod\`{e}le d'Einstein - de
Sitter} 
Conform\'{e}ment au mod\`{e}le standard de la Cosmologie
homog\`{e}ne isotropique (avec \emph{K} = 0), on a (c.f.[2], p.555
- 557):
$$ R(t) \propto t^{\frac{2}{3}} \hskip 3cm H(t) \propto
\frac{2}{3t}$$
$$ \hskip 1cm \lambda \propto \frac{1}{T} \propto R$$et par suite,
on obtient (en utilisant le r\'{e}sultat bien confirm\'{e} de
Stefan - Boltzmann)
$$ \rho_r(t) \propto T^4 \propto \frac{1}{R^4} \propto
\frac{1}{t^{\frac{8}{3}}}$$
$$ \rho_m(t) \propto T^3 \propto \frac{1}{R^3} \propto
\frac{1}{t^2}$$
ce qui est contradictoire au fait que $\rho_r+\rho_m=\rho$ est proportionnel \`a $\frac{1}{t^3}$ .\\
Par contre, notre mod\`{e}le montre clairement la
propri\'{e}t\'{e} fondamentale
$$ \rho \propto \frac{1}{R^3} \propto \frac{1}{t^3}$$qui est
conforme avec notre r\'{e}sultat
$$ \rho_r \propto \frac{1}{t^3},$$qui donne (jointe au r\'{e}sultat
de Stefan - Boltzmann)
$$ T^4 \propto \frac{1}{t^3} \hskip 2cm \mbox{ ou } \hskip 2cm T
\propto \frac{1}{t^{\frac{3}{4}}},$$
et, sur un petit intervalle de temps,
$$ \lambda \propto t \propto
\frac{1}{T^{\frac{4}{3}}}$$ 
et enfin
$$ \rho \propto \rho_r \propto \rho_m \propto T^4 \propto
\frac{1}{t^3}$$ qui est \'{e}videmment plus coh\'{e}rent.\\

Par ailleurs, le tenseur d'Einstein $T$ s'\'{e}crit, dans le cadre
de la cosmologie homog\`{e}ne isotropique, sous la forme ([4],
5.2.1)
$$ T_{ab} = \rho_m u_a u_b$$ o\`{u} $\rho_m$ est la densit\'{e}
moyenne de l'\'{e}nergie de masse. Or, dans le cadre de notre
mod\`{e}le, le champ de vecteur $u^a$ n'est autre que le champ de
vecteur $(\frac{\partial}{\partial t})^a$ et par suite on obtient
$$ T_{ab} = \rho_m dt^2  $$
D'autre part, l expression de la masse totale de l'Univers
\'{e}tablie dans le cadre de cette cosmologie est donn\'{e}e par
([4], 11.2.10)
$$ M = \frac{1}{4 \pi} \int_\Sigma R_{ab} n^a \xi^b dV = 2
\int_\Sigma(T_{ab} - \frac{1}{2} T g_{ab}) n^a \xi^b dV.$$ En
adaptant cette expression \`{a} notre mod\`{e}le, $\Sigma$ devient
$B(O,t), g_{ab}$ devient $h_t = dt^2 - g_t$ et $n^a$ s'identifie
\`{a} $(\frac{\partial}{\partial t})^a$ qui constitue
\'{e}galement une approximation raisonnable de $\xi^a$. D'o\`{u}
$T_{ab} n^a \xi^b = \rho_m, T = \rho_m$ et $g_{ab} n^a \xi^b = 1$
et par suite on obtient
$$ M = 2 \int_{B(O,t)} (\rho_m - \frac{1}{2}\rho_m) dV_t =
\int_{B(O,t)} \rho_m dV_t$$en accord avec notre d\'{e}finition de
la masse globale de l'Univers (incluant les trous noirs et la
masse de la mati\`{e}re invisible).\\\\

\subsection*{Comparaison avec la gravit\'{e} newtonienne}
En comparant notre \'{e}quation de la mati\`{e}re-\'{e}nergie
$$ \frac{\partial^2}{\partial t^2} E(t,X(t)) - \Delta E(t,X(t)) =
0$$et notre relation
$$ X^{''}(t) = \Gamma (t) = -\nabla^{g_e} E (t,X(t))$$avec les
\'{e}quations caract\'{e}risant la gravit\'{e} newtonienne (c.f.
[4],(4.4.17) et (4.4.21))
$$ \hskip 3cm \Delta \varphi = 4 \pi \rho_m \hskip 0.3cm \mbox{ (\'{E}quation de Poisson)}$$
$$ X^{''} = \Gamma = -\nabla^{g_e} \varphi ,$$on obtient (par
identification)
$$ \varphi(X(t)) = E(t,X(t))$$
$$ \Delta \varphi (X(t)) = \Delta E(t,X(t)) = 4\pi
\rho_m(X(t)).$$ 
Par cons\'{e}quent, on a
$$\frac{\partial^2E}{\partial t^2}(t,X(t)) = \Delta E(t,X(t)) = 4
\pi\rho_m(X(t)).$$ 
Or, la masse totale de l'Univers
(id\'{e}alis\'{e}) dans le cadre de la th\'{e}orie de la
gravit\'{e} newtonienne (r\'{e}adapt\'{e} \`{a} la relativit\'{e}
g\'{e}n\'{e}rale et \`{a} notre mod\`{e}le) est donn\'{e}e par
(c.f.[4],(11.2.2))
$$ M_N = \frac{1}{4 \pi} \int_{S(O,R)} \overrightarrow{\nabla
\varphi}.\overrightarrow{n} dS.$$ 
Ainsi, on a d'apr\`{e}s ce qui
pr\'{e}c\`{e}de
$$ M_N = \frac{1}{4 \pi} \int_{S(O,R)} \overrightarrow{\nabla E}.\overrightarrow{n}
dS$$qui est \'{e}gale (d'apr\`{e}s la loi de Gauss) \`{a}
$$ \frac{1}{4\pi} \int_{B(O,R)} \Delta E dX = \frac{1}{4\pi}
\int_{B(O,R)} 4\pi \rho_m\,dX = \int_{B(O,R)} \rho_m\,dX = M$$
o\`{u}
$\rho_m$ d\'{e}signe ici la densit\'{e} de la
mati\`{e}re de l'Univers \`{a} l'instant $t$ (en
prenant $R \simeq t$) et $M$ est la masse totale de la mati\`ere dans l'Univers d'apr\`{e}s notre mod\`{e}le.\\
Par ailleurs, en int\'{e}grant l'\'{e}galit\'{e} ci-dessus sur une
boule moyenne $B(I,r_0)$ ayant un volume \'egal \`a l'unit\'e de volume dans l'Univers \`{a} l'instant $t$, on
obtient
$$\int_{B(I,r_0)} \frac{\partial^2 E}{\partial t^2}(t,X_t)dX_t = \int_{B(I,r_0)} \Delta E(t,X_t) dX_t$$
$$ = 4 \pi \int_{B(I,r_0)} \rho_m (X_t)\,dX_t = 4 \pi
\rho_m(t) = 4\pi E_m(t)$$
o\`{u} $\rho_m(t) = E_m(t)$
d\'{e}signent ici la densit\'{e} moyenne de la mati\`ere (par
unit\'{e} de volume) de l'Univers \`{a} l'instant $t$. Ainsi, on a
$$ E^{''}(t) = 4\pi E_m(t) \hskip 1cm \mbox{ou} \hskip 1cm
\frac{1}{2}\rho_m^{''}(t) = 4 \pi \rho_m(t)$$
 et par
suite
$$ E_m(t) = \rho_m(t) = C e^{-2\sqrt{2\pi}t}$$\emph{C}
pouvant \^{e}tre d\'{e}termin\'{e}e \`{a} partir des valeurs
connues $t_0$ et $\rho_0$. En effet, la relation
$$E_m(t_0) = C e^{-2 \sqrt{2\pi}t_0} = \rho_0$$implique
$$C = \rho_0 e^{2 \sqrt{2\pi}t_0}\;\; ;$$ ce qui donne
$$\rho_m(t) = \rho_0 e^{2 \sqrt{2\pi}(t_0 - t)}.$$

\noindent{\bf Remarque:} L'expression ci-dessus montre que $\rho_m(t)$ ainsi que $\rho(t)$ tendent exponentiellement vers $0$ tandis qu'en fait $\rho(t)$ tend vers $0$ proportionnellement \`a $\frac{1}{t^3}$ lorsque $t$ tend vers $+\infty$. Ceci prouve que le potentiel gravitationnel newtonien, qui explique bien les lois de gravitation d'un corps statique isol\'e, n'explique pas exactement la gravitation macroscopique cosmique. Cette d\'eviation de la d\'ecroissance de $\rho(t)$ par rapport \`a sa d\'ecroissance exacte signifie un ralentissement r\'eel de l'expansion qui est d\^u, \`a notre avis, aux forces de liaisons gravitationnelles entre \'etoiles, galaxies et amas de galaxies au sein de l'Univers qui conduisent \`a l'existence des ``constantes" cosmologiques. Une autre raison pour cette d\'eviation est le fait qu'on souvent travaill\'e, le long de cette \'etude, avec le laplacien  euclidien $\Delta_{g_e}$ \`a la place de l'op\'erateur de Laplace-Beltrami $\Delta_{g_t}$. Les r\'esultats ainsi obtenus sont de bonnes approximations sur le plan cosmologique mais ne pourrons pas donner exactement le r\'eel comportement de $\rho(t)$.\\

\noindent Par ailleurs, il est bien connu que, malgr\'{e} le fait qu'au sein
de la th\'{e}orie newtonienne, la notion de densit\'{e} de
l'\'{e}nergie gravitationnelle est d\'{e}finie par $\rho_G = -
\frac{1}{8 \pi} | \nabla^{g_e} \varphi |^2$, on a bien du mal
\`{a} d\'{e}finir une chose pareille au sein de la th\'{e}orie de
la relativit\'{e} g\'{e}n\'{e}rale classique. Dans le cadre de
notre th\'{e}orie, la quantit\'{e} qui constitue une bonne
candidate \`{a} jouer le r\^{o}le de densit\'{e} de l\'{e}nergie
g\'{e}n\'{e}rale incluant la densit\'{e} de l'\'{e}nergie
gravitationnelle (\`{a} l'instant $t$) est
$$\rho _G = - \frac{1}{8 \pi} | \nabla^{g_t} E_t (X) |^2 = -
\frac{1}{8 \pi} | \nabla^{g_t} \varphi | ^2$$ et, qu'en toute
logique, on a
$$ \rho_G = - \rho_m = 2 \rho_r = - 2\rho = - \frac{\Delta
\varphi}{4 \pi}.$$ 
Rappelons que $g_t$ est ici notre m\'{e}trique physique qui refl\`{e}te toute la
consistance physique de l'Univers et qui est caract\'{e}ris\'{e}e
par notre tenseur global de la mati\`{e}re-\'{e}nergie $T^*_{ab}$
et que $\nabla^{g_t}$ est ici le gradient, par rapport \`{a}
$g_t$, de la distribution de la mati\`{e}re-\'{e}nergie $E_t(X)$
\`{a} l'instant $t$ sur $B(O,t)$.\\\\
\textbf{Remarque:} La force de gravit\'{e} globale est, dans le
cadre de la relativit\'{e} g\'{e}n\'{e}rale classique,
proportionnelle \`{a} $\rho + 3P = \rho + \rho_r$ qui est, dans le
cadre de notre mod\`{e}le, nulle; ce qui explique et confirme que
l'expansion de l'Univers est, \`{a} partir d'un certain temps,
uniforme et
permanent.\\\\
Ceci implique
$$ - \frac{1}{8 \pi} | \nabla^{g_t} \varphi|^2 = - \frac{1}{4 \pi}
\Delta \varphi$$ ou
$$ \Delta \varphi = \frac{1}{2} | \nabla^{g_t} \varphi|^2.$$
Cette relation implique que, sur une trajectoire quelconque
$X(t)$, on a
$$ \Delta \varphi(X(t)) = \frac{1}{2} | \nabla ^{g_t} \varphi
(X(t))|^2$$ ou
$$ \Delta_x E_t(X(t)) = \frac{1}{2} | \nabla ^{g_t} E_t
(X(t))|^2$$ ce qui est compatible avec notre mod\`{e}le puisque
lorsque $X(t)$ d\'{e}signe un mouvement libre (i.e. $X(t)$ est une
g\'{e}od\'{e}sique pour $g_t$) alors on a
$$ E_t(X(t)) = E(t,X(t)) = \mbox{cte}$$
et
$$ \nabla ^{g_t} E_t(X(t)) = F_{g_t}(X(t)) = \nabla^{g_t}_{X'(t)}
X'(t) = \widetilde{\Gamma}(t) = 0.$$

\subsection*{Remarques}
 Les d\'{e}viations des valeurs trouv\'{e}es ci
- dessus (pour $t_0, r_0, E$,...) vis \`{a} vis des valeurs
approximatives pr\'{e}vues par le mod\`{e}le standard d'Einstein -
de Sitter, par exemple, peuvent \^{e}tre expliqu\'{e}es
essentiellement \`{a} l'aide des deux facteurs suivants:\\\\
$1^\circ$) La valeur g\'{e}n\'{e}ralement admise du param\`{e}tre
de Hubble $H_0$ est incorrect pour deux raisons:\\
La premi\`{e}re est l'utilisation de quelques notions
relativistes, dont on a d\'{e}j\`{a} \'{e}tabli la d\'{e}faillance
dans un article (pas encore publi\'{e}) ant\'{e}rieur (cf [1]), du
temps propre et des formules relativistes utilis\'{e}es pour
d\'{e}terminer le
param\`{e}tre du redshift $z$.\\
La seconde raison est le fait de se baser sur des mesures
effectu\'{e}es moyennant des galaxies visibles dont le
positionnement, les distances mutuelles et les vitesses relatives
sont loins de repr\'{e}senter fid\`{e}lement le param\`{e}tre
d'expansion. Nous ne sommes pas au centre de l'Univers et il y a
des galaxies qui se trouvent et se d\'{e}placent en dehors de
notre horizon. De plus, l'espace de l'Univers s'\'{e}tant bien au del\`{a} de toutes les galaxies et les agglom\'{e}rations mat\'{e}\\
L'\'{e}norme diff\'{e}rence entre les deux estimations de
l'\'{e}nergie moyenne des densit\'{e}s de masse est d\^{u}e \`{a}
la grande diff\'{e}rence entre les deux estimations
de la taille de l'Univers.\\\\
$2^\circ$) Le param\`{e}tre de courbure utilis\'{e}
g\'{e}n\'{e}ralement dans la premi\`{e}re \'{e}quation de
Friedmann est suppos\'{e} constant. Or ceci est absurde puisque la
notion fondamentale de courbure (qui, rappelons le, refl\`{e}te et
caract\'{e}rise la distribution de la mati\`{e}re - \'{e}nergie
qui est l'essence m\^{e}me de l'Univers r\'{e}el) est
essentiellement dynamique et \'{e}volutive (localement et
globalement) puisque, selon la th\'{e}orie bien confirm\'{e}e de
l'expansion, l'Univers n'est pas r\'{e}duit ni \`{a}
$\mathbb{R}^3$ ni \`{a} un domaine fixe de $\mathbb{R}^3$ mais
plut\^{o}t, conform\'{e}ment \`{a} notre mod\`{e}le, \`{a} une
boule \emph{B}(\emph{O},\emph{t}) toujours en expansion.\\
C'est pour cette raison qu'on a commenc\'{e} par utiliser la
premi\`{e}re \'{e}quation de Friedmann afin de montrer la
n\'{e}cessit\'{e} de l'utilisation d'un param\`{e}tre de courbure
(macroscopique et global) d\'{e}pendant du temps caract\'{e}risant
ainsi une m\'{e}trique (\`{a} la fois locale et globale)
d\'{e}pendante elle m\^{e}me du temps, bien que la
d\'{e}termination de $t_0$ n\'{e}cessite uniquement l'utilisation
de la seconde \'{e}quation
de Friedmann.\\\\
$3^\circ$) Remarquons aussi que la diff\'{e}rence de signe entre
les densit\'{e}s $\rho_m$ et $\rho_r$ est fondamentalement d\^{u}e
au fait que la premi\`{e}re est associ\'{e}e \`{a} la force
attractive de la gravit\'{e} tandis que la seconde est
associ\'{e}e \`{a} la pression qui engendre une force ayant une
nature essentiellement oppos\'{e}e. Ce sont les deux forces
fondamentales de la Nature, \`{a} savoir la force de gravitation
et la force de la pression radiationnelle
\'{e}lectromagn\'{e}tique. Ainsi, la relation
$$\rho + 3P = \rho_m + \rho_r +3P = 0$$ajoute une dimension
nouvelle au probl\`{e}me de la conservation de l'\'{e}nergie
cosmique. En effet, si on attribue \`{a} l'expression $3P$ le
qualificatif (de densit\'{e} moyenne) d'\'{e}nergie n\'{e}gative
et \`{a} l'expression $\rho + 3P$ le qualificatif (de densit\'{e}
moyenne) d'\'{e}nergie
g\'{e}n\'{e}ralis\'{e}e globale, on peut \'{e}noncer\\

L'\'{e}nergie g\'{e}n\'{e}ralis\'{e}e globale de l'Univers est
\'{e}ternellement nulle (c.f. [1]).\\\\
Ce principe rappelle le principe de la conservation du "momentum"
qui, appliqu\'{e} \`{a} l'Univers tout
entier (voir paragraphe 6), s'\'{e}nonce\\

Le "momentum" global de l'Univers est
\'{e}ternellement nul.\\\\
N\'{e}anmoins, remarquons que, dans notre mod\`{e}le, on ne peut
parler de pression et d'\'{e}nergie n\'{e}gative et par suite
d'\'{e}nergie g\'{e}n\'{e}ralis\'{e}e globale nulle (dans le cas
o\`{u} l'Univers \'{e}tait r\'{e}duit originellement \`{a} une
\'{e}nergie $E_0$ concentr\'{e}e en un point) qu'apr\`{e}s le Big
Bang. Par cons\'{e}quent, notre mod\`{e}le ne cautionne pas les
th\'{e}ories qui pr\'{e}tendent que l'Univers est sorti de
litt\'{e}ralement rien.\\Signalons aussi que le sens du terme
d'\'{e}nergie n\'{e}gative utilis\'{e} ici est compl\`{e}tement
diff\'{e}rent de celui qu'on lui attribue au sein de la
th\'{e}orie d'inflation et de la "gravit\'{e} n\'{e}gative".
\\\\
\section{Constantes fondamentales de la Physique\\ moderne}
\subsection*{La constante de la Statistique quantique}

On va \'{e}tablir ici quelques relations impliquant plusieurs
constantes fondamentales de la physique et montrant qu'un grand
nombre parmi elles d\'{e}pendent du temps (et de la
temp\'{e}rature). Ces relations conduisent \`{a} l'unification des
forces fondamentales ainsi qu'\`{a} l'unification de toutes les
branches de la physique: Relativit\'{e} g\'{e}n\'{e}rale i.e.
Cosmologie, Th\'{e}orie quantique, \'{E}lectr-omagn\'{e}tisme,
Thermodynamique et la M\'{e}canique de Newton - Lagrange -
Hamilton.\\

Reprenons la relation (33), i.e.
$$ E = \frac{c^5 C_0}{2 G_0}$$
o\`{u}
$$ C_0 = K_0 t_0^3 + t_0 = \frac{4}{9} t_0 + \frac{8 \pi G_0
a_0}{3 c^2}.$$ Ceci implique
$$ G_0 = \frac{ c^5 C_0}{2E} = \frac{c^5}{2E}(\frac{4}{9} t_0 +
\frac{8 \pi G_0 a_0}{3 c^2})$$
$$ = \frac{2}{9} \frac{c^5 t_0}{E} + \frac{4 \pi}{3}
c^3 \frac{G_0 a_0}{E}\;\;\;\;\;\;\;$$
$$ = \frac{2}{9} \frac{c^5 t_0}{E} + \frac{4 \pi}{3}
c^3 \frac{G_0 t_0^3}{E} \frac{a_0}{t_0^3}\;\;\;\;\;$$
$$ = \frac{2}{9} \frac{c^5 t_0}{E} - \frac{4 \pi c^3
t_0^3}{3} G_0 \frac{\rho_0}{E}\;\;\;\;\;$$
$$ = \frac{2}{9} \frac{c^5 t_0}{E} -G_0.\;\;\;\;\;\;\;\;\;\;\;\;\;\;\;\;\;$$
D'o\`{u}
$$ 2G_0 = \frac{2}{9} \frac{c^5 t_0}{E}$$et
$$ G_0 = \frac{c^5 t_0}{9 E} ( = \frac{2.365 \times {10}^{19}
\times243 \times {10}^{40}}{9 \times 9.57 \times {10}^{70}} = 6.67
\times {10}^{-11})$$Ceci \'{e}tant pour $t_0 \gg $ 1 arbitraire,
on obtient, pour $t \gg$ 1:
$$\;\;\;\;\;\;\;\;\;\;\;\;\hskip 1cm G = \frac{c^5 t}{9 E} \hskip 1cm \mbox { et } \hskip 1cm E =
\frac{c^5 t}{9 G}
\;\;\;\;\;\;\;\;\;\;\;\;\;\;\;\;\;\;\;\;\;\;\;\hskip 1cm
(35)$$D'autre part, on a
$$ \rho_m = -2 \rho_r = 2 \rho = \frac{6 E}{4 \pi c^3 t^3}\hskip 0.1cm,$$
ce qui implique, en se basant sur la statistique quantique (c.f.
[4] p.108),
$$ \frac{6 E}{4 \pi c^3 t^3} = \sum_{i = 1} ^{n} \alpha_i g_i
\frac{\pi^2 (K_B \hskip 0.05cm T)^4}{30 \overline{h}^3 \hskip
0.05cm
c^5}$$o\`{u} la constante de Boltzmann est not\'{e}e ici $K_B$.\\
D'o\`{u}
$$  \frac{3 E}{2 \pi c^3 t^3} = \sum_{i = 1}^{n} \alpha_i g_i
\frac{\pi^2}{30 c^5} \frac{(K_B \hskip 0.05cm
T)^4}{\overline{h}^3}$$Ainsi, on a
$$ \frac{\frac{3 c^5 t}{9 G}}{2 \pi c^3 t^3} = \frac{c^2}{6 \pi G
t^2} = (\sum_{i = 1}^{n} \alpha_i g_i \frac{\pi^2}{30 c^5})
\frac{(K_B \hskip 0.05cm T)^4}{\overline{h}^3}$$qui s'\'{e}crit
$$ \frac{5 c^7}{\pi G t^2} = (\sum_{i = 1}^{n} \alpha_i g_i \pi^2)
\frac{(K_B \hskip 0.05cmT)^4}{\overline{h}^3}$$ou
$$ \frac{1}{G t^2} = (\sum_{i = 1}^{n} \frac{\alpha_i g_i \pi^3}{5
c^7}) \frac{(K_B \hskip 0.05cmT)^4}{\overline{h}^3}$$
$$ =: A \frac{(K_B\hskip 0.05cm
T)^4}{\overline{h}^3} ; \;\;\;\;\;\;\;\;$$ce qui donne
$$ \hskip 3cm G = \frac{1}{A} \frac{\overline{h}^3}{(K_B \hskip 0.05cmT)^4}
\frac{1}{t^2} \hskip 5.1cm (36)$$o\`{u}
$$ A = \frac{1}{G} \frac{\overline{h}^3}{(K_B \hskip0.05cm T)^4}
\frac{1}{t^2}$$peut \^{e}tre calcul\'{e} \`{a} partir des valeurs
actuelles des constantes impliqu\'{e}es correspondant \`{a} $t_0$
= 2.365 $\times {10}^{19}$.\\

Ainsi, on a (d'apr\`{e}s la relation (35)):
$$ \frac{c^5 \hskip0.05cm t}{9 E} = \frac{1}{A} \frac{\overline{h}^3}{(K_B \hskip0.05cm
T)^4} \frac{1}{t^2}; $$ce qui donne
$$ \hskip 3cm  \frac{\overline{h}^3}{(K_B \hskip 0.05cmT)^4} = \frac{A c^5}{9 E} t^3 = A G
t^2 ,\hskip 4.2cm (37) $$ Par ailleurs, en d\'{e}signant par
$\alpha$ le facteur classique de la force
\'{e}lectromagn\'{e}tique et par $K_E$ la constante
\'{e}lectromagn\'{e}tique d\'{e}sign\'{e}e usuelllement par
\emph{k}, on obtient
$$ \alpha = \frac{K_E \hskip 0.05cm e^2}{\overline{h}\hskip0.05cm c} = \frac{K_E \hskip 0.05cm e^2}{c}
\times(\frac{9 E}{(K_B \hskip 0.05cm T)^4 A c^5
t^3})^{\frac{1}{3}}$$
$$ \hskip 2.7cm = K_E \hskip 0.05cm e^2 \hskip 0.1cm(\frac{9E}{Ac^8})^{\frac{1}{3}}\hskip 0.1cm (K_B \hskip 0.05cm
T)^{-\frac{4}{3}}\hskip 0.1cm t^{-1}\;\;\;\;\;\;\;\;\;\hskip 3cm
(38)$$et
$$ \alpha = \frac{K_E \hskip 0.05cm e^2}{c} ((K_B \hskip 0.05cm T)^4\hskip 0.1cm A G t^2)^{-\frac{1}{3}}\;\;\;\;\;\;\;\;\;\;\;\;\;\;\;\;$$
$$ \hskip 2.7 cm = \frac{K_E \hskip 0.05cm e^2}{c A^{\frac{1}{3}}} \hskip 0.1cm (K_B \hskip 0.05cm
T)^{-\frac{4}{3}} \hskip 0.1cm G^{-\frac{1}{3}} \hskip 0.1cm
t^{-\frac{2}{3}} .\;\;\;\;\;\;\;\;\;\;\;\;\hskip 3cm (39)$$

\subsection*{Relations fondamentales}

\`{A} partir de ces relations, on peut obtenir un certain nombre
de relations pr\'{e}cisant la d\'{e}pendance des constantes
fondamentales les unes en fonction des autres et en fonction du
temps.\\
En effet la relation (36) implique
$$\hskip 2cm  \frac{G (K_B T)^4}{\overline{h}^3} = \frac{1}{A}\hskip 0.1cm \frac{1}{t^2} =
\frac{C_1}{t^2} \hskip 2 cm \mbox{ avec } \hskip 0.5cm C_1 =
\frac{1}{A} \hskip 1.5cm (40)$$ La relation (37) implique
$$\hskip 1cm \frac{\overline{h}^3}{(K_B T)^4} = \frac{Ac^5}{9E}
t^3 = C_2 t^3 \hskip 1.5cm \mbox{ avec } \hskip 0.5cm C_2 =
\frac{Ac^5}{9E}$$ou
$$ \hskip 4cm \frac{\overline{h}}{(K_B T)^{\frac{4}{3}}} =
C_2^{\frac{1}{3}} t \hskip 6cm (41)$$La relation (38) implique
$$\alpha t = k e^2 (\frac{9E}{Ac^8})^{\frac{1}{3}} (K_B
T)^{-\frac{4}{3}}\;\;\;\;\;\;\;\;\;\;\;\;\;\;\;\;\;\;\;\;\;\;\;\;\;\;\;\;\;\;\;$$
$$ \hskip 3cm = \frac{C_3}{(K_B T)^{\frac{4}{3}}}\hskip 2.7cm \mbox{
avec } \hskip 0.8cm C_3 = k e^2 (\frac{9E}{Ac^8})^{\frac{1}{3}}$$
ou
$$\hskip 4cm \alpha (K_B T)^{\frac{4}{3}} = \frac{C_3}{t} \hskip 6cm
(42)$$Cette relation jointe \`{a} la relation
$$G = \frac{c^5}{9E} t =: C_0 t$$qui implique $t$ =
$\frac{G}{C_0}$ donne
$$ \alpha (K_B T)^{\frac{4}{3}} = \frac{C_3 C_0}{G}$$ et par suite
$$ \hskip 4cm \alpha G (K_B T)^{\frac{4}{3}} = C_3 C_0 =: C_5 \hskip 4cm
(43)$$avec
$$ C_5 = k e^2(\frac{9E}{Ac^8})^{\frac{1}{3}} \frac{c^5}{9E} = k e^2
(9E)^{\frac{1}{3}}\times(9E)^{-1}\times(Ac^8)^{-\frac{1}{3}}c^5$$
\\
$$ = \frac{k e^2
c^{\frac{7}{3}}}{A^{\frac{1}{3}}(9E)^{\frac{2}{3}}}\hskip 7.5cm
$$qui, jointe \`{a} (41) implique
$$ \alpha G \frac{\overline{h}}{C_2^{\frac{1}{3}} t} = C_5$$et par
cons\'{e}quent
$$ \hskip 3.6cm\alpha G \overline{h} = C_5 C_2^{\frac{1}{3}} t = C_6 t \hskip
4cm (44)$$avec
$$C_6 = \frac{k e^2 c^4}{9E}$$
(v\'{e}rification: $\alpha \overline{h}$ = $C_6 \frac{t}{G}$ =
$C_6 \frac{9E}{c^5}$ = $\frac{ke^2c^4}{9E}\frac{9E}{c^5}$ =
$\frac{ke^2}{c}$ qui est une constante absolue bien que $\alpha$ et $\overline{h}$ d\'ependent toutes les deux du temps $t$).\\\\
Finalement, les relations (40) et (44) impliquent respectivement
$$\alpha G \overline{h} \hskip0.2cm = \hskip 0.2cm \alpha \frac{\overline{h}^3}{A(K_B
T)^4\hskip 0.05cm t^2} \overline{h} \hskip 0.2cm = \hskip 0.2cm
\alpha (\frac{\overline{h}}{K_BT})^4 \frac{1}{At^2}$$ et
$$\alpha G \overline{h} = \frac{k e^2 c^4 t}{9E}$$
D'o\`{u}
$$E=\frac{k e^2 c^4 t}{9\alpha\overline{h}G}=\frac{k e^2 c^4 t}{9\frac{ke^2}{c}G}=\frac{c^5 t}{9G},$$
ce qui est conforme \`a la relation obtenue pr\'ec\'edemment en utilisant la densit\'e radiationnelle \'egale \`a l'oppos\'e de la densit\'e de la mati\`ere-\'energie:
$$\rho(t)=\int_{B(I,r_0)}E_t(X)\,dX.$$
De m\^eme
$$ \hskip 2cm\alpha (\frac{\overline{h}}{K_BT})^4 = \frac{A}{9E} k e^2 c^4
t^3 \hskip 4.2cm (45)$$
\\
(v\'{e}rification: $\alpha$ ($\frac{\overline{h}}{K_BT})^4$ =
$\frac{1}{G} \frac{\overline{h}^3}{(K_B T)^4} \frac{1}{t^2} \times
\frac{1}{9E} k e^2 c^4 t^3$ = $\frac{1}{9E} k e^2 c^4
\frac{\overline{h}^3}{G(K_BT)^4} t$\\\\
ce qui donne
$$\alpha
\overline{h} = \frac{1}{9E} k e^2 \frac{c^4 \hskip0.05cm t}{G} = k
e^2 \frac{c^4 \hskip 0.05cm t}{9E}
\frac{1}{\frac{c^5\hskip0.05cmt}{9E}}  = \frac{k e^2}{c}).$$
Ainsi, on a 
$$\alpha\overline{h} =\frac{ke^2}{c}= \frac{2,304\times 10^{-19}}{3\times 10^8}\simeq 7,68\times 10^{-28}.$$
Remarquons que les valeurs classiques de $\alpha$ et $\overline{h}$ donnent 
$$\alpha\overline{h} = \frac{1,06\times 10^{-34}}{137}\simeq 7,737\times 10^{-37}$$
ce qui est visiblement contradictoire.\\
Notons aussi que la relation $E_t(X)=h(t)f(t)$ implique 
$$\int_{B(I,r_0)}h(t)f(t)\, dX=\int_{B(I,r_0)}E_t(X)\, dX$$
qui donne
$$h(t)f(t)=\rho(t)$$
et
$$E=\int_{B(O,t)}E_t(X)\,dX=\int_{B(O,t)}h(t)f(t)\,dX=\int_{B(O,t)}\rho(t)\,dX.$$
Le fait que $\rho(t)\propto \frac{1}{t^3}$ et $h(t)\propto t$ implique que $f(t)\propto\frac{1}{t^4}\propto T^{\frac{16}{3}}$.\\
Si maintenant $\mu$ est la valeur propre moyenne sur la boule unit\'e du probl\`eme de Dirichlet classique (i.e. sur $B(O,1),g_e))$ associ\'e \`a l'\'equation de la mati\`ere-\'energie, alors la relation
$$f(t)\propto\frac{\rho(t)}{h(t)}\propto\frac{1}{t^4}\quad\mbox{implique}\quad f(t)^2\propto\frac{1}{t^8}$$ 
et
$$\mu=Cf(t)^2t^8\quad\mbox{$\Leftrightarrow$}\quad f(t)=\frac{1}{C}\frac{\sqrt{\mu}}{t^4}.$$ 
Donc la relation liant $\mu$ \`a $f_\mu(t)$ obtenue en supposant que 
$$e(\mu)=E_\mu(t,X(t))=h_\mu(t)f_\mu(t)=\rho(t)$$
est constant (i.e. $f(t)=\frac{1}{2\pi}\frac{\sqrt{\mu}}{t}$) n'est pas exacte \`a l'\'echelle cosmique. Ceci est d\^u, comme on l'a d\'ej\`a signal\'e \`a plusieurs reprises, au refroidissement moyen permanent de l'Univers. La relation  
$$\alpha\overline{h}=7.68\times 10^{-28}$$ 
implique
 $$\alpha h=2\pi\times 7.68\times 10^{-28}$$ 
qui est de l'ordre de $10^{-27}$ et par suite $h$ est, \`a  pr\'esent, de l'ordre de $10^{-25}$, ce qui implique
\begin{eqnarray*}
\mu&\simeq& C\frac{(6.4\times 10^{-14})^2}{(10^{-25})^2}\times(2.365\times 10^{19})^8\\
&\simeq& C\times 10^{180}.
\end{eqnarray*}
De m\^eme
$$h(t_0)\simeq bt_0\simeq 10^{-25}$$
implique que $b$ est de l'ordre de $10^{-44}$.\\

\noindent En r\'{e}sum\'{e}, on a les relations fondamentales suivantes
$$ T \propto \frac{1}{t^{\frac{3}{4}}}, \hskip 1cm \rho \propto T^4 \propto \frac{1}{t^3} \propto \frac{1}{V}
 \hskip 1cm \mbox{ et } \hskip 1cm T \propto
 \frac{1}{V^{\frac{1}{4}}}$$(o\`{u} \emph{V} est le volume de
 l'Univers \`{a} l'instant \emph{t})
$$ \alpha \hskip0.1cm \overline{h} = \frac{K_E\hskip0.05cm
e^2}{c}$$
$$ \frac{\overline{h}^3}{(K_B\hskip0.05cm T)^4} \propto t^3 \hskip 1cm
\mbox{ et } \hskip 1cm \overline{h} \propto (K_B \hskip0.05cm
T)^{\frac{4}{3}}\hskip0.05cm t \sim
K_B\hskip0.05cm^{\frac{4}{3}}$$

$$\;\;\; E = \frac{c^5 \hskip0.05cm t}{9G} \hskip1cm \mbox{ et }
\hskip1cm G = \frac{c^5}{9E} t \propto \frac{c^5}{9E}
\frac{1}{T^{\frac{4}{3}}}.$$
La d\'ependance effective de la longueur d'onde $\lambda$ en fonction de la temp\'erature cosmique $T$ peut \^etre exprim\'ee par:
$$\lambda\propto\frac{1}{f}\propto t^4\propto\left(\frac{1}{T^{\frac{4}{3}}}\right)^4=\frac{1}{T^{\frac{16}{3}}}\propto V^{\frac{4}{3}}.$$ 
Quant au param\`{e}tre de courbure
$K(t)$, on a d'apr\`{e}s les relations (32) et (34),
$$ K(t) = -\frac{5}{9t^2} + \frac{8 \pi G a}{3 c^2 t^3}\hskip 5cm$$
$$ = -\frac{5}{9t^2} + \frac{8 \pi}{3c^2 t^3}(-\frac{c^2}{6\pi}t +
3 \frac{EG}{4\pi c^3}) \hskip 1.2cm$$
$$ = -\frac{5}{9t^2} + \frac{8 \pi}{3c^2 t^3}(-\frac{c^2}{6\pi}t +
\frac{3}{4 \pi c^3} \frac{c^5}{9}t) \hskip 0.8 cm$$
$$ = -\frac{5}{9t^2} + \frac{8 \pi}{3c^2 t^3}(-\frac{c^2}{6\pi}t +
 \frac{c^2}{12 \pi}t) \hskip 1.3cm$$
$$ = -\frac{5}{9t^2} -\frac{2}{9t^2} = -\frac{7}{9t^2} \hskip 2.9cm$$
Enfin on a, d'une part
$$ \alpha = \frac{K_E e^2}{c \overline{h}} \propto \frac{K_E
e^2}{c K_B^{\frac{4}{3}} }$$et d'autre part
$$ \alpha = \frac{K_E\hskip0.05cm e^2}{c\hskip0.05cm\overline{h}}
= \frac{K_E\hskip0.05cm e^2}{c(K_B\hskip0.05cm
T)^{\frac{4}{3}}\hskip0.05cm t} = \frac{K_E\hskip0.05cm
e^2}{c(K_B\hskip0.05cm T)^{\frac{4}{3}}} \times \frac{c^5}{9GE} =
\frac{c^4 K_E\hskip0.05cm e^2}{9(K_B\hskip0.05cm
T)^{\frac{4}{3}}\hskip0.05cm GE}$$ou
$$ \alpha G \propto \frac{c^4 K_E\hskip0.05cm e^2}{9(K_B\hskip0.05cm
T)^{\frac{4}{3}}\hskip0.05cm E}.$$ Notons que, puisque
$\overline{h} \propto t$, la relation (36) implique que $K_B T$
est constante et il est facile de montrer que les relations
pr\'{e}c\'{e}dentes impliquent
$$\alpha \propto \frac{1}{t} \hskip 0.2cm , \hskip 0.2cm G \propto t \hskip 0.2cm \mbox{et}\hskip 0.2cm K_B
\propto t^{\frac{3}{4}}.$$ 
{\bf Remarque}: Etant donn\'e que, pour $T=300K$, on a $K_BT\simeq 0.02585$eV, on a actuellement $K_B\simeq\frac{0.02585}{300}\simeq 8.62\times 10^{-5}$ et par suite la constant $K_BT$, o\`u $T$ es la temp\'erature cosmique, est \'egale \`a $8.62\times 10^{-5}\times 2.74\simeq 2.361\times 10^{-4}$eV.\\

Notons enfin que, dans le cadre de notre mod\`{e}le, on a
$$R_{t_0}(t) = \frac{t}{t_0} \mbox{  et  } \frac{dR_{t_0}}{dt} =
\frac{1}{t_0}$$et l'\'{e}quation de Friedmann devient
$$ \frac{1}{t_0^2} = \frac{ 8 \pi \rho G R^2}{3 c^2} - K(t)$$ce
qui donne, pour $t = t_0$
$$\frac{1}{t_0^2} = \frac{8 \pi \rho_0 G_0}{3c^2} + \frac{7}{9t_0^2}
= \frac{8 \pi}{3c^2} \frac{3E}{4 \pi c^3 t_0^3} \frac{c^5 t_0}{9E}
+ \frac{7}{9t_0^2}$$
$$ = \frac{2}{9t_0^2} + \frac{7}{9t_0^2} \hskip 4.7cm$$
Ceci montre, de nouveau, la validit\'{e} de notre mod\`{e}le et la
n\'{e}cessit\'{e} de la d\'{e}pendence du temps pour les
constantes $K$ et $G$.\\

\subsection*{Remarques}
$1^\circ$) Les relations ci - dessus montrent que seules les
constantes fondamentales \emph{E}, \emph{A}, \emph{c}, $K_B T$ et
$K_E \hskip0.05cm e^2$ sont ind\'{e}pendantes du temps, les autres
constantes \emph{G}, $\overline{h}, K,$ $\alpha$ et la courbure
$K$ d\'{e}pendent du temps et sont reli\'{e}es entre elles
ainsi qu'aux autres quatre premi\`{e}res.\\

En prenant \emph{c} = 1 et en tenant compte des autres relations,
on r\'{e}alise que seuls \emph{E}, $K_E \hskip0.05cm e^2$, $K_B T$
et \emph{t} ont une existence intrins\`{e}que r\'{e}duisant ainsi
l'Univers \`{a} trois
\'{e}l\'{e}ments de base:\\

L'\'{e}nergie originelle, l'\'{e}lectromagn\'{e}tisme et le
temps.\\
Ce dernier facteur n'est autre que la distance et par suite
l'\'{e}tendue, d\'{e}terminant ainsi, avec le deuxi\`{e}me
facteur, l'expansion. L' Univers est donc essentiellement
l'\'{e}nergie
originelle en expansion.\\\\
$2^\circ$) En d\'{e}signant $K_B \hskip0.05cm T$ =  $\frac{2}{3}
\langle E_c \rangle$, o\`{u} $\langle E_c \rangle$ est
l'\'{e}nergie cin\'{e}tique g\'{e}n\'{e}ralis\'{e}e moyenne, par
$E_\ast$ et en rempla\c{c}ant la constante
\'{e}lectromagn\'{e}tique $K_E$ par \emph{k}, alors les relations
pr\'{e}c\'{e}dentes donnent lieu, pour \emph{c} = 1, aux relations
suivantes:
$$ \alpha \hskip0.05cm \overline{h} = k e^2$$

$$ \;\;\;\overline{h}  \propto E^{\frac{4}{3}}_\ast \hskip0.1cm t \hskip 1cm
\mbox{ et } \hskip1cm \alpha \hskip0.05cm E^{\frac{4}{3}}_\ast
\hskip0.1cm t  \propto k e^2$$

$$ G = \frac{t}{9E} \hskip 1cm \mbox{ et } \hskip 1cm \overline{h} = 9 E^{\frac{4}{3}}_\ast \hskip0.1cm EG$$

$$ \alpha = \frac{ke^2}{9 E^{\frac{4}{3}}_\ast \hskip0.1cm
E}\frac{1}{G} \hskip 1cm \mbox{ ou } \hskip 1cm \alpha G =
\frac{ke^2}{9E E^{\frac{4}{3}}_\ast}.$$Ces relations montrent,
entre autre, l'unit\'{e}
des forces fondamentales de la nature.\\\\
$3^\circ$) La relation $A = \frac{\overline{h}^3}{(K_BT)^4}
\frac{1}{t^2}$ montre la puissance ultime de la Statistique
quantique. Les autres relations montrent, d'une part, que la
quantisation tout azimut est non justifi\'{e}e (\emph{G} et
$\overline{h}$, par exemple, d\'{e}pendent contin\^{u}ment du
temps) et, d'autre part, que la quantisation issue d'une
exp\'{e}rimentation prolong\'{e}e et pr\'{e}cise trouve sa
l\'{e}gitimit\'{e} globale en s'unifiant avec la Physique et la
M\'{e}canique th\'{e}orique. Il s'agit ici de l'unification
retrouv\'{e}e de la Physique quantique avec la relativit\'{e}
g\'{e}n\'{e}rale toutes les deux unifi\'{e}es avec la
M\'{e}canique de Newton - Lagrange - Hamitlon. En particulier, on
a ainsi montr\'{e}
 que l'\'{e}volution de l'Univers est essentiellement d\'{e}crite par la th\'{e}orie de la relativit\'{e} g\'{e}n\'{e}rale
 (revis\'{e}e) d'Einstein.\\\\
$4^\circ$) Ces m\^{e}mes relations jointes aux r\'{e}sultats
\'{e}tablis au paragraphe 9, ainsi qu'aux autres paragraphes, ne
font qu'instituer les bases de l'unification de toutes les
branches de la Physique: L'\'{E}lectromagn\'{e}tisme, la
Relativi\'{e} g\'{e}n\'{e}rale (i.e. la Cosmologie), la
Thermodynamique, la Physique et la M\'{e}canique quantiques et la
Physique des particules avec la M\'{e}canique de Newton - Lagrange
- Hamilton.
\subsubsection*{Note}
Cet article a \'{e}t\'{e} \'{e}crit le long de la p\'{e}riode:
 Janvier 2007-Mars 2009. Plusieurs parties ont \'{e}t\'{e}
"publi\'{e}es" aux Arxiv de Math\'{e}matiques.\\

\section{Commentaires et probl\`{e}mes ouverts}
\subsection*{R\'eflexions autour de notre mod\`ele et quelques autres}

Notre mod\`{e}le physico - math\'{e}matique global permet de
donner des r\'{e}pon-\\ses \`{a} un certain nombre de
probl\`{e}mes ouverts ainsi que quelques clarifications et
pr\'{e}cisions les concernant. De m\^{e}me, il permet de
reformuler quelques autres probl\`{e}mes et donner de nouvelles
perspectives dans la
direction de leur r\'{e}solution.\\

Les points cruciaux qui ont conduit \`{a} ces deux
possibilit\'{e}s sont:\\

\noindent $1^\circ)$ La r\'{e}futation th\'{e}orique de certaines
interpr\'{e}tations du deuxi\`{e}me postulat de la relativit\'{e}
restreinte et le remplacement des formules relativistes
approximatives par des formules qui refl\`{e}tent fid\`{e}lement
les lois de la nature \`{a} une pr\'{e}cision qui pourrait
\^{e}tre am\'{e}lior\'{e}e et quantifi\'{e}e le plus correctement
possible (voir paragraphes $4$ et $7$). Ces modifications
devraient se g\'{e}n\'{e}raliser \`{a} tous les recoins de la Physique moderne.\\\\
2$^\circ$) La clarification logique et math\'{e}matique des
limites th\'{e}oriques et exp\'{e}ri-\\mentales de la th\'{e}orie
quantique et de la nature conjoncturelle du principe d'incertitude
de Heisenberg malgr\'{e} l'efficacit\'{e} pratique de ces outils
et surtout celle des \'{e}quations de Schr\"{o}dinger (paragraphes
$7$ et $8$). N\'{e}anmoins la th\'{e}orie quantique doit \^{e}tre
utilis\'{e}e pour r\'{e}soudre tous les probl\`{e}mes qui ne se
pr\^{e}tent pas \`{a} une mod\'{e}lisation-id\'{e}alisation
permettant de les r\'{e}soudre \`{a} l'aide de la M\'{e}canique et de la Physique classiques.\\\\
3$^\circ$) La g\'{e}om\'{e}trisation de l'Univers tridimensionnel
\`{a} l'aide de la m\'{e}trique physique $g_t$ et
l'\'{e}tablissement de l'\'{e}quation de la mati\`{e}re -
\'{e}nergie ($E^*$) ainsi que la g\'{e}om\'{e}trisation de
l'espace - temps quadrimensionnel \`{a} l'aide de la m\'{e}trique
d'\'{e}volution $h$ = $dt^{2} - g_{t}$, apr\`{e}s le
r\'{e}tablissement de la relation naturelle de l'espace - temps.
La prise en compte de l'\'{e}volution avec le temps de toutes les
grandeurs est assur\'{e}e \`{a} l'aide de ces deux m\'{e}triques.
(paragraphes $4,5$ et $6$) La m\'{e}trique d'Einstein dans le vide
tient seulement compte du champ gravitationnel et le tenseur
d'Einstein caract\'{e}rise quelques champs de mati\`{e}re et
champs \'{e}lectromagn\'{e}tiques particuliers. Notre m\'{e}trique
et notre tenseur tiennent compte de tous les aspects et effets de
la mati\`{e}re-\'{e}nergie: champ gravitationnel global (y compris
celui des trous noirs), toutes les formes de champ de mati\`{e}re,
champ \'{e}lectromagn\'{e}tique global, radiations cosmiques,
pression et temp\'{e}rature, toutes les \'{e}volutions \'{e}nerg\'{e}tiques, les interactions et les singularit\'{e}s.\\\\
4$^\circ$) La r\'{e}futation logique (et \`{a} post\'{e}riori
physique) du premier principe cosmologique de Hubble selon lequel
n'importe quelle galaxie pourrait \^etre consid\'er\'ee comme
\'etant au centre de l'Univers. Cette r\'{e}futation s'appuie
solidement sur notre mod\`{e}le qui a r\'{e}v\'{e}l\'{e} sa
coh\'{e}rence et sa compatibilit\'{e} extr\^{e}me avec les lois de
la physique et de la M\'{e}canique parfaitement \'{e}tablies.
Cette r\'{e}futation, jointe au premier point, a permis de
modifier la th\'{e}orie de la relativit\'{e} g\'{e}n\'{e}rale
d'Einstein, les \'{e}quations de Friedmann - Einstein et la
Cosmologie homog\`{e}ne macroscopique d'Einstein - Hubble - de
Sitter - Friedmann de mani\`{e}re \`{a} prouver clairement, une
fois r\'{e}adapt\'{e}es \`{a} notre mod\`{e}le, qu'elles
d\'{e}crivent correctement l'\'{e}volution (i.e. l'expansion) de
l'Univers. Cette r\'{e}adaptation, jointe \`{a} la preuve que
toutes les constantes fondamentales (\`{a} l'exception de $E$,
$K_B T$, $k e^2$, $c$ et $A$) d\'{e}pendent du temps (ce qui
donne, entre autre, une nouvelle dimension \`{a} la
quantification), conduit \`{a} l'unification de toutes les
branches de la physique (paragraphes $10,11$ et $12$).\\\\
5$^\circ$) L'\'{e}quation de la mati\`{e}re - \'{e}nergie et
l'utilisation de l'op\'{e}rateur de Dirac ont permis de donner une
nouvelle approche \`{a} la physique des particules fondamentales
qui a une chance (\`{a} condition d'\^{e}tre appuy\'{e}e par des
r\'{e}sultats exp\'{e}rimentaux) de servir de base \`{a} cette
branche de la Physique qui est, \`{a} l'heure actuelle,
indissociable de la Cosmologie et qui a des applications
primordiales pour l'avenir de l'humanit\'{e} (paragraphe $9$).\\\\

Le long de cet article, on a fourni des r\'{e}ponses claires \`{a}
un grand nombre de questions ouvertes (comme, \`{a} titre
d'exemple, celles indiqu\'{e}es \`{a} la fin de [2] ou celles qui
sont \'evoqu\'ees \`a travers l'excellent livre panoramique
\'ecrit sous la direction de Paul Davies savemment intitul\'e: La
Nouvelle Physique) ainsi que des r\'{e}ponses possibles \`{a}
d'autres probl\`{e}mes ouverts et on a pos\'{e} d'autres questions
et \'emis d'autres hypoth\`{e}ses. Cependant, concernant le point
5$^\circ$), on va souligner les remarques
suivantes:\\\\
1$^\circ$) Dans le cadre de notre mod\`{e}le, les particules
mat\'{e}rielles fondamentales sont au nombre de neuf (auxquels on
ajoute les trois neutrinos):
$$
e\hskip 1cm u \hskip1cm d \hskip1cm s  \hskip1cm c \hskip 1cm b
\hskip 1cm t\hskip 1cm \mu\hskip 1cm \tau\hskip 1cm (\nu_e\hskip
1cm \nu_\mu\hskip 1cm \nu_\tau).
$$
Chacune d'elle existe d'une mani\`ere intrins\`eque avec deux spins diff\'erents et chaque quark existe intrins\`equement avec trois couleurs diff\'erentes.\\
Elles constituent, avec leurs antiparticules, les trois vecteurs
d'\'{e}nergie pure repr\'{e}sent\'{e}s par $\Gamma_1, \Gamma_2,
\Gamma_3$ et par d'autres vecteurs $\Gamma$ qui ont chacun une
polarit\'e donn\'ee et qui sont porteurs, en puissance, de toutes
les particules fondamentales et leur donnent naissance. Ces
derniers forment tous les \'{e}tats li\'{e}s plus ou moins stables
nomm\'{e}s hadrons (baryons et m\'{e}sons). Les neutrinos sont les
partenaires directs ou indirects d'un grand nombre d'interactions
(faibles, fortes et \'{e}lectromagn\'{e}tiques), de
d\'{e}sint\'{e}gration, de collisions, de synth\`{e}ses
nucl\'{e}aires et d'annihilations. Ils sont produits
(originellement et continuellement) lors de ces interactions,
essentiellement pour les rendre possibles et compatibles avec les
lois de conservation. Des exemples significatifs de production de
ces particules ainsi que d'interactions les impliquant sont
reproduits sch\'{e}matiquement ci - dessous (c.f. [2]):
$$W^+ \rightarrow e^+ + \nu_e \hskip 2cm W^- \rightarrow
e^- + \overline{\nu_e}$$
$$\pi^+ \rightarrow \mu^+ + \nu_\mu \hskip 2cm \pi^- \rightarrow
\mu^- + \overline{\nu_\mu}$$
$$K^+ \rightarrow \mu^+ + \nu_\mu \hskip 2cm K^- \rightarrow \mu^-
+ \overline{\nu_\mu}$$
$$\tau^- \rightarrow \nu_\tau + e^- + \overline{\nu_e}$$
$$s \rightarrow u + W^- \rightarrow u + e^- +
\overline{\nu_e}$$
$$c \rightarrow s + W^+ \rightarrow s + e^+ + \nu_e$$
$$c \rightarrow s + W^- \rightarrow s + \mu^+ + \nu_\mu$$
$$b \rightarrow c + W^- \rightarrow c + e^- +
\overline{\nu_e}$$
$$b \rightarrow c + W^- \rightarrow c + \mu^- +
\overline{\nu_\mu}$$
$$b \rightarrow c + W^- \rightarrow c + \tau^- +
\overline{\nu_\tau}$$
$$\nu_\mu + N \rightarrow \mu^- + \mbox{hadrons}$$
$$\overline{\nu_\mu} + N \rightarrow \mu^+ + \mbox{hadrons}$$
$$\nu_\mu + d \rightarrow W^- \rightarrow \mu^- + u$$
$$\overline{\nu_\mu} + u \rightarrow W^+ \rightarrow \mu^+ +
u$$
$$\nu_\mu + q \rightarrow Z^0 \rightarrow \nu_\mu + q$$
$$\overline{\nu_\mu} + q \rightarrow Z^0 \rightarrow \overline{\nu_\mu} +
q$$
$$\nu_\mu + e^- \rightarrow \nu_\mu + e^-$$
$$\mu^- \rightarrow e^- + \overline{\nu_e} + \nu_\mu$$
$$\mu^+ \rightarrow e^+ + \nu_e + \overline{\nu_\mu}$$
$$\tau^- \rightarrow e^- + \overline{\nu_e} + \nu_\mu$$
$$\tau^- \rightarrow \mu^- + \overline{\nu_\mu} + \nu_\tau$$
$$\tau^- \rightarrow \nu_\tau + \mbox{hadrons}$$
$$\tau^+ \rightarrow e^+ + \nu_e + \overline{\nu_\tau}$$
$$\tau^+ \rightarrow\overline{\nu_\tau} + \mbox{hadrons}$$
Ces derni\`eres relations, comme beaucoup d'autres, montrent que
les deux familles leptoniques $\mu$ et $\tau$ sont form\'{e}es par
des interactions entre neutrinos et divers types de particules
(fondamentales et autres) ainsi que par d'autres interactions et
par des d\'{e}sint\'{e}grations diverses. Elles sont produites
avec une dur\'{e}e de vie extr\^{e}mement courte, ce qui appuie le
fait que, bien qu'elles soint fondamentales (i.e. impliqu\'{e}es
dans des solutions de l'\'equation d'onde associ\'ee \`a
l'op\'erateur de Dirac), elles sont potentiellement composites.
Elles sont \'{e}lectriquement charg\'{e}es et leur
d\'{e}sint\'{e}gration rapide donne naissance (\`{a} c\^{o}t\'{e}
des stables neutrinos et des instables hadrons) \`{a} des
\'{e}lectrons qui sont stables ou \`{a} des positrons qui
interagissent et s'annihilent tr\`{e}s
rapidement. C'est \`a peu pr\`es le cas des $5$ quarks les plus massifs qui, bien qu'ils soient des particules fondamentales, ils ne sont pas absolument stables et finissent par donner naissance au quark $u.$ \\\\
2$^\circ$) Il est bien connu que les neutrinos sont des particules
gauches (i.e. ayant une h\'{e}licit\'{e} n\'{e}gative) et que les
antineutrinos sont des particules droites (ayant une
h\'{e}licit\'{e} positive). Ces propri\'{e}t\'{e}s sont \`{a}
mettre en rapport avec les deux polarisations des ondes
\'{e}lectromagn\'{e}tiques et avec l'existence de deux types
d'\'{e}lectrons de spins oppos\'{e}s $e_{\frac{1}{2}}$ et
$e_{-\frac{1}{2}}$. Elles peuvent \^{e}tre consid\'{e}r\'{e}es
comme des caract\'{e}ristiques intrins\`{e}ques de ces particules.
La d\'{e}sint\'{e}gration $\beta$ de la particule $W^-$ produit un
\'{e}lectron polaris\'{e} \`a gauche et un antineutrino qui ne
peut \^etre polaris\'e qu'\`{a} droite. La collision -
annihilation $p - \overline{p}$ ou plut\^ot $q - \overline{q}$
d'un quark gauche et d'un antiquark droit produit (par
l'interm\'{e}diaire d'une particule $W$) un positron droit et un
neutrino qui ne peut \^{e}tre polaris\'{e} qu'\`{a} gauche. Faut -
il pour autant
parler de violation de parit\'{e}?\\\\
3$^\circ$) Notre mod\`{e}le est bas\'{e} sur la notion de champ.
Il ne n\'{e}cessite pas l'existence de charges fortes (ni
d'ailleurs de charge gravitationnelle associ\'{e}e \`{a} la force
de gravit\'{e}) ni d'interm\'{e}diaires pour v\'{e}hiculer ces
charges; les gluons sont en fait des particules semblables aux
photons et comme ces derniers ne portent pas de charges et servent
essentiellement en tant que m\'{e}ssagers des interactions fortes
entre les quarks et les nucl\'{e}ons (comme le sont les photons
pour les interactions \'{e}lectromagn\'{e}tiques entre particules
charg\'{e}es \'{e}lectriquement). \`{A} la diff\'{e}rence des
photons qui, d'une part, forment une mer en \'{e}volution
permanente \`{a} l'int\'{e}rieur des atomes et, d'une autre part,
se propagent partout dans l'Univers, les gluons forment uniquement
une mer de particules (parmi d'autres) \`{a} l'int\'{e}rieur des
hadrons (et des noyaux) qui \'{e}volue et se transforme
perpetuellement subissant et provoquant toute sorte
d'int\'{e}raction. Par contre, les gluons sont fondamentalement
diff\'{e}rents des particules massives $W$ et $Z^0$ qui sont \`{a}
la base des int\'{e}ractions faibles d\`{e}s leur formation
jusqu'\`{a} leur d\'{e}sint\'{e}gration et les effets qui en r\'esultent.\\
Signalons que, dans notre mod\`{e}le, les gravitons n'existent
pas, les boules de glu non plus. Le champ gravitationnel existe
bel et bien; il courbe l'espace et son action n'a pas besoin ni de
charges ni d'interm\'{e}diaires, il a une port\'{e}e illimit\'{e}e
dans l'Univers. Les int\'{e}ractions fortes existent uniquement
\`{a} l'int\'{e}rieur des hadrons et des noyaux; ces
int\'{e}ractions sont m\'{e}diatis\'{e}es par les gluons.\\\\
4$^\circ$) Les seules particules non hadroniques stables au sein
de notre mod\`{e}le sont les \'{e}lectrons, les photons, les
neutrinos et le quark $u.$ Le seul hadron vraissemblablement
stable est le proton. Les autres hadrons (\`{a} l'exception du
neutron) ont une existence \'{e}ph\'{e}m\`{e}re. Ceci consolide le
bien fond\'{e} de notre hypoth\`{e}se sur la non existence de
charges fortes et de forces fortes particuli\`{e}res. Les
\'{e}tats de liaison au sein des hadrons sont assur\'{e}s par les
attractions r\'{e}sultant, d'une mani\`{e}re \'{e}ph\'{e}m\`{e}re
et \'{e}pisodique, des forces \'{e}lectromagn\'{e}tiques et de la
gravitation.\\Ce fait est \'{e}tay\'{e}, \`{a} titre d'exemple,
par la comparaison suivante des trois \'{e}tats de liaison
suivants: l'\'{e}tat $u \hskip 0.05cm u \hskip 0.05cm d$ pour le
proton, l'\'{e}tat $u \hskip 0.05cm d \hskip 0.05cm d$ pour le
neutron et l'\'{e}tat $u \hskip 0.05cm u \hskip 0.05cm u$ pour le
$\Delta^{++}$. Les deux premiers ont un spin \'{e}gal \`{a}
$\frac{1}{2}$ et le troisi\`{e}me a un spin \'{e}gal \`{a}
$\frac{3}{2}$. Ceci montre que pour les deux premiers seulement
deux des quarks constituants ont des spins align\'{e}s, tandis que
les trois spins sont align\'{e}s au sein du troisi\`{e}me. De plus
le premier \'{e}tat est constitu\'{e} de deux quarks charg\'{e}s
positivement (+$\frac{2}{3}$) et d'un quark charg\'{e}
n\'{e}gativement (-$\frac{1}{3}$), le deuxi\`{e}me est
constitu\'{e} de deux quarks charg\'{e}s n\'{e}gativement
(-$\frac{1}{3}$) et d'un quark charg\'{e} positivement
(+$\frac{2}{3}$) tandis que le troisi\`{e}me est constitu\'{e} de
trois quarks charg\'{e}s positivement (+$\frac{2}{3}$). Ces deux
facteurs contribuent au fait que l'\'{e}tat $\Delta^{++}$ est
extr\^{e}mement instable compar\'{e} aux deux autres et a une
dur\'{e}e de vie infinimement petite. Le fait que le proton a une
dur\'{e}e de vie extr\^{e}mement plus longue que celle du neutron
pourrait s'expliquer par le fait que le quark qui est
antialign\'{e} avec les deux autres au sein du proton est le quark
\emph{u} qui a une charge sup\'{e}rieure \`{a} celle du quark
\emph{d} qui est antialign\'{e} avec les deux autres quarks au
sein du neutron rendant ainsi l'attraction
\'{e}lectromagn\'{e}tique \`{a} l'int\'{e}rieur du proton plus
importante que celle \`{a} l'int\'{e}rieur du neutron (cette
attraction \'{e}tant inexistante \`{a} l'int\'{e}rieur du
$\Delta^{++}$). La diff\'{e}rence de masse entre le quark \emph{d} et
le quark \emph{u} et par cons\'{e}quent entre le neutron et le
proton joue un r\^{o}le d\'{e}terminant en ce qui concerne la
diff\'{e}rence de stabilit\'{e}; le neutron a la possibilit\'{e}
de se d\'{e}sint\'{e}grer pour donner naissance \`{a} un proton
que le proton n'a pas. La transformation du quark \emph{d} en un
quark \emph{u} au sein du neutron n'a pas de contraintes majeures.
Le quark \emph{u} ne peut pas se transformer
naturellement en un autre quark. En r\'esum\'e, l\'equilibre \'energ\'etique entre l'\'energie potentielle (gravitationnelle et \'electromagn\'etique), l'\'energie cin\'etique (vibrationnelle et rotationnelle) et l'\'energie de masse qui est \'etabli (\`a tr\`es courte distance) entre les trois quarks $u,u$ et $d$ au sein du proton est extr\`emement plus stable que celui qui est \'etabli entre les quarks $u,d$ et $d$ au sein du neutron qui est, lui m\^eme extr\^emement plus stable que celui qui est \'etabli parmi les quarks et antiquarks au sein des autres hadrons.\\\\
5$^\circ$) La non existence de trois charges diff\'{e}rentes et de
trois ou huit gluons charg\'{e}s (bien que l'existence de chacun
des quarks avec trois couleurs diff\'{e}rentes a \'{e}t\'{e} bien
prouv\'{e} et l'existence de plusieurs types de gluons
diff\'{e}remment color\'{e}s et mix\'{e}s n'est pas exclue)
n'emp\^{e}che pas que les interactions fortes pourraient \^{e}tre
associ\'{e}es \`{a} la sym\'{e}trie $SU$(3). De m\^{e}me, les
int\'{e}ractions faibles pourraient, elles aussi, \^{e}tre
associ\'{e}es \`{a} la sym\'{e}trie $SU(2).$ Par ailleurs, les
interactions faibles et les interactions \'electromagn\'etiques
pourraient \^{e}tre unifi\'{e}es, \`{a} l'aide d'une th\'{e}orie
de jauge associ\'{e}e \`{a} $SU$(2) $\times$ \emph{U}(1), au sein
d'une th\'{e}orie \'{e}lectrofaible. Il est \'{e}galement possible
de pouvoir construire une th\'{e}orie de grande unification
bas\'{e}e sur le groupe $SU$(5) et de se poser des questions sur
les conditions sp\'{e}cifiques concernant les diff\'{e}rentes
brisures de sym\'{e}trie. Tout cela nous permettrait peut \^{e}tre
de construire la table de toutes les particules qui pourraient
jamais exister \`{a} partir de nos $24$ particules fondamentales
et de caract\'{e}riser toutes les particules qualifi\'{e}es de
particules de jauge, y compris celles
du genre particules de Higgs.\\
D'autre part, on pourrait c\'{e}der \`{a} la charme
math\'{e}matique de la th\'{e}orie de la supersym\'{e}trie et de
lui associer une th\'{e}orie de supergravit\'{e}. Mais l'existence
de gravitons, de particules de Higgs et de leurs partenaires
supersym\'{e}triques (gravitinos et Higgsinos) est exclue du cadre
de notre mod\`{e}le. De m\^{e}me, il faut avouer que la
compactification de l'espace \`{a} cinq dimensions de Kaluza -
Klein en un espace - temps \`{a} quatre dimensions apparentes (et
une cinqui\`{e}me invisible correspondant \`{a}
l'\'{e}lectromagn\'{e}tisme) et que la th\'{e}orie du tout (TOE)
de Cremmer et Julia et son association \`{a} la th\'{e}orie de la
supersym\'{e}trie $N = 8$ (avec ses pyrgons) sont compl\`{e}tement
\'{e}trang\`{e}res \`{a} notre
mod\`{e}le.\\
Il y a aussi la th\'{e}orie des cordes supersym\'{e}triques \`{a}
dix dimensions qui est associ\'{e}e \`{a} une sym\'{e}trie de
jauge poss\'{e}dant un groupe de jauge \emph{G} de rang 16 (qui
pourrait \^{e}tre $SO$(32)$/Z_2$ ou \emph{E8} $\times$ \emph{E8})
et qui reproduit $496$ particules de Yang-Mills dont $480$
solitons. L'association des beaux r\'{e}sultats topologiques de
Witten ainsi que d'autres ingr\'{e}dients math\'{e}matiques \`{a}
cette th\'{e}orie ne constitue pour nous qu'une fascinante
gymnastique intellectuelle d'une esth\'{e}tique exceptionnelle
qui, au bout de longues confrontations th\'{e}oriques et
exp\'{e}rimentales et de modifications ad\'{e}quates, pourrait
contribuer \`{a} \'{e}clairer le dernier bout du chemin de la
connaissance des lois
ultimes de la nature.\\\\
Signalons enfin que l'inflation, l'\'{e}nergie du vide et sa
polarisation n'ont pas de place dans notre mod\`{e}le. Le vide qui
pullule de toute sorte de particules r\'{e}elles mat\'{e}rielles
ou immat\'{e}rielles qui se cr\'{e}ent, s'annihilent,
int\'{e}ragissent et se d\'{e}sint\`{e}grent plus ou moins
rapidement n'est pas r\'{e}ellement vide. Ces particules peuvent
r\'{e}ellement exister et participer \`{a} des diff\'{e}rentes
transformations \'{e}nerg\'{e}tiques et peuvent aussi \^{e}tre
sp\'{e}cifiquement polaris\'{e}es, mais pourrait-on alors parler
de vide polaris\'{e}? L'\'energie n\'egative (du vide) ne peut
s'expliquer que par la pression radiationnelle (r\'esultant d'une
mer radiationnelle l\`a o\`u elle existe) qui agit dans le sens de
l'antigravit\'{e} et de l'expansion. Il ne s'agit donc pas de la
notion (bizarre) de gravit\'{e} n\'{e}gative associ\'{e}e \`{a} un
(pr\'{e}sum\'{e}) faux vide comme il en est question au sein des th\'{e}ories de l'inflation.\\\\
6$^\circ)$ Une fois l'hypoth\`ese de l'existence de l'\'energie originelle concentr\'ee en un point est admise, notre mod\`ele explique le processus de la cr\'eation et de la formation de l'Univers \`a l'aide des trois d\'edoublement suivants: \\
\begin{itemize}
\item Le d\'edoublement de l'\'energie lors de la propagation en
deux polarisations. \item Le d\'edoublement
mati\`ere-antimati\`ere accompagn\'e du d\'edoublement des charges
\'electriques positives et n\'egatives.
\item Le d\'edoublement du spin de l'\'electron et les autres particules qui existent avec deux spins oppos\'es.\\
\end{itemize}
Ces trois ph\'enom\`enes se sont produits en liaison \'etroite
avec la temp\'erature: Temp\'erature = Energie (via les
fr\'equences) = pression (radiationnelle).
L'ex-\\pansion-propagation, les interactions et l'\'evolution
ult\'erieure peuvent \^etre expliqu\'ees et ob\'eissent \`a des
lois bien pr\'ecises. Les
deux processus capitaux sont:\\

\noindent$1.\;$ La propagation, li\'ee au d\'edoublement de la polarisation, des ondes \'electro-\\magn\'etiques \`a partir de l'\'energie originelle.\\

\noindent$2.\;$ La cr\'eation de la mati\`ere-antimati\`ere et plus particuli\`erement des \'electrons-positrons et des quarks-antiquarks u. La cr\'eation des autres quarks-antiquarks et des autres leptons-antileptons avec sans doute les neutrinos-antineutrinos a conduit \`a l'\'etat qualifi\'e de soupe de quarks et de leptons qui a pr\'ec\'ed\'e la formation des hadrons et de tout le reste.\\

Ce sch\'ema indique que les particules ultra-fondamentales (avec leurs antiparticules) se r\'eduisent \`a trois types: Le photon (avec ses deux polarisations) qui n'est autre qu'une particule d\'energie en mouvement, l'\'electron (avec ses deux spins oppos\'es) et le quark u. Les autres particules fondamentales finissent toutes par donner naissance \`a ces trois types de particules auxquels s'ajoutent les fameux neutrinos.\\
Le monde r\'eel est constitu\'e essentiellement des cinq
particules les plus stables: Photons, \'{e}lectrons, neutrinos,
protons et neutrons. Ces deux derniers sont form\'es de quarks u
et d; l'\'etat de liaison uud \'etant plus stable que l'\'etat udd
puisque le quark d peut se transformer naturellement en un quark
u, d'une part, et grace \`a la plus grande attraction des deux quarks u \`a l'int\'erieur du proton que celle des deux quarks d \`a l'int\'erieur du neutrons \`a cause de leur plus grande charge et leur mouvement oppos\'e ainsi que leur spin oppos\'e. Les autres leptons et hadrons sont extr\`emement moins stables
et ont une vie \'eph\'em\`ere. L'\'etat de liaison uu (ou dd) n'existe pas
et l'\'etat uuu ne peut exister que d'une fa\c con
ultra-\'eph\'em\`ere. La disparition de l'antimati\`{e}re pourrait
s'expliquer par l'extr\^{e}me instabilit\'{e} de toutes les
particules form\'{e}es partiellement d'antiparticules
fondamentales d'une part et par l'absorption des antineutrinos par
un grand nombre de protons pour donner naissance \`{a} des
neutrons qui sont relativement instables et moins nombreux que les
protons, d'une autre part.
Les antiparticules s'annihilent tr\`{e}s rapidement et finissent par former des photons et des gluons.

\subsection*{R\'eflexions autour de l'Au del\`a du Big Bang}
Il r\'esulte de notre mod\'elisation physico-math\'ematique de l'expansion de l'Univers (publi\'ee aux ``Arxiv" sous le titre ``The expanding universe and energy problems") qu'originellement il y avait l'espace tridimensionnel th\'eorique virtuel ind\'efiniment \'etendu dans toutes les directions et que le choix d'un point quelconque $O$ en tant qu'origine de l'espace l'identifie \`a $\mathbb{R}^3$. Il y avait aussi une quatri\`eme dimension repr\'esentant l'\'ecoulement continu du temps que le choix d'un origine correspondant au temps $t=0$ l'identifie \`a $\mathbb{R}=]-\infty,+\infty[$. Comme toute notion impliquant la notion de l'infini, elle rel\`eve de la m\'etaphysique puisque l'infini, que ce soit spatial ou temporel, est, bien qu'il soit d\'efini math\'ematiquement, ne peut pas \^etre atteint physiquement et reste quelque chose de fonci\`erement abstrait. A partir de l\`a, notre mod\`ele sh\'ematise l'Univers au temps d\'esign\'e par $t=0$ \`a l'aide d'une quantit\'e d'\'energie \'eternelle bien d\'etermin\'ee concentr\'ee en $O$ dont le calcul \`a l'aide de nos unit\'es conventionnelles fixe \`a $E_0=9.57\times 10^{70}$ J. La distribution de la mati\`ere-\'energie \`a travers l'Univers physique \`a l'instant $t>0$, d\'esign\'ee par $E_t(X)=E(t,X)$, se r\'eduit alors, pour le temps d\'esign\'e par $t=0$, \`a $E_0(X)=\delta E_0$ o\`u $\delta$ est la mesure de Dirac \`a l'origine de l'espace $O$. Il r\'esulte \'egalement de notre mod\`ele que l'Univers \`a l'instant $t>0$ est mod\'elis\'e par
$$U(t)=\left(B(O,R(t)),g_t\right),$$
o\`u $B(O,R(t))$ est la boule de rayon euclidien $R(t)$ et $g_t$ est une m\'etrique riemannienne d\'efinie sur $B(O,R(t))$ qui refl\`ete, \`a tout instant $t$, l'influence de la distribution de la mati\`ere-\'energie remplissant l'espace $B(O,R(t))$. La m\'etrique r\'egularis\'ee $g_t$ sera appel\'ee la m\'etrique r\'eelle physique.\\

En consid\'erant la forme volume canonique $dv_{g_t}$ associ\'ee \`a la m\'etrique $g_t$ et en d\'efinissant la mesure r\'egularis\'ee $\nu_t=E(t,X)\,dX$, on obtient alors
$$dv_{g_t}=dX-\nu_t=(1-E(t,X))\,dX.$$
Ainsi, $dv_{g_t}$ mesure le volume r\'eel physique dans l'espace physique $B(O,R(t))$ et $\nu_t$ mesure le d\'efaut de volume, caus\'e par l'existence de la consistance physique d'un domaine $D$ de $B(O,R(t))$ (i.e. l'existence de la distribution $E(t,X)$ dans ce domaine ainsi que l'influence de cette distribution \`a travers tout l'Univers physique sur lui), pour que le volume physique de $D$ soit \'egal au volume euclidien de $D$ conventionnellement mesur\'e par la mesure de Lebesgue $dX$ dans le cas o\`u ce domaine est absolument vide de mati\`ere, sous toutes ses formes, et de ses effets.

A partir du temps que l'on a consid\'er\'e comme $t=0$, l'Univers physique se forme \`a l'aide de la propagation, \`a travers l'espace th\'eorique $\mathbb{R}^3$, des ondes \'electromagn\'etiques ob\'eissant \`a l'\'equation des ondes 
$$\frac{1}{v^2(t)}\frac{\partial^2}{\partial t^2}E(t,X)-\Delta E(t,X)=0$$ 
o\`u $v(t)$ est la vitesse de propagation des ondes, qui est extr\`emement petite au tout d\'ebut \`a cause de l'immensit\'e de la gravit\'e centrale caus\'ee par la densit\'e immense de la mati\`ere-\'energie pour $t$ tr\`es voisin de $0$. En fait, ce que l'on a consid\'er\'e comme correspondant au temps initial $t=0$ s\'etend sur un intervalle de temps qui pourrait \^etre tr\`es grand et m\^eme infini. Cet intervalle de temps pourrait \^etre qualifi\'e par l'au del\`a de l'\`ere de Planck \`a partir de laquelle $R(t)$ devient assez significatif tout en restant extr\`emement petit. Le long de cet intervalle, on peut consid\'erer que $R(t)\cong0$ gr\^ace \`a l'extr\`eme petitesse de la vitesse $v(t)$ tout au long de cet intervalle. Ensuite, l'intensit\'e de la gravit\'e d\'ecro\^it au fure et \`a mesure que $t>0$ cro\^it et $v(t)$ augmente progressivement pour devenir, plus tard, tr\`es proche de $c=1$ et qui tendra fermement vers $1$ lorsque $t$ tendrait vers $+\infty$. Durant cet intervalle, que nous qualifierons aussi par l'\`ere du Big Bang, la vitesse $v(t)$ et le rayon $R(t)$ \'etaient extr\`emement petits, la fr\'equence des ondes \'etait extr\`emement grande, leur longueur d'onde \'etait extr\`emement petite et la temp\'erature de l'Univers \'etait extr\`emement \'elev\'ee. En consid\'erant que, tout au long de cet intervalle de temps, $E(t,X)=\delta E_0$, on a en fait suppos\'e que $v(t)\equiv0$, $R(t)\equiv0$, $\lambda(t)\equiv0$, $\omega=+\infty$, $T=+\infty$ et que la densit\'e de la mati\`ere-\'energie ainsi que la courbure de l'espace physique, r\'eduit \`a un point, \'etaient $+\infty$. Ceci pourrait \^etre justifi\'e par le fait que $R(t)$ \'etait alors infiniment petit compar\'e au rayon actuel $R(t_0)$ o\`u $t_0\simeq 2.365\times 10^{19}$s d\'esigne le temps \'ecoul\'e \`a partir du temps o\`u la vitesse $v(t)$ devient proche de $1$. La situation pr\'ec\'edente o\`u tout pourrait se r\'eduire \`a $0$ ou $\pm\infty$ (\`a l'exception de l\'energie $E_0$), pourrait exister r\'eellement bien avant l'intervalle d\'esign\'e par $t=0$. Elle pourrait exister \`a un instant fini $t_1<0$, qu'il est impossible de d\'eterminer, ou s\'etendre sur un intervalle fini $[t_1,0[$ ou alors sur un intervalle infini qui serait d\'esign\'e, en utilisant nos conceptions math\'ematiques, par $]-\infty,0[$. Nous allons qualifier ce point ou cet intervalle par l'au del\`a du Big Bang. Durant les deux intervalles qualifi\'es successivement par l'au del\`a de l'\`ere de Planck (ou l'\`ere du Big Bang) qui correspond \`a $t=0$ et par l'au del\`a du Big Bang qui correspondrait \`a $\{t_1\}$, $[t_1,0[$ ou $]-\infty,0[$, l'Univers physique (dynamique pour le premier et statique pour le second) est suppos\'e se r\'eduire \`a une seule dimension qui correspond \`a celle du temps; les trois dimensions spatiales \'etant in\'existantes pour le second intervalle et pouvant \^etre consid\'er\'ees comme in\'existantes pour le premier. Ce r\'esultat r\'esulterait \'egalement \`a partir de la th\'eorie des cordes.\\

Une autre possibilit\'e pourrait \^etre s\'erieusement envisag\'ee dans le cas o\`u $t_1$ est fini, qui pourrait \^etre pris comme origine du temps, $t_1=0$, et dans le cas o\`u l'\'energie totale $E_0$ de l'Univers \'etait concentr\'ee uniquement au temps $t_1=0$ en un point O consid\'er\'e comme \'etant l'origine de l'espace virtuel \`a trois dimensions $\mathbb{R}^3$. On pourrait alors consid\'erer l'\'evolution de l'Univers dans les deux directions du temps: les temps positifs, $t>0$, et les temps n\'egatifs, $t<0$.\\
Ainsi, il r\'esulterait de la discussion de la section ``Temps universel et temps propre'' que, si $\tau$ d\'esigne le temps propre de l'expansion, on pourrait consid\'erer la boule $B(O,r_0)$ repr\'esentant le quasi-trou noir originel. $B(O,r_0)$ coupe alors toutes les directions de l'espace $\mathbb{R}^3$ (i.e. les axes passant par O) en des intervalles $]-r_0,r_0[$.\\
 D'apr\`es la caract\'erisation des m\'etriques des trous noirs on a, dans $B\setminus O$, $\tau=|t|$ puisque $\|X'(t)\|_{g_0}=0$ et en $t=0$ on a $\tau=t=0$ puisque $\tau=1-v_{g_0}^2=1-\|X'(0)\|_{g_0}=1-1=0$ en consid\'erant que l'\'energie totale de l'Univers est l'unit\'e d'\'energie.\\
Il s'en suit que
$$\tau'=\left\{\begin{array}{ll}
\;\;1&\;\;\mbox{pour}\;t>0,\\ 
{}\\ 
-1&\;\;\mbox{pour}\;t<0,\\
\end{array}
\right.
$$  
et par suite
$$\tau'=H(t)-\check{H}(t)\qquad\mbox{o\`u}\qquad \check{H}(t)=H(-t).$$
Etant donn\'e que $<\tau'',\varphi>=-<\tau',\varphi'>=-\left(\int_0^\infty\varphi'(t)\,dt-\int_{-\infty}^0\check{\varphi}'(t)\,dt\right)=-\left(\int_0^\infty\varphi'(t)\,dt+\int_{+\infty}^0\varphi'(t)\,dt\right)=-(-\varphi'(0)+\varphi'(0))=0$, pour tout $\varphi\in C^\infty_c(\mathbb{R}^3)$ et le long de toutes les directions de $\mathbb{R}^3$, on obtient
$$\tau''=\delta_{\mathbb{R}^+}+\delta_{\mathbb{R}^-}=\Gamma^{+}+\Gamma^{-}=F^++F^-=0$$
et $F^+=-F^-$ et $\tau'$ est la vitesse euclidienne $v_e$ tandis que $\tau$ est la distance euclidienne au sein de la boule $B(O,r_0)$.\\
Si l'on consid\`ere, \`a partir du temps $t_1=0$ le temps n\'egatif contin\^ument d\'ecroissant vers $-\infty$ et le temps positif contin\^ument croissant vers $+\infty$, on obtient le c\^one de l'espace-temps form\'e de deux semi-c\^ones oppos\'es; le premier correspondant aux temps positifs et le second aux temps n\'egatifs. Le premier, \'etudi\'e le long de cet article, correspond \`a la mati\`ere caract\'eris\'ee par une densit\'e mat\'erielle et une densit\'e d'\'energie positives et par une densit\'e radiationnelle et une pression n\'egatives qui constitue l'Univers physique r\'eel dans lequel nous vivons tous. Quant au second correspondrait \`a l'antimati\`ere caract\'eris\'ee par une densit\'e antimat\'erielle et une densit\'e d'\'energie n\'egatives et par une densit\'e radiationnelle et une pression positives qui constitue l'Univers virtuel au sujet duquel nous ne connaissons pas grand chose. Si on admet les relations $\displaystyle E_0=\lim_{t\rightarrow0}h(t)f(t)$ et $\displaystyle v_0=\lim_{t\rightarrow0}\lambda(t)f(t)$ alors, puisque $\displaystyle \lim_{t\rightarrow0}h(t)=0$ et $E_0$ et $v_0$ sont finis, on a $\displaystyle \lim_{t\rightarrow0}f(t)=+\infty$ et $\displaystyle \lim_{t\rightarrow0}\lambda(t)=0$; ce qui est conforme aux r\'esultats obtenus pr\'ec\'edemment.\\
Dans cette \'eventualit\'e pr\'ecise, on a $B(O,t_0)=B(O,r_0)$ et c'est l'intervalle $]-t_0,t_0[$ qui repr\'esente l'intervalle du temps que l'on a d\'esign\'e par l'intervalle qui correspond \`a $t=0$. L'\'evolution de l'Univers r\'eel, comme l'Univers virtuel d'ailleurs, d\'emarre \`a partir de $t=t_0$ provoqu\'e par l'apparition des particules mat\'erielles en m\^eme temps que les particules antimat\'erielles qui ont, comme on le sait, des charges \'electriques oppos\'ees produisant des champs de forces oppos\'es.\\
Le c\^one de l'espace et du temps peut alors \^etre esquiss\'e, en se bornant \`a $3$ dimensions, sous la forme de la figure $16$. Finalement, nous pensons qu'il est possible de d\'eterminer le volume de $B(O,r_0)$, une fois calcul\'e $E_0$ (en le comparant \`a d'autres trous noirs ordinaires), et par suite de d\'eterminer $t_0$. Il est \'egalement possible de d\'eterminer la loi qui r\'egit, \`a partir de $t_0$, l'\'evolution des vitesses (euclidienne et relative \`a $g_t$) du propagation des ondes \'electromagn\'etiques (i.e. les vitesses de l'expansion) toujours en \'etudiant les trous noirs.\\
Notons $\tau_t$ et $\tau_e$ les temps propres de l'expansion par rapport aux deux m\'etriques $g_t$ et $g_e$ ainsi que $\gamma_t:=(1-v_t^2)^{-1/2}$ et $\gamma_e:=(1-v_e^2)^{-1/2}$ o\`u $v_t:=\|X'(t)\|_{g_t}$ et $v_e:=\|X'(t)\|_{g_e}$. Nous avons
$$\frac{\tau_t^2}{t^2}=\frac{1}{\gamma_t^2}=1-v_t^2\qquad\mbox{et}\qquad \frac{\tau_e^2}{t^2}=\frac{1}{\gamma_e^2}=1-v_e^2.$$
par suite, on a
$$\tau_t^2\gamma_t^2=t^2=\tau_e^2\gamma_e^2\qquad\mbox{et}\qquad \gamma_t^2(1-v_t^2)=1=\gamma_e^2(1-v_e^2).$$
D'o\`u
$$\frac{\tau_t^2}{\tau_e^2}=\frac{\gamma_e^2}{\gamma_t^2}=\frac{1-v_t^2}{1-v_e^2}$$
et lorsque $g_t$ d\'ecro\^it de $g_e$ \`a $g_0=0$, $v_t$ d\'ecro\^it de $v_e$ \`a $0$ et $\displaystyle \frac{\tau_t^2}{\tau_e^2}=\frac{\gamma_e^2}{\gamma_t^2}$ croissent de $1$ \`a $\displaystyle \frac{1}{1-v_e^2}$.\\
Dans $B\setminus O$, on a $\tau_t=t$ puisque $v_t=\|X'(t)\|_{g_t}=0$, $\tau_e=0$ puisque $v_e=\|X'(t)\|_{g_e}=1$, $\gamma_t=1$ et $\gamma_e=+\infty$.\\

Ainsi, notre mod\`ele aboutit \`a la conclusion que l'Univers r\'eel originel se r\'eduit \`a un unique gigantesque trou noir (comme il r\'esulterait \`a partir de la th\'eorie des boucles) \`a la diff\'erence pr\`es que l'espace physique n'existait pas originellement tandis que, pour un trou noir ordinaire \`a l'instant $t>0$, on a  $E(t,X)=\delta e$ sur une boule $B(I,r)\subset B(O,R(t))$ qui existe bel et bien physiquement et \`a l'int\'erieur de laquelle la gravit\'e est extr\`emement intense et la distance physique entre deux points est pratiquement nulle. Quant \`a l'existence de l'\'energie originelle $E_0$ (pour $t=-\infty$ ou $t=t_1$) concentr\'ee en un point $O$ et \`a l'instant originel \`a partir duquel commence le processus de l'expansion ainsi que la raison pour laquelle $v_{g_t}(t)$ cesse d'\^etre nulle, si jamais elle \'etait nulle \`a un certain instant fini, tout ceci rel\`eve de la m\'etaphysique ou de la r\'eligion  puisqu'il \'evoque les notions du nul et de l'infini qui se situent au del\`a de notre conception de la r\'ealit\'e physique qui est \`a l'origine de notre conscience. Reste que les lois physiques de l'Univers $U(t)$ \`a partir du moment o\`u il est mod\'elis\'e par $(B(O,R(t)),g_t)$ (i.e. pour $t\geq t_0$) sont tout \`a fait accessibles \`a notre compr\'ehension, \`a nos moyens techniques, qui \'evoluent continuellement, et \`a notre intelligence. En effet, notre mod\`ele, \'etablit rigoureusement un certain nombre de lois math\'ematiques et physiques coh\'erentes auxquelles ob\'eit l'\'evolution de notre Univers tout en unifiant toutes les branches de la Physique: Relativit\'e g\'en\'erale, Th\'eorie quantique, Thermodynamique, Electromagn\'etisme et M\'ecanique de Newton-Lagrange-Hamilton. En particulier, notre mod\`ele confirme que la Relativit\'e g\'en\'erale (l\'eg\`erement modifi\'ee) instaur\'ee par Einstein gouverne bel et bien l'\'evolution de notre Univers et la Cosmologie.\\

A partir de l\`a, nous consid\'erons que la d\'ecouverte et la compr\'ehension des lois de la Nature et de l'\'evolution ult\'erieure de l'\'energie, de la mati\`ere et de l'Univers constituent le domaine de la Science, l'existence de l'\'energie premi\`ere et la raison premi\`ere du d\'edoublement et du mouvement constituent le domaine de la M\'etaphysique, tandis que la reflexion sur les retomb\'ees de ces lois et de cette \'evolution sur l'humanit\'e constitue le domaine de la Philosophie et de la raison humaine.\\\\

\textbf{e.mail}\\
hmoukadem@hotmail.com

\end{document}